\documentclass[final,5p,times,twocolumn]{elsarticle}

\usepackage{amsmath}
\usepackage{amsfonts}
\usepackage{amssymb}
\usepackage{bm}
\usepackage{enumerate}
\usepackage{epsfig}
\pdfoutput=1

\journal{Journal of Instrumentation}

\begin{document}

\title{Studies of single-photoelectron response and of performance in magnetic field of a H8500C-03 photomultiplier tube}

\newcommand*{\DUKE}{Duke University, Durham, North Carolina 27708}
\newcommand*{\JLAB}{Thomas Jefferson National Accelerator Facility, Newport News, Virginia 23606}

\author {S.P.~Malace} 
\address{\DUKE}
\address{\JLAB}

\author {B.D.~Sawatzky} 
\address{\JLAB}

\author {H.~Gao} 
\address{\DUKE}

\date{\today}

\begin{abstract}

We studied the single-photoelectron detection capabilities of a multianode photomultiplier 
tube H8500C-03 and its performance in high magnetic field. Our results show that the 
device can readily resolve signals at the single photoelectron level making it 
suitable for photon detection in both threshold and ring imaging Cerenkov detectors. 
We also found that a large longitudinal magnetic field, up to 300 Gauss, induces a change in 
the relative output of at most 55\% for an edge pixel, and of at most 15\% for a central pixel. 
The H8500C-03 signal loss in transverse magnetic fields it is significantly more pronounced 
than for the longitudinal case. Our studies of single photoelectron reduction 
in magnetic fields point to the field induced misfocusing of the photoelectron extracted 
from the photocathode as primary cause of signal loss. With appropriate shielding this PMT 
could function in high magnetic field environments.

\end{abstract}
\maketitle


\section{Motivation}\label{sec:motivation}

The Cerenkov radiation emitted by a 
dielectric medium at the passage of charged particles 
has been used successfully for many decades as means of 
particle identification in physics experiments. The most common devices 
utilized to detect Cerenkov photons have been photomultiplier 
tubes (PMTs). With the increasing complexity of experimental setups, a 
variety of PMTs have been developed to meet the demands. In particular, 
a special category of PMTs that would function in high magnetic field 
environments have allowed for Cerenkov detectors to be used in the 
proximity of high-field focusing magnets.

A Solenoid Large Intensity Device (SoLID) that would require threshold 
Cerenkov detectors for positive identification of electrons and pions 
it is conceptually designed to be used in experiments at Jefferson 
Laboratory in experimental Hall A \citet{solid}. The apparatus consists of a 
focusing solenoid with maximum magnetic field of approximately 1.5 Tesla 
combined with an open geometry detector package that would utilize calorimeters 
and Cerenkov detectors for particle identification, gas electron multiplier (GEM) 
chambers for tracking, and multi-gap resitive plate chambers (MRPCs) for time-of-flight 
measurements. The solenoid residual field at the location of the photon 
detector to be used for Cerenkov radiation measurements is expected to reach 
200 Gauss. Stringent requirements for PMTs suitable for the SoLID Cerenkov 
detectors include the ability to operate in this high magnetic field, good resolution 
for single-photoelectron detection, and suitability for tiling to cover a large 
active area (up to 36 inch$^{2}$).

One possible choice for photon detector for the SoLID Cerenkov counters 
is the 2 inch 64 anode Hamamatsu photomultiplier tube H8500C-03. It has 
been designed to function in non-negligible magnetic field environments and 
Hamamatsu's measurements of signal degradation in magnetic fields up to 100 Gauss 
show this PMT to be particularly stable when exposed to longitudinal fields 
(the longitudinal orientation would be perpendicular to the face of the PMT). 
This is of great importance as the longitudinal magnetic field component is 
experimentally the hardest to shield. 
Additionally, H8500C-03 has a square profile with a photocathode coverage of 
89\% which makes it suitable for tiling. The largest tile currently 
envisioned for SoLID Cerenkov counters would be made of 9 H8500C-03 PMTs. 
The SoLID Cerenkov detectors are {\it NON-ring-imaging} and do not 
require each of the 64 pixels 
of H8500C-03 to be instrumented separately. Instead, groups of 16 anodes are expected to 
be ganged together and the summed output digitized. The SoLID requirement of good 
resolution for single-photoelectron detection must be maintained however, 
so the impact of pixel to pixel gain non-uniformity on the summed reponse must be tested 
and characterized.

Our paper is structured as indicated below. Section \ref{sec:expsetup} describes the experimental 
setup we used and the readout of the PMT we tested. In Section \ref{sec:singlePEmeas} we present 
the single-photoelectron measurements we performed on individual pixels and on 
two groups of 16 pixels and the extraction of the H8500C-03 resolution 
by using a well-established PMT response function; we also show our gain 
measurements for several pixels which confirmed the gain non-uniformities 
outlined in the Hamamatsu data and we use the output-matching method to 
improve the single-photoelectron detection resolution for groups of pixels. 
In Section \ref{sec:magFieldMeas} we present our results of the PMT test in high magnetic field. 
We measured and quantified the signal degradation of H8500C-03 when 
exposed to transverse and longitudinal magnetic field orientations with magnitudes 
up to 300 Gauss. The tests were performed on both small signals, at the 
single photoelectron level, and on large signals, 30 to 50 photoelectrons. 
Section \ref{sec:conclusions} summarizes our results.

\section{Experimental Setup}\label{sec:expsetup}

\begin{figure}[htbp]
\vspace*{-0.1in}
\centering
\begin{tabular}{c}
\includegraphics[width=8.8cm]{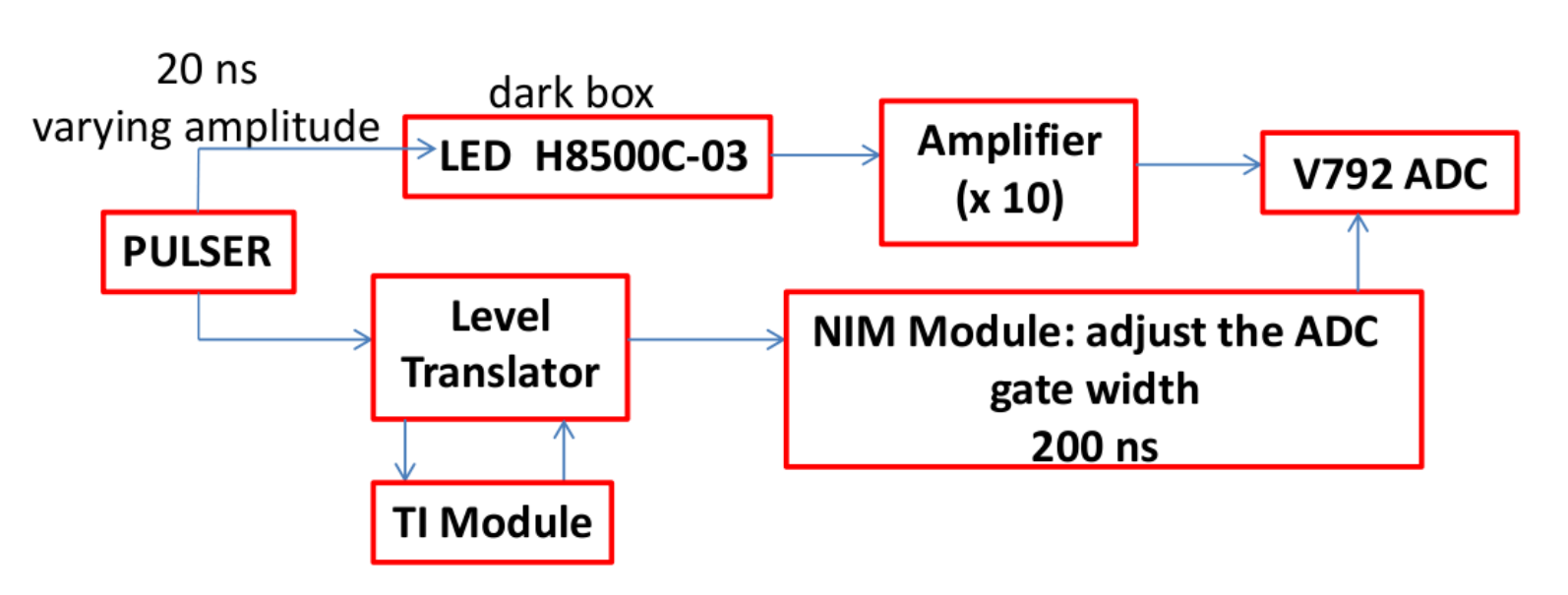}
\end{tabular}
\linespread{0.5}
\caption[]{
{} Diagram of the circuit used for the single-photoelectron and magnetic 
field measurements.}
\label{test_diagram}
\end{figure}

All measurements were performed at Thomas Jefferson Laboratory 
from August to December 2012. We only tested one H8500C-03 PMT unit.
A sketch of our setup is shown in Fig.~\ref{test_diagram}.

The H8500C-03 PMT was placed inside a dark box together with a 
green Light Emitting Diode (LED) to be used as a source of photons. 
A pulser was used to power the LED 
with a pulse width of 20 ns and of varying amplitude from 0.85 to 4 V. 
The same pulser was used to trigger the data acquisition. The signal 
from the PMT was passed through a low-noise 10X amplifier and then 
was sent to a CAEN v792 charge integrating Analog-to-Digital Converter (ADC). 
The software used for data acquisition was CODA 2.5, developed at Jefferson Lab. 
Pedestal events were 
taken before each run to monitor the stability of our system. For pedestal 
runs the LED was not powered thus only the background signal from the PMT 
in the absence of light along with any electronic circuit noise was recorded 
by the ADC. For the entire duration of the experiment 
the pedestal peak position and its width have been very stable, with a 
standard deviation of the pedestal distributions less than 6 ADC channels. 
For the magnetic field measurements we used additionally an iron-core ``C'' magnet with its 
power supply and a gaussmeter to monitor the magnitude of the field. 
More details on the experimental setup for the magnetic field tests will be 
given in Section \ref{sec:magFieldMeas}.

\begin{figure}[htbp]
\vspace*{-0.1in}
\centering
\begin{tabular}{c}
\includegraphics[width=8.8cm]{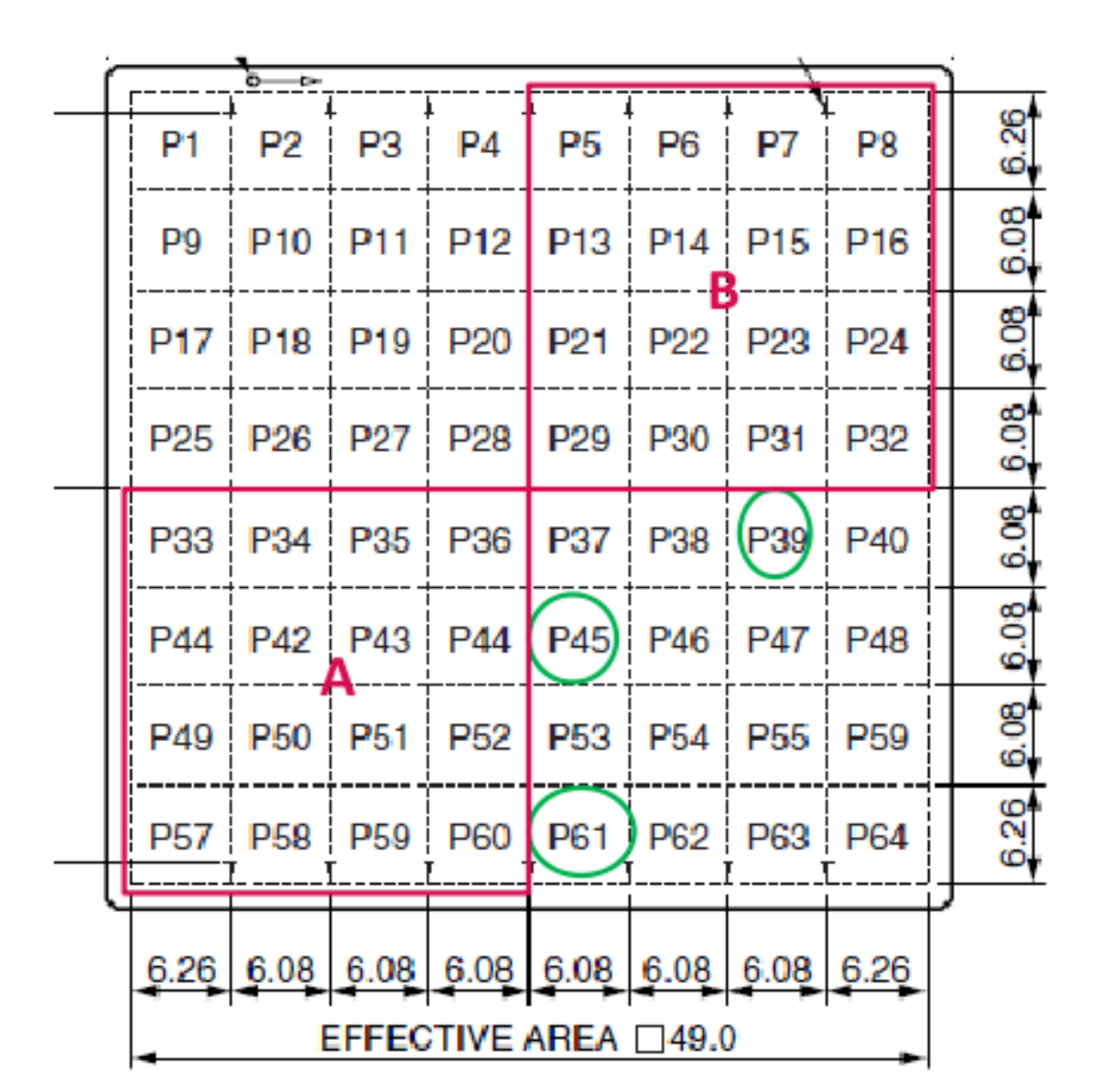}
\end{tabular}
\linespread{0.5}
\caption[pmt_face]{
{} Pixel layout of a H8500C-03. The individual pixels tested, 39, 45, 61, 
are circled in green. The two groups of 16 pixels labeled ``quad A'' and 
``quad B'' are outlined by red squares.}
\label{pmt_face}
\end{figure}

A schematic of the PMT division in pixels is shown in Fig.~\ref{pmt_face}. 
We recorded the output from three individual pixels, 39, 45 and 61, and from 
2 groups of 16 pixels, ``quad A'' and ``quad B'', as shown. 
Two pixels were selected to evaluate whether an 
edge pixel (61) behaves differently under a magnetic field 
than a more central pixel (45). Pixel 39 was chosen because it had one of the lowest 
relative gains in the Hammamatsu gain map shipped with the device. 
Pixel 61 had the highest gain of all 64 pixels, 
while 45 and 39 were labeled with a relative output 
of 79\% and 54\%, respectively, when compared to the highest output. 
The signal from quad A represents the sum over 16 pixels with relative outputs 
ranging from 55\% to 97\% while quad B gets contribution from pixels 
with relative gains between 48\% and 86\%. For the SoLID application a 
four quad division of each PMT would be a reasonable choice as output.

\section{Single-photoelectron Measurements}\label{sec:singlePEmeas}

The methodology of single-photoelectron identification relies on the 
quantum nature of the process of electron extraction from the PMT 
photocathode where at the microscopic level an electron (also called 
photoelectron) is extracted from the photocathode via the photoelectric 
effect by one incident photon. 
In reality not every single absorbed photon will release a photoelectron. 
The probability of photoemission is quantified by the quantum efficiency 
defined as the ratio of output electrons to incident photons which depends 
on several parameters specific to the photocathode material such as 
the reflection coefficient, full absorption coefficient of photons, the 
work function and also on the energy of the incident photon. 
Typically, materials used as visible range photocathodes have quantum 
efficiencies below 30\% which means that for a given number of photons 
incident on the photocathode only a relatively small fraction will be 
converted into electrons.

In our experiment a trigger was formed every time a pulse was sent to 
the LED. The pulse amplitude was reduced until, on average, only
0 (\emph{i.e.} pedestal events) to one photoelectron was being produced at the
photocathode per pulse. The yield of photons was then 
varied in small steps by adjusting the pulse amplitude in order to change 
the fraction of single photoelectron events to pedestal events in a run. 
The expectation was that the single photoelectron distribution will peak at 
the same location on the ADC histogram when compared to the 
pedestal position since for a fixed PMT high voltage the gain of the 
PMT is independent of the magnitude of incoming photons yield.

\begin{figure}[htbp]
\vspace*{-0.1in}
\centering
\begin{tabular}{c}
\vspace{-0.2in}
\includegraphics[width=11cm]{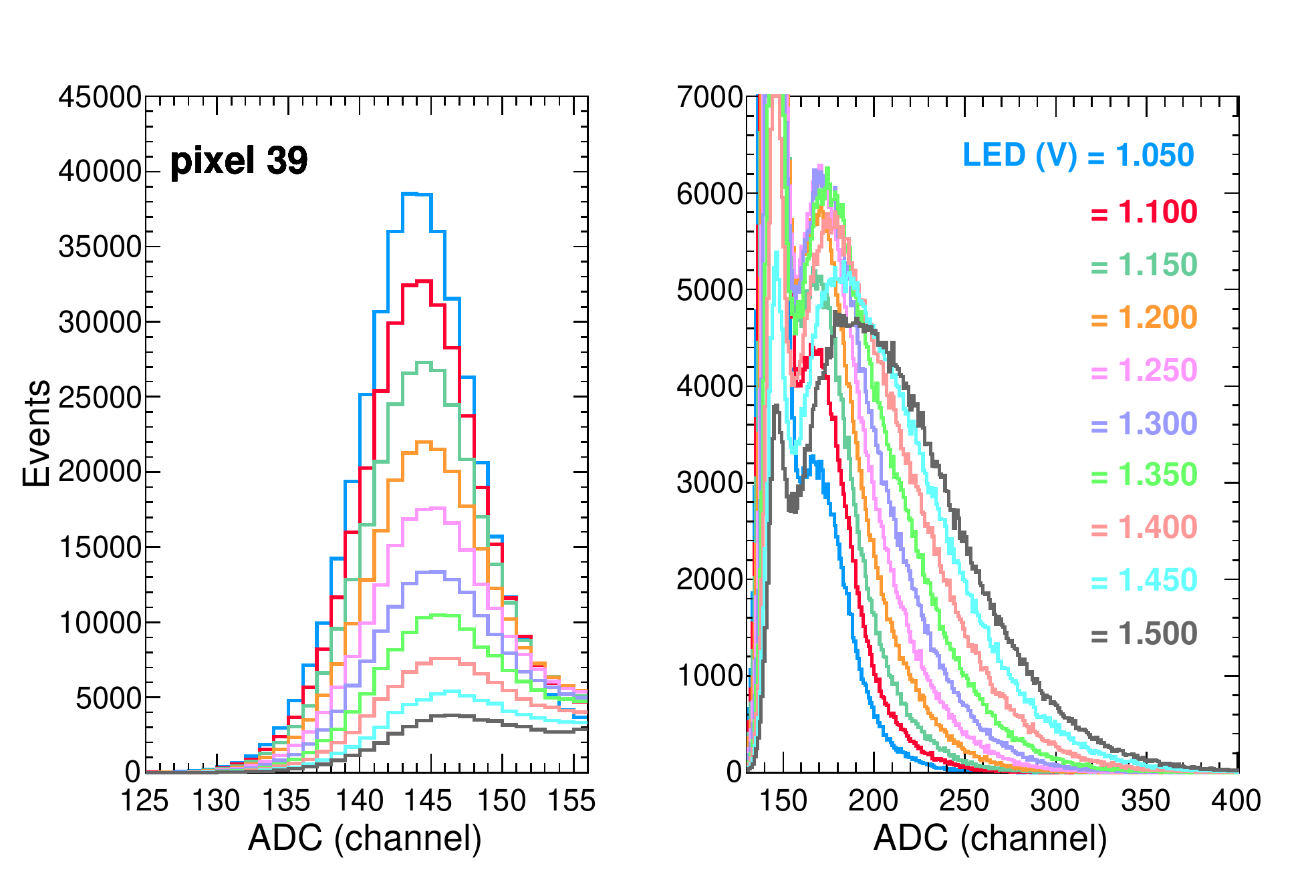} \\
\vspace{-0.2in}
\includegraphics[width=11cm]{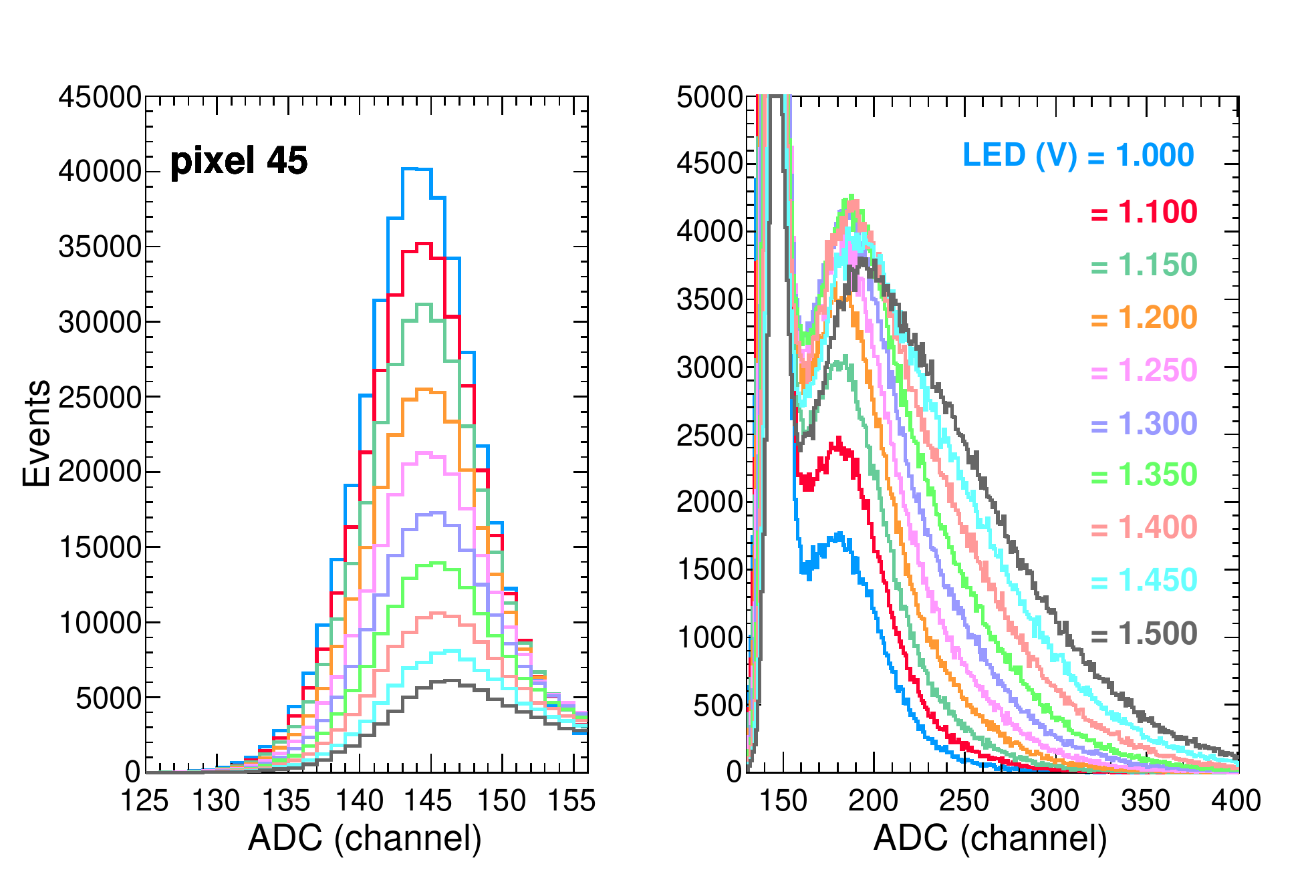} \\
\includegraphics[width=11cm]{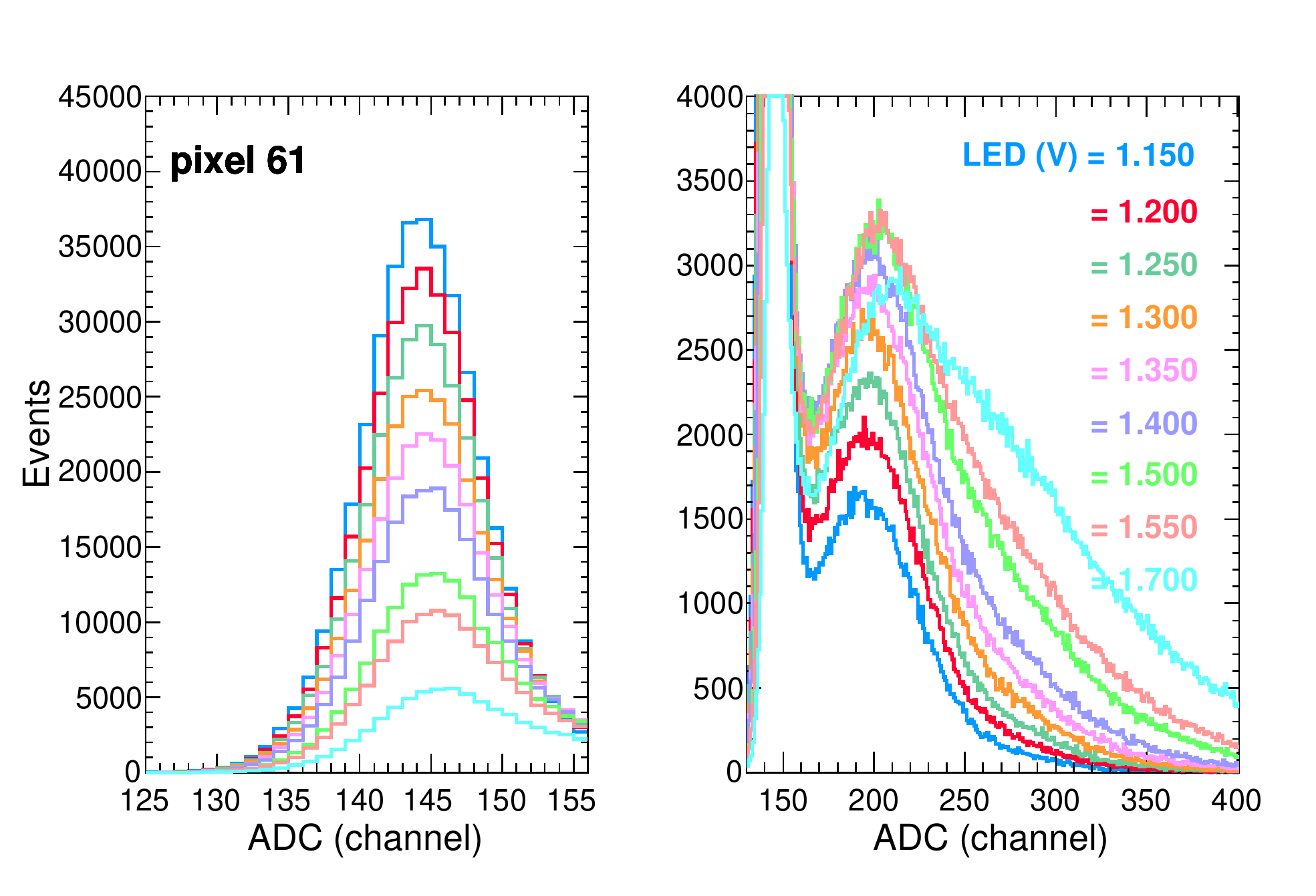}
\end{tabular}
\linespread{0.5}
\caption[spe-on-pixels-39-45-61]{
{} Single-photoelectron measurements on three individual pixels (top, middle, bottom).  
The left panels show the pedestal region only, while the right panels display the full spectrum, 
but with a y-axis range selected to allow the signal components to be seen clearly. Different curves 
represent the PMT response to a varying yield of incoming photons.}
\label{spe-on-pixels-39-45-61}
\end{figure}

Our measurements taken on pixels 39, 45 and 61 are shown in 
Fig.~\ref{spe-on-pixels-39-45-61}. The amplitude of the pulse 
that powered the LED was varied between 1--1.7 V in small 
steps. For the lowest amplitude pulse the single photoelectron 
peak is visible in all pixels but the predominant contribution 
to the run comes from pedestal events. As the yield of photons per trigger 
increases, the fraction of single photoelectron events within a 
run becomes larger. For the highest pulse amplitude, 2 photoelectron 
events are produced in a non-negligible fraction as well. It can 
be seen qualitatively that the peak of the single photoelectron 
distribution is separated from the pedestal peak by approximately 
the same number of ADC channels per pixel. Based on Hamamatsu's 
pixel gain map, pixel 39 has the lowest gain of the three pixels 
under measurement while 61 the highest. This is clearly visible 
in Fig.~\ref{spe-on-pixels-39-45-61} when looking at the pedestal 
to single photoelectron peak separation. For the highest gain pixel, 
61, the measurement taken with a LED setting of 1.7 V shows 1 and 
2 photoelectron peaks as distinguishable features of the ADC distribution.

\begin{figure}[htbp]
\vspace*{-0.1in}
\centering
\begin{tabular}{c}
\vspace{-0.2in}
\includegraphics[width=11cm]{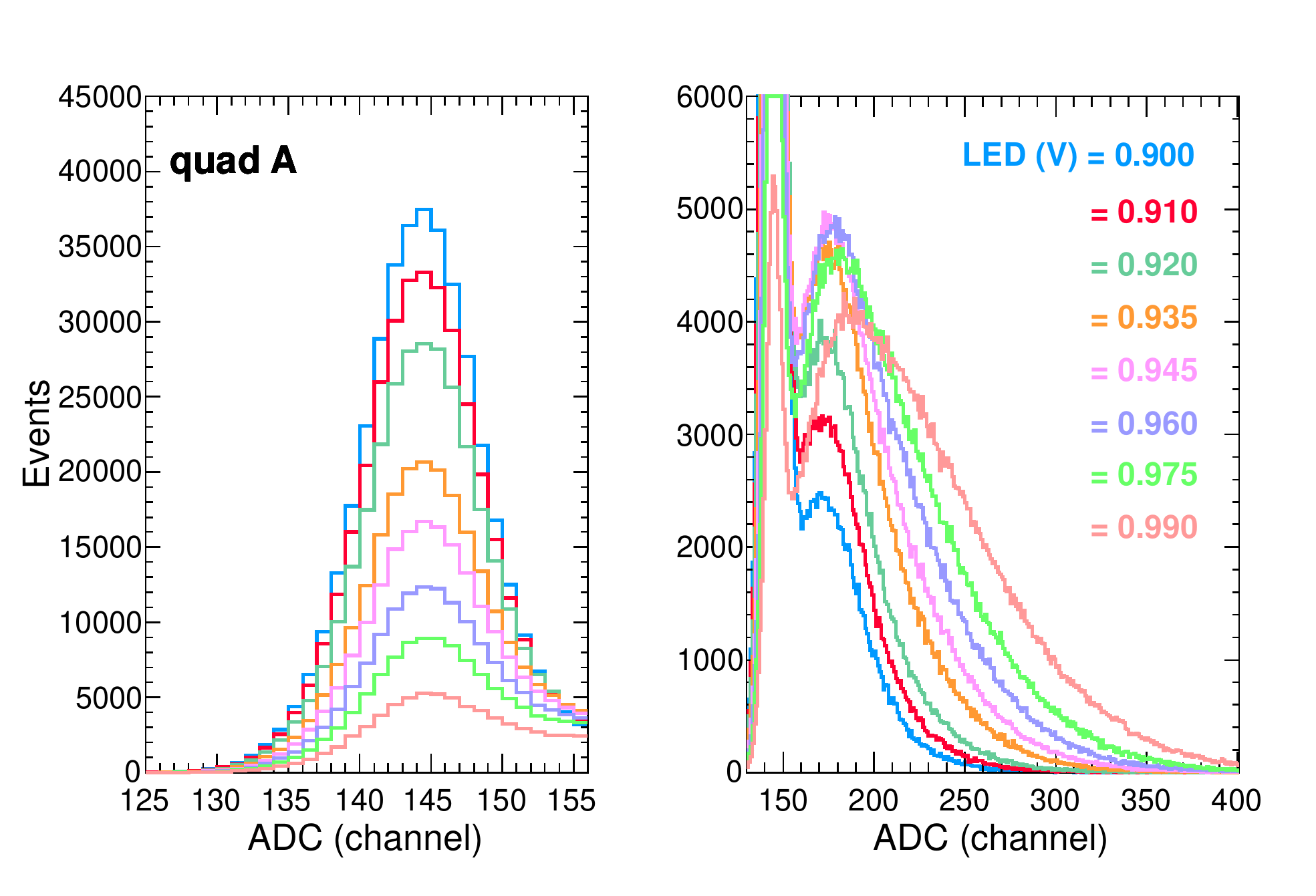} \\
\includegraphics[width=11cm]{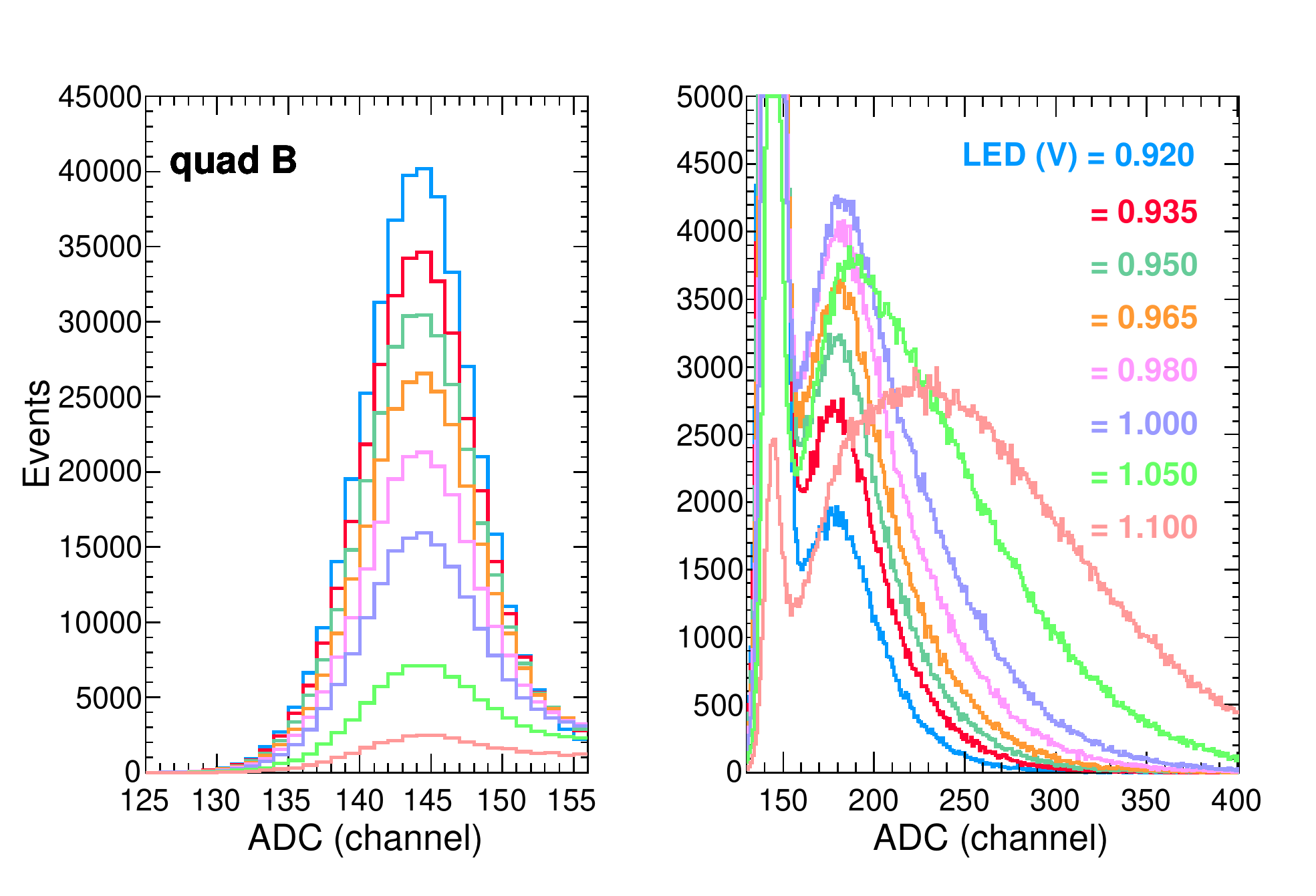} 
\end{tabular}
\linespread{0.5}
\caption[spe-on-quad-a-quad-b]{
{} Single-photoelectron measurements on 2 groups of 16 pixel sums.  
The left panels show the pedestal region only, while the right panels display the full spectrum, 
but with a y-axis range selected to allow the signal components to be seen clearly. Different curves 
represent the PMT response to a varying yield of incoming photons.}
\label{spe-on-quad-a-quad-b}
\end{figure}

Of marked interest for SoLID's application is the single-photoelectron 
response of groups of pixels. 
Figure~\ref{spe-on-quad-a-quad-b} shows our 
measurements on two groups of 16 pixels labeled quad A and quad B 
as indicated in Fig.~\ref{pmt_face}. We use the same procedure for 
single-photoelectron detection as for single pixels but 
we start with a lower LED voltage (\emph{i.e.} lower yield of photons) 
since we need only one pixel from the group of 16 to respond per event. 
Given that each pixel of the group has a different gain, the charge 
resulting from a single photoelectron extracted from the photocathode 
can be different per event if different pixels respond. Thus the 
resulting ADC distributions shown in Fig.~\ref{spe-on-quad-a-quad-b} 
can be viewed as a superposition of single pixel ADC distributions with 
possibly different gains. Nevertheless, our measurements show that the single-photoelectron 
signal is still clearly identifiable when summing over the 
output of 16 pixels.

\subsection{Single-photoelectron Characterization}

The ADC single-photoelectron spectra have been analyzed using a PMT 
response function developed by Bellamy and collaborators~\citet{bellamy} 
which parametrizes the photoconversion process at the photocathode and the 
electron collection and amplification through the dynode chain. The conversion 
of photons into electrons at the photocathode is described by a Poisson 
distribution while the response of the multiplicative dynode chain is 
parametrized using a Gaussian distribution. The PMT response function 
accounts also for various background processes which would generate 
additional charge like thermoelectron emission from the photocathode 
and/or dynode chain, leakage current in the anode circuit, etc. The 
background is parametrized by a combination of Gaussian and exponential 
functions. The realistic response function is then given by:
\begin{equation}   
F = \sum_{n=0}^{\infty} \frac{\mu^n e^{-\mu}}{n!} \times [(1 - w) G_{n}(x - Q_{0})] + w I_{G_{n} \bigotimes E} (x - Q_{0})
\end{equation}
where

\begin{equation}
\begin{split}
I_{G_{n} \bigotimes E} (x - Q_{0}) = \int_{Q_{0}}^{x} G_{n}(x' - Q_{0}) \alpha exp[-\alpha(x - x')]dx') \\
= \frac{\alpha}{2} exp[-\alpha(x - O_{n} - \alpha \sigma_{n}^2)] \times [erf(\frac{Q_{0} - Q_{n} - \sigma_{n}^{2} \alpha}{\sigma_{n}\sqrt{2}}) \\
+ sign(x - Q_{n} - \sigma_{n}^{2} \alpha) \times erf(\frac{x - Q_{n} - \sigma_{n}^2 \alpha}{\sigma_{n} \sqrt{2}})] 
\end{split}
\end{equation}
with $Q_{n} = Q_{0} + nQ_{1}$ and $\sigma_{n} = \sqrt{\sigma_{0}^2 + n\sigma_{1}^2}$. 
Here $Q_{0}$ is the pedestal and $erf(x)$ is the error function. There are 
seven fit parameters: $Q_{0}$ and $\sigma_{0}$ define the pedestal position and width, $w$ 
and $\alpha$ describe the discrete background and the remaining three parameters, 
$Q_{1}$, $\sigma_{1}$ and $\mu$ characterize the spectrum of the real signal. 
Of the latter three parameters, $\mu$ is proportional to the intensity of the 
light source, and $Q_{1}$ and $\sigma_{1}$ characterize the amplification process 
of the dynode system.

\begin{figure}[htbp]
\vspace*{-0.2in}
\centering
\begin{tabular}{cc}
\includegraphics[width=8cm]{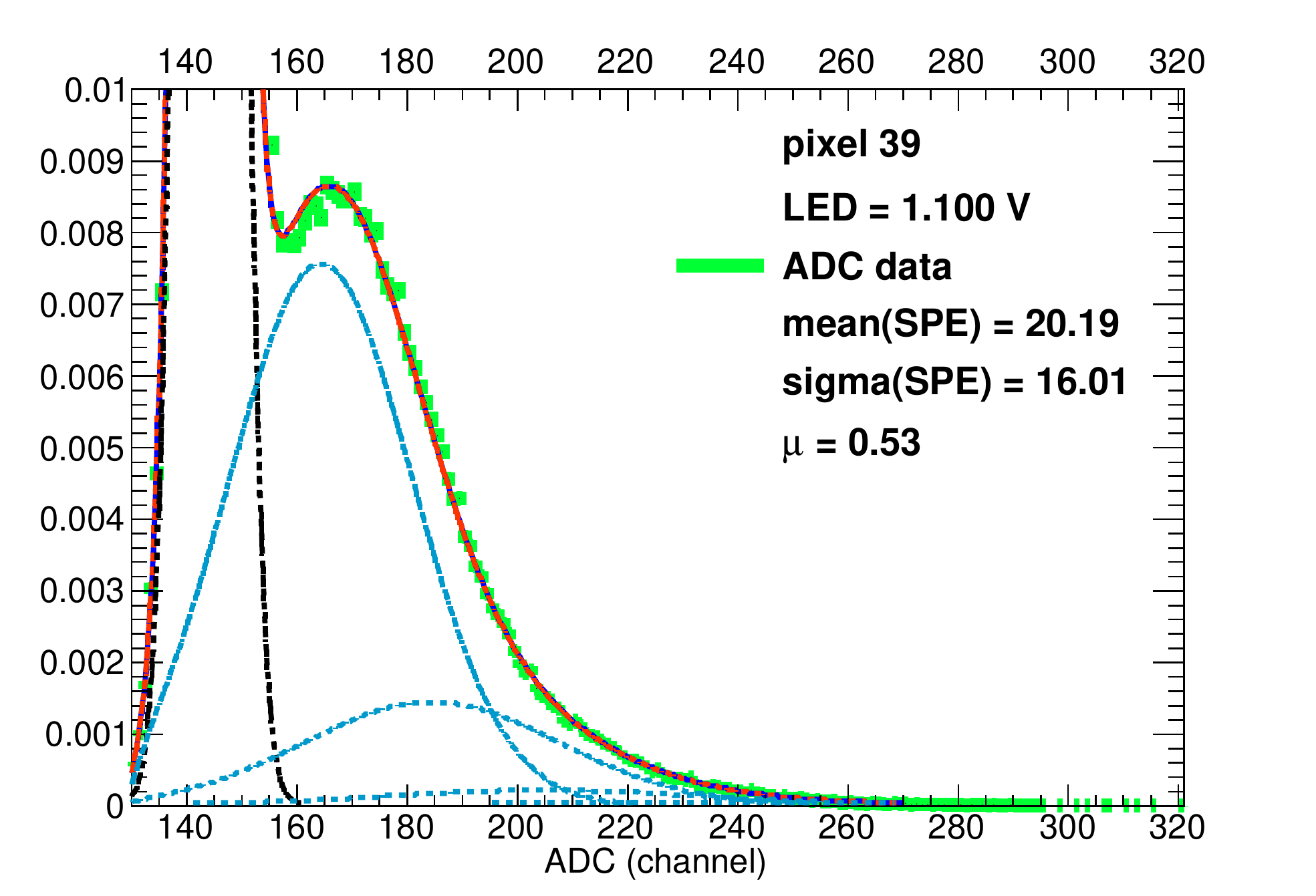}
&
\includegraphics[width=8cm]{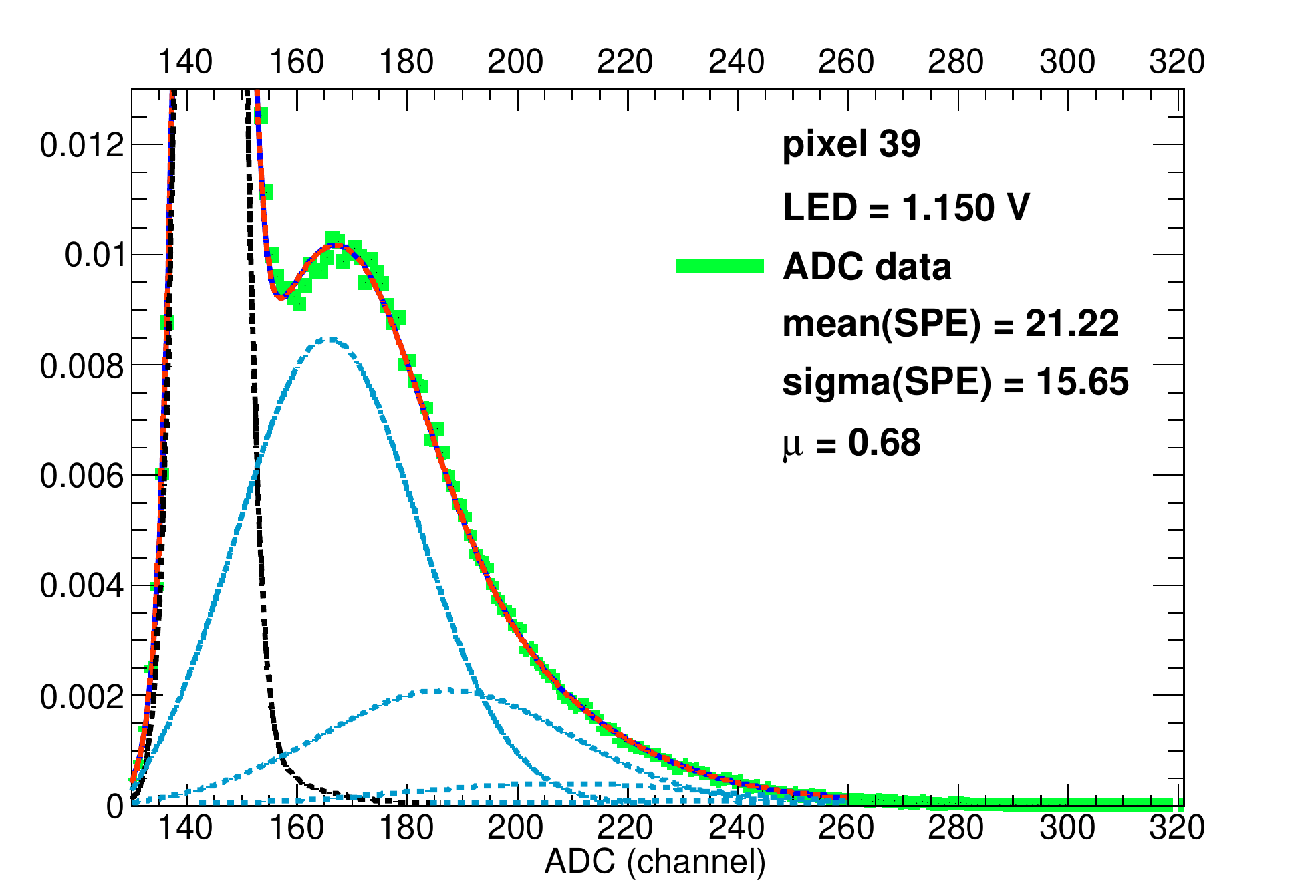} \\
\includegraphics[width=8cm]{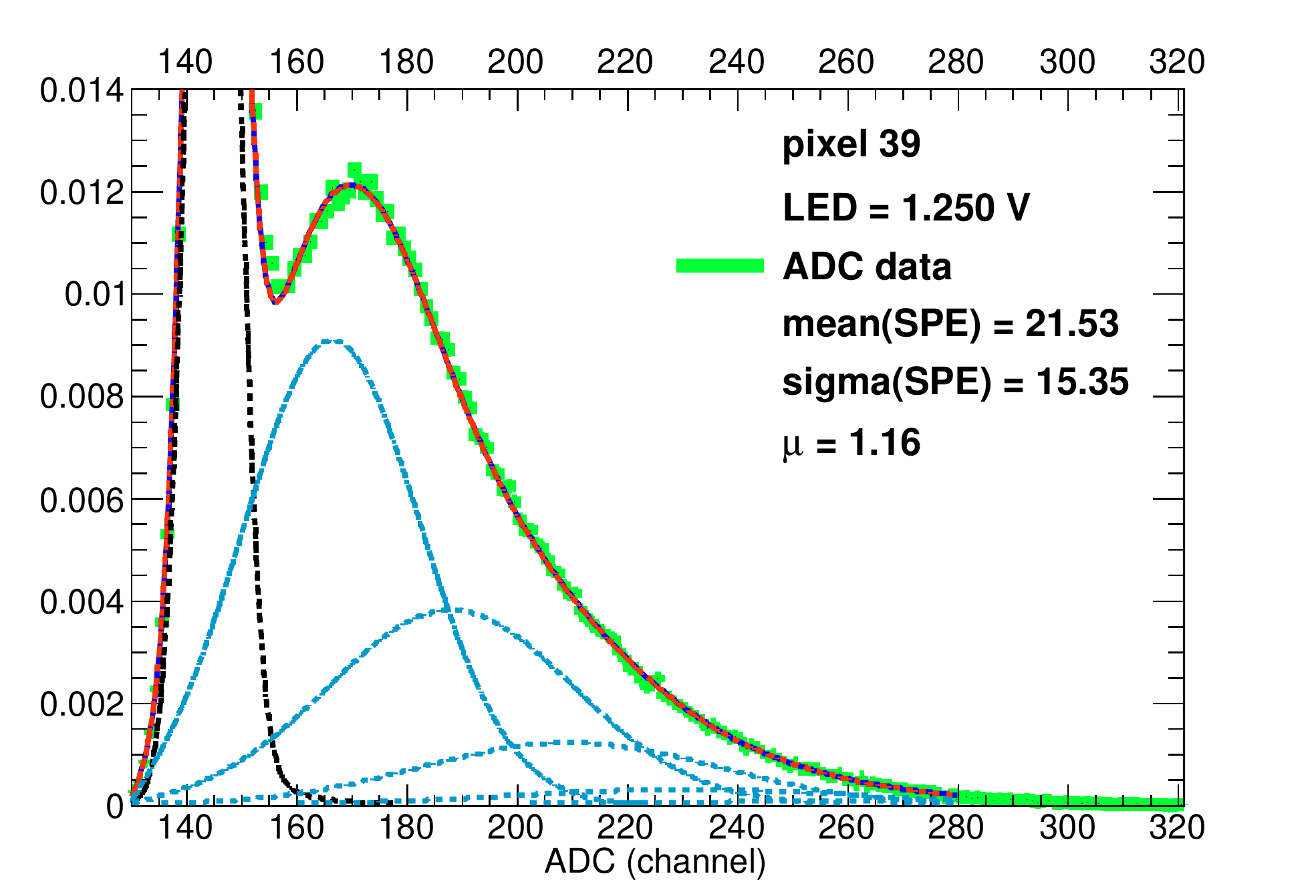}
&
\includegraphics[width=8cm]{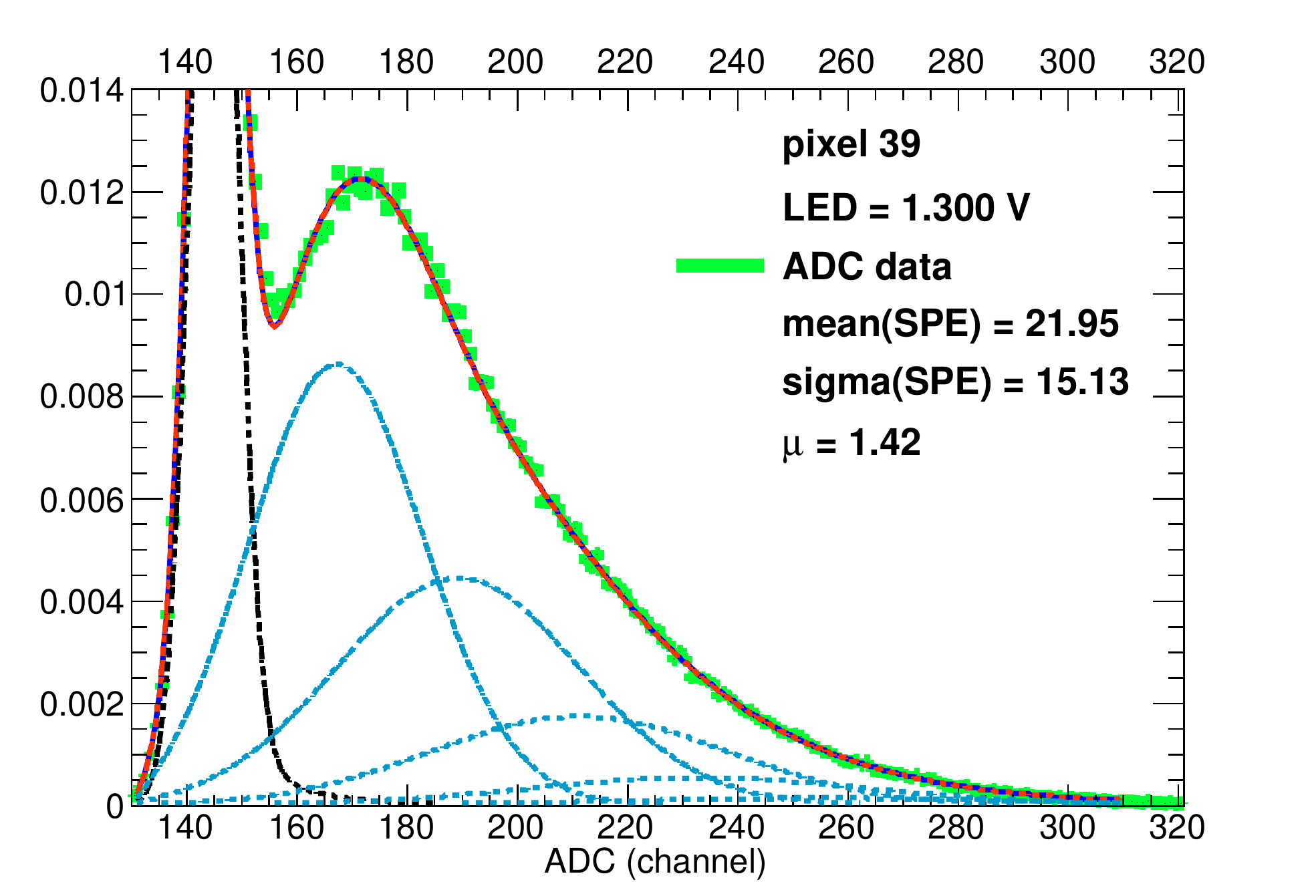} \\
\includegraphics[width=8cm]{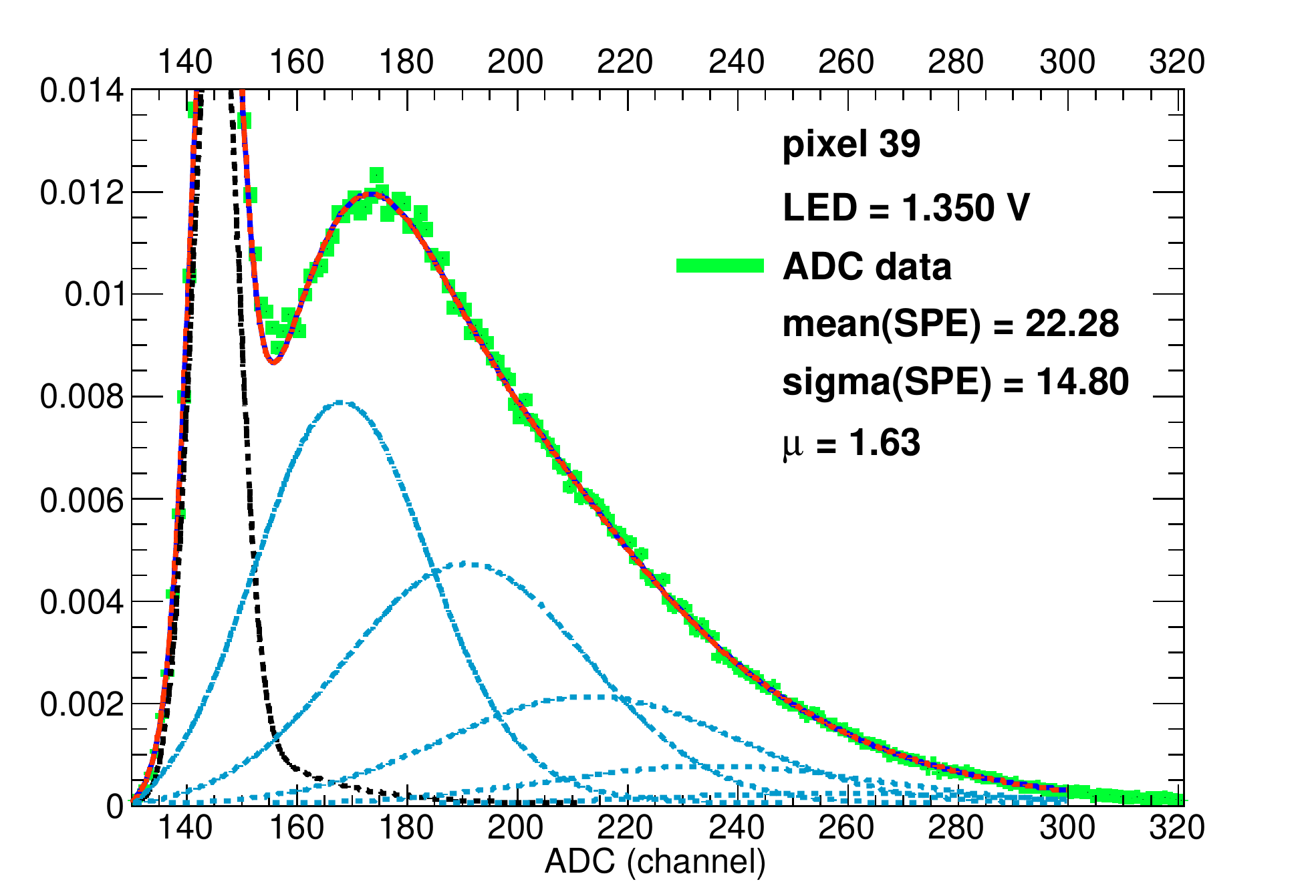}
&
\includegraphics[width=8cm]{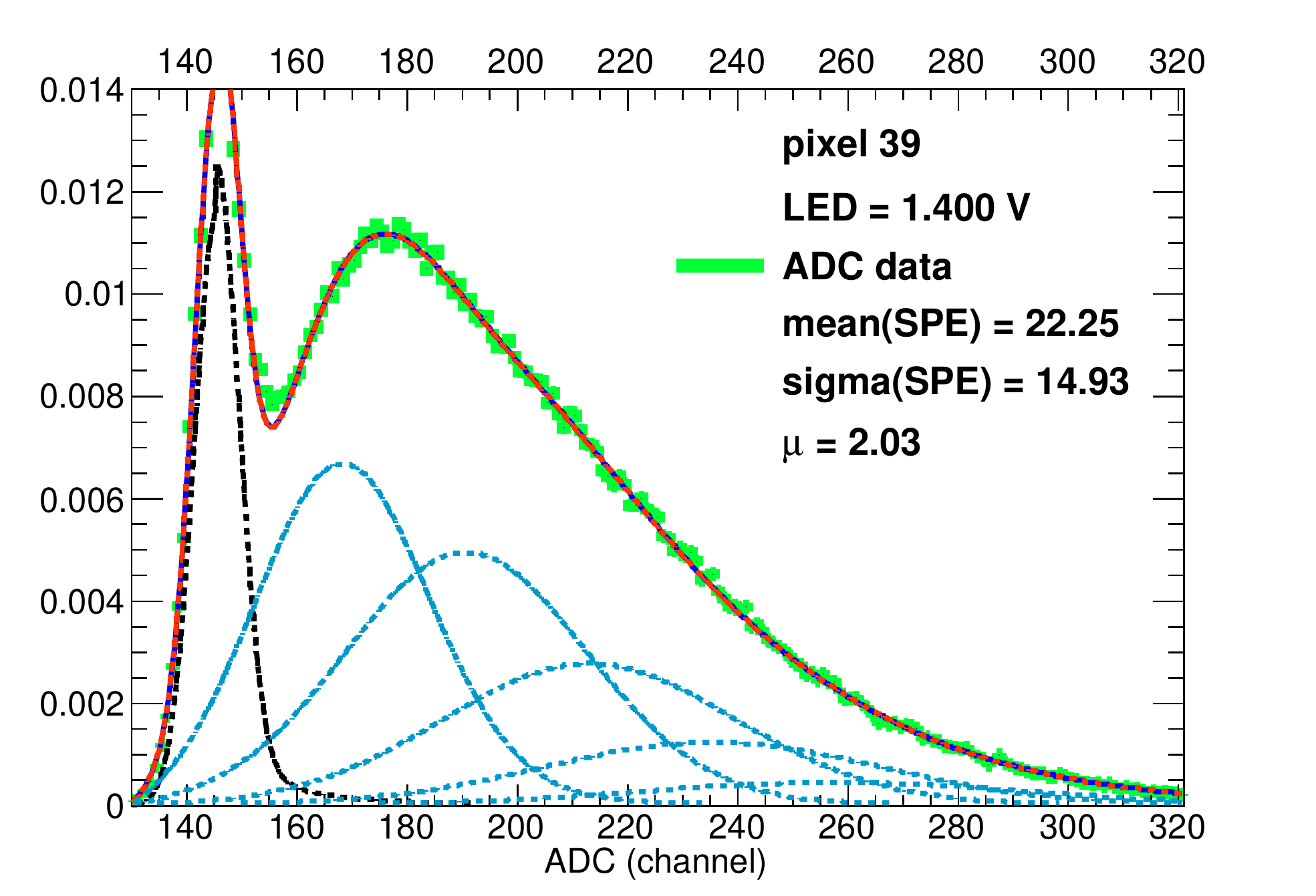} \\
\includegraphics[width=8cm]{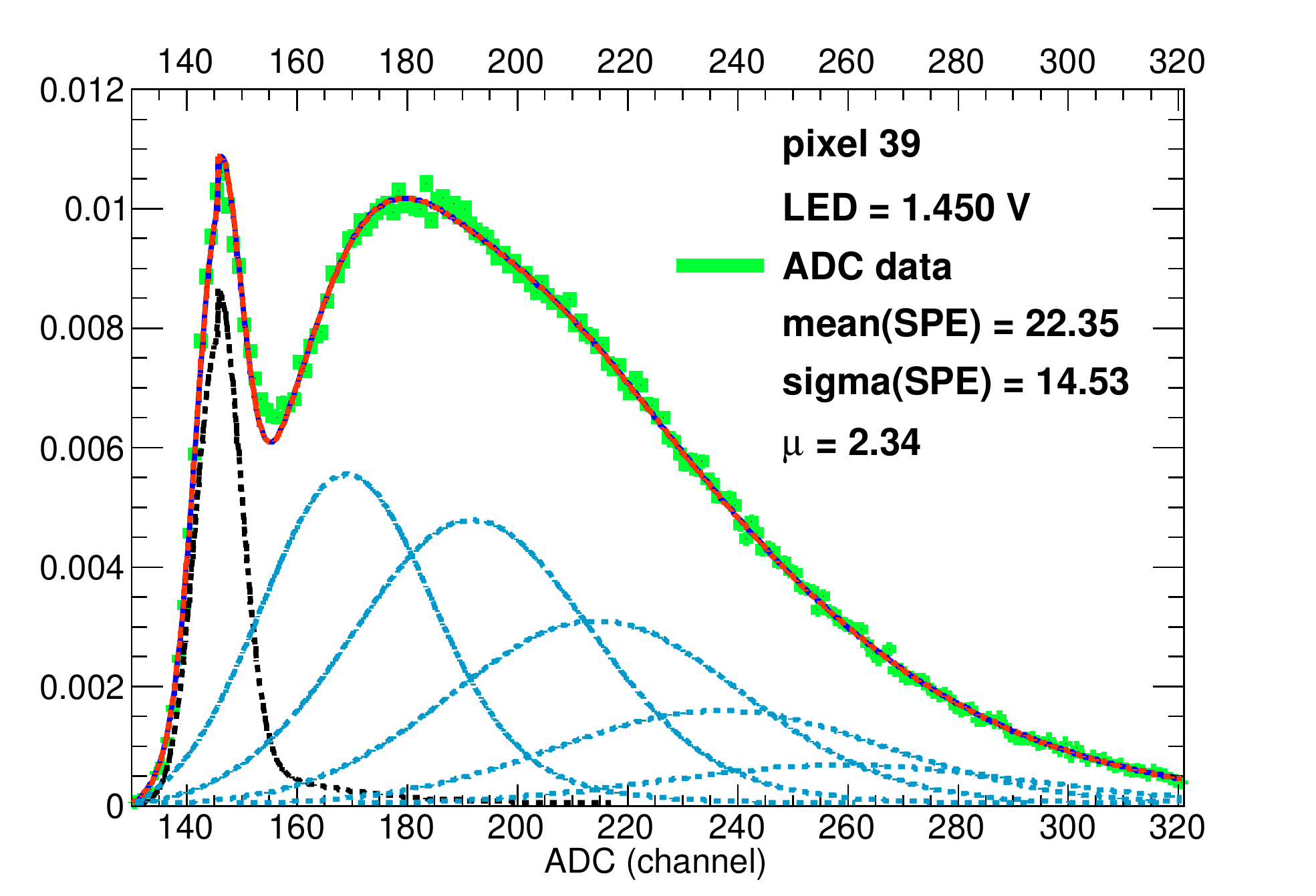}
&
\includegraphics[width=8cm]{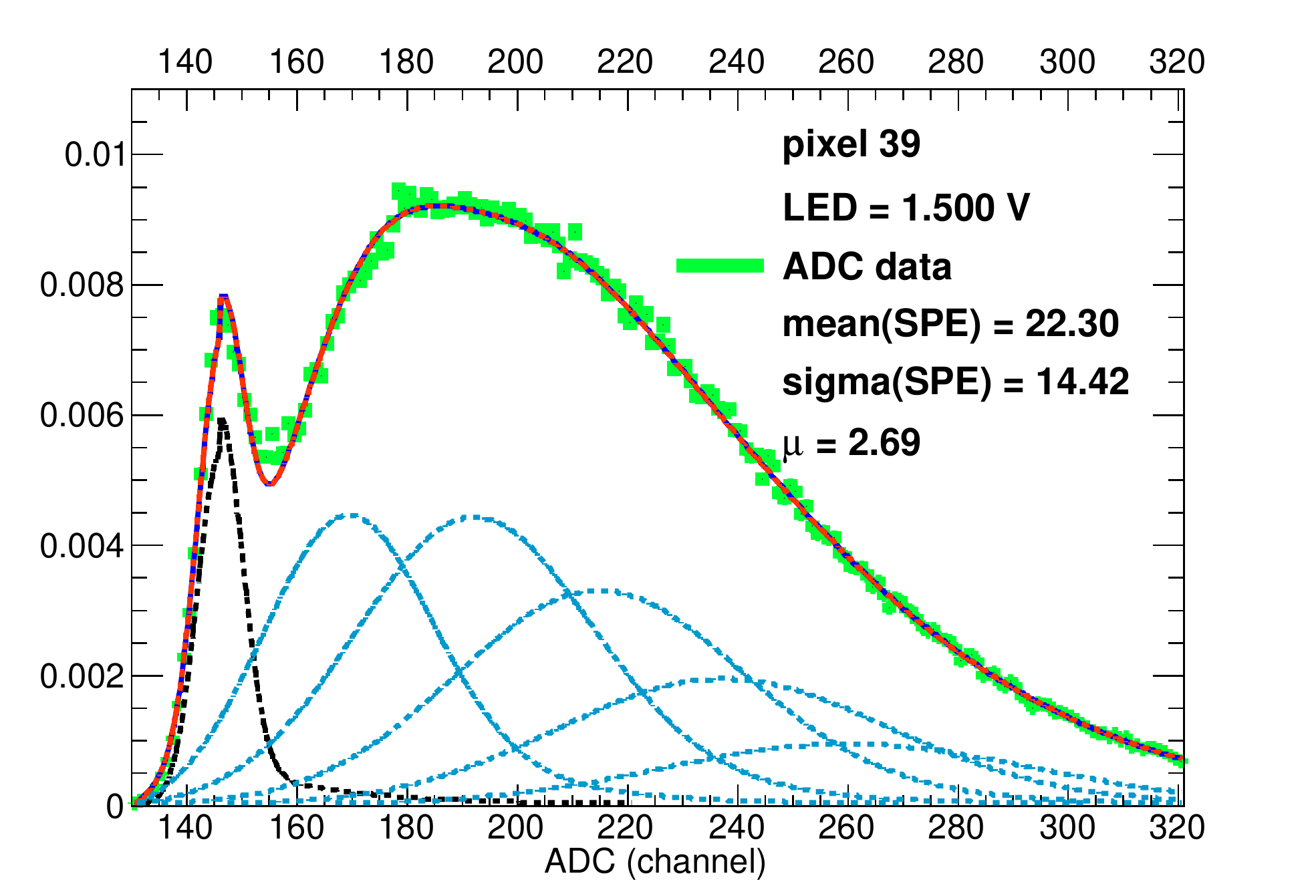}
\end{tabular}
\linespread{0.5}
\caption[]{
{Fits to the ADC distributions recorded from pixel 39 responding to varying 
yields of incoming photons (see text for more details). The green points represent 
the data, the red continuous curve shows the fit to data while the black dashed and 
blue dotted curves depict the background and real signal contributions to the 
total fit.} }
\label{spe-fit-pixel-39}
\end{figure}

\begin{figure}[htbp]
\vspace*{-0.2in}
\centering
\begin{tabular}{cc}
\includegraphics[width=8cm]{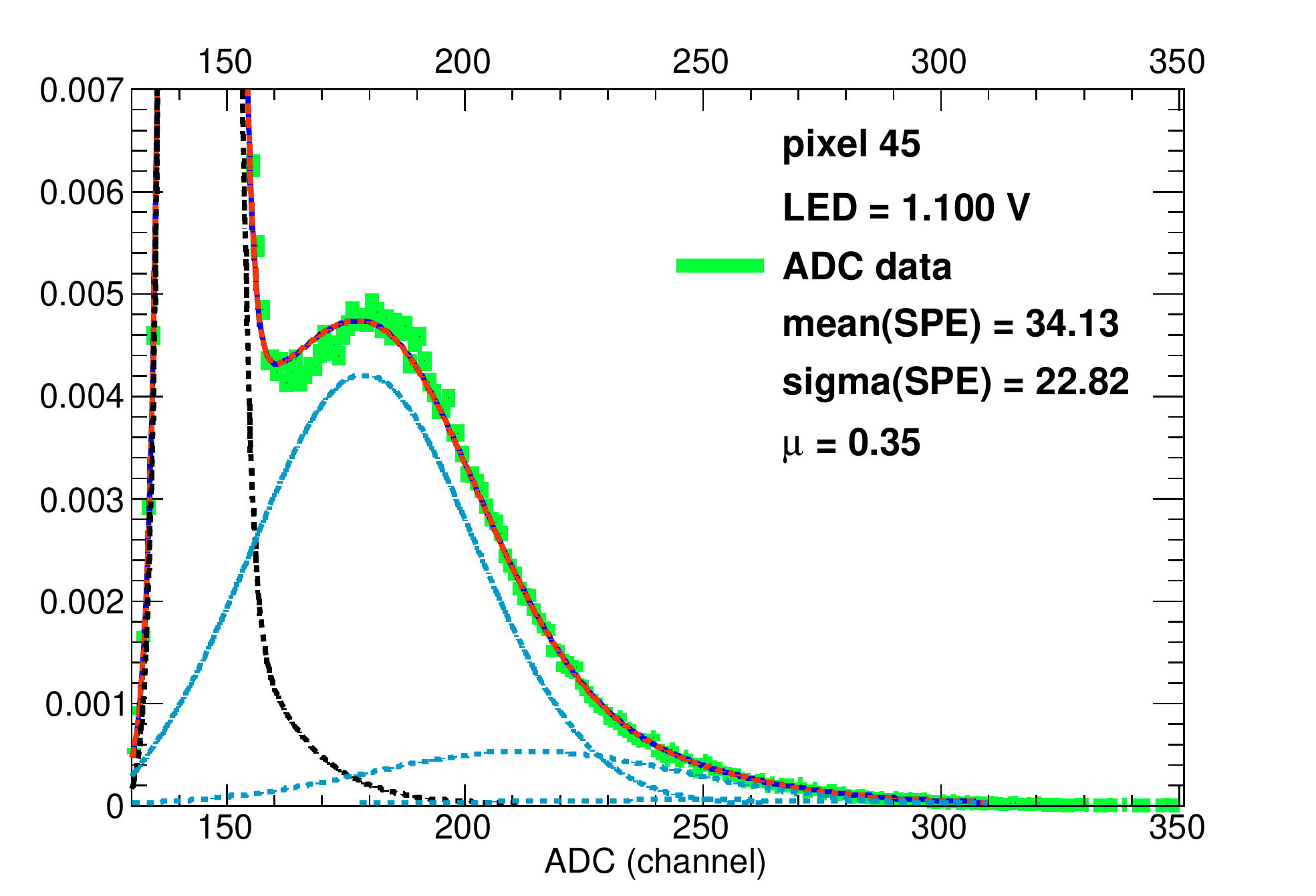}
&
\includegraphics[width=8cm]{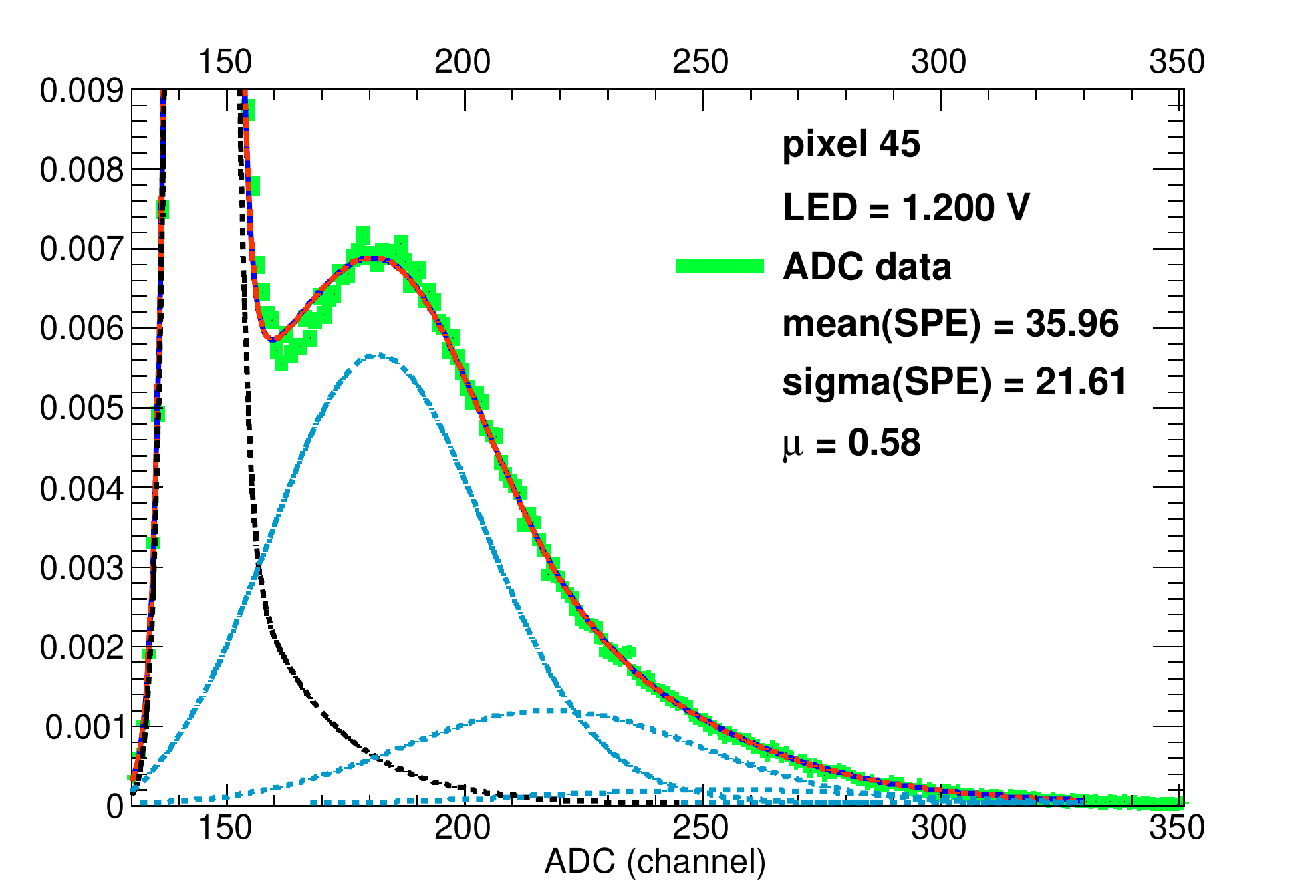} \\
\includegraphics[width=8cm]{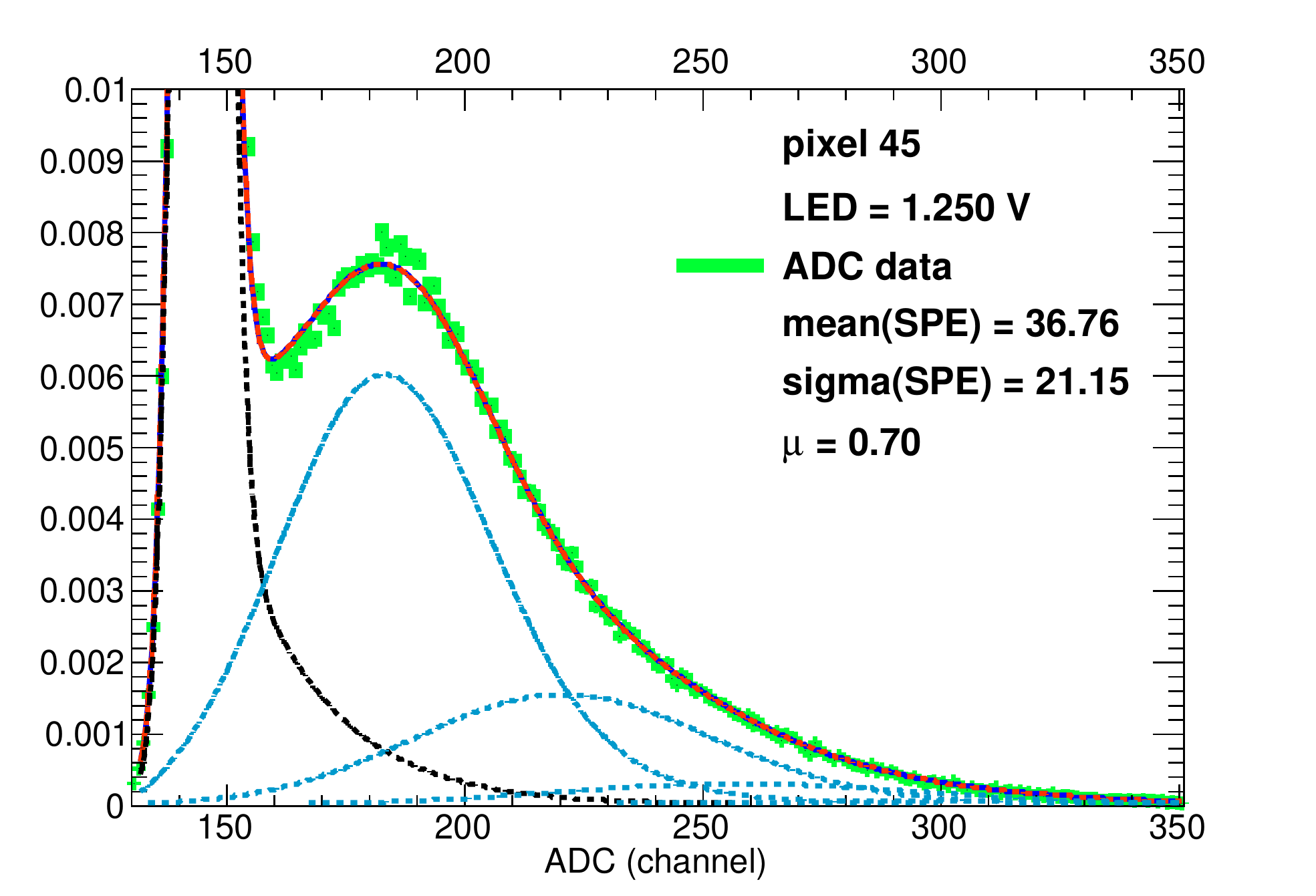}
&
\includegraphics[width=8cm]{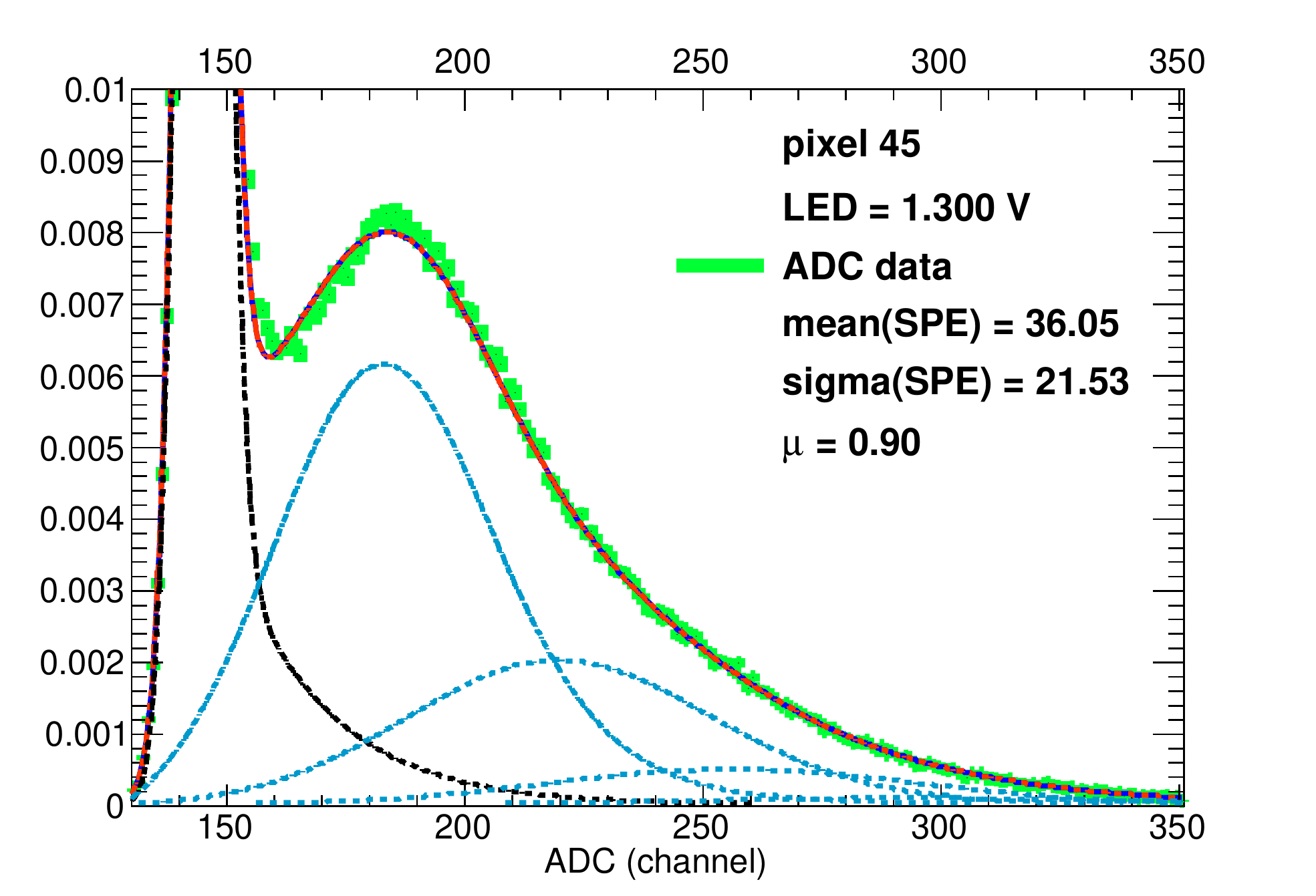} \\
\includegraphics[width=8cm]{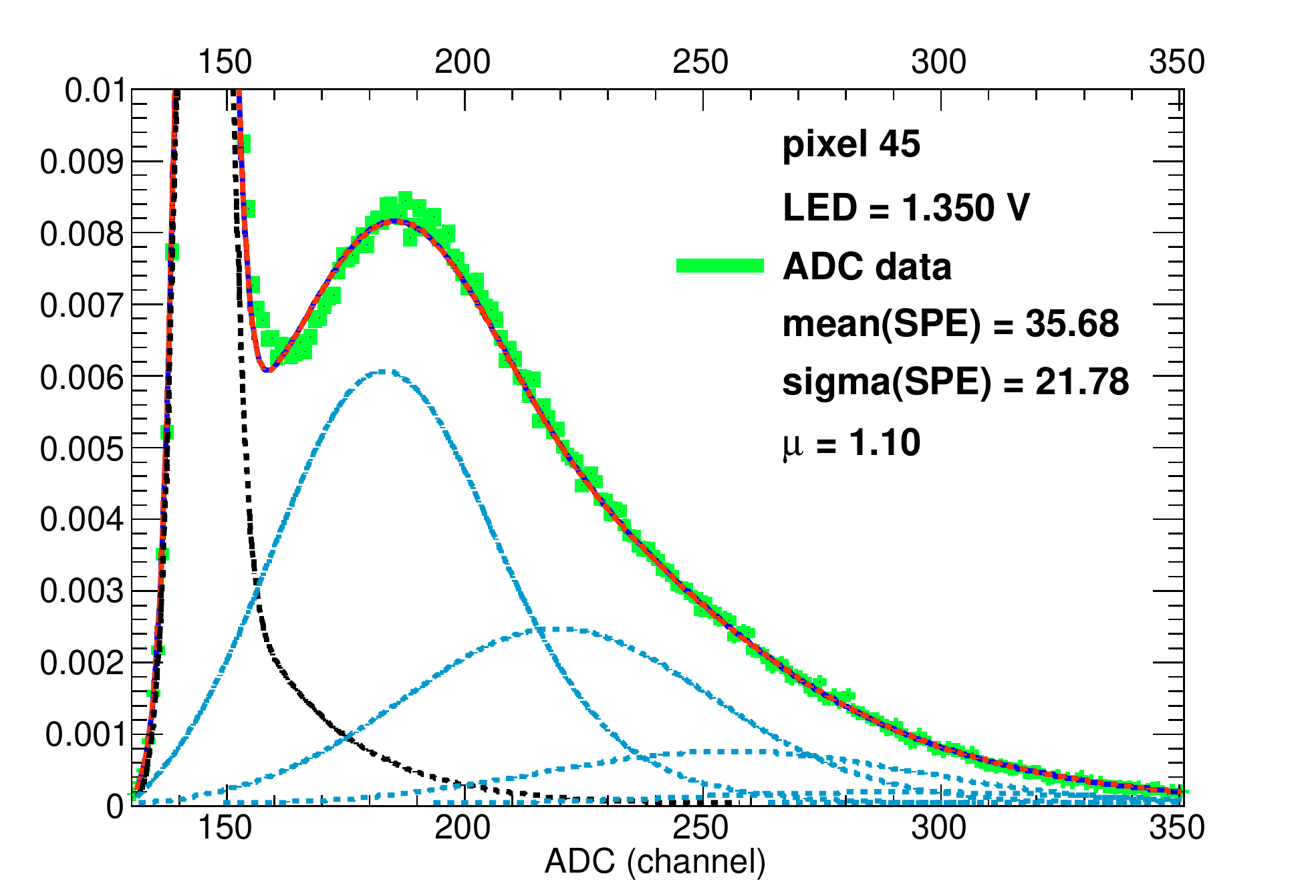}
&
\includegraphics[width=8cm]{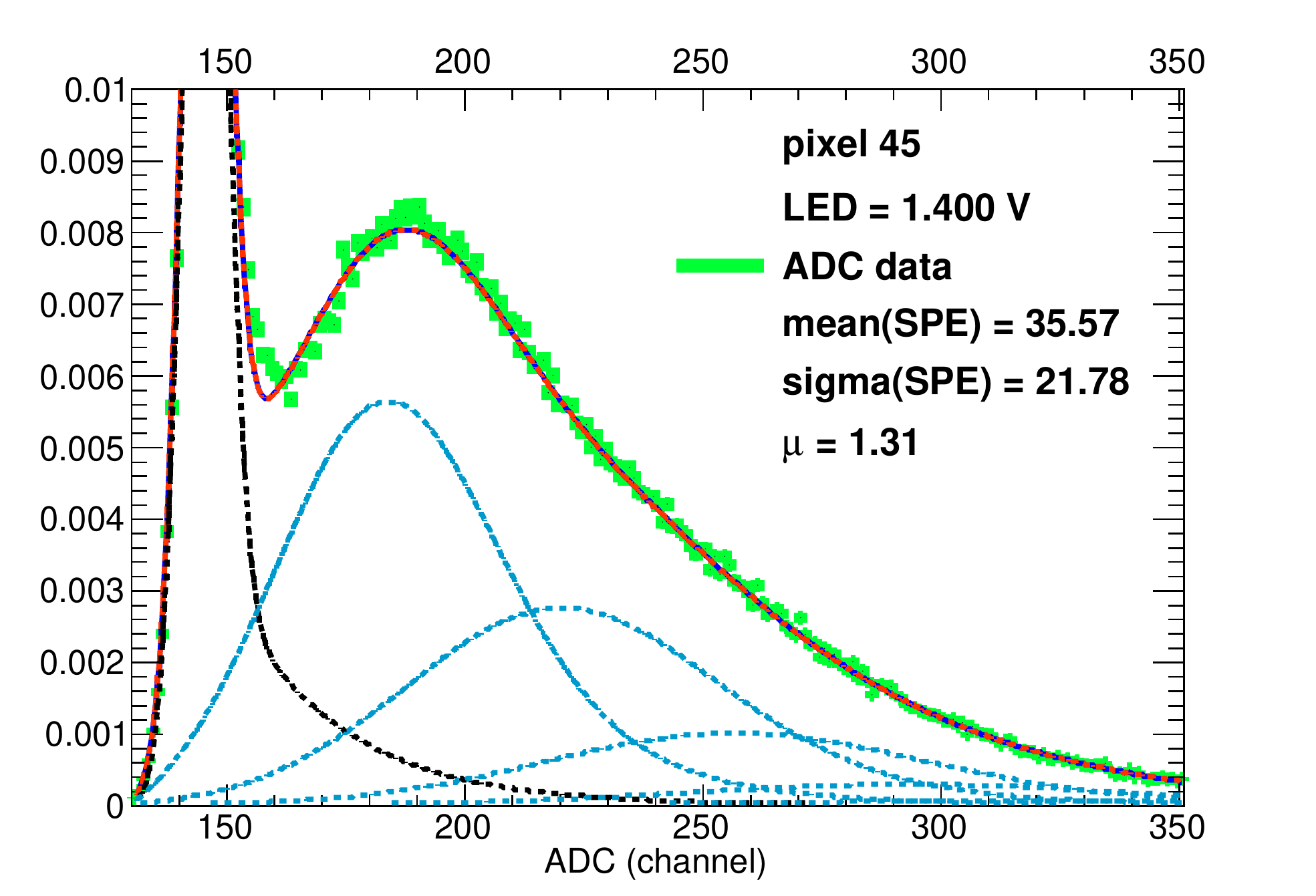} \\
\includegraphics[width=8cm]{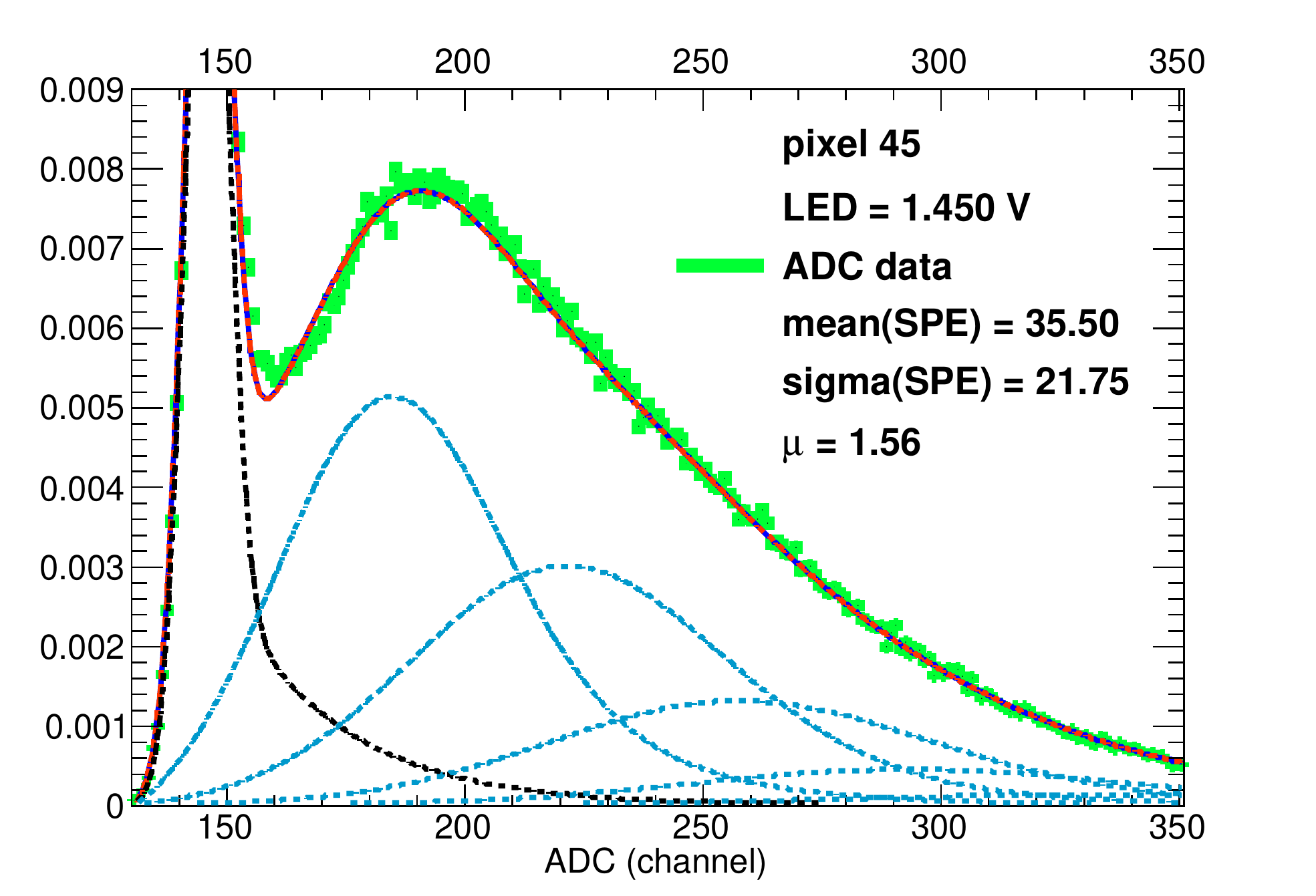}
&
\includegraphics[width=8cm]{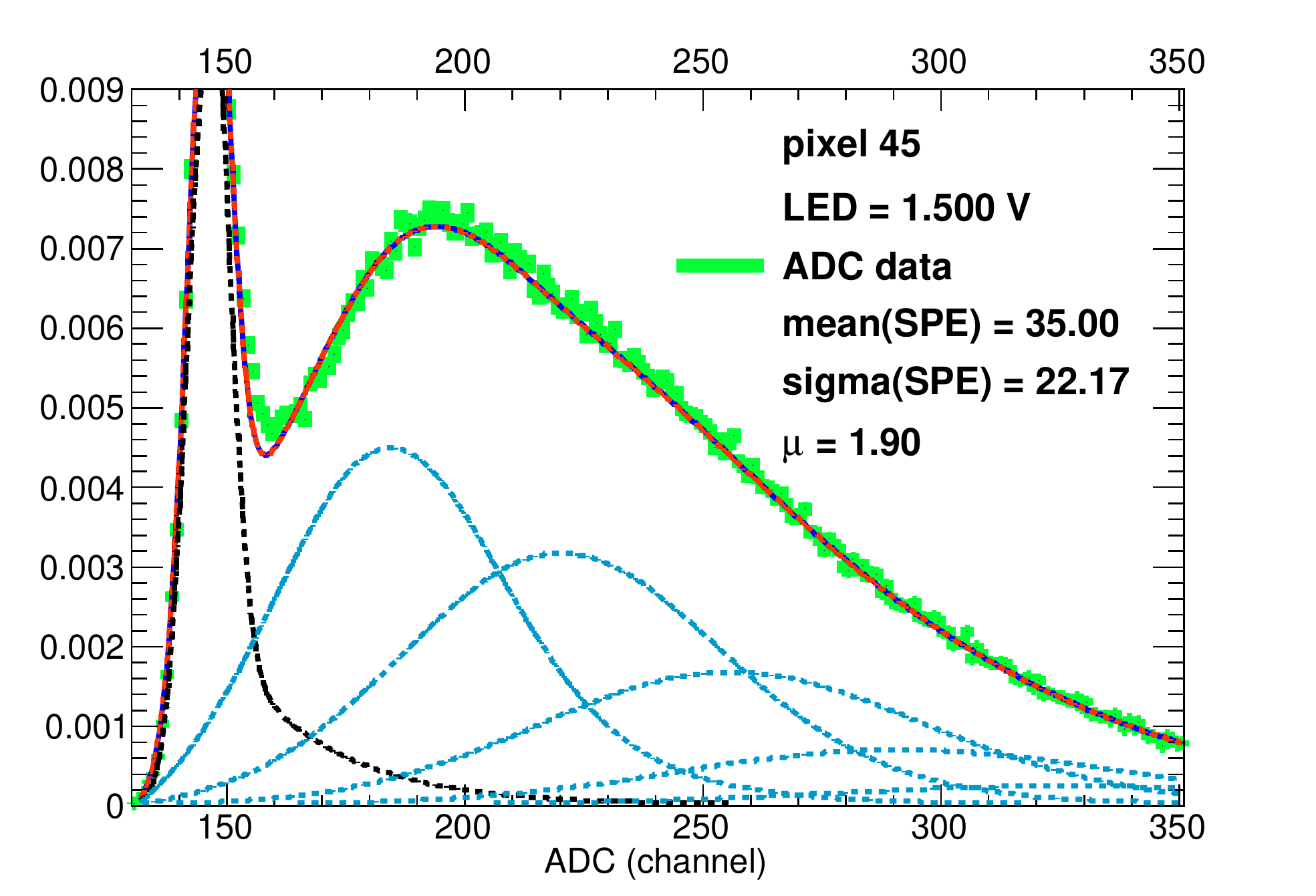}
\end{tabular}
\linespread{0.5}
\caption[]{
{Fits to the ADC distributions recorded from pixel 45 responding to varying 
yields of incoming photons (see text for more details). The green points represent 
the data, the red continuous curve shows the fit to data while the black dashed and 
blue dotted curves depict the background and real signal contributions to the 
total fit.} }
\label{spe-fit-pixel-45}
\end{figure}

\begin{figure}[htbp]
\vspace*{-0.2in}
\centering
\begin{tabular}{cc}
\includegraphics[width=8cm]{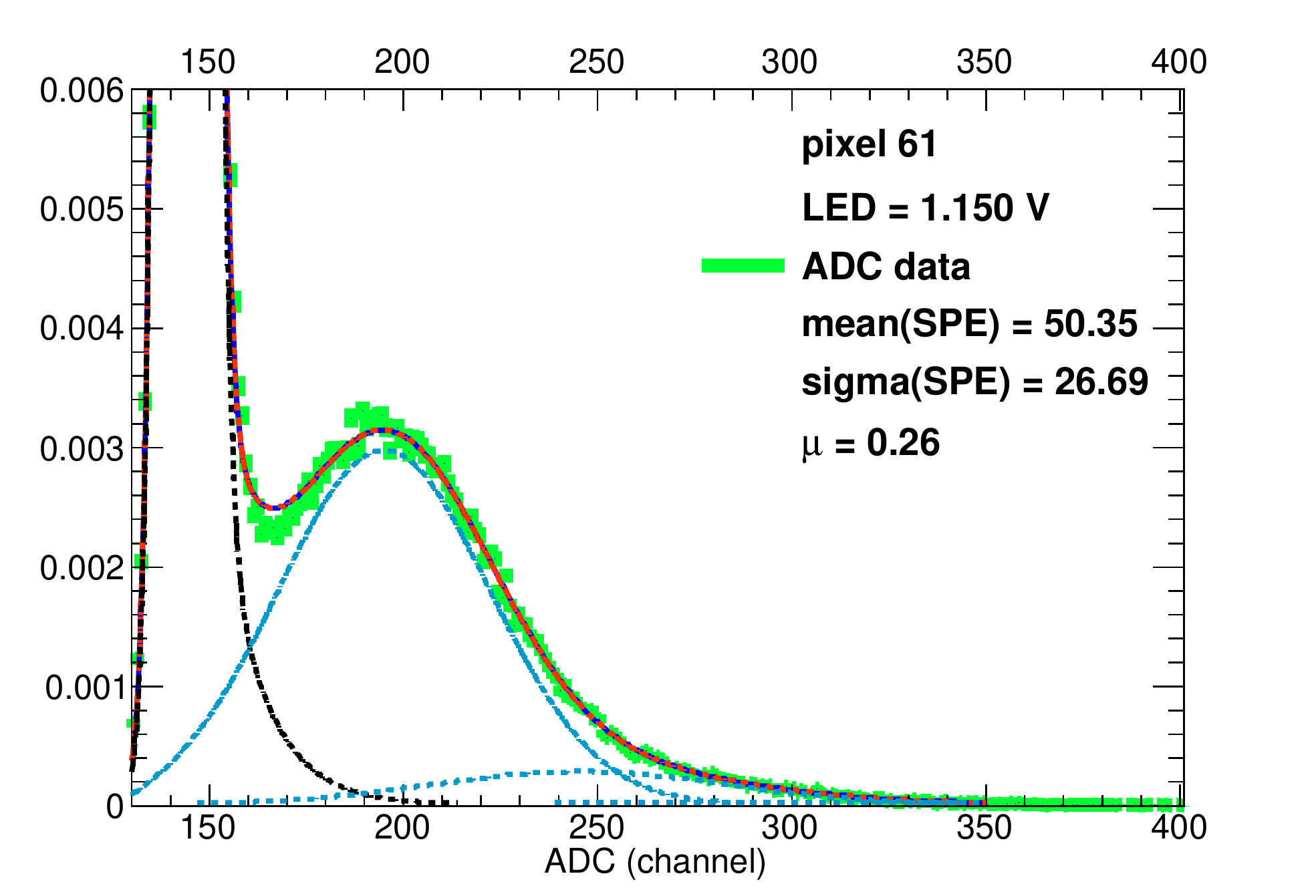}
&
\includegraphics[width=8cm]{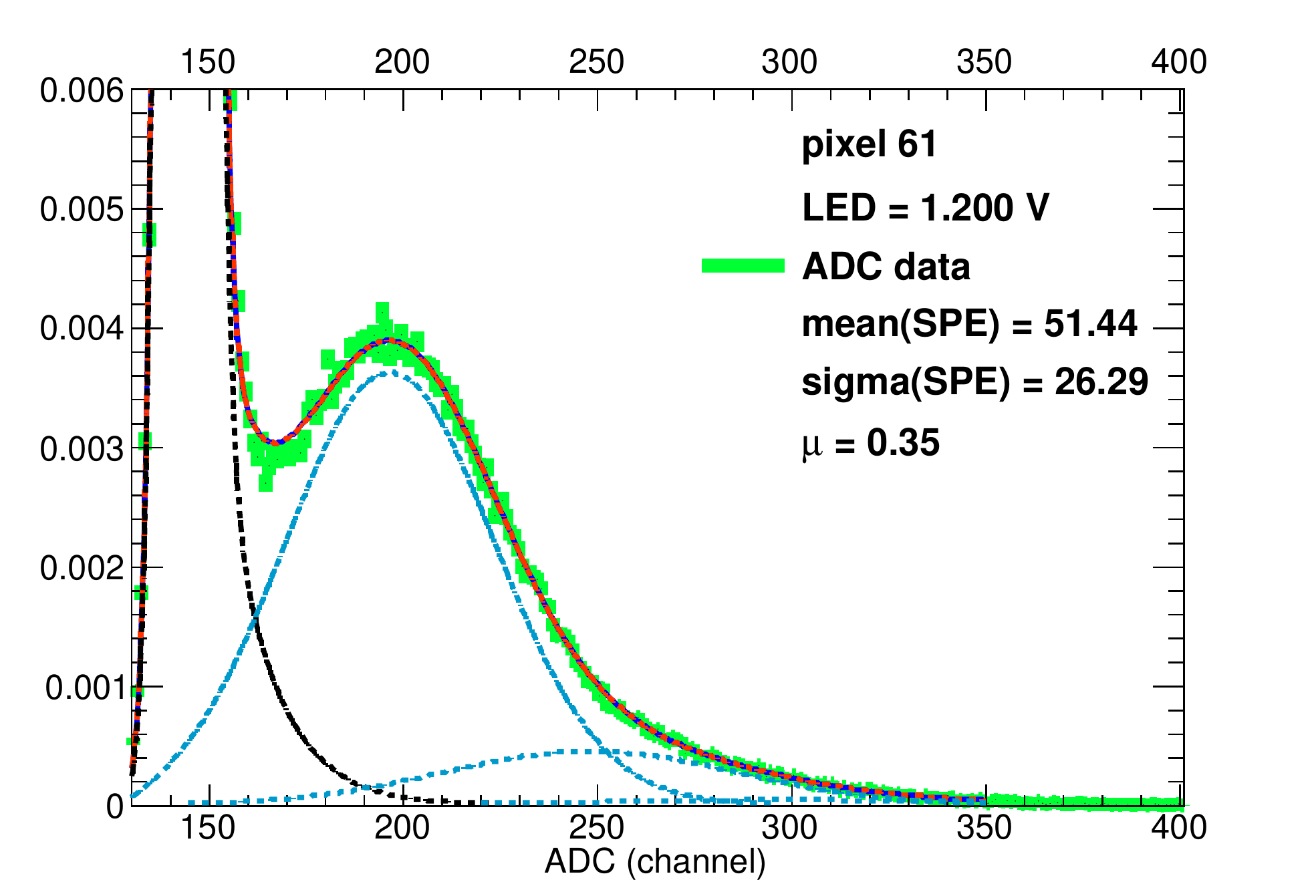} \\
\includegraphics[width=8cm]{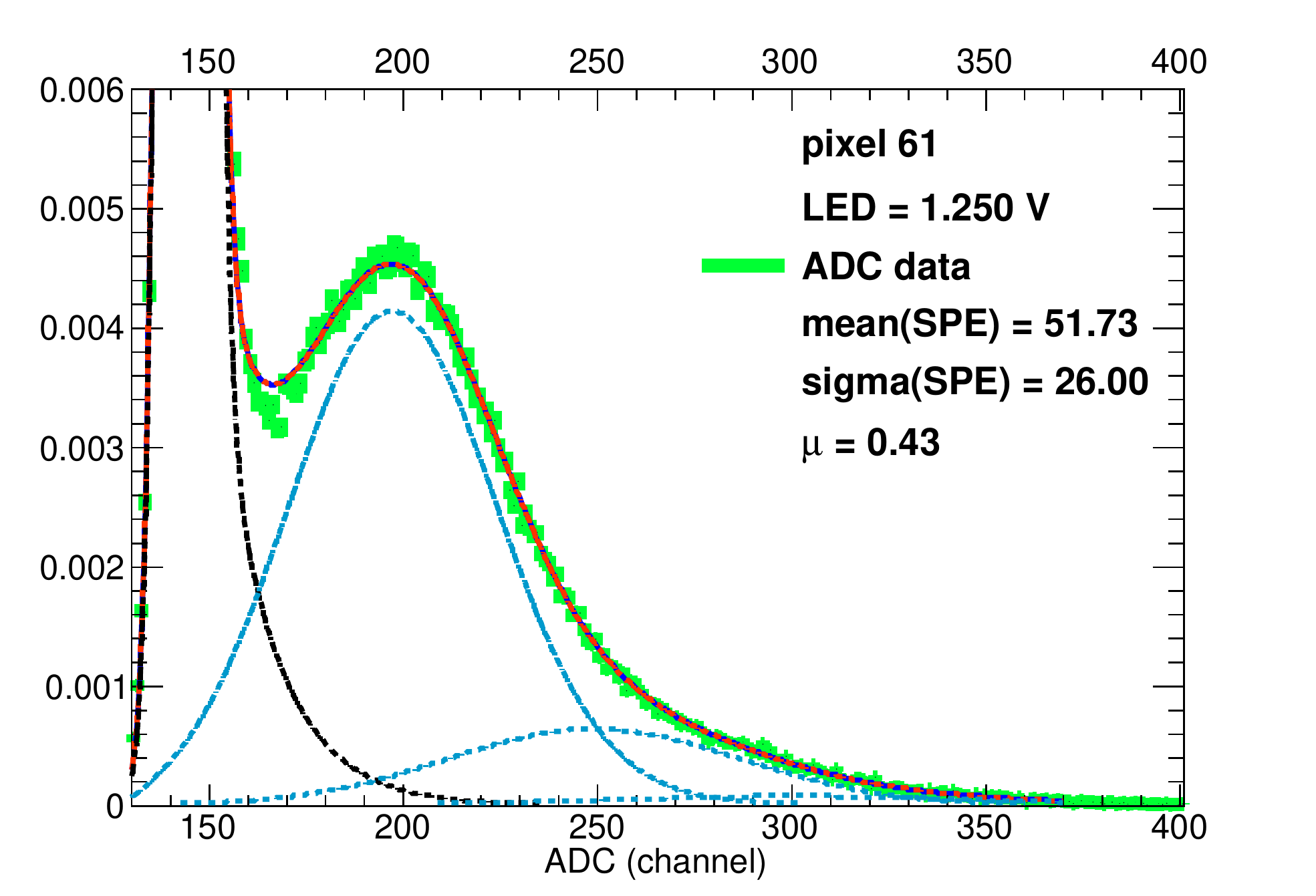}
&
\includegraphics[width=8cm]{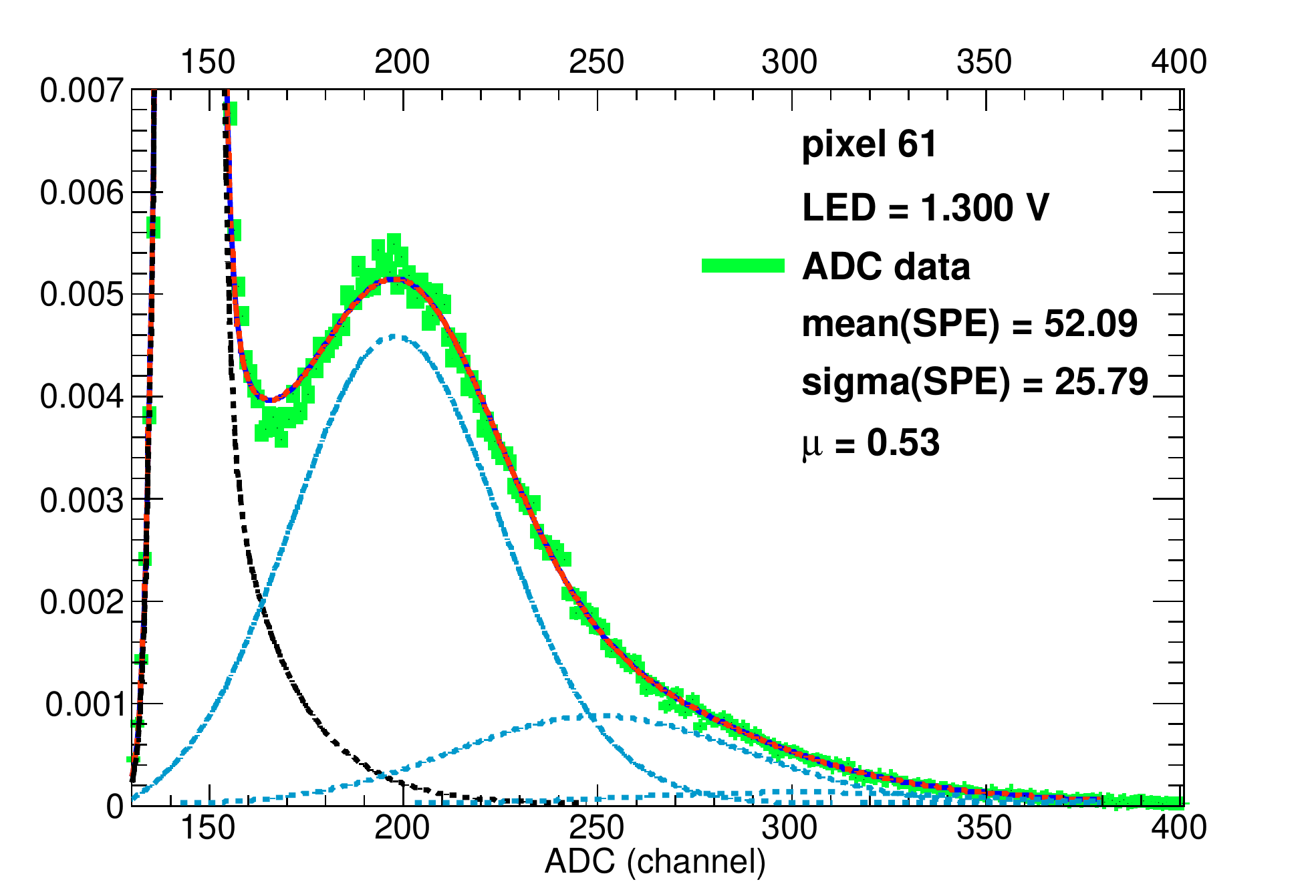} \\
\includegraphics[width=8cm]{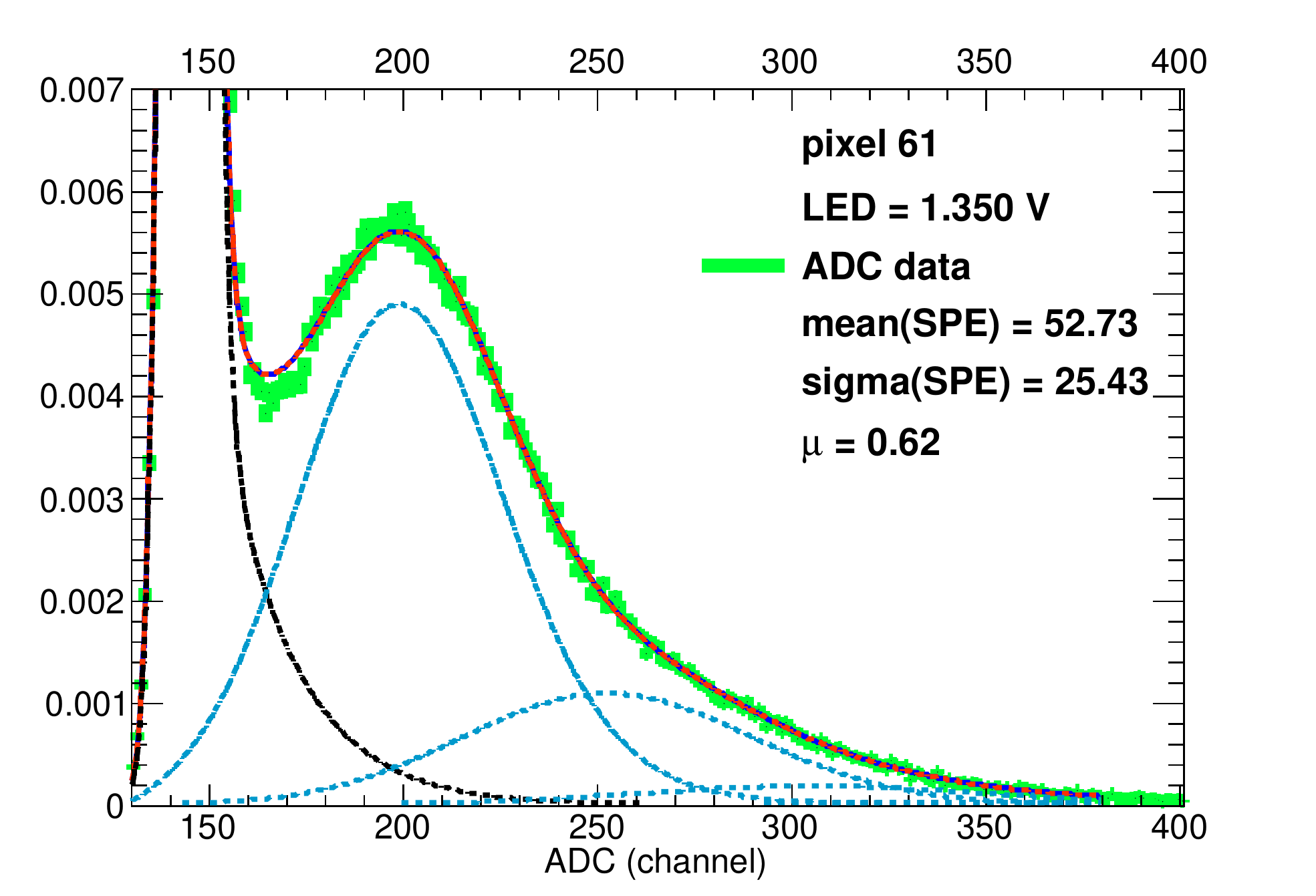}
&
\includegraphics[width=8cm]{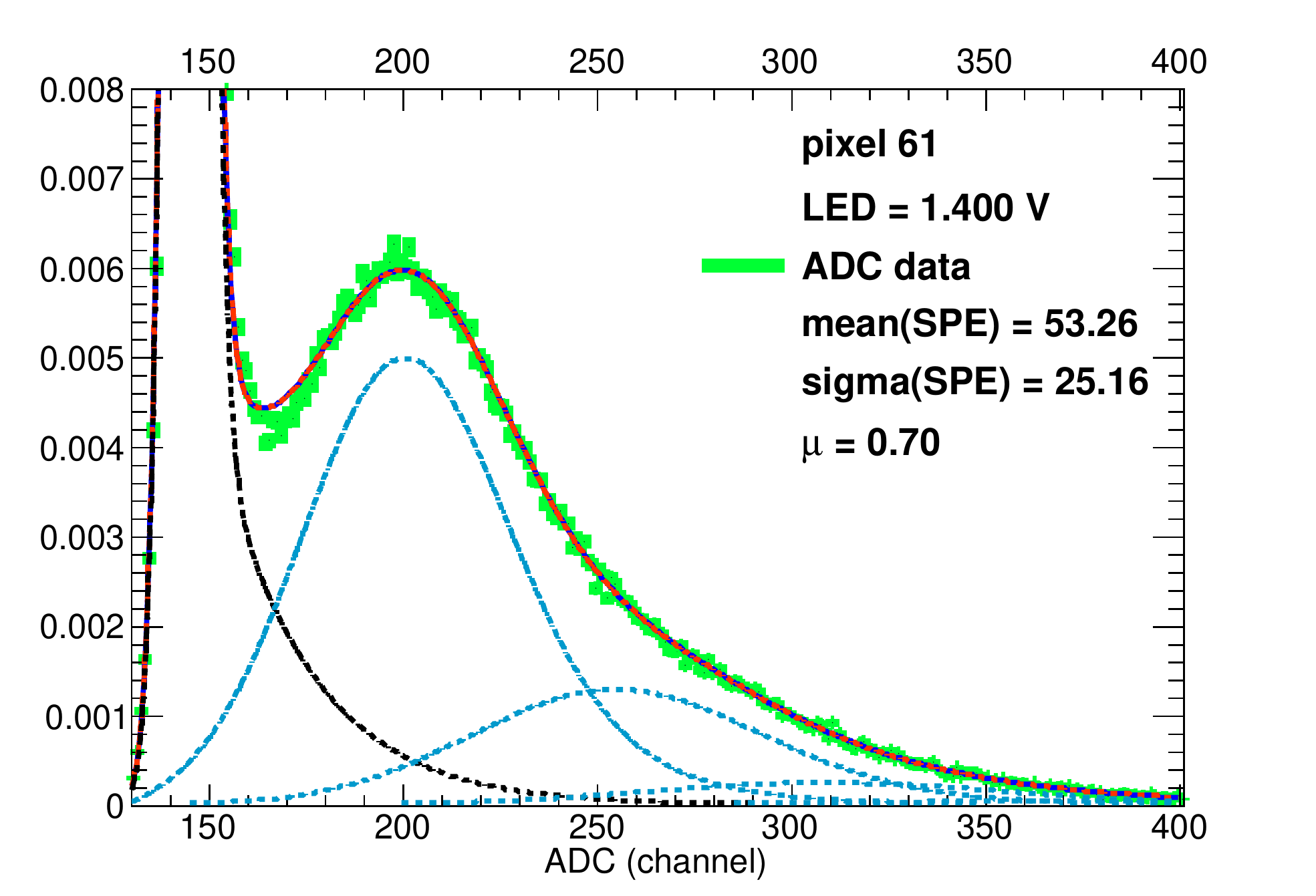} \\
\includegraphics[width=8cm]{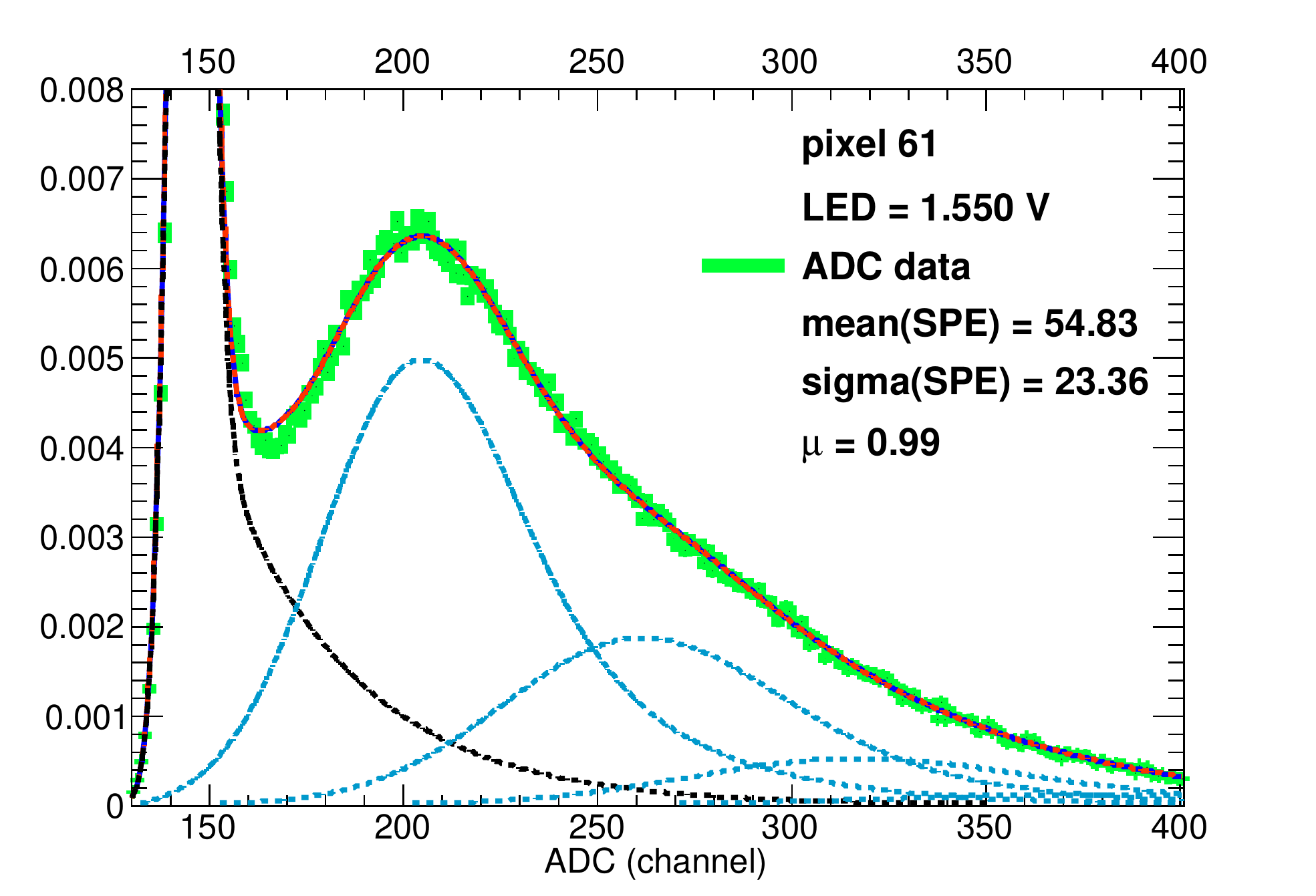}
&
\includegraphics[width=8cm]{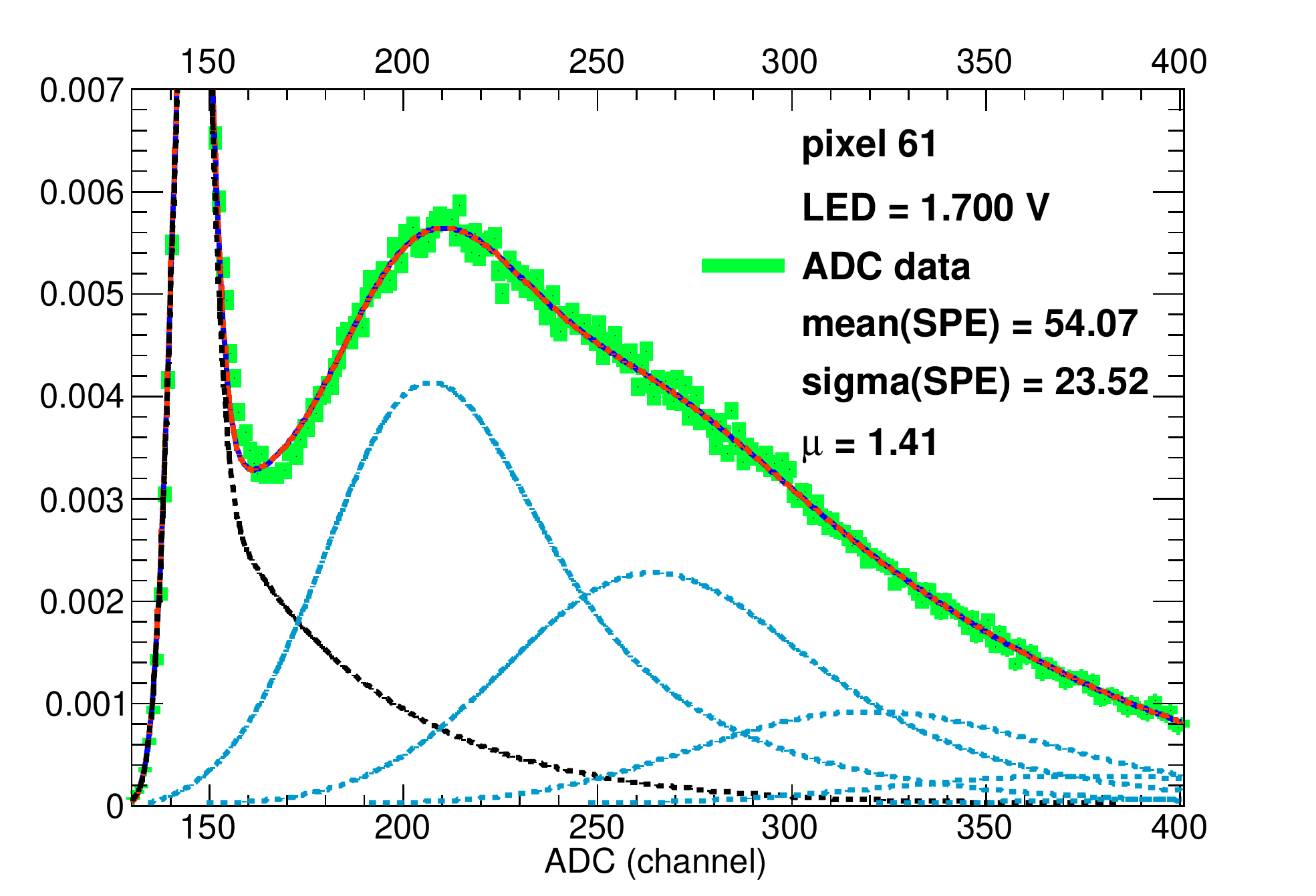}
\end{tabular}
\linespread{0.5}
\caption[]{
{Fits to the ADC distributions recorded from pixel 61 responding to varying 
yields of incoming photons (see text for more details). The green points represent 
the data, the red continuous curve shows the fit to data while the black dashed and 
blue dotted curves depict the background and real signal contributions to the 
total fit.} }
\label{spe-fit-pixel-61}
\end{figure}

\begin{figure}[htbp]
\vspace*{-0.2in}
\centering
\begin{tabular}{cc}
\includegraphics[width=8cm]{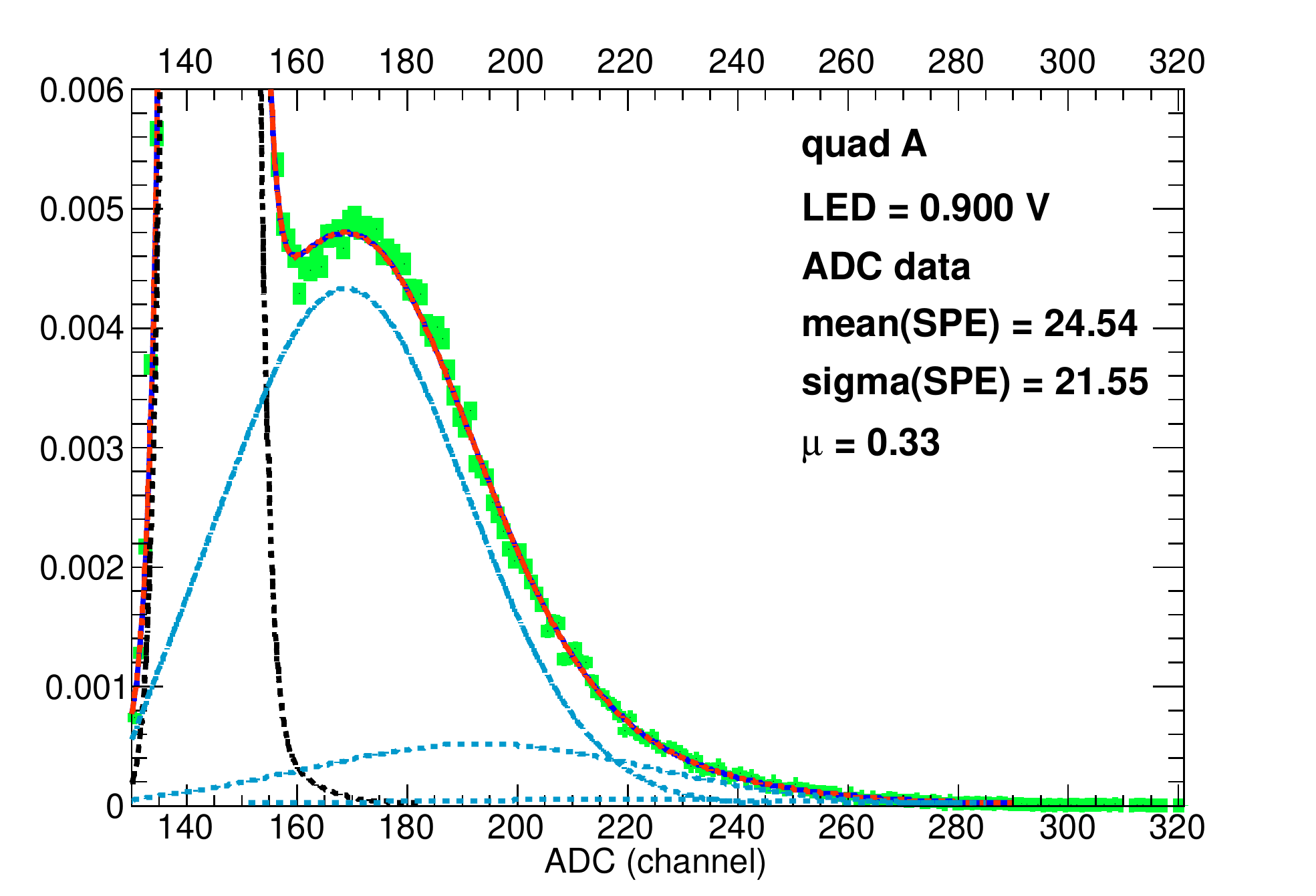}
&
\includegraphics[width=8cm]{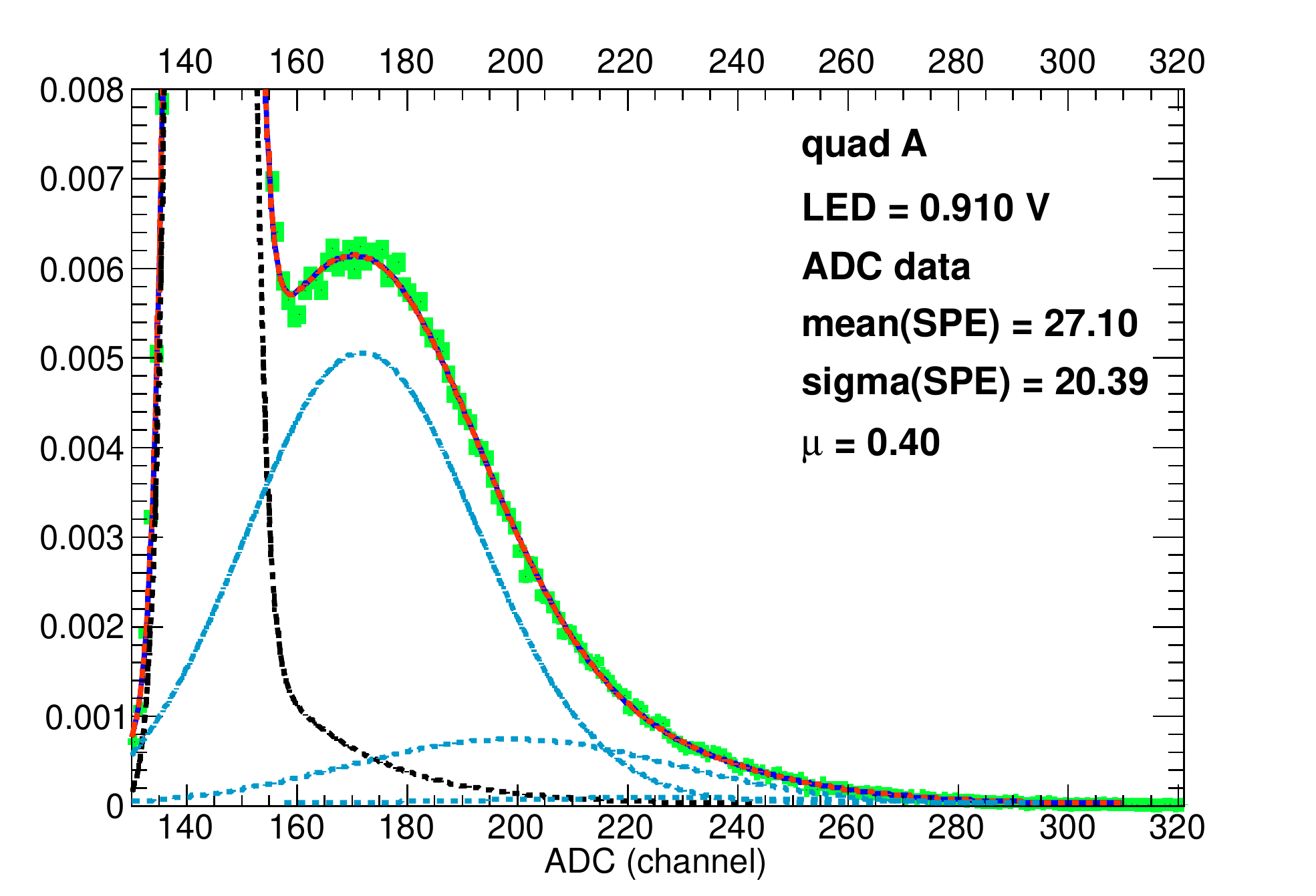} \\
\includegraphics[width=8cm]{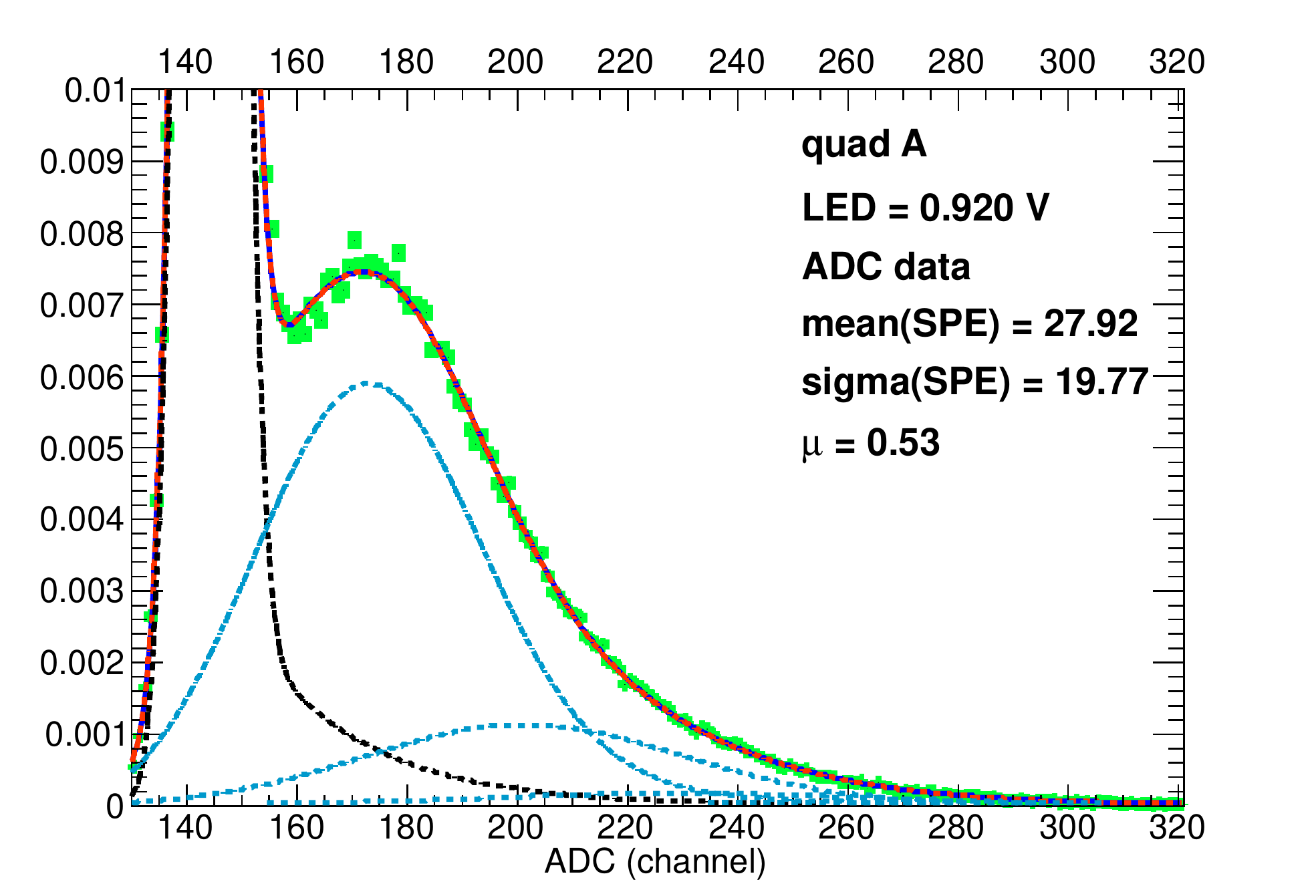}
&
\includegraphics[width=8cm]{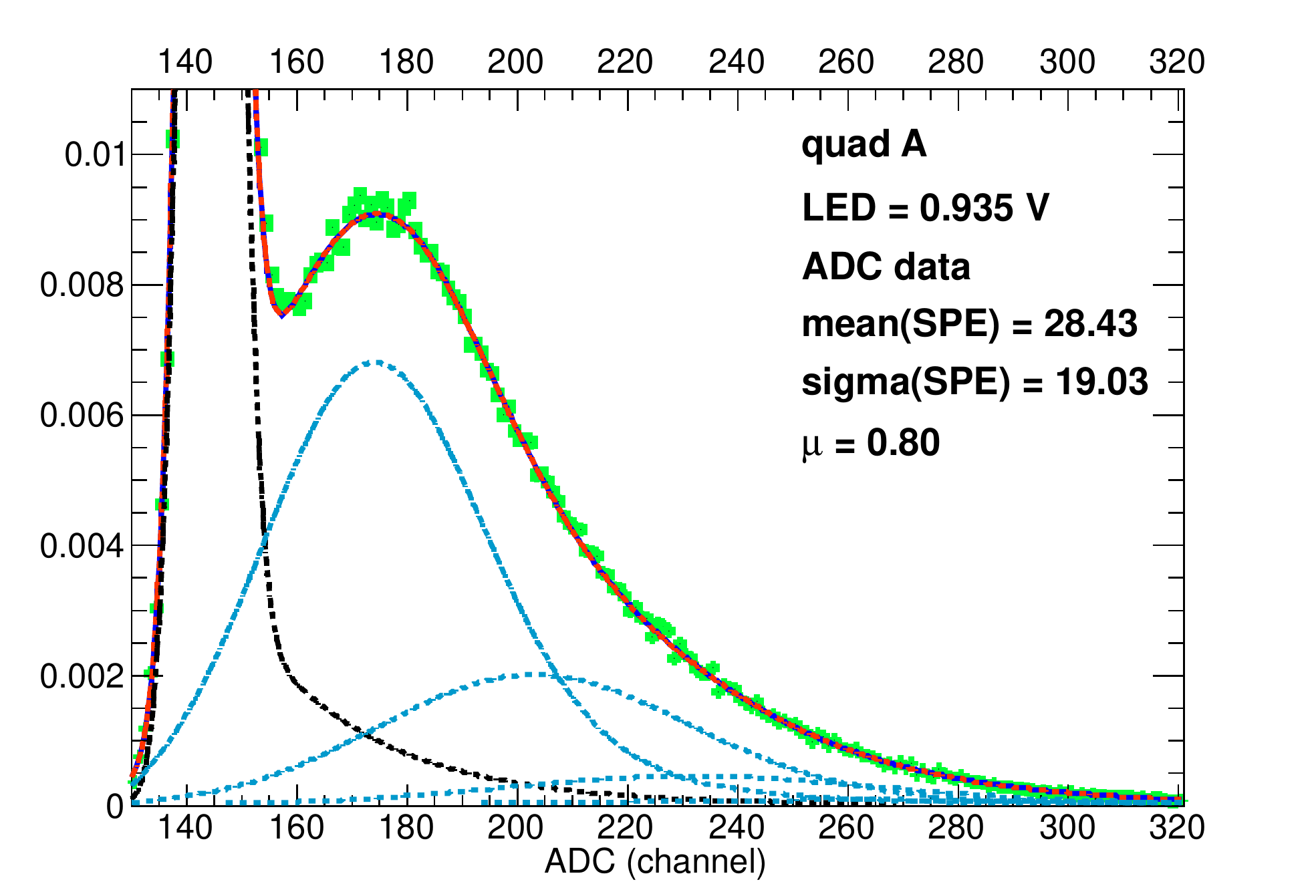} \\
\includegraphics[width=8cm]{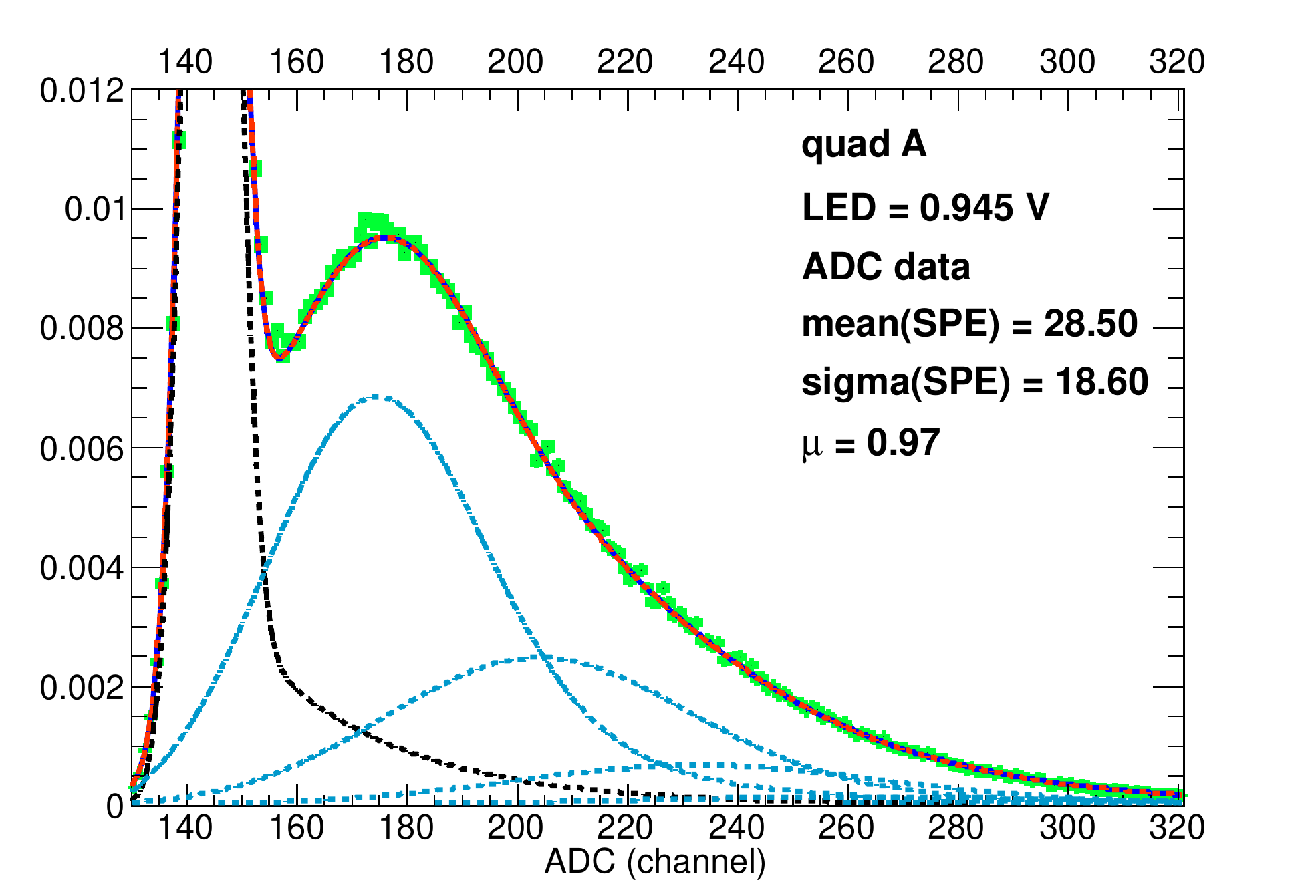}
&
\includegraphics[width=8cm]{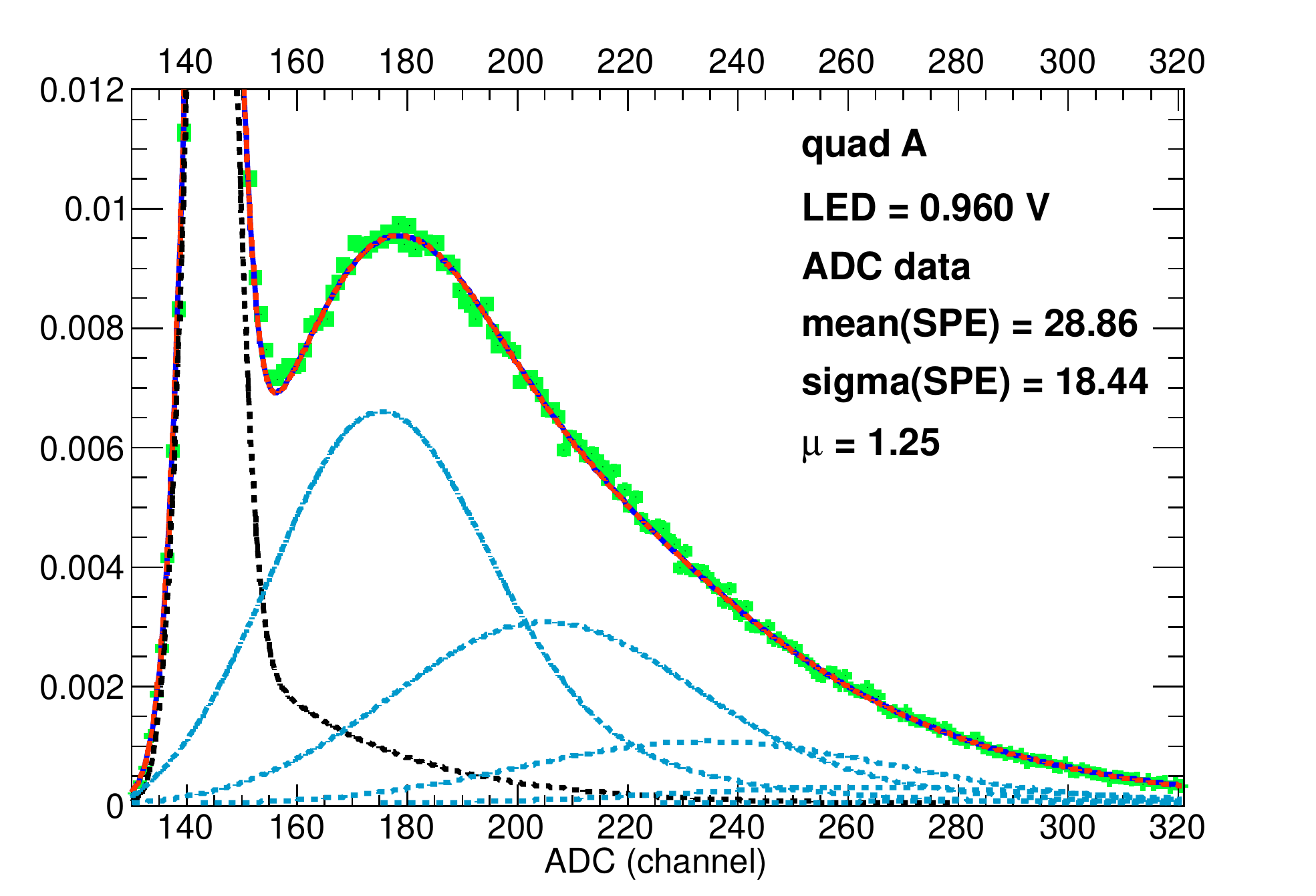} \\
\includegraphics[width=8cm]{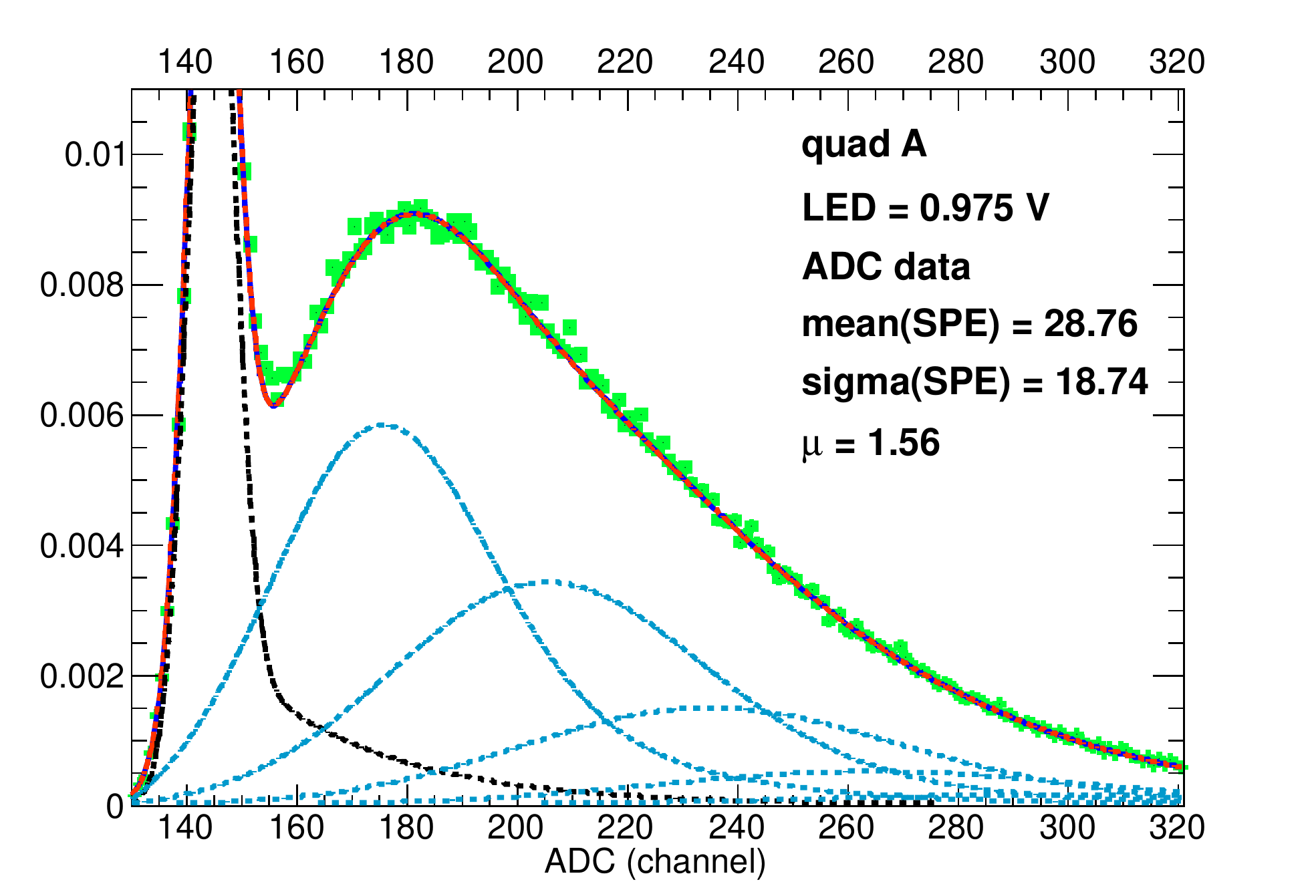}
&
\includegraphics[width=8cm]{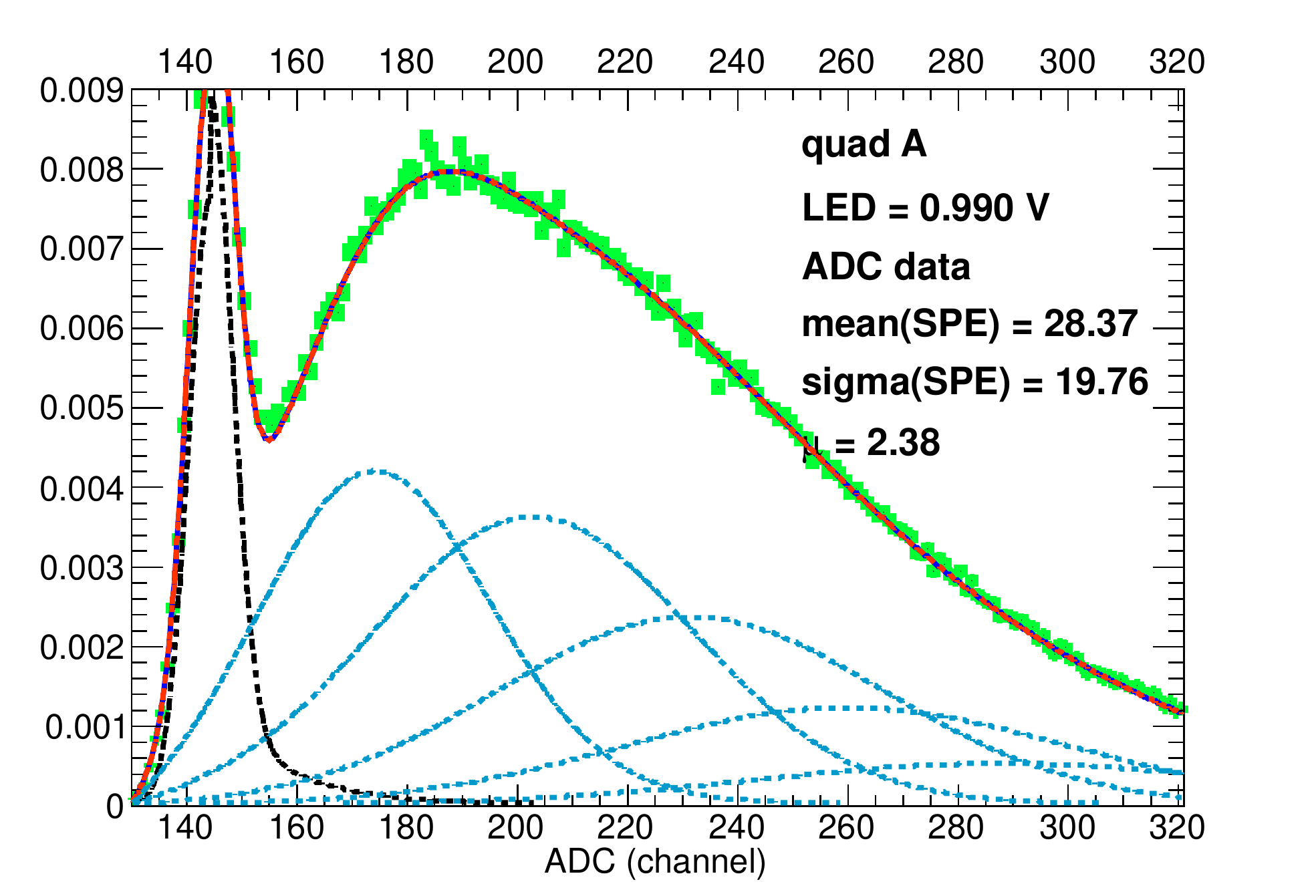}
\end{tabular}
\linespread{0.5}
\caption[]{
{Fits to the ADC distributions recorded from quad A responding to varying 
yields of incoming photons (see text for more details). The green points represent 
the data, the red continuous curve shows the fit to data while the black dashed and 
blue dotted curves depict the background and real signal contributions to the 
total fit.} } 
\label{spe-fit-quad-a}
\end{figure}

\begin{figure}[htbp]
\vspace*{-0.2in}
\centering
\begin{tabular}{cc}
\includegraphics[width=8cm]{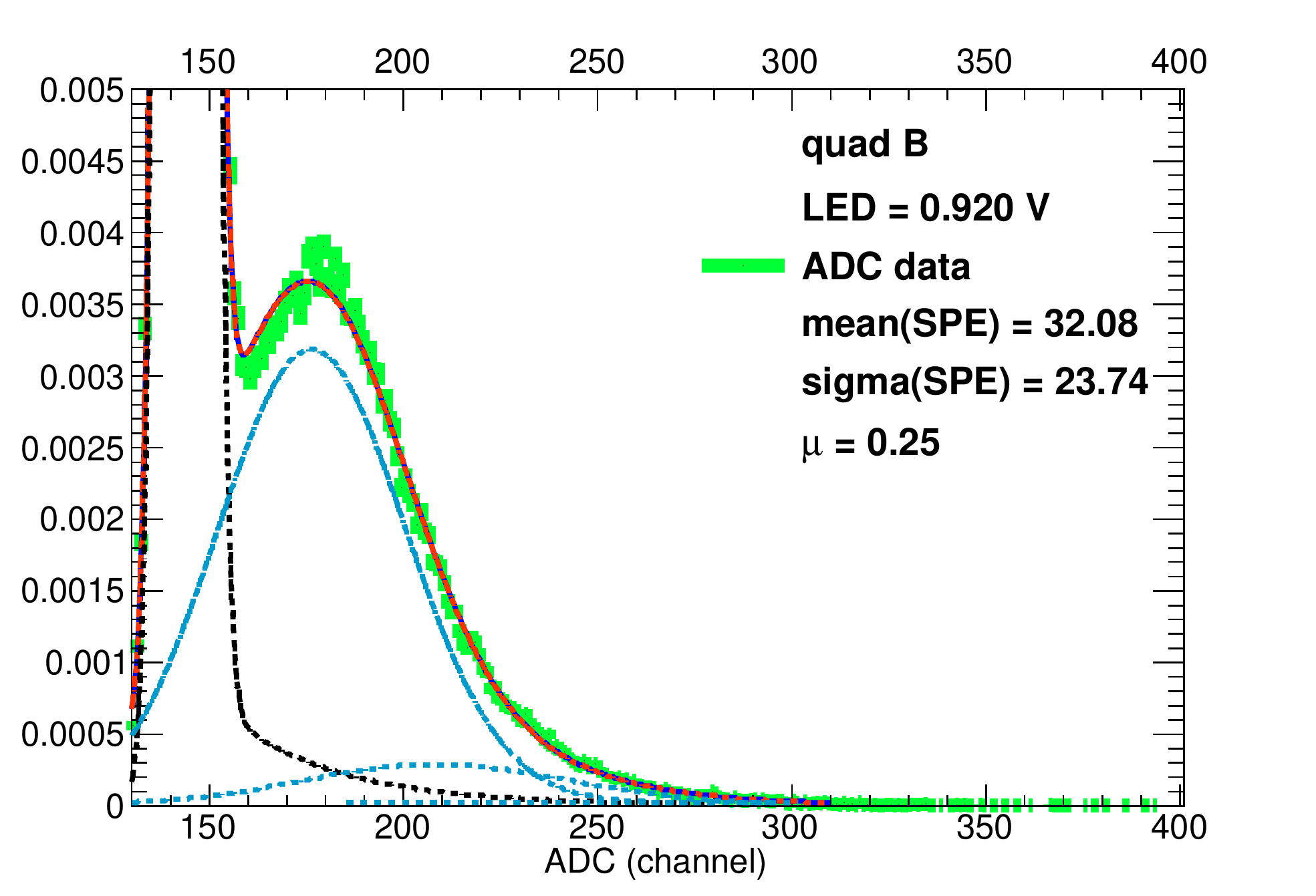}
&
\includegraphics[width=8cm]{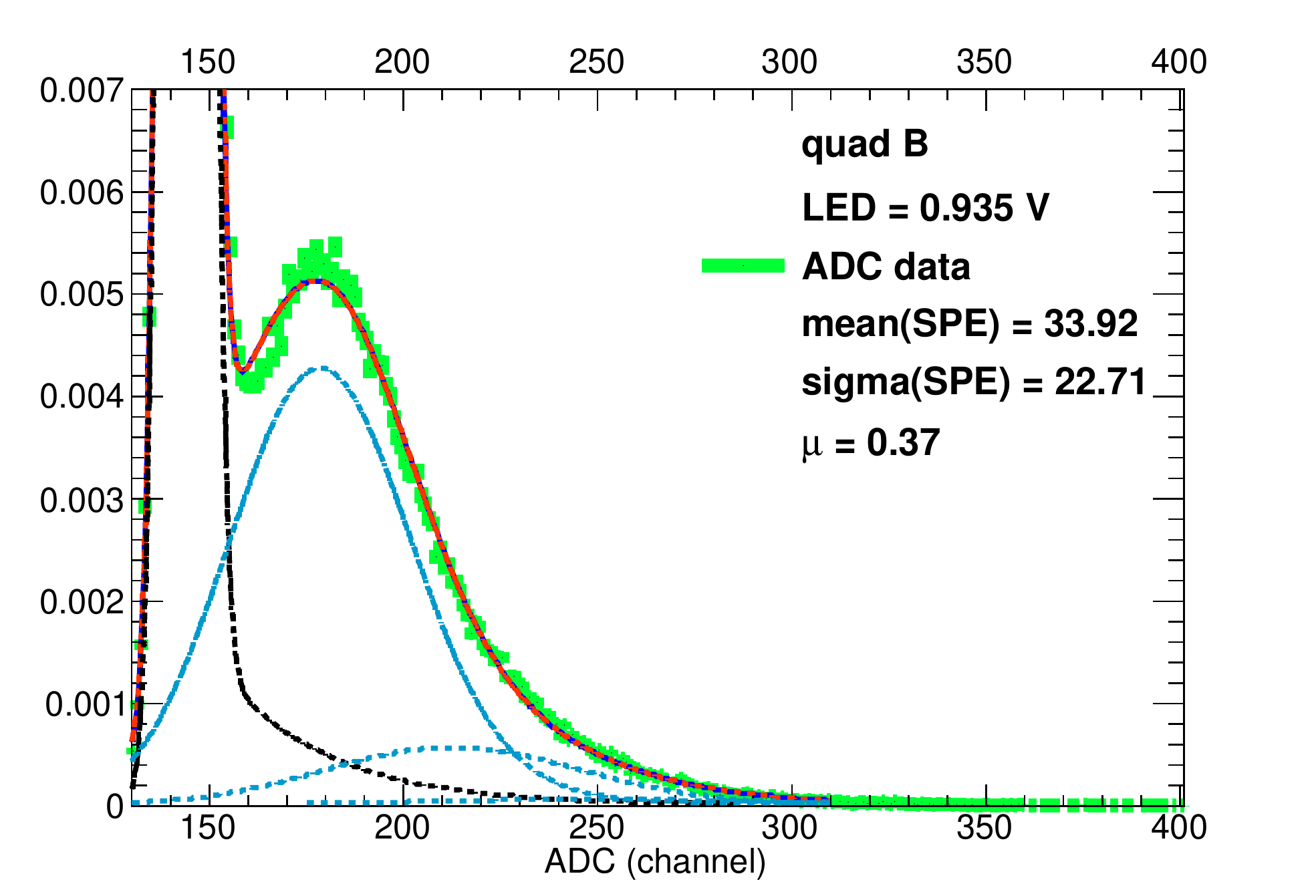} \\
\includegraphics[width=8cm]{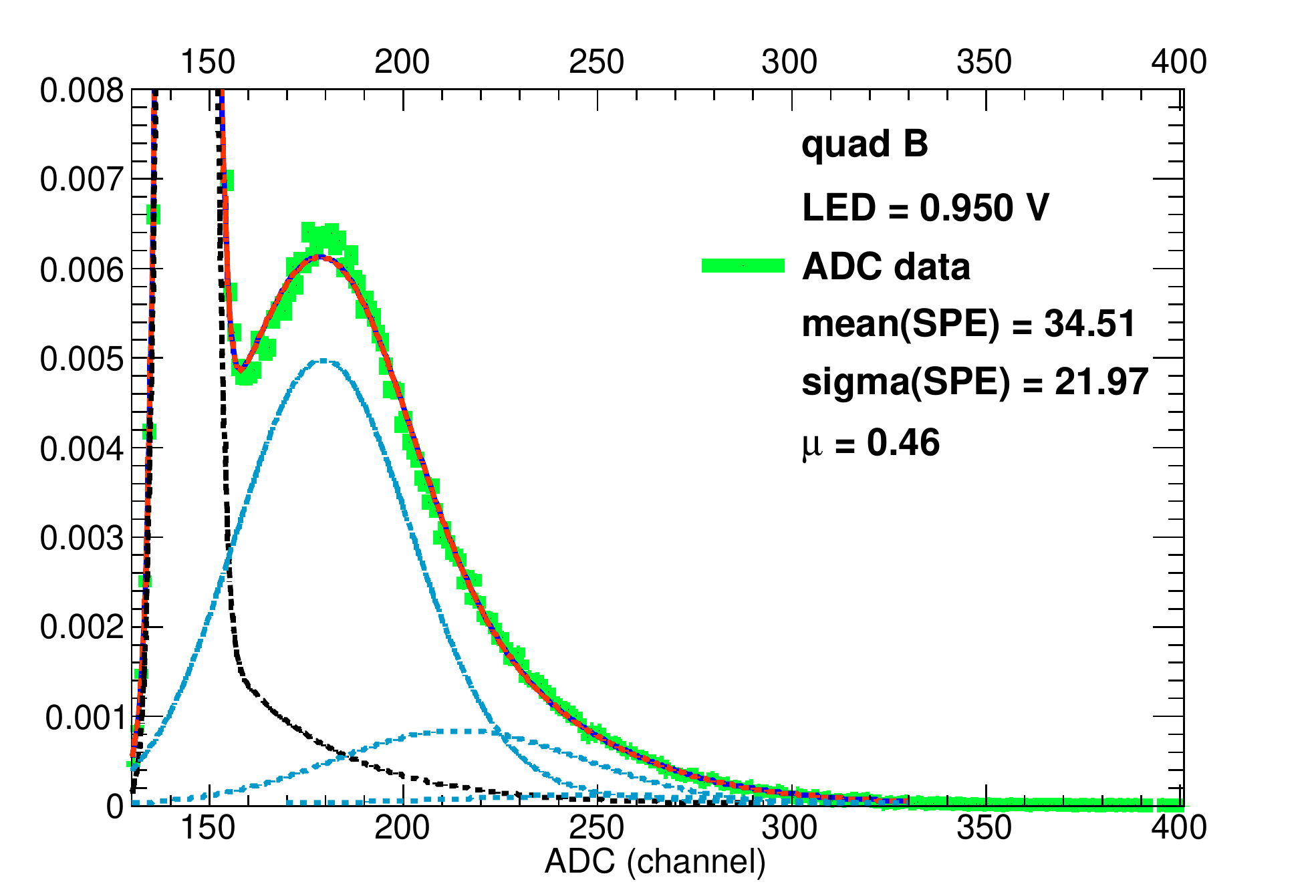}
&
\includegraphics[width=8cm]{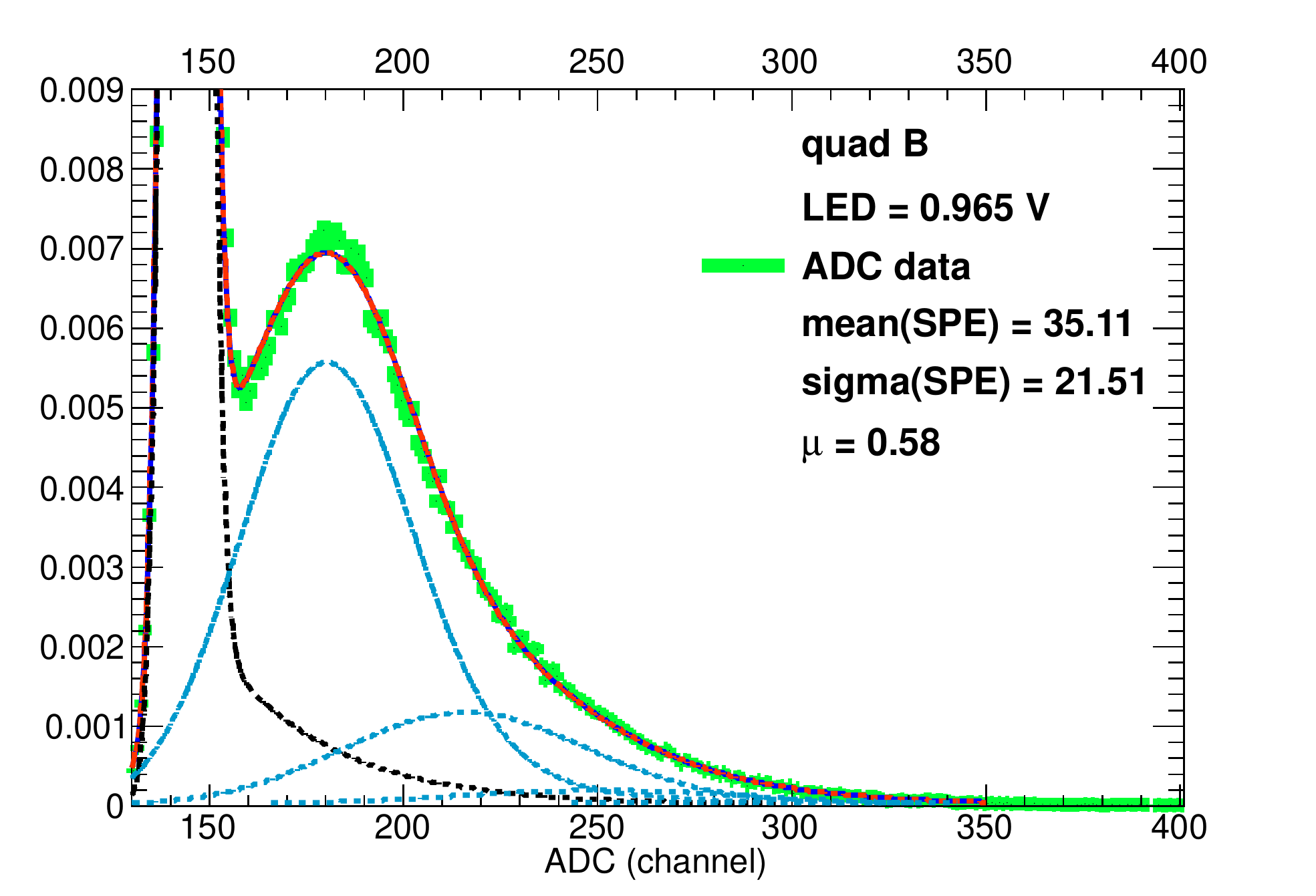} \\
\includegraphics[width=8cm]{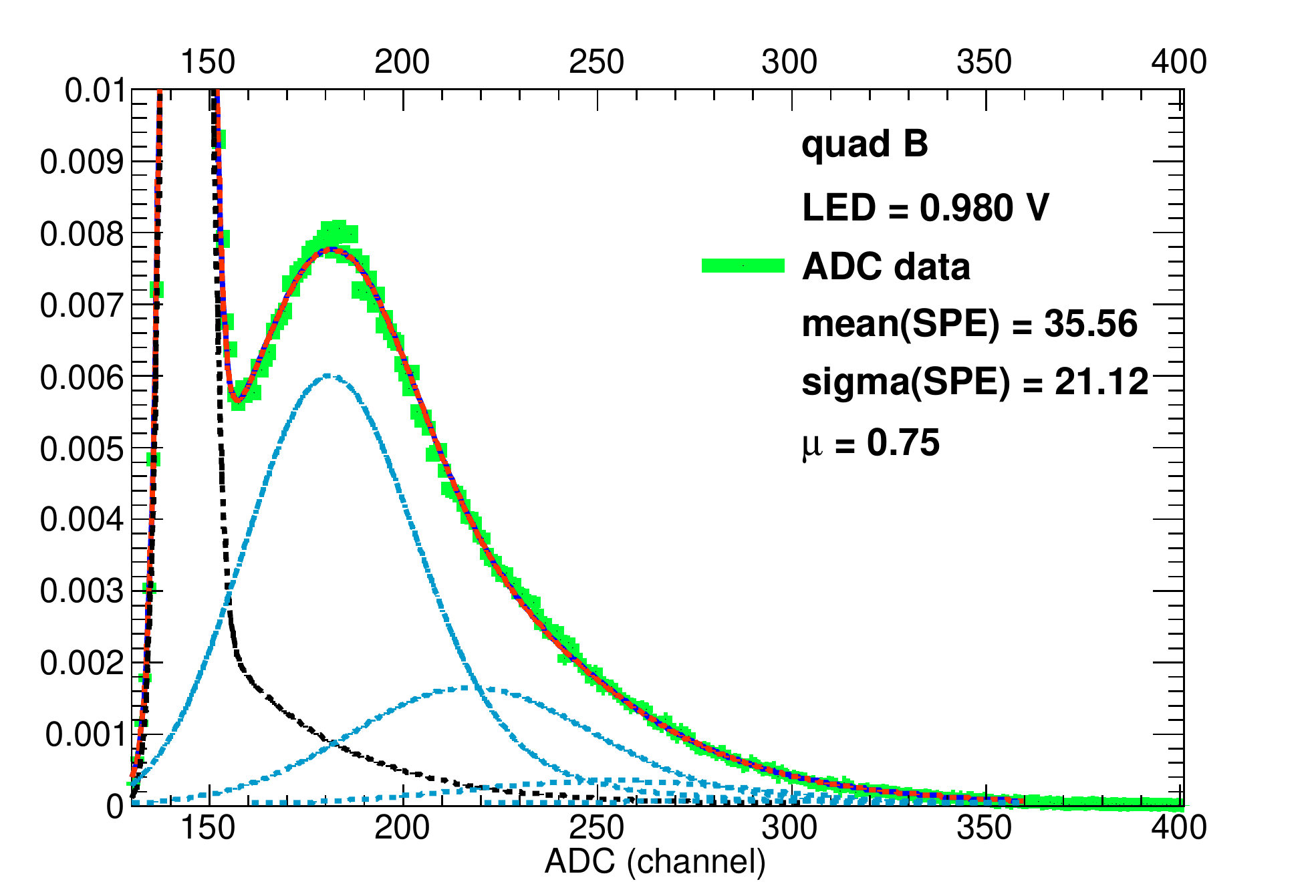}
&
\includegraphics[width=8cm]{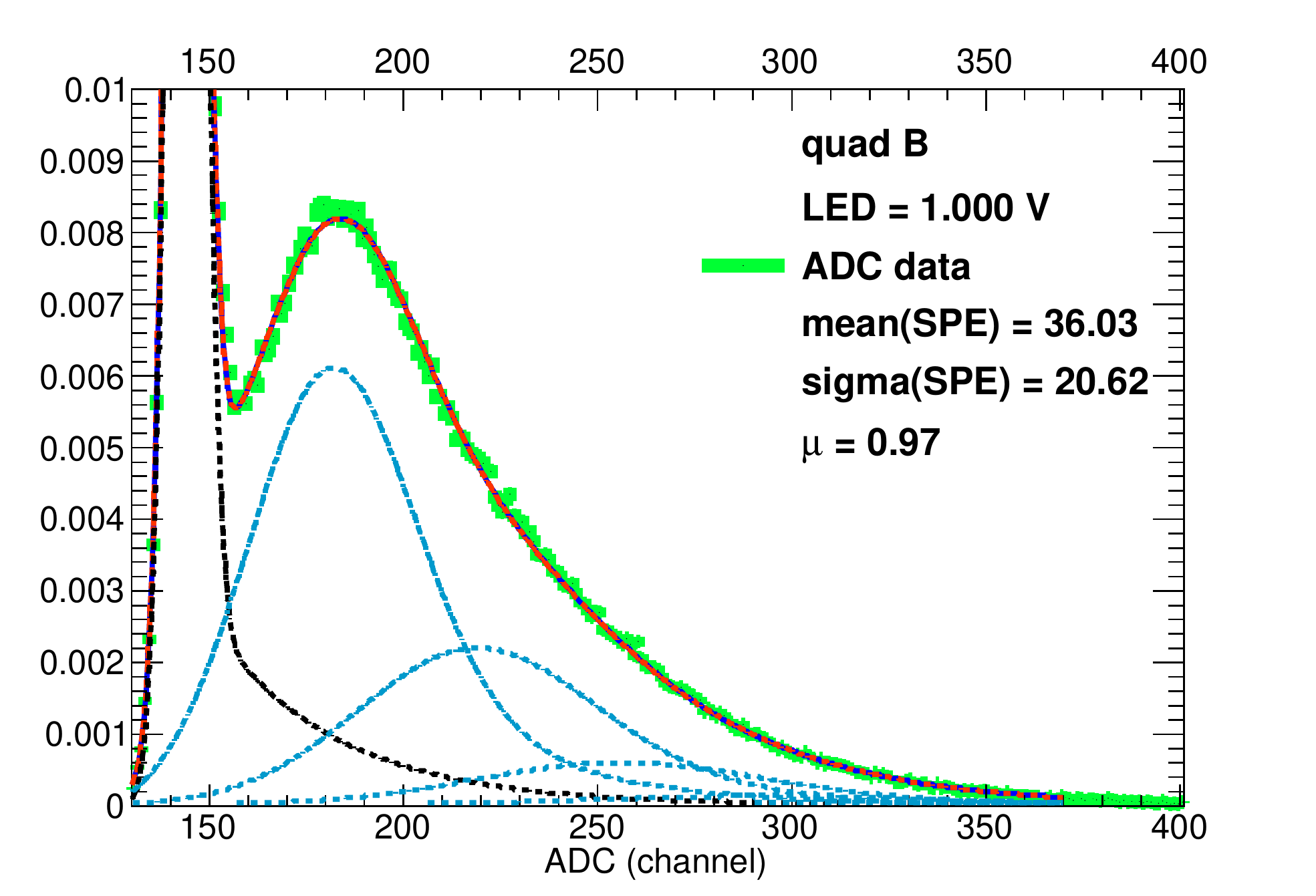} \\
\includegraphics[width=8cm]{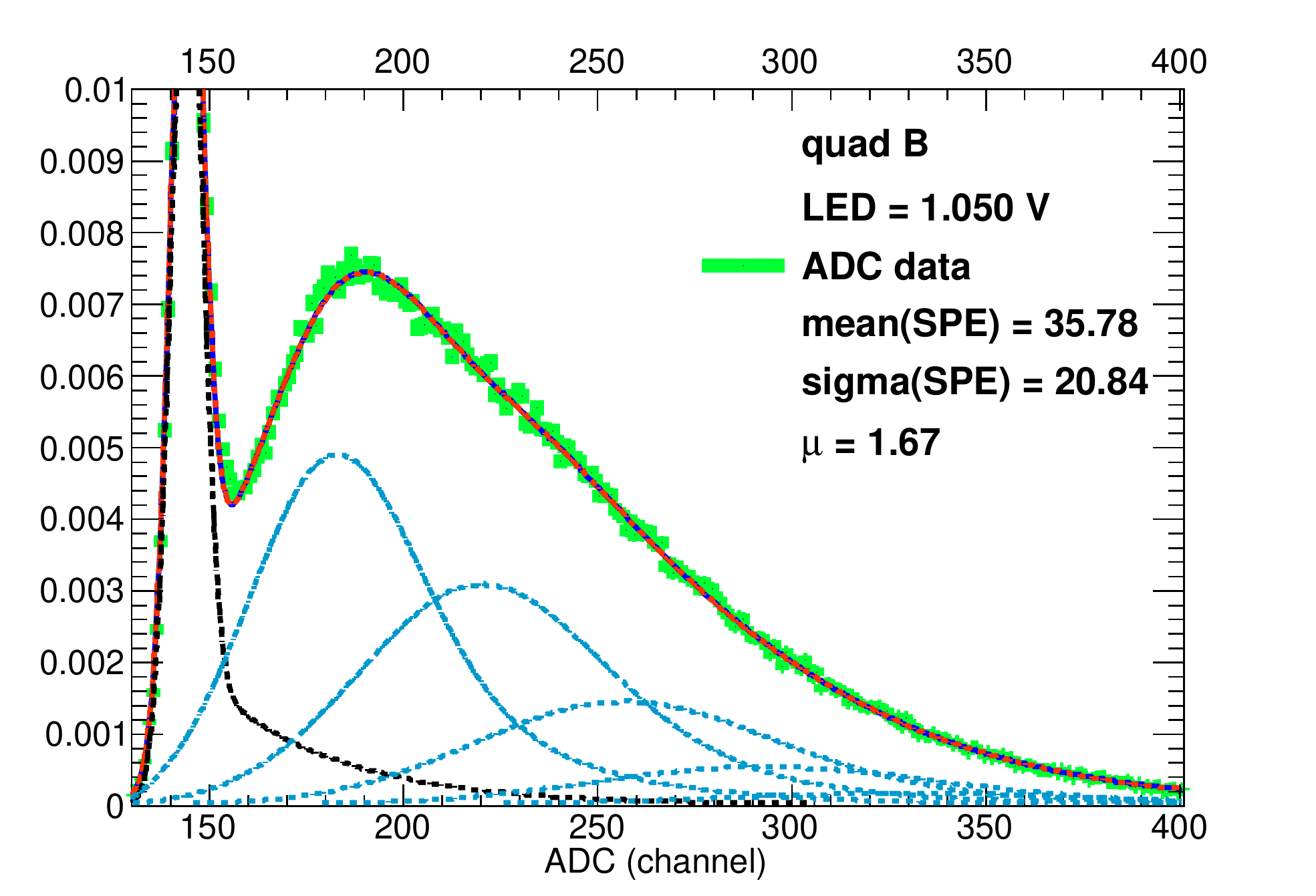}
&
\includegraphics[width=8cm]{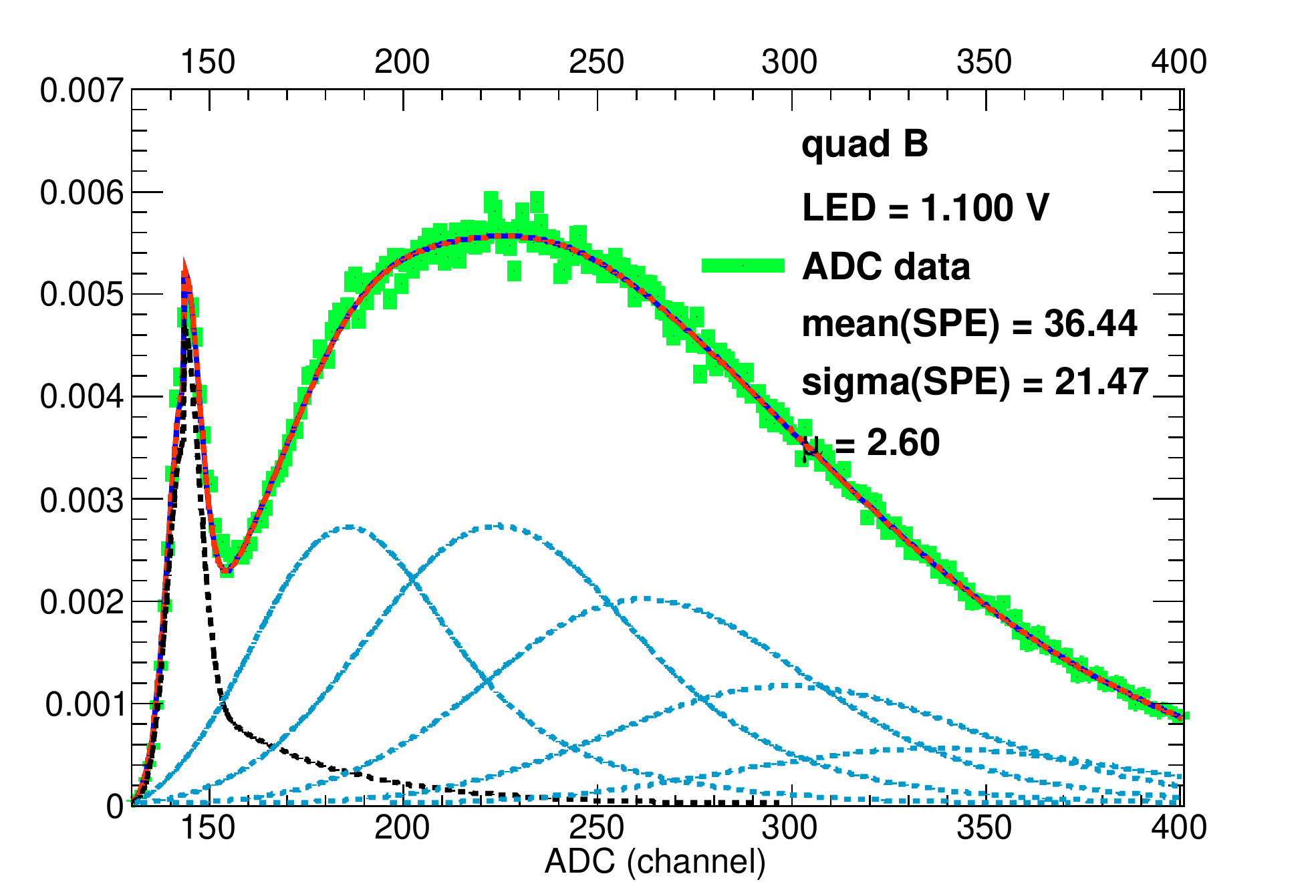}
\end{tabular}
\linespread{0.5}
\caption[]{
{Fits to the ADC distributions recorded from quad B responding to varying 
yields of incoming photons (see text for more details). The green points represent 
the data, the red continuous curve shows the fit to data while the black dashed and 
blue dotted curves depict the background and real signal contributions to the 
total fit.} }
\label{spe-fit-quad-b}
\end{figure}

Our fits of the single photoelectron ADC spectra are shown in 
Figs.~\ref{spe-fit-pixel-39}, \ref{spe-fit-pixel-45}, \ref{spe-fit-pixel-61}, 
\ref{spe-fit-quad-a} and \ref{spe-fit-quad-b} for pixel and quad 
measurements. Each of the 8 panels in Figs.~\ref{spe-fit-pixel-39} to 
\ref{spe-fit-quad-b} shows the ADC distribution for one LED setting 
along with the fit where various contributions have been plotted on the same graph. 
The solid red curve displays the full fit function as given by Eq. 1. The 
black dashed curve shows the background contribution only and the blue dotted 
curves represent the photoelectron distributions coming from the real signal. 

The location of the single photoelectron peak (in ADC channels above pedestal) 
\textit{mean(SPE)} which is a measure of the gain and the standard deviation of the single 
photoelectron distribution as extracted from the fit, $sigma(SPE)$ 
are also shown together with the Poisson mean describing the average number of photoelectrons, $\mu$. 
For each of the pixels and quads we show fits to 
the ADC distributions corresponding to various LED settings. Assuming 
the fit function characterizes reliably the PMT response, the expectation is that 
for an increasing LED voltage per pixel or quad \textit{$mean(SPE)$} and $sigma(SPE)$ 
will remain constant while the average number of photoelectrons, $\mu$, 
will increase. This pattern is indeed observed for all the pixels and quads 
studied. Also the charge corresponding to a single photoelectron for each of 
the three individual pixels confirms the gain per pixel measurements provided by 
Hamamatsu where pixel 39 has the lowest gain of the pixels tested while 45 and 
61 have higher gains (in this order). 

\begin{figure}[htbp]
\vspace*{-0.1in}
\centering
\begin{tabular}{cc}
\vspace{-0.3in}
\hspace{0.in}
\includegraphics[width=8cm]{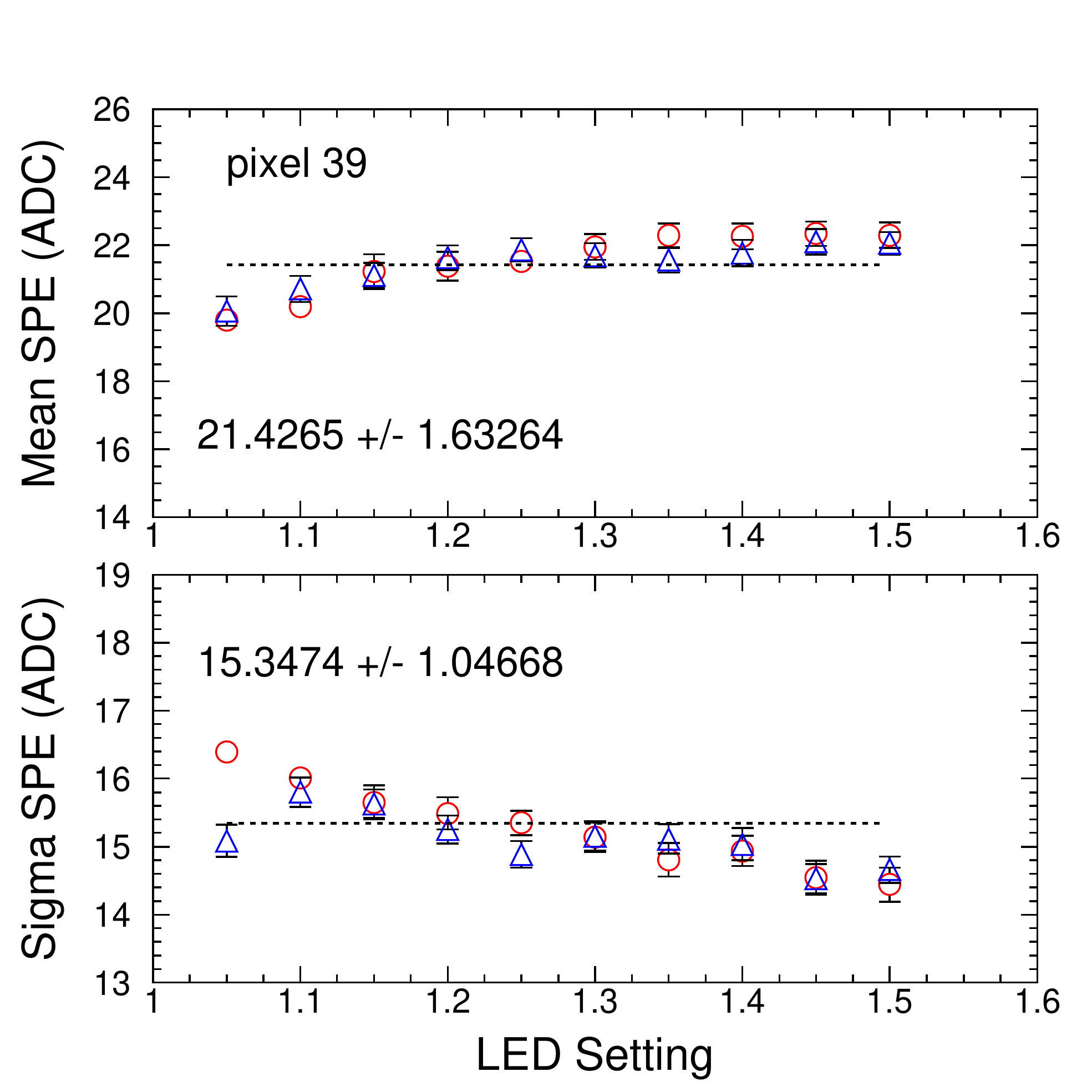}
&
\hspace{-0.2in}
\includegraphics[width=8cm]{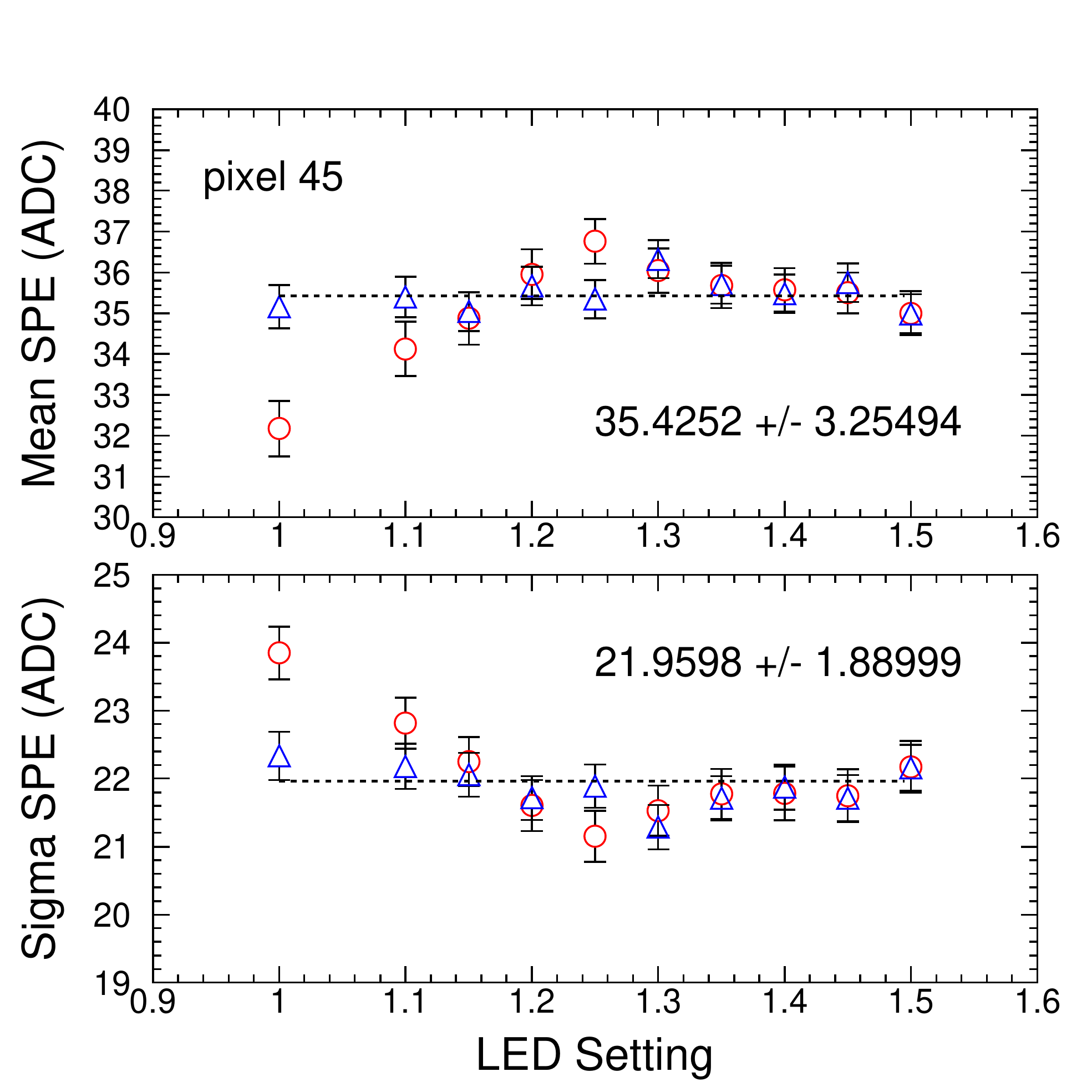} \\
\vspace{-0.3in}
\hspace{0.in}
\includegraphics[width=8cm]{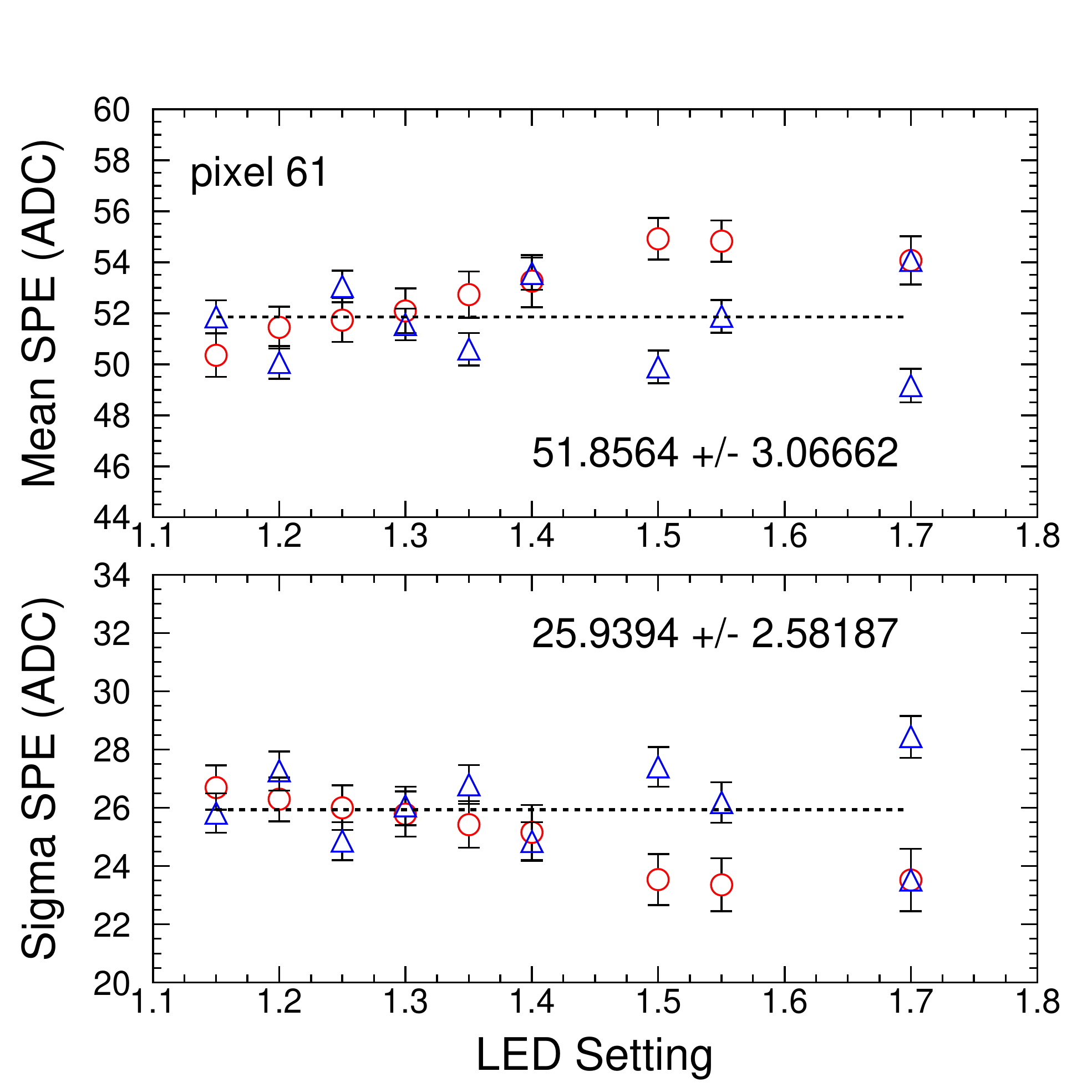}
&
\hspace{-0.2in}
\includegraphics[width=8cm]{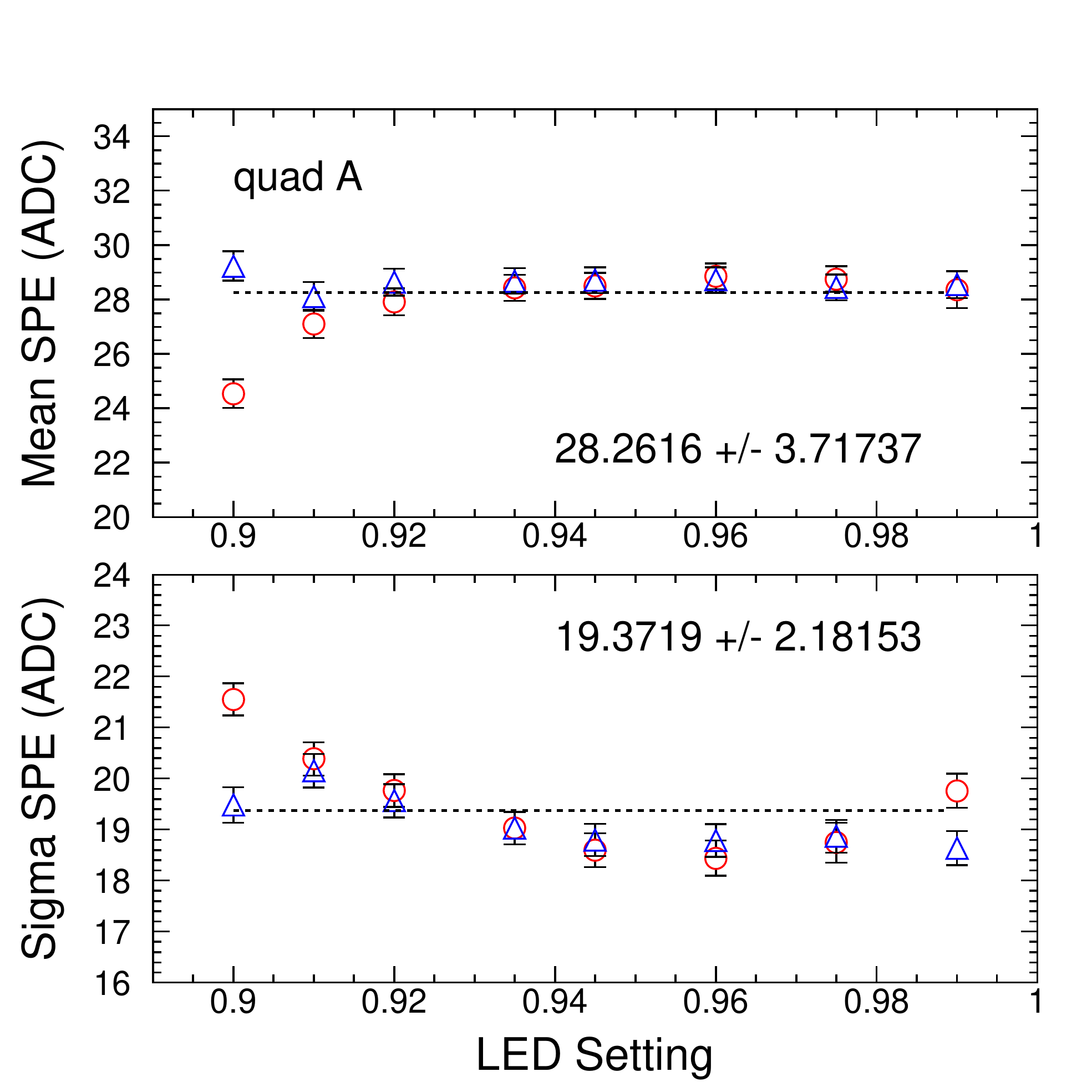} \\
\hspace{0.in}
\includegraphics[width=8cm]{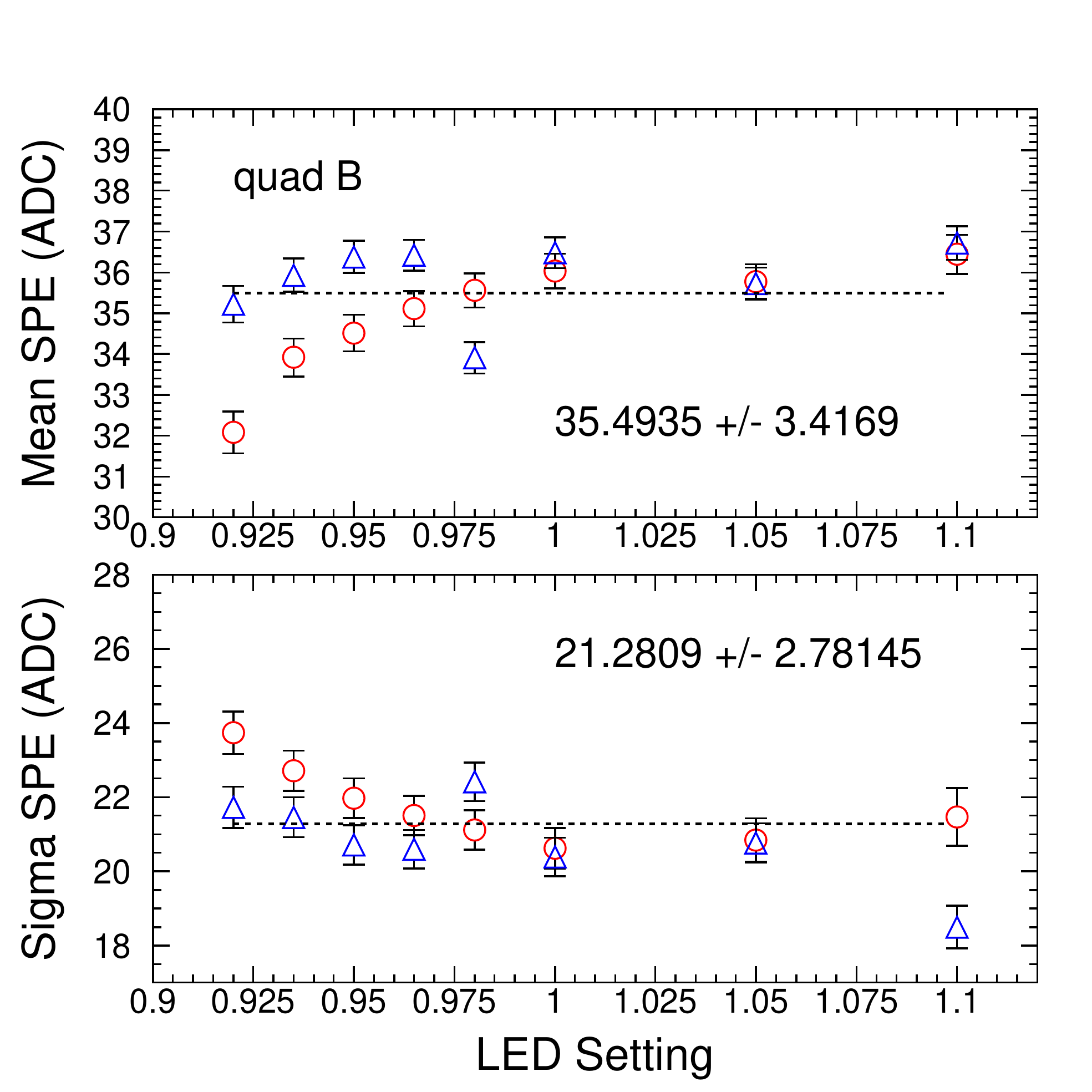}
&
\hspace{-0.2in}
\includegraphics[width=8cm]{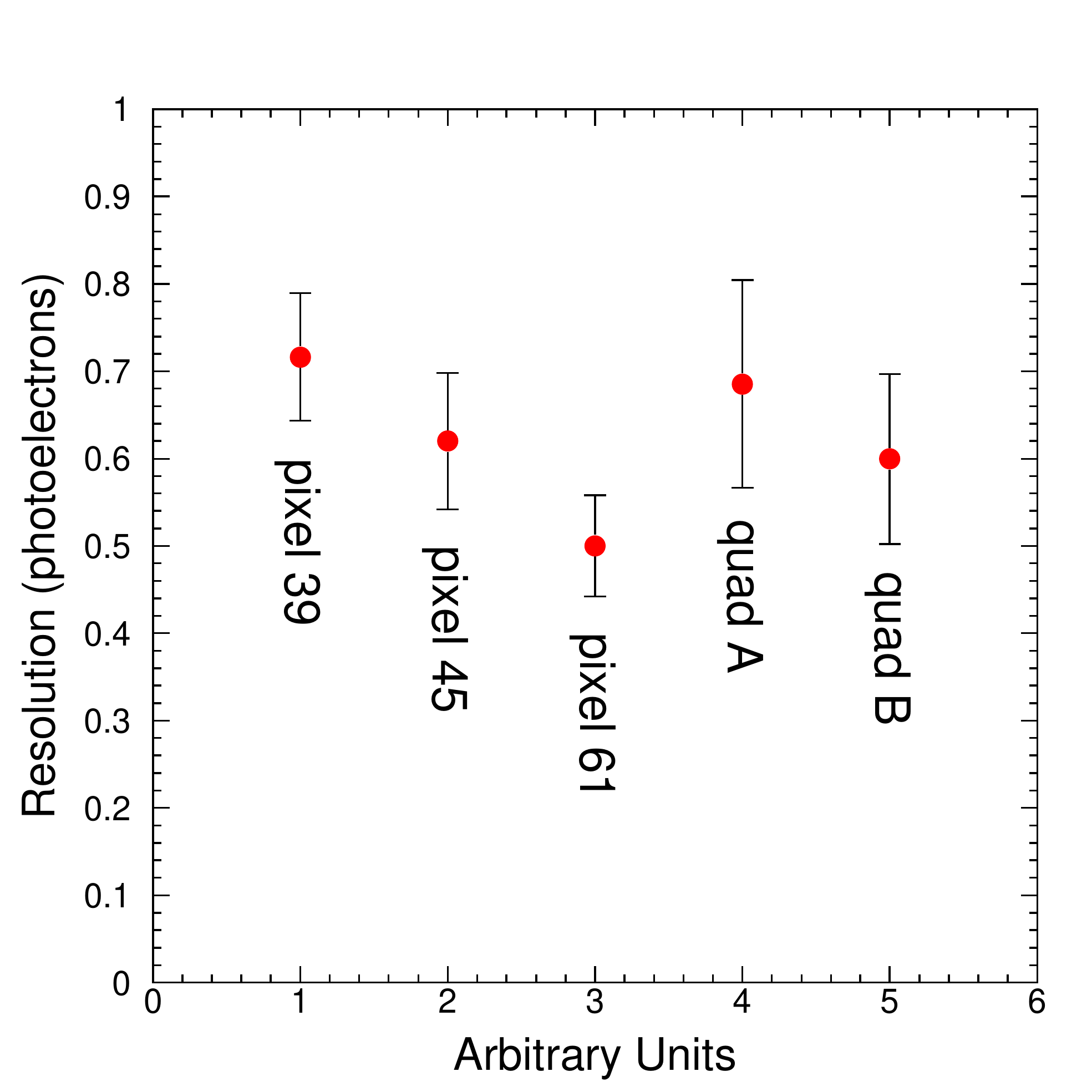}
\end{tabular}
\linespread{0.5}
\caption[]{
{Summary of fit results to single photoelectron ADC distributions from pixels 
39, 45, 61 and quads A and B. The extracted resolution for single photoelectron detection 
is also shown.} }
\label{summary_spe}
\end{figure}

We summarize our fit results in Fig.~\ref{summary_spe} where we plot \textit{$mean(SPE)$} 
and $sigma(SPE)$ \emph{v.s.} LED pulse voltage, characterizing 
the single photoelectron distribution as extracted from the fit for all LED settings 
for the three individual pixels and for the two quads. For all outputs we 
performed two sets of fits. In one case, we fit allowing all 
parameters to change during the minimization process; these results are 
represented in Fig.~\ref{summary_spe} by the empty red circles. We also fit 
fixing the background parameters, $w$ and $\alpha$, and the results 
are shown by the blue empty triangles. To extract the average mean and 
standard deviation characterizing the single photoelectron distribution 
for pixels and quads we then perform a linear fit to \textit{$mean(SPE)$} and 
$sigma(SPE)$ using a zero order polynomial (dashed black line in 
Fig.~\ref{summary_spe}). We take the largest deviation of the 
measurements from this fit as the uncertainty of the quantity extracted. 
Both the average mean and standard deviation of the single photoelectron 
distribution together with their associated uncertainties are shown 
Fig.~\ref{summary_spe} for all pixels and quads tested. The uncertainty on the 
mean is mostly within 10\% (slightly larger for quad A) and of a similar 
magnitude for the standard deviation.

We calculate the resolution for single photoelectron detection from the 
fit results as the ratio of the standard deviation and the mean. As seen 
in Fig.~\ref{summary_spe}, bottom right panel, pixel 39 with the lowest 
gain has the lowest resolution of about 0.7 photoelectrons while for 
the highest gain pixel, 61, we obtain $\approx$0.5 photoelectrons. For 
quads the resolution is again below 1 photoelectron. The demonstrated performance 
of group of pixels is quite satisfactory for applications like SoLID 
where the single photoelectron identification is crucial. Overall, although 
the H8500C-03 has a typical gain several orders of magnitude lower 
than PMTs conventionally used for Cherenkov light detection, it can clearly resolve 
single photoelectron signals with a resolution of less than one 
photoelectron.

\subsection{Gain Measurements}

\begin{figure}[htbp]
\vspace*{-0.2in}
\centering
\begin{tabular}{cc}
\includegraphics[width=9.cm]{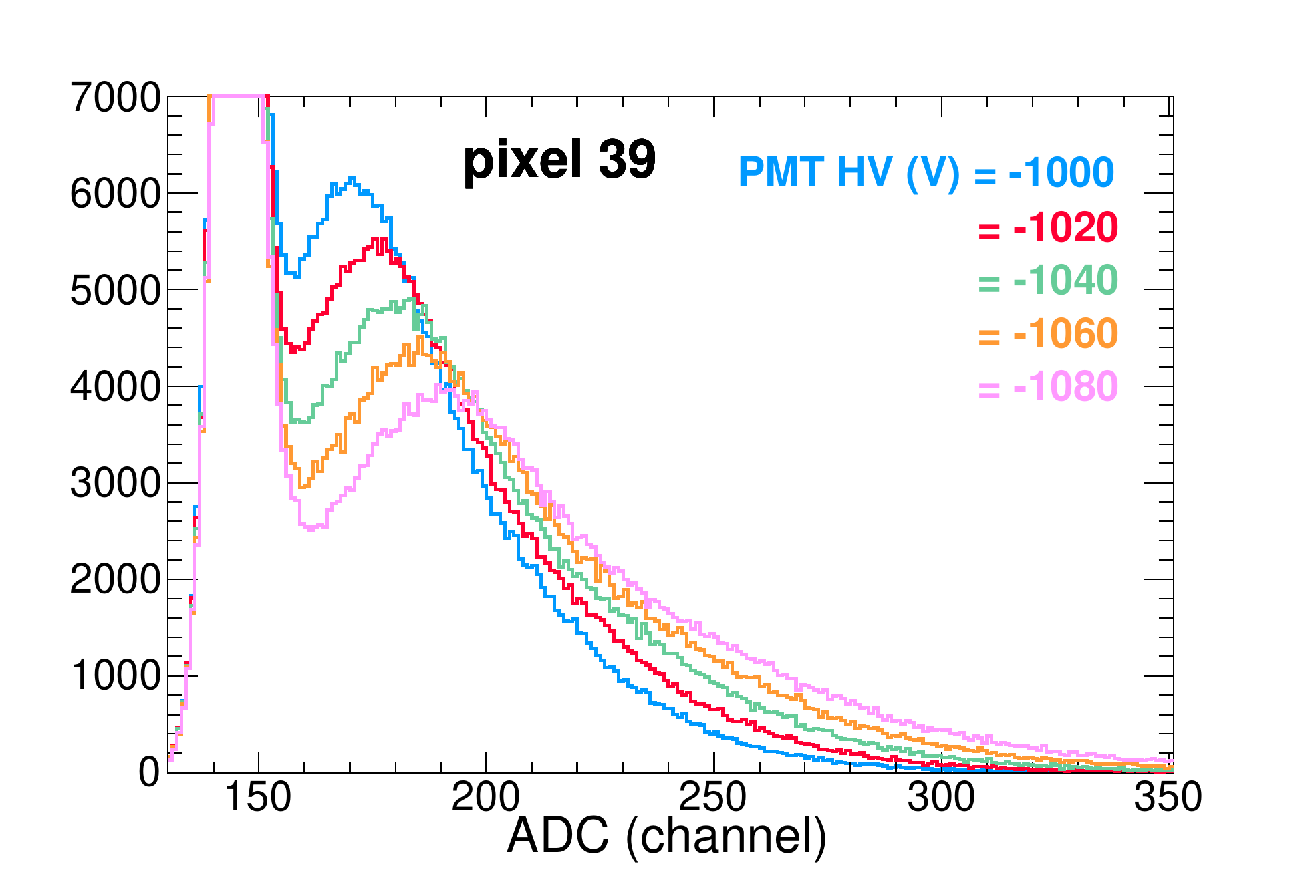}
&
\hspace{-0.5in}
\includegraphics[width=8.8cm]{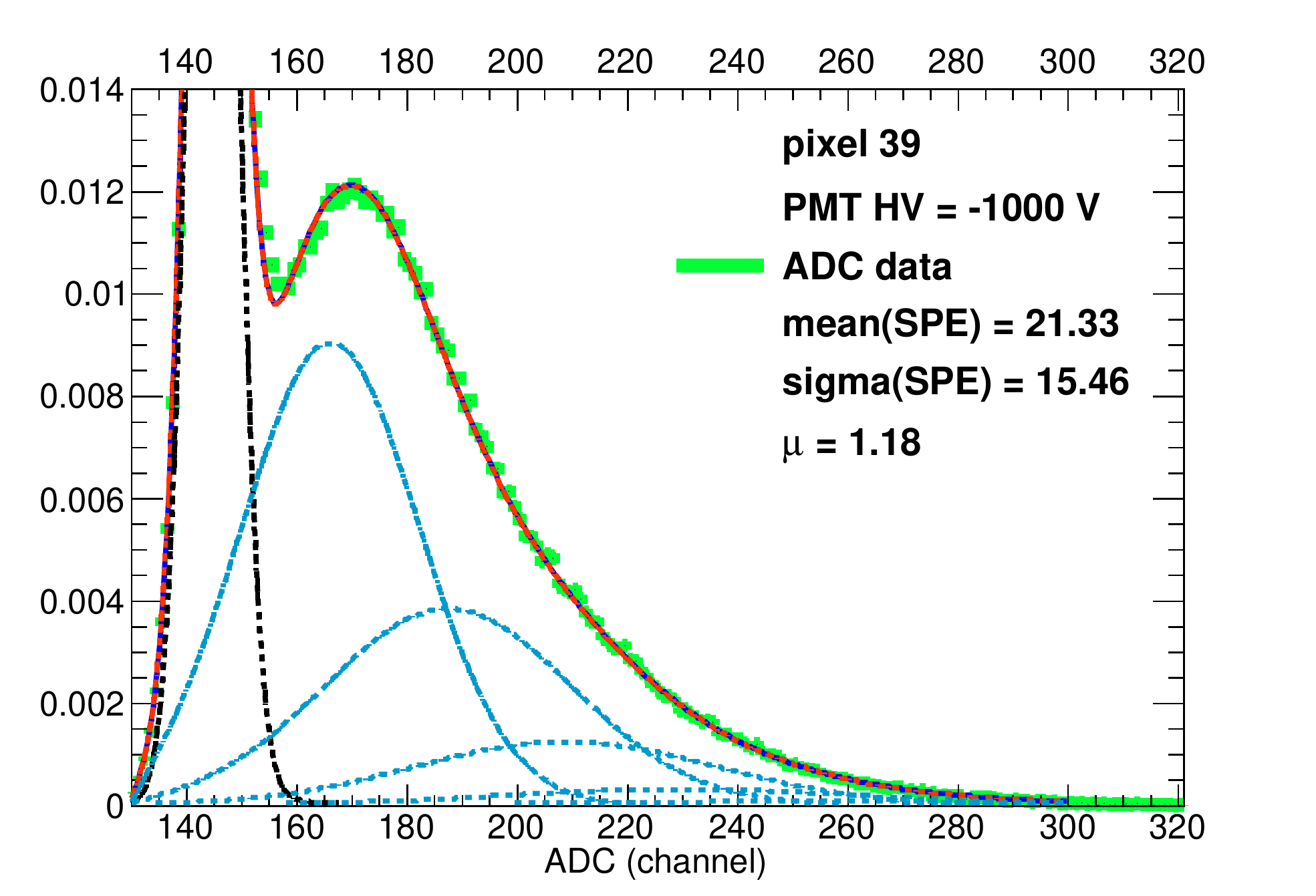} \\
\includegraphics[width=8.8cm]{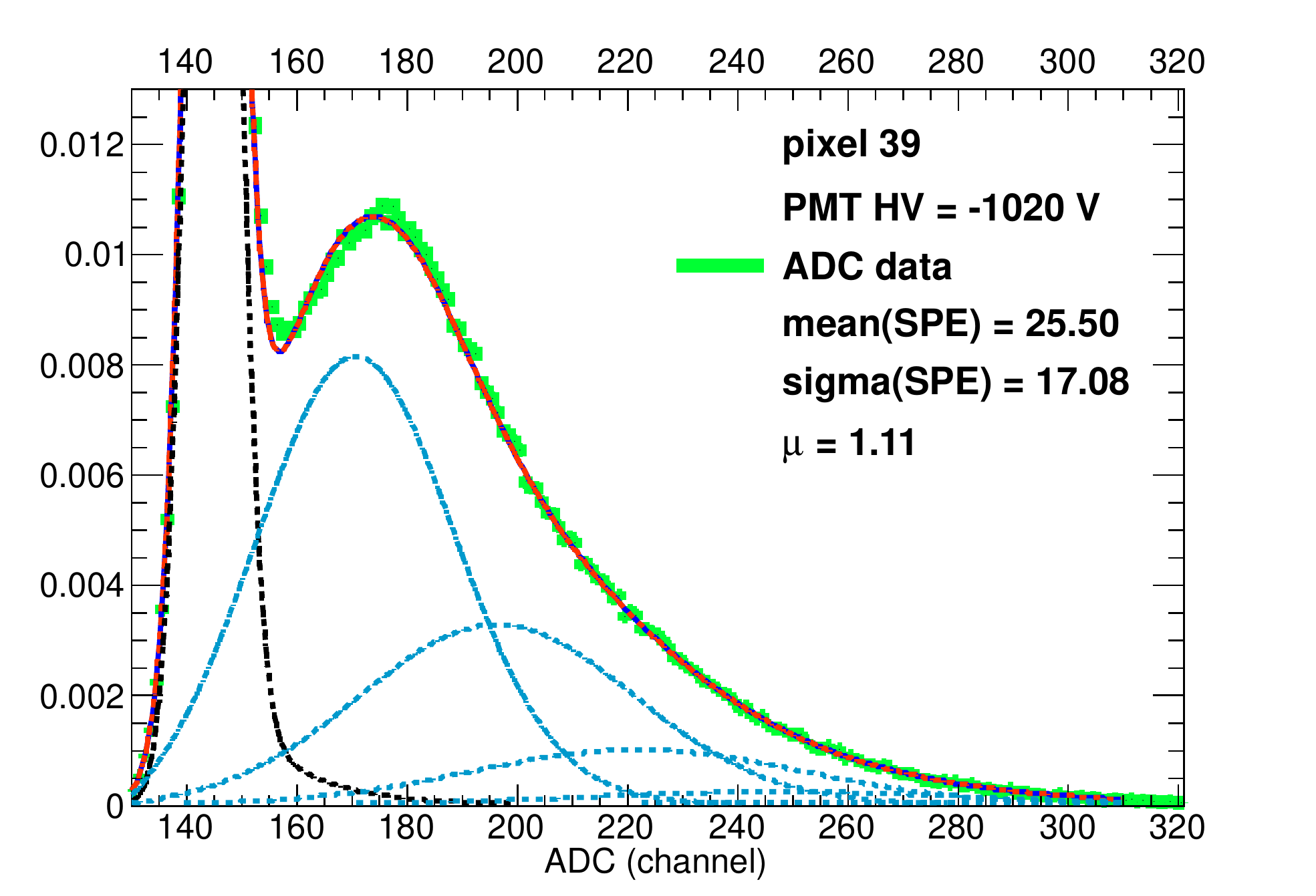}
&
\hspace{-0.5in}
\includegraphics[width=8.8cm]{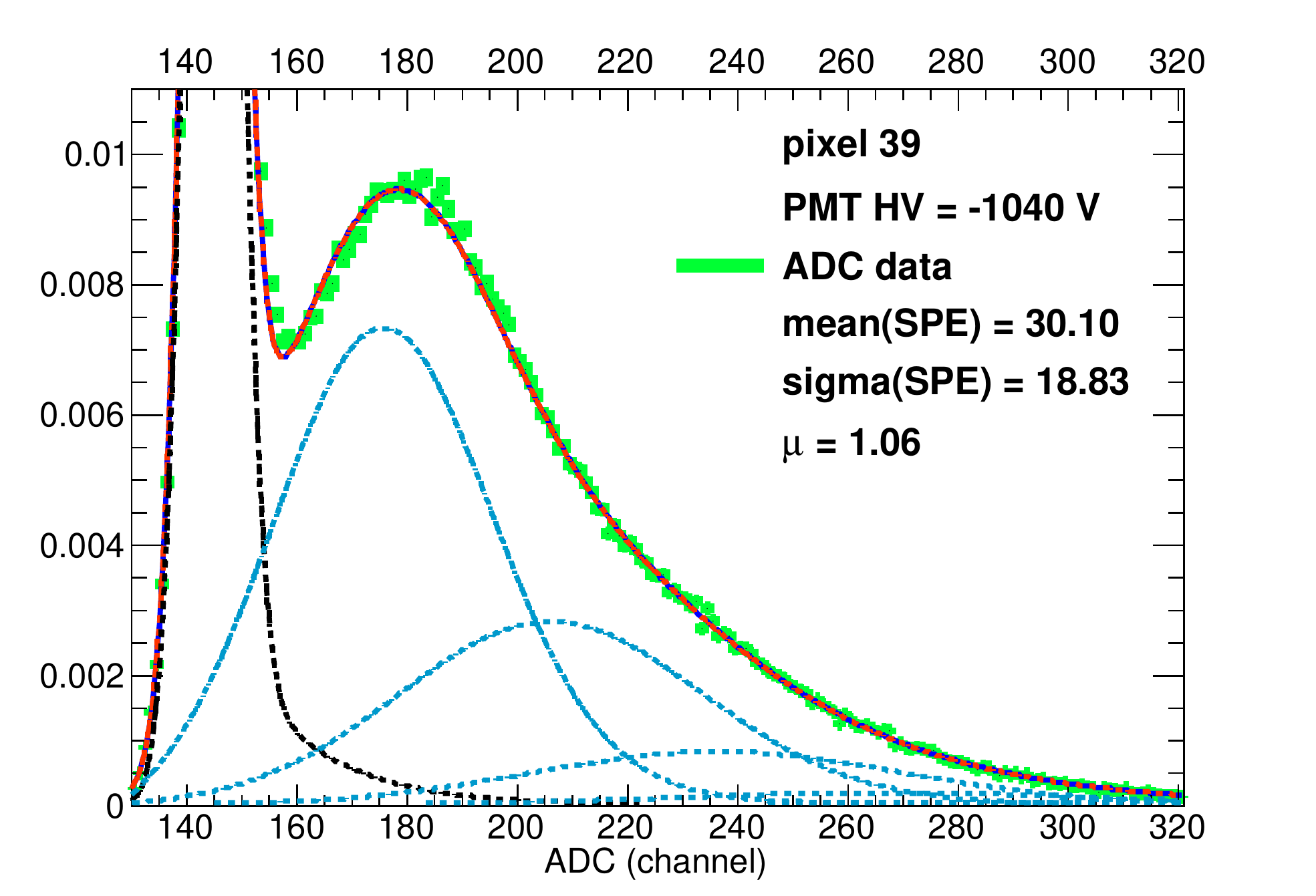} \\
\includegraphics[width=8.8cm]{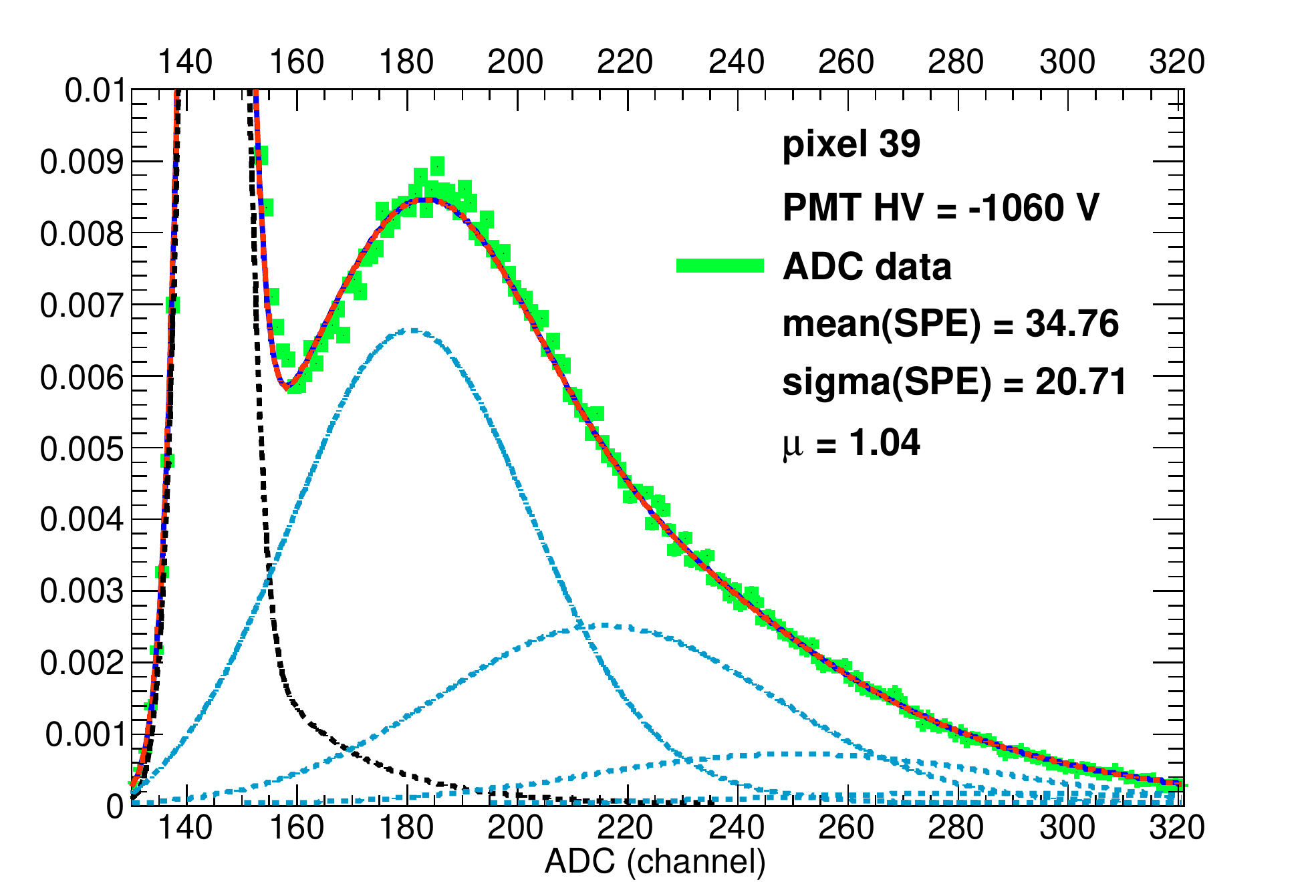}
&
\hspace{-0.5in}
\includegraphics[width=8.8cm]{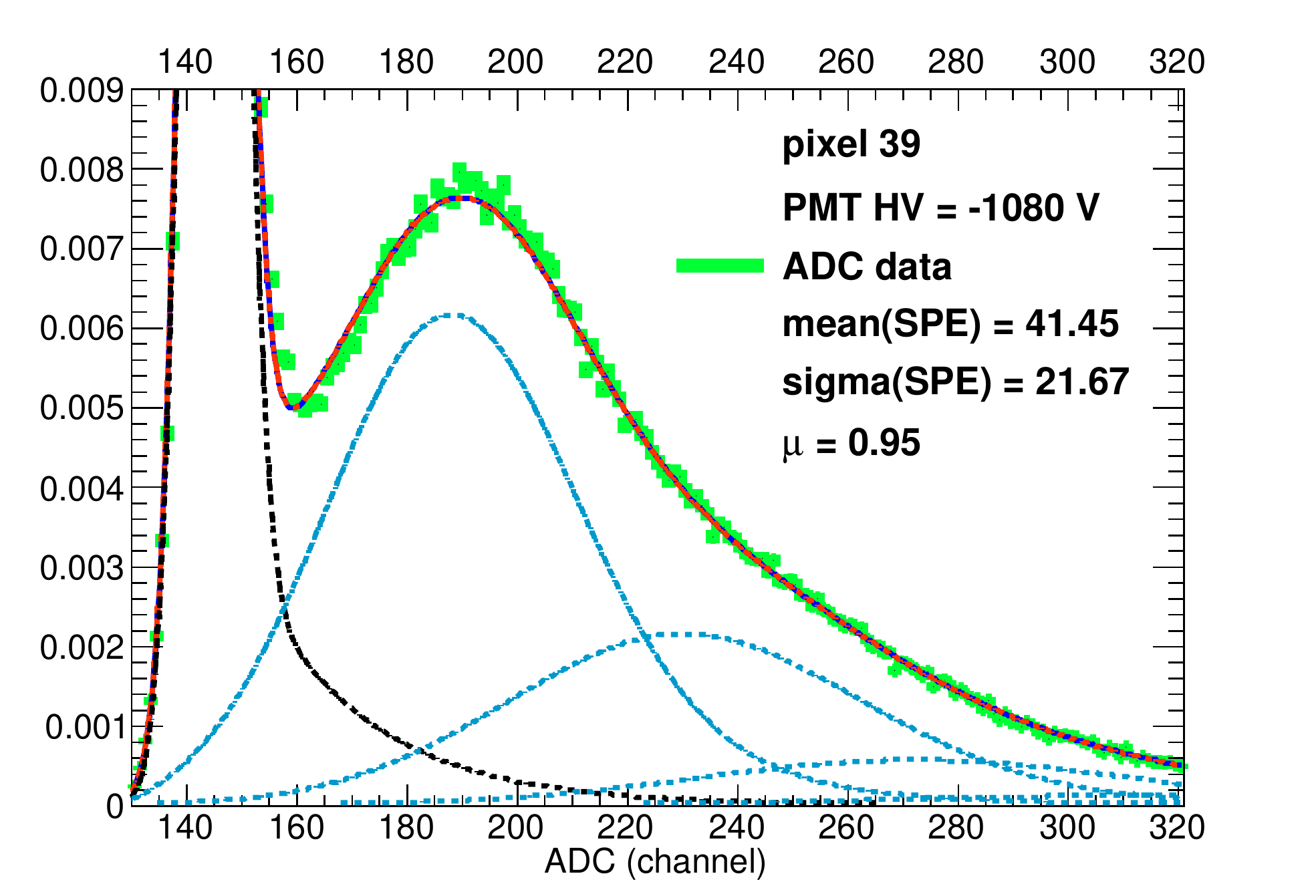} 
\end{tabular}
\linespread{0.5}
\caption[]{
{High voltage scan for pixel 39. The high voltage of the PMT was varied between 
-1000 V (nominal operating value) and -1080 V in steps of 20 V and the single 
photoelectron ADC distribution was recorded. The LED setting was optimized for 
single photoelectron detection and then kept 
constant during each high voltage scan. As expected, the pixel gain 
increases with an increasing high voltage and this is clearly indicated by 
the increase in the number of ADC channels above the pedestal (\emph{i.e.} 
integrated charge) corresponding to the single photoelectron peak.} }
\label{hv_scan_39}
\end{figure}

\begin{figure}[htbp]
\vspace*{-0.2in}
\centering
\begin{tabular}{cc}
\includegraphics[width=9.cm]{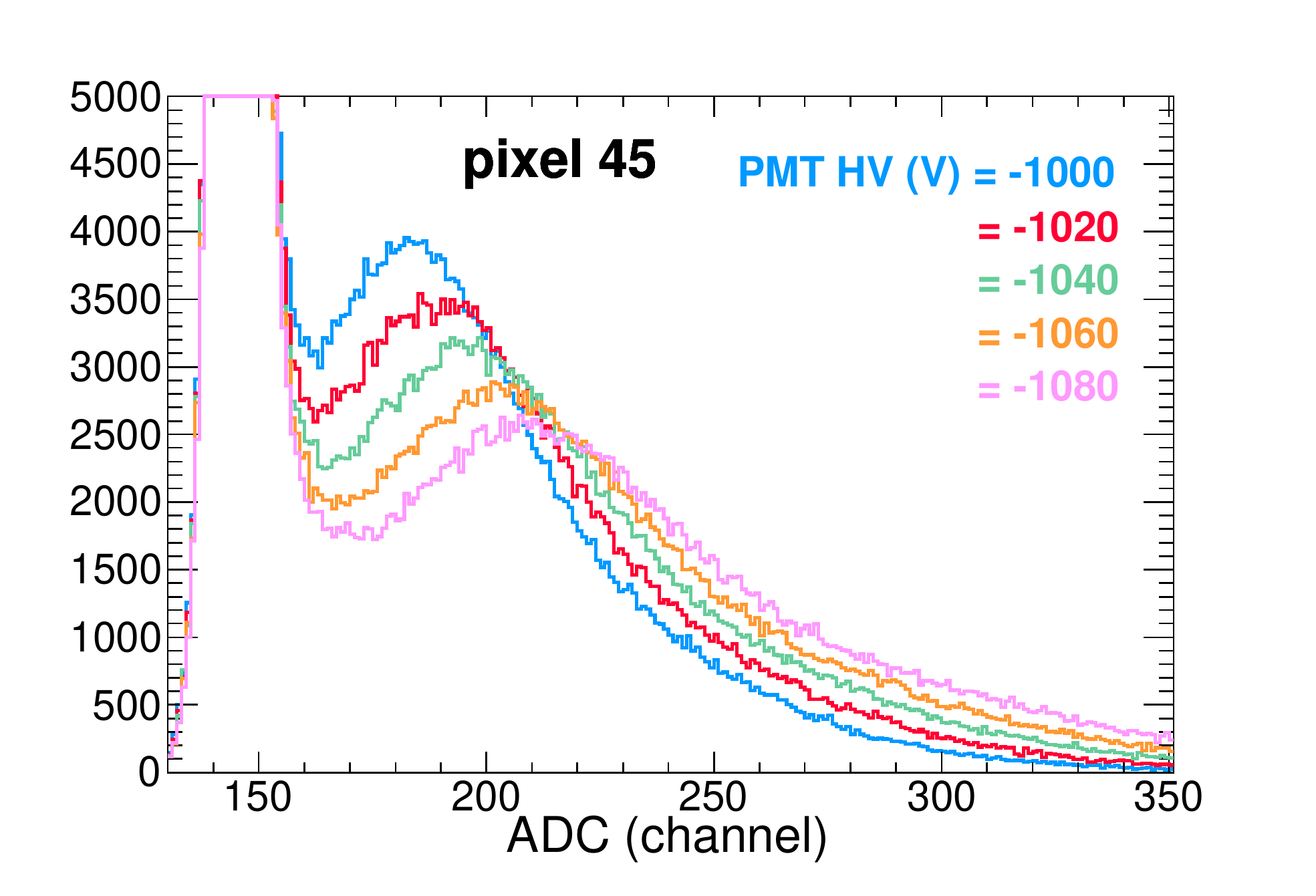}
&
\hspace{-0.5in}
\includegraphics[width=8.8cm]{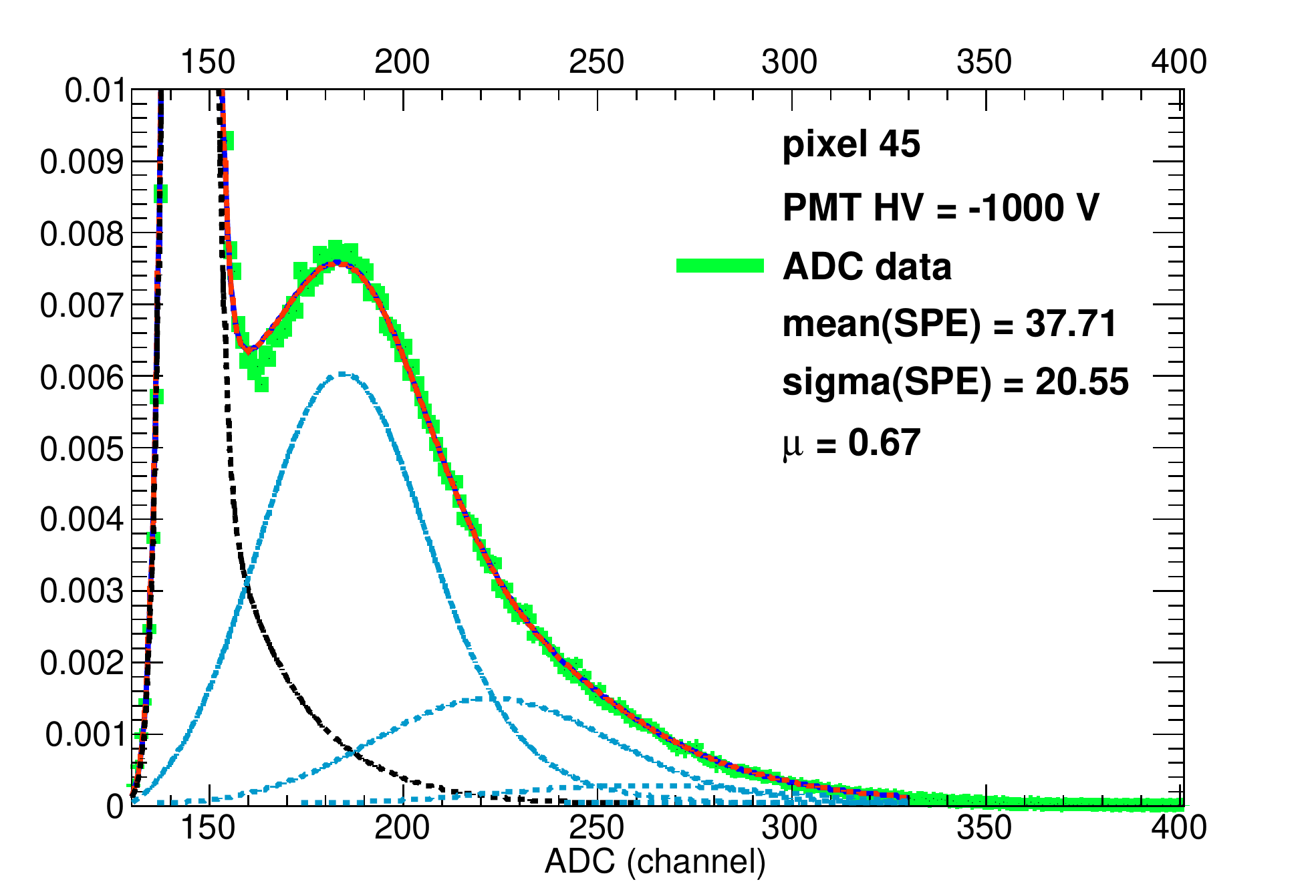} \\
\includegraphics[width=8.8cm]{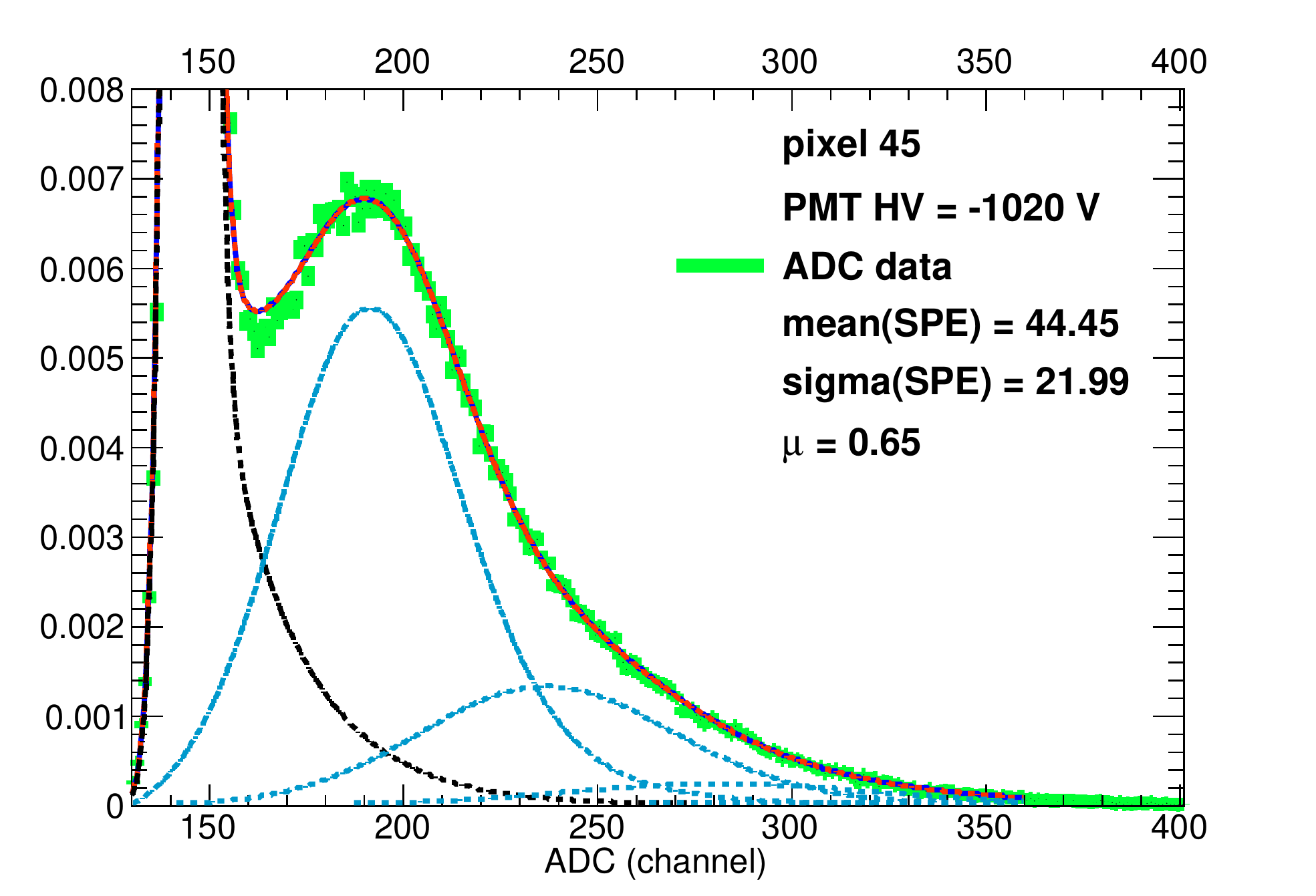}
&
\hspace{-0.5in}
\includegraphics[width=8.8cm]{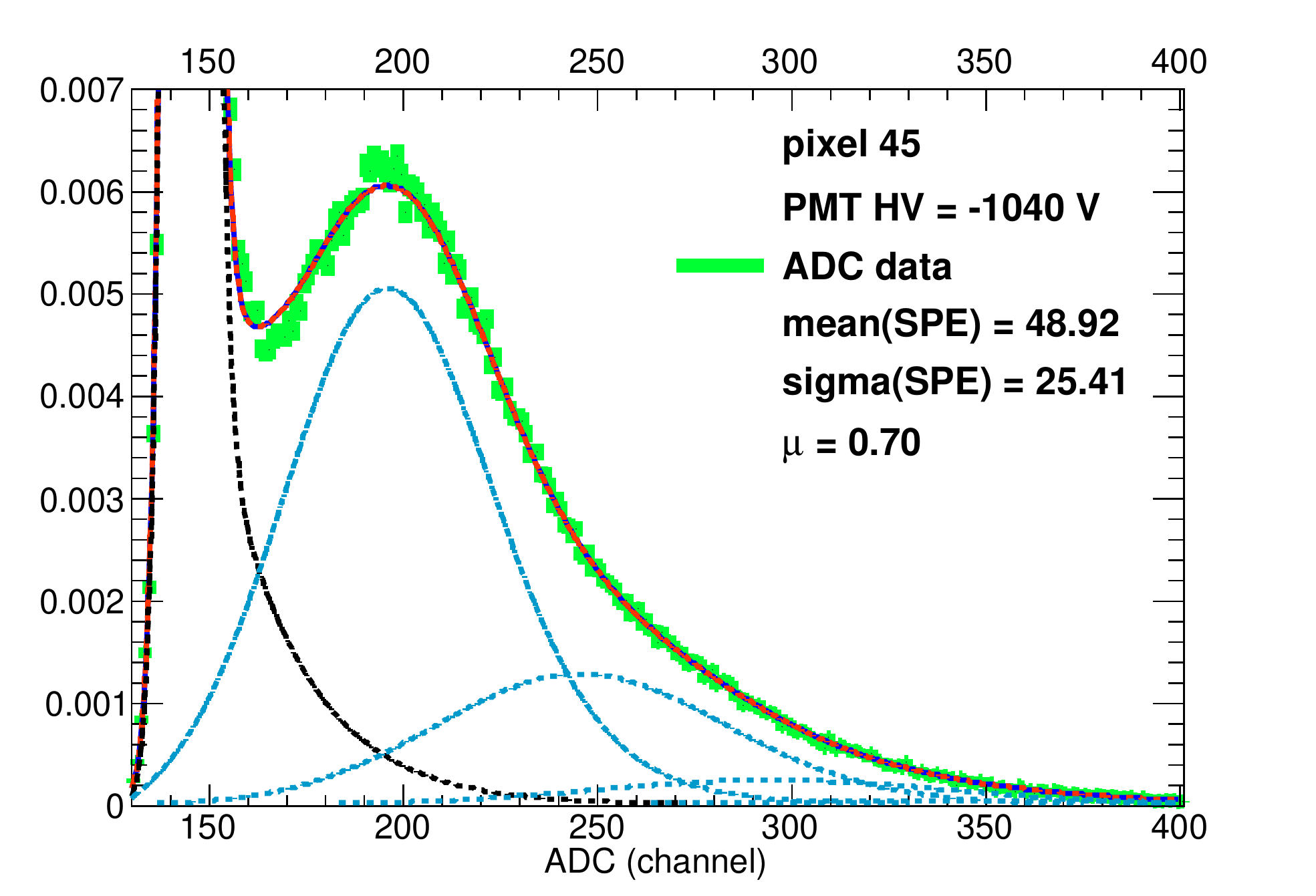} \\
\includegraphics[width=8.8cm]{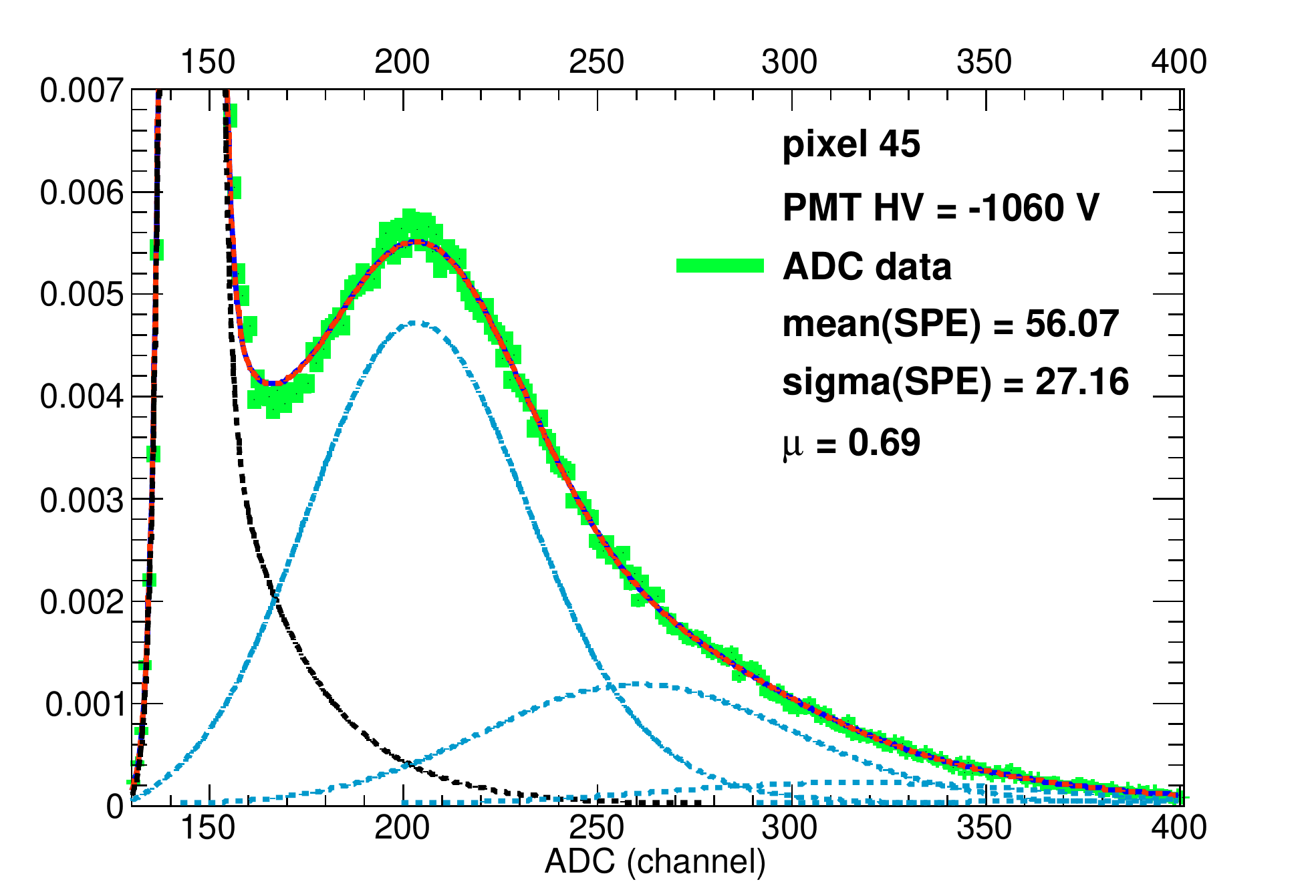}
&
\hspace{-0.5in}
\includegraphics[width=8.8cm]{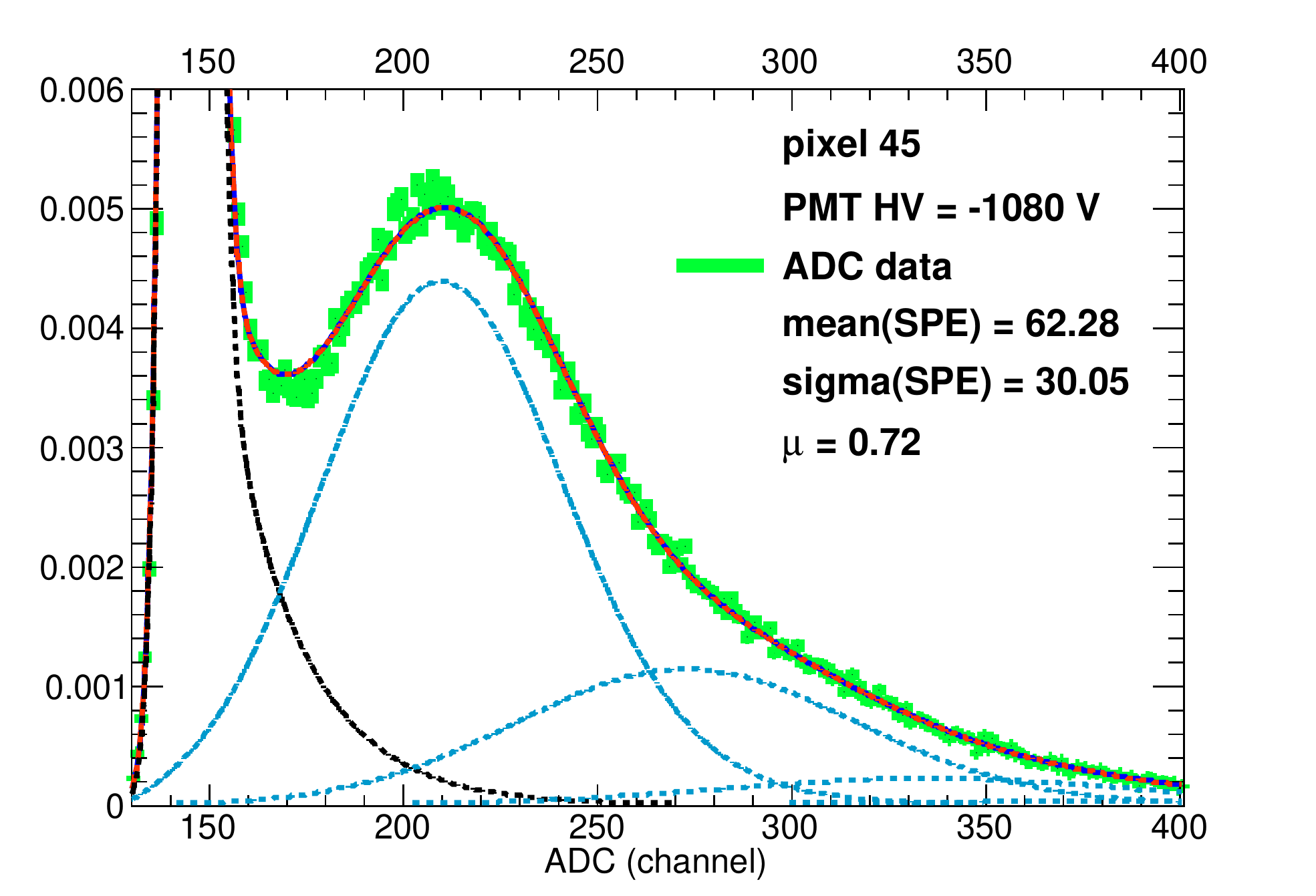} 
\end{tabular}
\linespread{0.5}
\caption[]{
{High voltage scan for pixel 45. The high voltage of the PMT was varied between 
-1000 V (nominal operating value) and -1080 V in steps of 20 V and the single 
photoelectron ADC distribution was recorded. The LED setting was optimized for 
single photoelectron detection and then kept 
constant during each high voltage scan. As expected, the pixel gain 
increases with an increasing high voltage and this is clearly indicated by 
the increase in the number of ADC channels above the pedestal (\emph{i.e.} 
integrated charge) corresponding to the single photoelectron peak.} }
\label{hv_scan_45}
\end{figure}

\begin{figure}[htbp]
\vspace*{-0.2in}
\centering
\begin{tabular}{cc}
\includegraphics[width=9.cm]{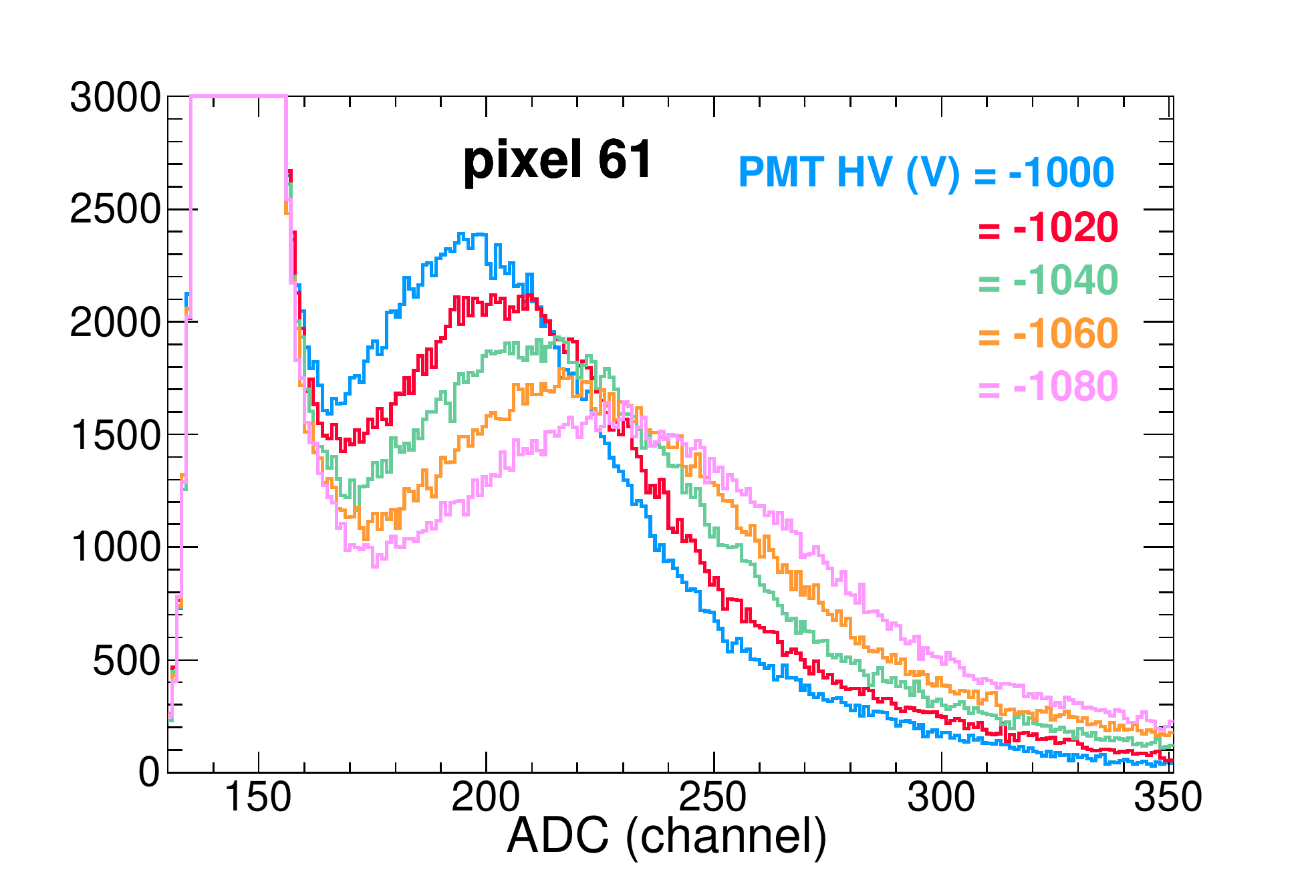}
&
\hspace{-0.5in}
\includegraphics[width=8.8cm]{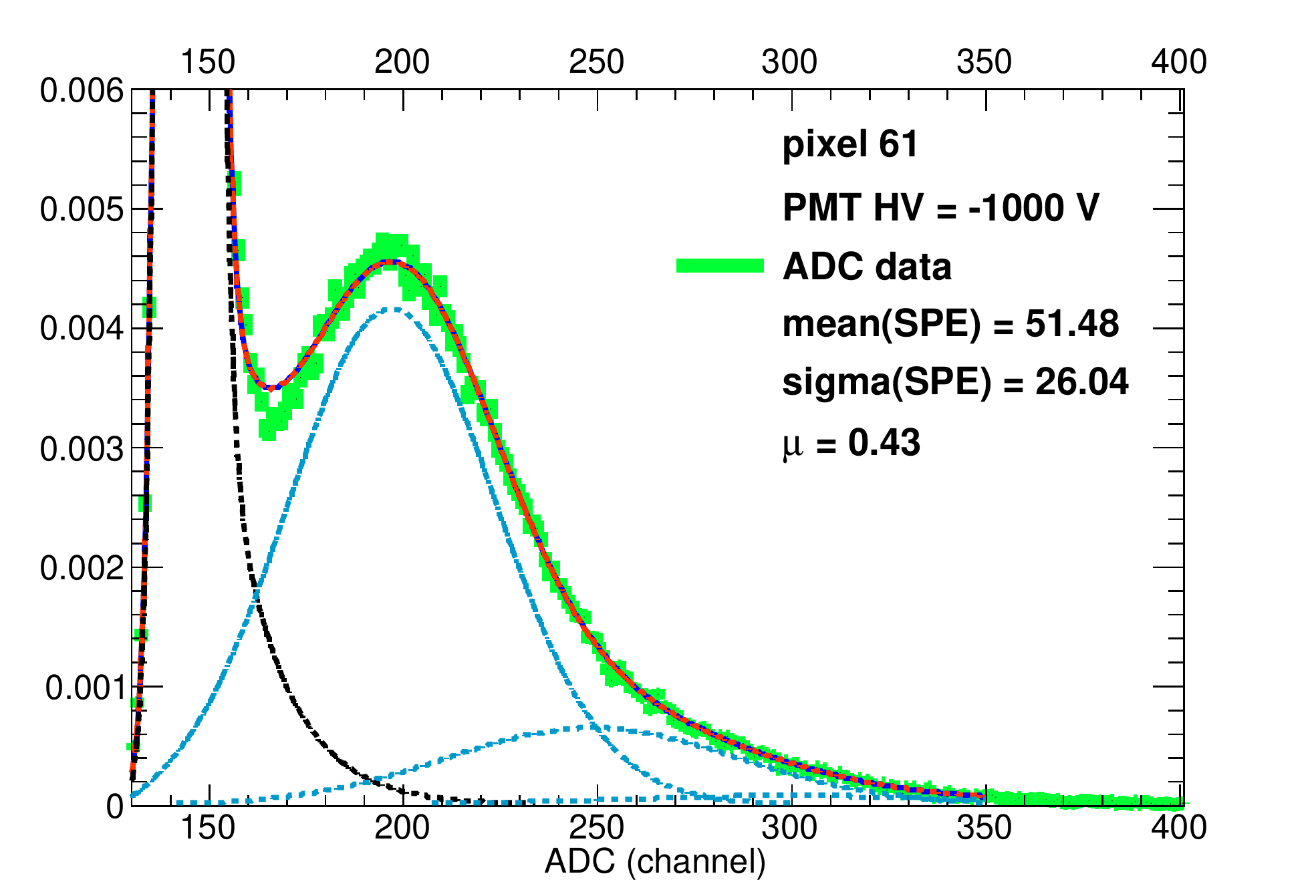} \\
\includegraphics[width=8.8cm]{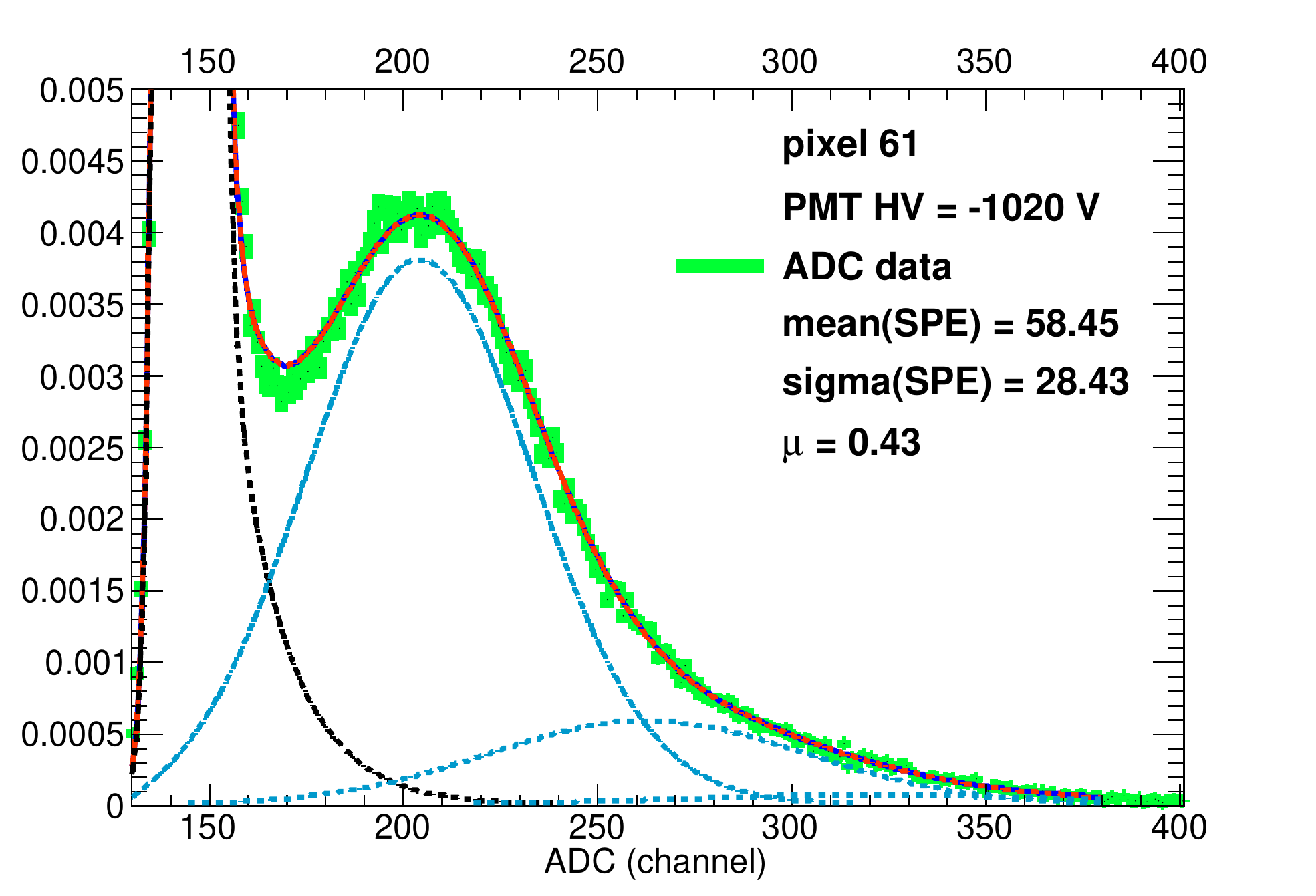}
&
\hspace{-0.5in}
\includegraphics[width=8.8cm]{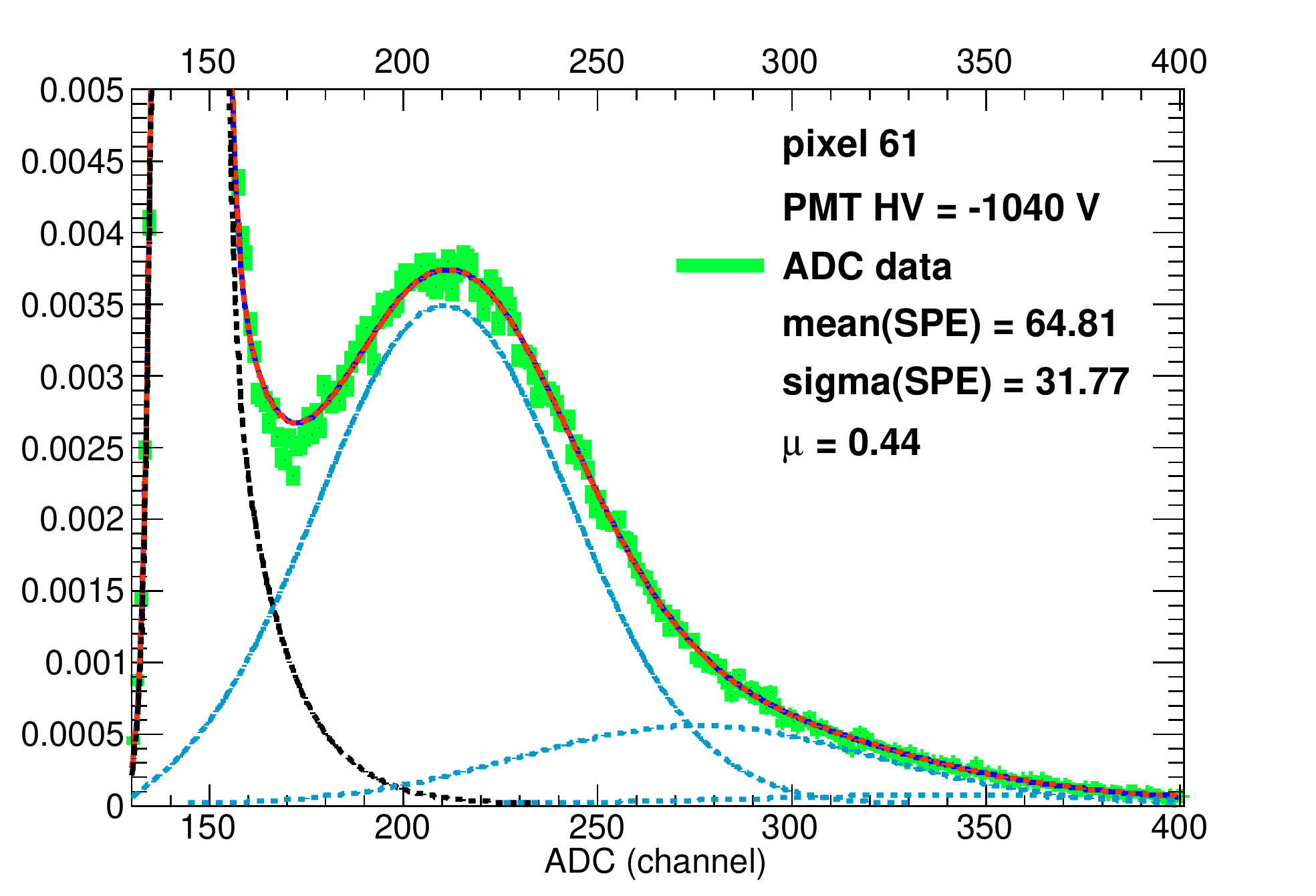} \\
\includegraphics[width=8.8cm]{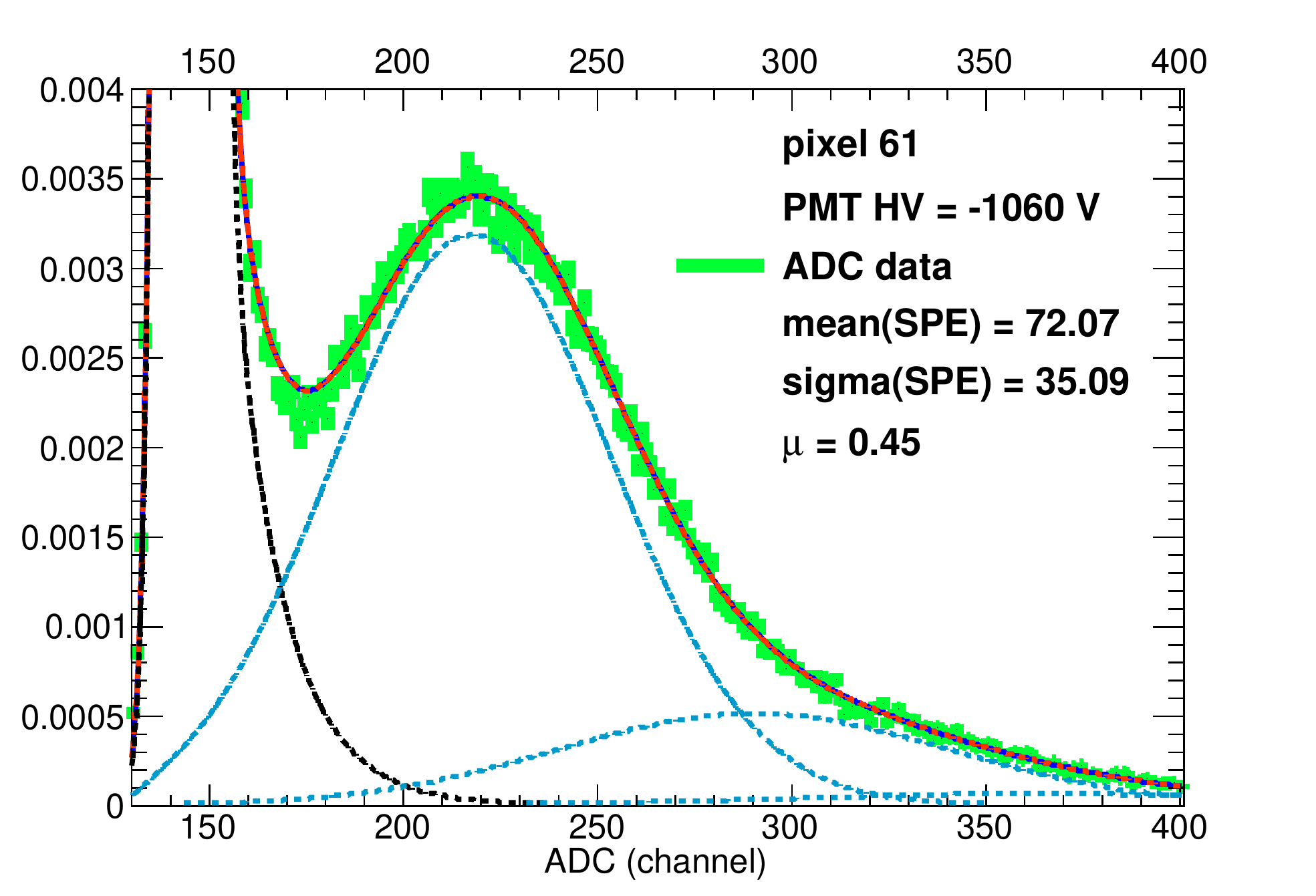}
&
\hspace{-0.5in}
\includegraphics[width=8.8cm]{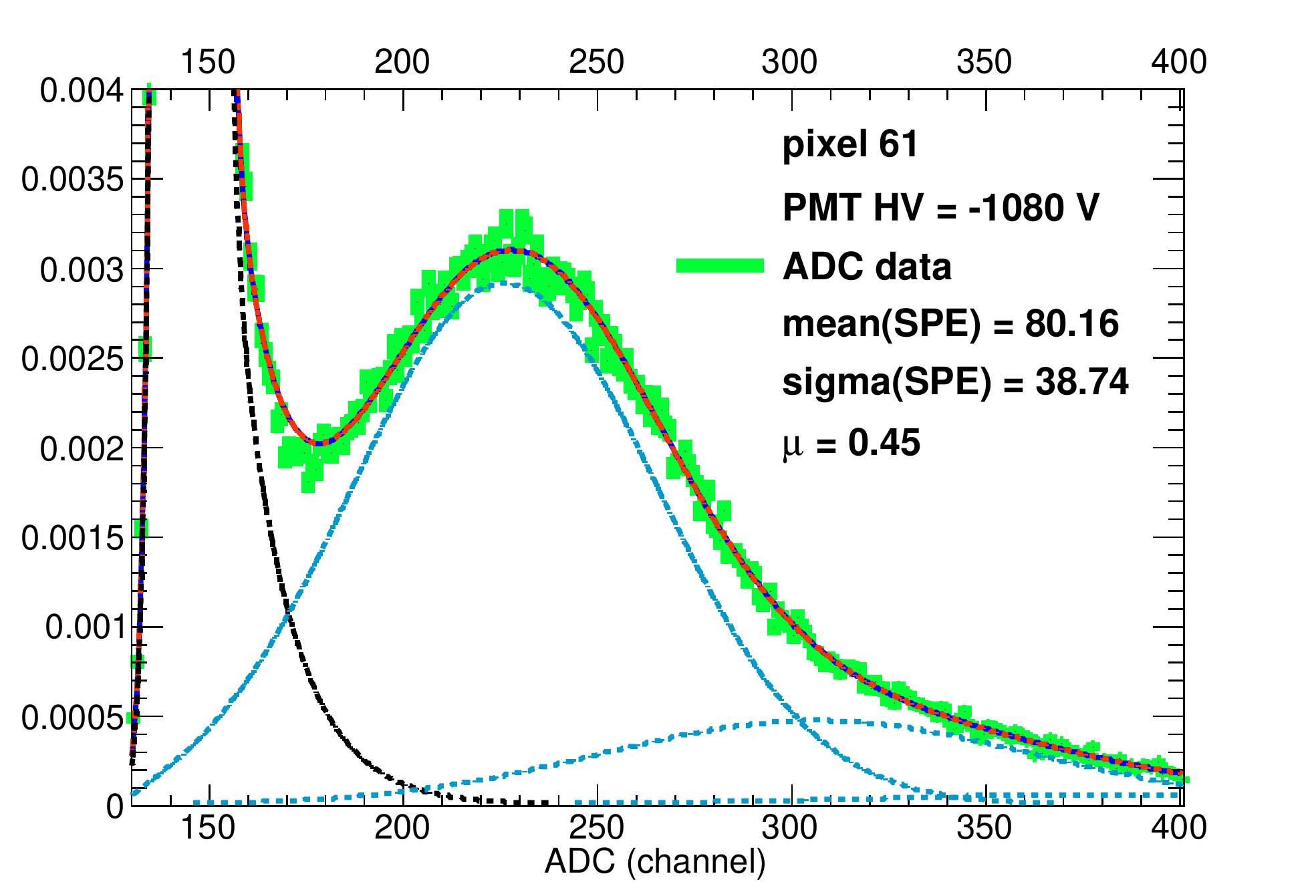} 
\end{tabular}
\linespread{0.5}
\caption[]{
{High voltage scan for pixel 61. The high voltage of the PMT was varied between 
-1000 V (nominal operating value) and -1080 V in steps of 20 V and the single 
photoelectron ADC distribution was recorded. The LED setting was optimized for 
single photoelectron detection and then kept 
constant during each high voltage scan. As expected, the pixel gain 
increases with an increasing high voltage and this is clearly indicated by 
the increase in the number of ADC channels above the pedestal (\emph{i.e.} 
integrated charge) corresponding to the single photoelectron peak.} }
\label{hv_scan_61}
\end{figure}

We took measurements to determine the gain of each of the three pixels 
tested. Once the single photoelectron signal is identified the gain 
was extracted using:

\begin{equation}
Q_\mathit{ADC} = \frac{\mathrm{Charge \left. per \right. bin}_\mathit{ADC} \times \mathit{mean(SPE)}}{
\mathit{external \left. amplification \right. factor}}
\label{gain}
\end{equation}
The CAEN V792 ADC we used has a resolution of 100 picoCoulomb per bin. The 
PMT signal was amplified by a factor of 10 and \textit{$mean(SPE)$}, the number of 
ADC channels above the pedestal corresponding to the single photoelectron 
peak, was extracted 
by fitting the ADC distributions as explained in the previous subsection. 
We measured the boost in the PMT gain by increasing the PMT high voltage from 
a nominal value of -1000 V to -1080 V in steps of 20 V (the maximum safe 
high voltage value is -1100 V according to Hamamatsu's specifications). 
The ADC distributions for pixels 39, 45 and 61 for varying PMT high 
voltage settings are shown 
in Figs.~\ref{hv_scan_39}, \ref{hv_scan_45} and \ref{hv_scan_61} together 
with our fits. During the 
high voltage scan the LED setting was kept constant for each pixel. Thus, 
as the PMT high voltage increases it is expected that the pixel gain will 
increase but the average number of photoelectrons will remain constant. Indeed, 
as seen in Figs.~\ref{hv_scan_39}, \ref{hv_scan_45} and \ref{hv_scan_61}, 
with each 20 V increase in the PMT high voltage the pixel gain grows by 
$\approx$ 10-15\% as indicated by \textit{$mean(SPE)$} while the average number of 
photoelectrons given by $\mu$ remains the same.

\begin{figure}[htbp]
\vspace*{-0.1in}
\centering
\begin{tabular}{c}
\includegraphics[width=11cm]{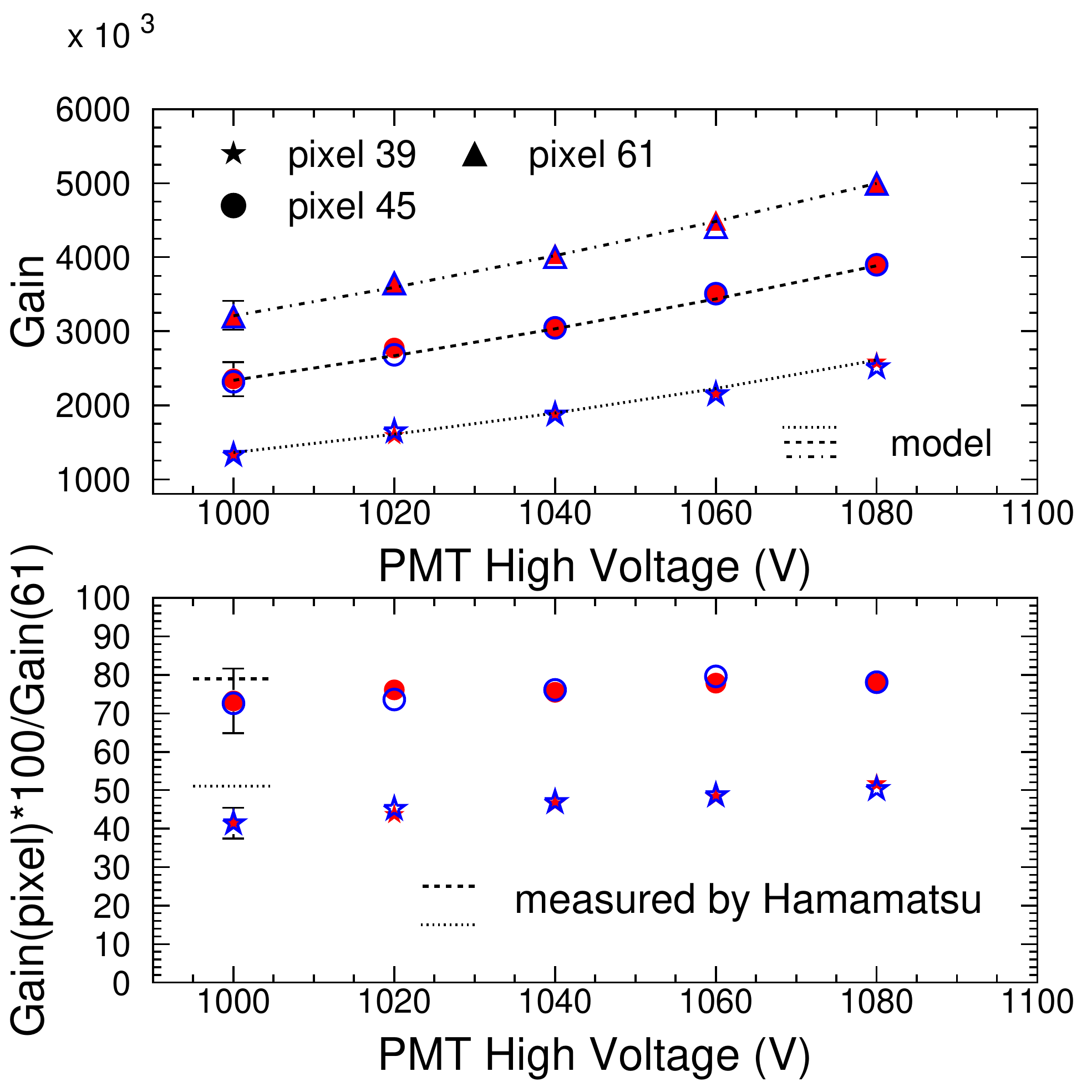} 
\end{tabular}
\linespread{0.5}
\caption[]{
{{\bf Top:} Gain as a function of the PMT high voltage for pixels 39, 45 and 61. A 
model calculation is also shown (curves). See text for more details. {\bf Bottom:} Measured 
output from pixels 39 and 45 relative to the highest output pixel (61). The values 
provided by Hamamatsu at a PMT voltage of -1000 V are also shown as short dashed/dotted lines.}}
\label{gain_vs_hv}
\end{figure}

The gain extracted according to Eq.~\ref{gain} for each of the three pixels 
tested is shown in Fig.~\ref{gain_vs_hv}, top panel, together with a model that 
parametrizes the change in gain with voltage of a 12 dynode stage PMT as 
$\frac{a^{12}}{13^{k \times 12}} \times V^{k \times 12}$ where $a$ is a 
constant and $k$ is determined by the structure and material of the 
dynode. The curves shown in Fig.~\ref{gain_vs_hv}, top panel, represent 
the model prediction when using the coefficients $a$ and $k$ as extracted 
from measurements taken at -1000 V and -1080 V. We also calculated the 
relative gain for pixels 39 and 45 with respect to the highest gain pixel, 
61, and our data, shown in Fig.~\ref{gain_vs_hv}, bottom panel, are in 
agreement with the measurements from Hamamatsu at -1000~V (dashed and 
dotted lines).

\subsection{Output Matching}

\begin{figure}[htbp]
\vspace*{-0.1in}
\centering
\begin{tabular}{cc}
\includegraphics[width=8cm]{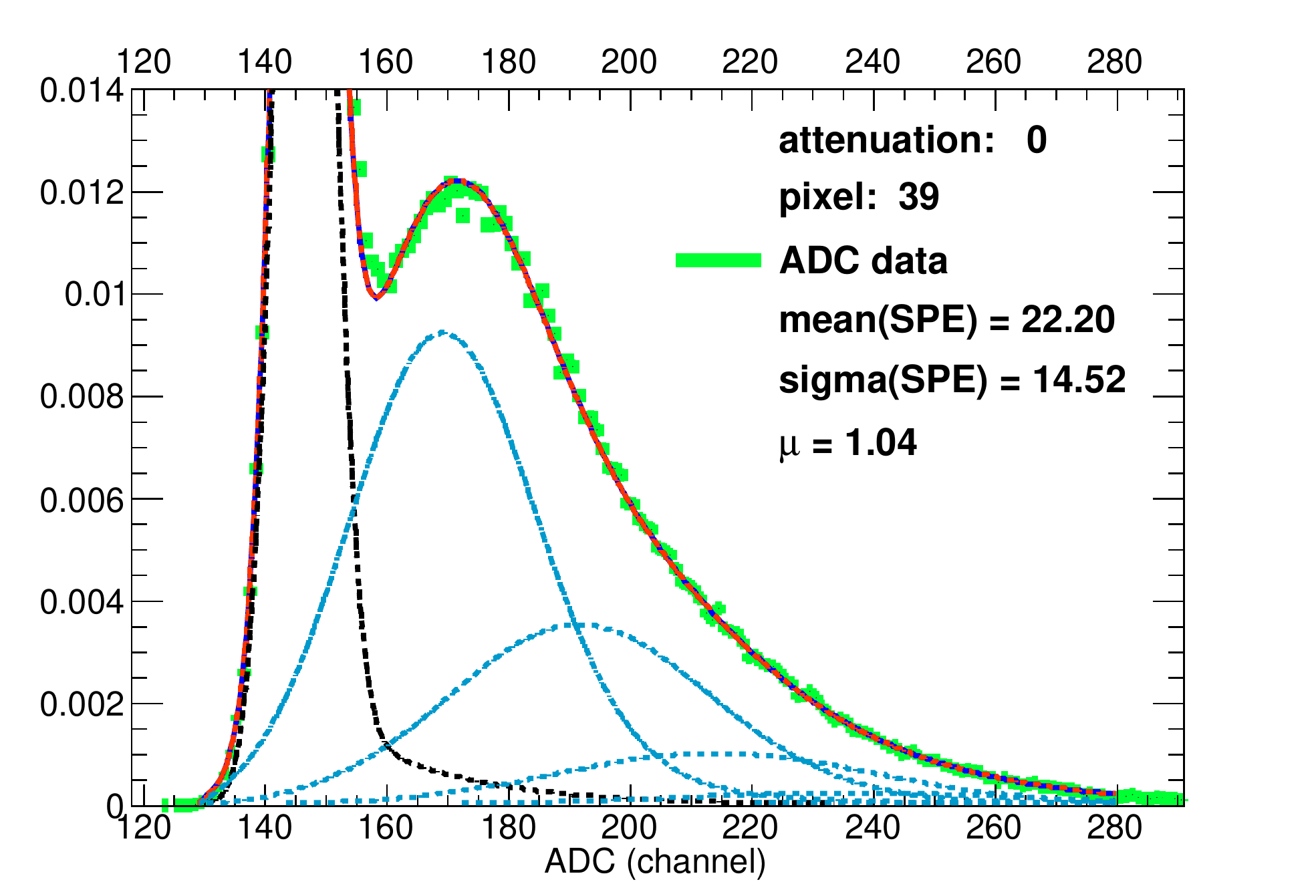}
&
\includegraphics[width=8cm]{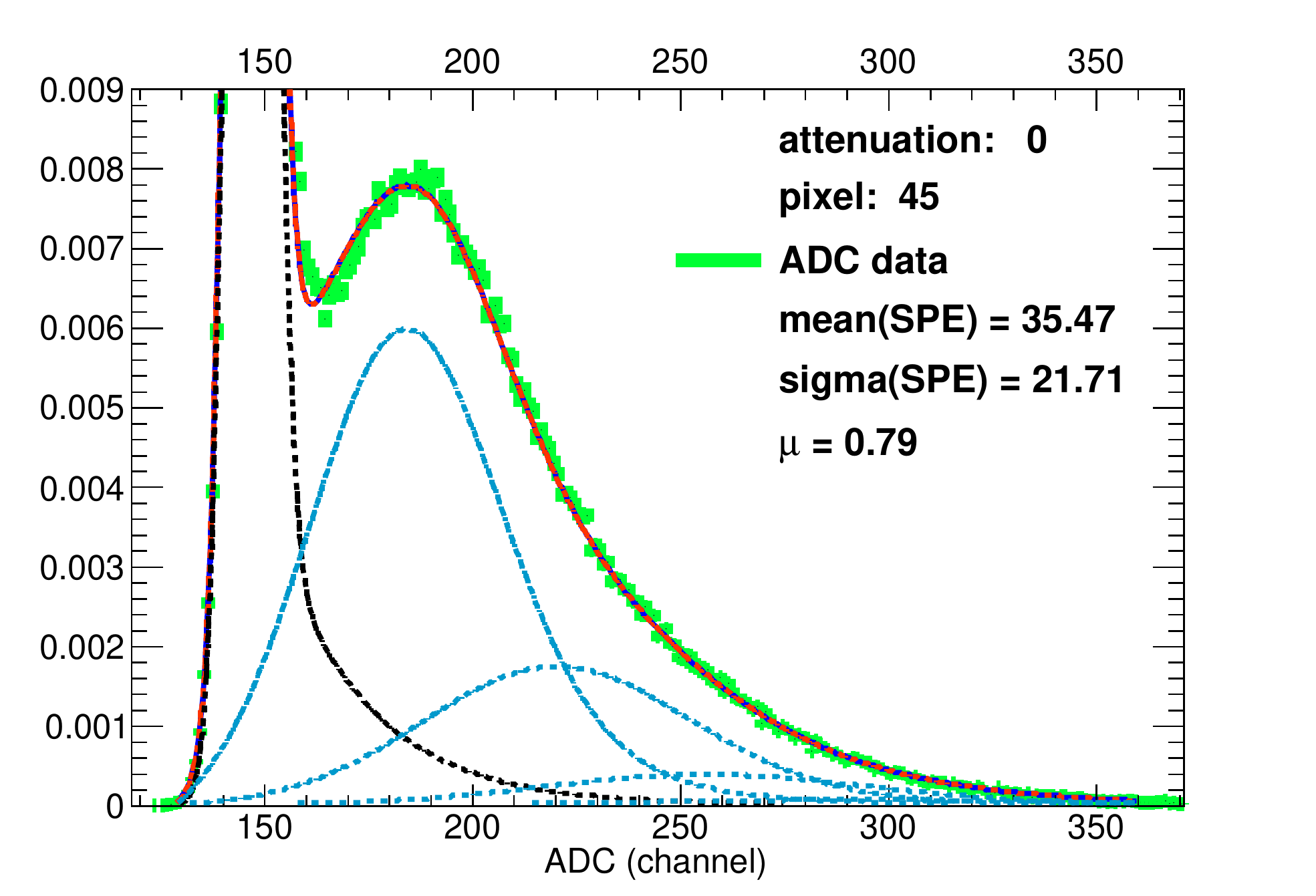} \\
\includegraphics[width=8cm]{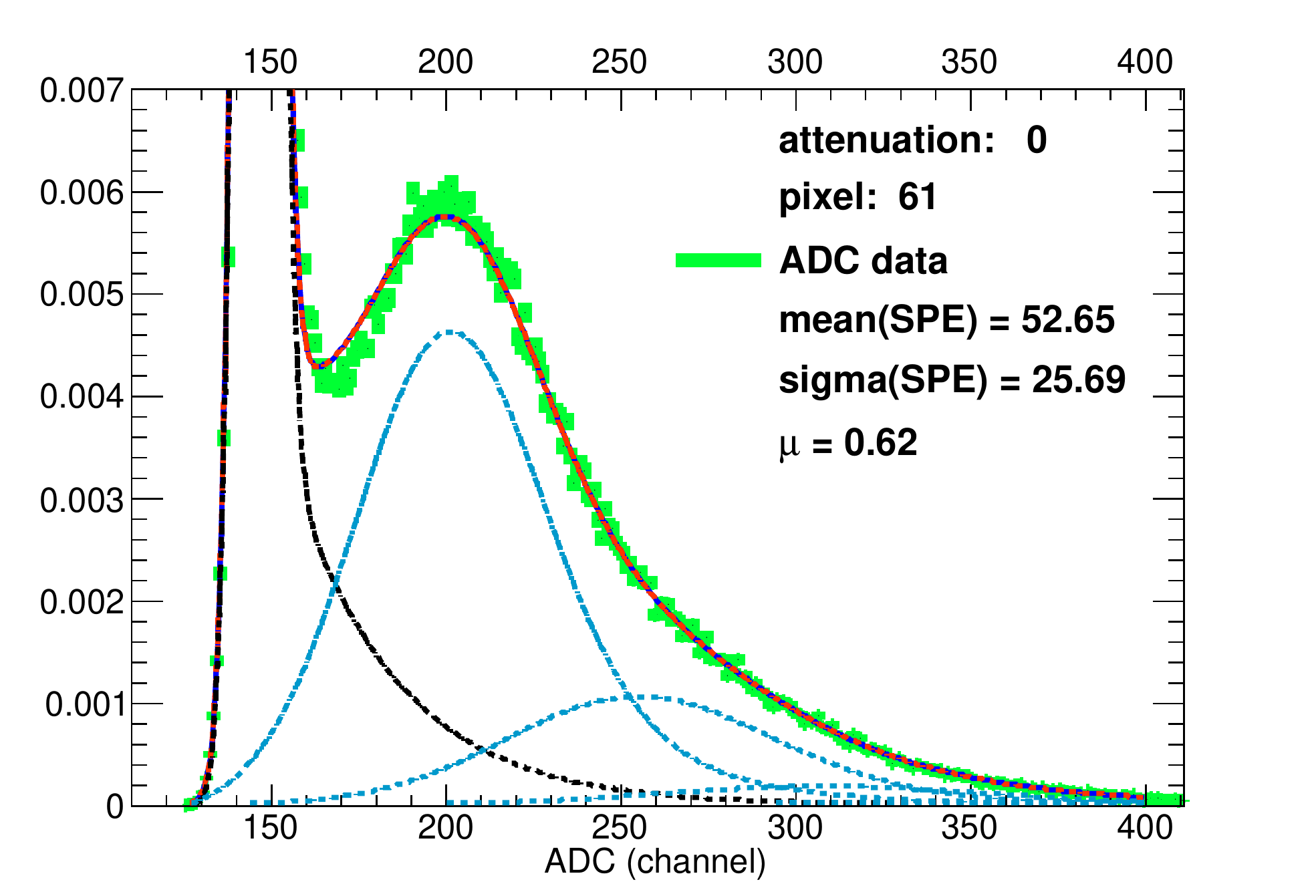}
&
\includegraphics[width=8cm]{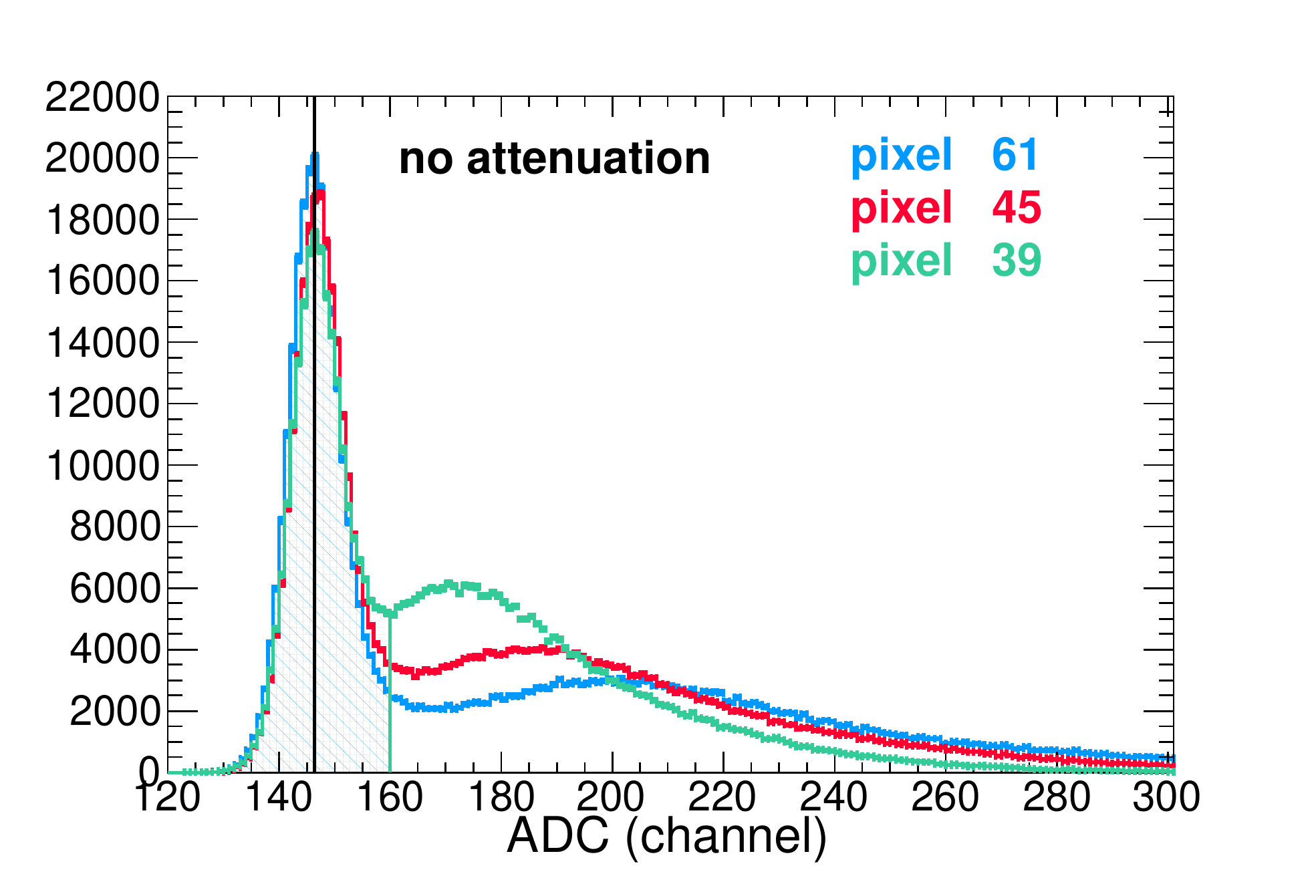} \\
\includegraphics[width=8cm]{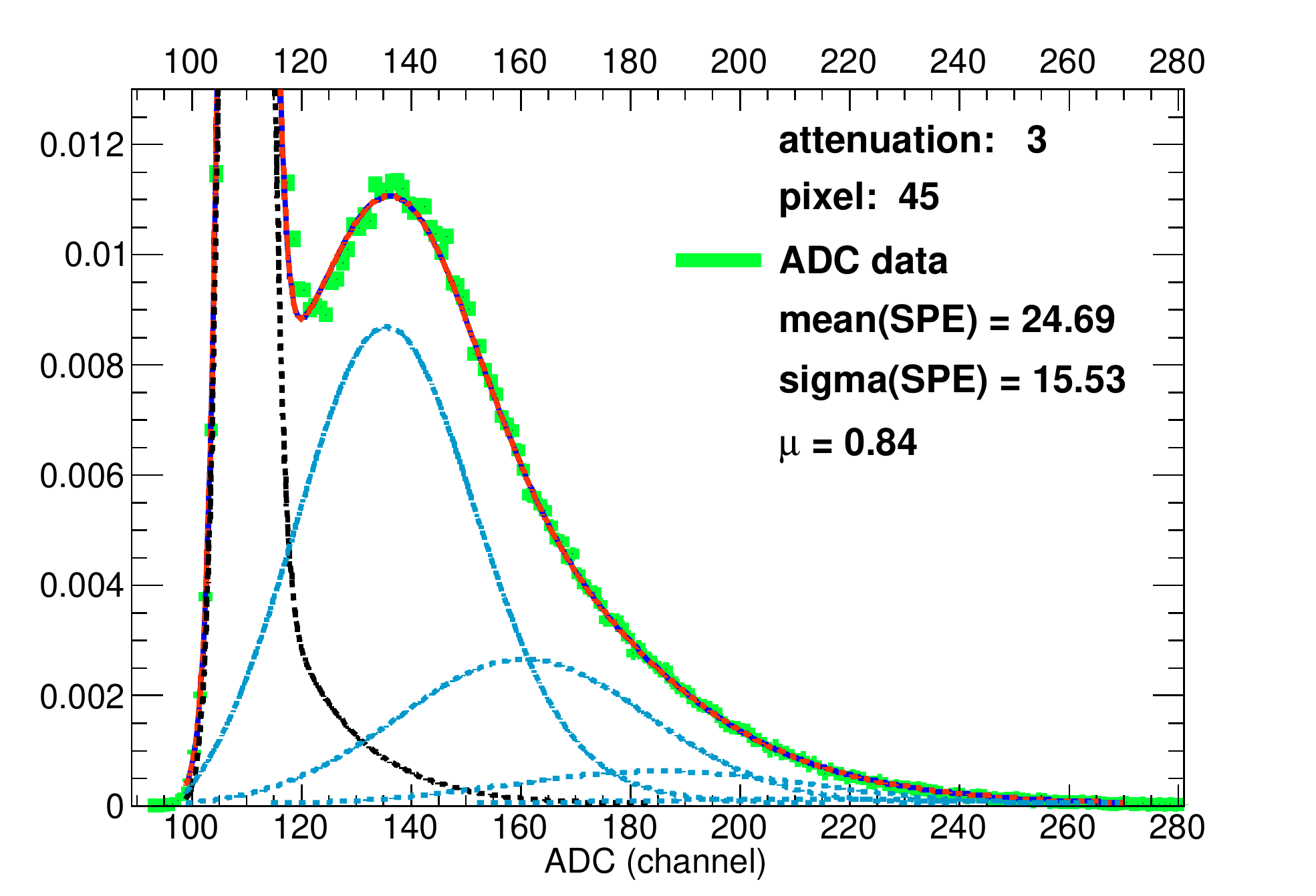}
&
\includegraphics[width=8cm]{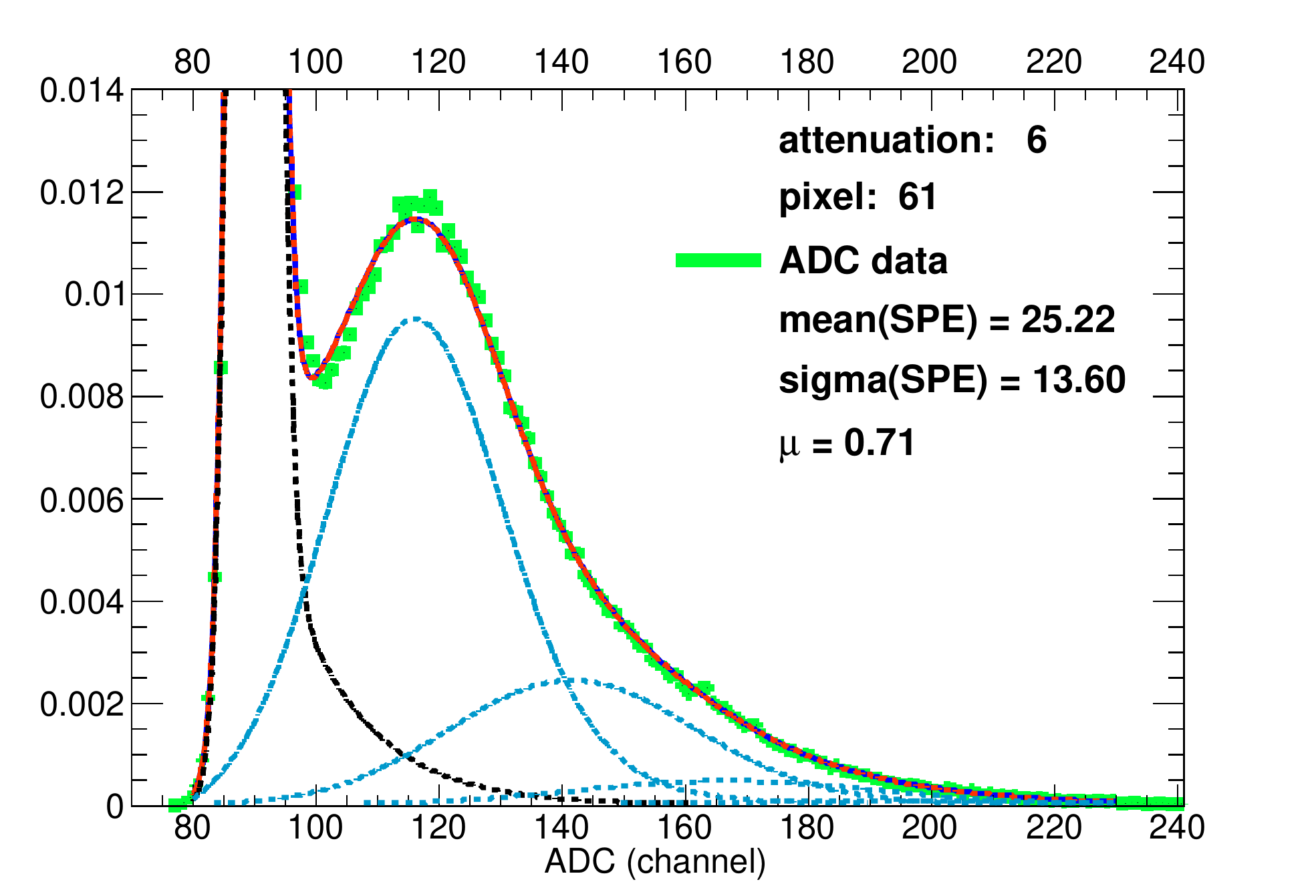} \\
\includegraphics[width=8cm]{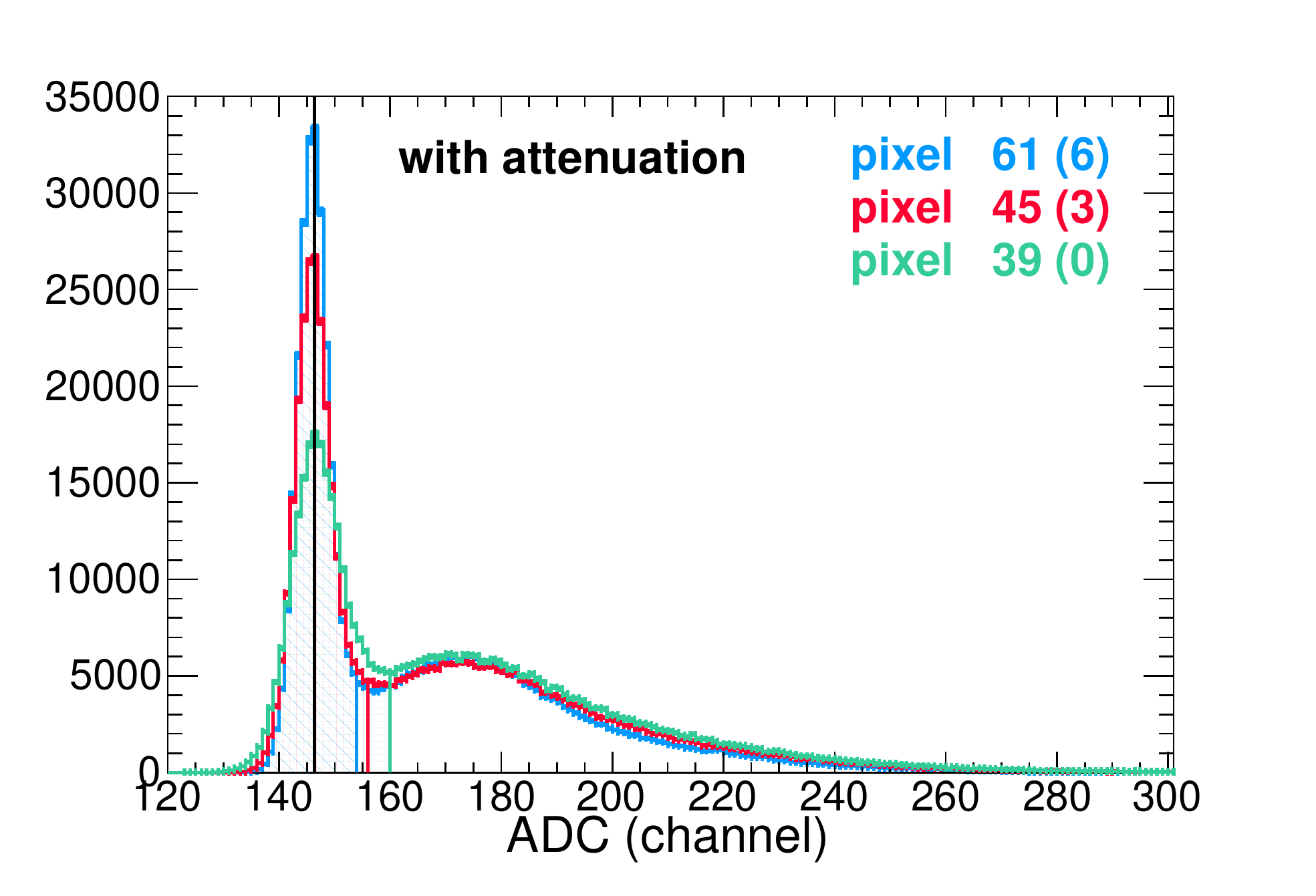}
&
\includegraphics[width=8cm]{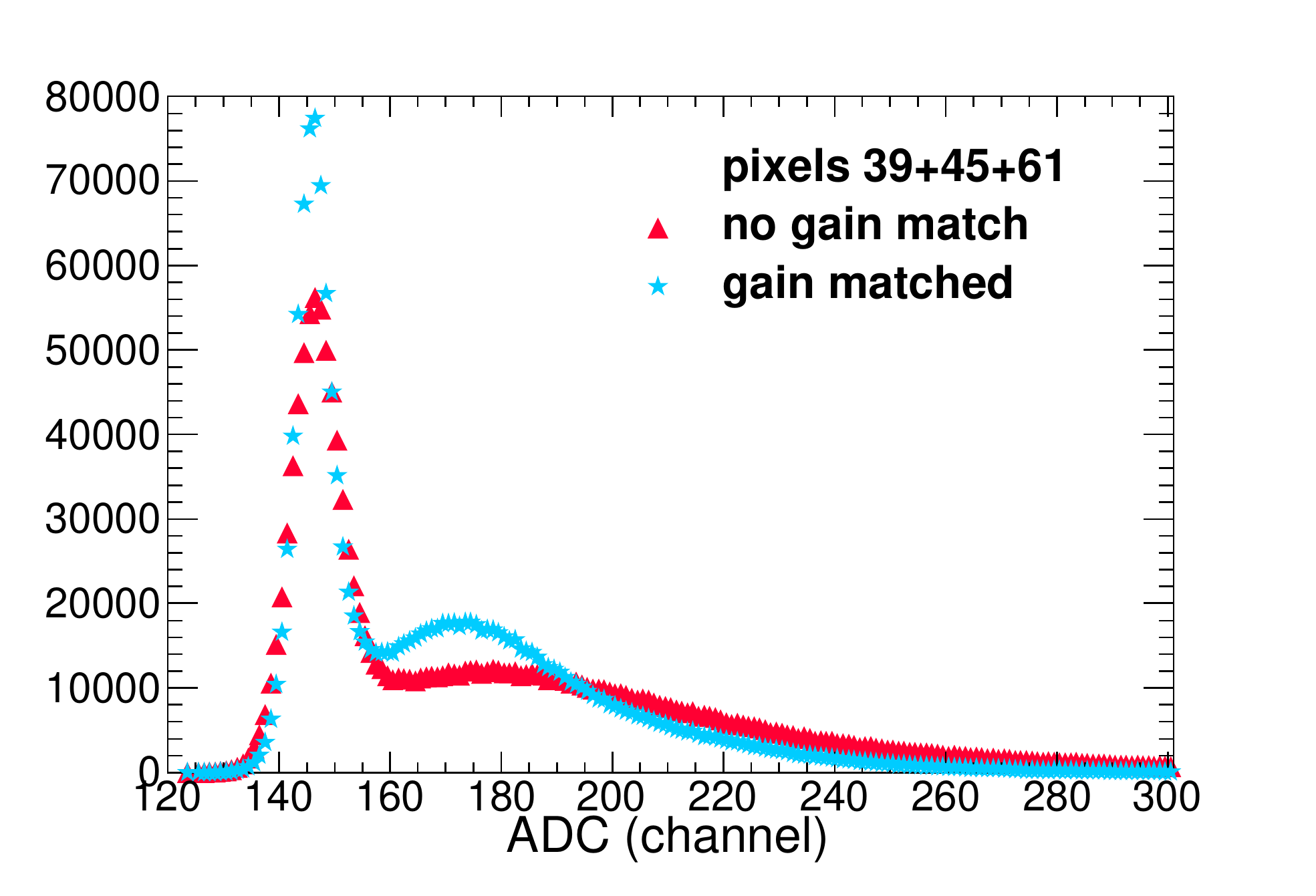}
\end{tabular}
\linespread{0.5}
\caption[]{
{ADC histograms associated with output matching of pixels 45 and 61 to that of pixel 39 (see text for details).} }
\label{match_gain}
\end{figure}

As pointed out in previous subsections, gain non-uniformities 
up to a factor of two or more between pixels are common and this can result in a decrease in 
the single photoelectron detection resolution when outputs from several pixels 
are summed. The effect can be corrected for by normalizing the 
output of pixels with higher gains to the output of the lowest gain pixel. 
This may be achieved by adding a simple resistor-based voltage divider into 
the readout circuit so that the output of the higher gain pixels is attenuated 
to match that of the lowest gain pixel. We followed this procedure to match 
the output of pixels 45 and 61 to that of pixel 39 and we show our results 
in Fig.~\ref{match_gain}. The first two panels on the left and the first 
panel on the right display the single photoelectron ADC distributions for 
pixels 39, 45 and 61 together with a fit which shows the three pixels 
to have varying gains, up to 60\%, as indicated by the fit 
parameter \textit{$mean(SPE)$}. To 
emphasize the effect we superimposed the ADC distributions of the three 
pixels on the second right panel. We passed the outputs from 
pixels 45 and 61 through a rotary attenuator and the attenuated signals 
as recorded by the ADC are shown in the third left and right panels, 
respectively. We used 3 and 6 db attenuation for pixel 45 and 61 to 
match the output of pixel 39. After attenuation, the output of the three 
pixels is very similar, within 10\%, as indicated by the fit parameter 
\textit{$mean(SPE)$}. The bottom left panel shows the superposition of the 
ADC distributions from the three pixels after attenuation and it can be 
seen that the single photoelectron peak is positioned at the same location 
with respect to the pedestal. Lastly, in the bottom right panel we show the 
sum of the ADC spectra of all three pixels before (red triangles) and 
after (blue starts) output matching. The improvement in the 
resolution for single photoelectron detection is evident.      

\section{Magnetic Field Measurements}\label{sec:magFieldMeas}

\begin{figure}[htbp]
\vspace*{-0.1in}
\centering
\begin{tabular}{cc}
\hspace{-1.3in}
\includegraphics[clip,width=9.cm]{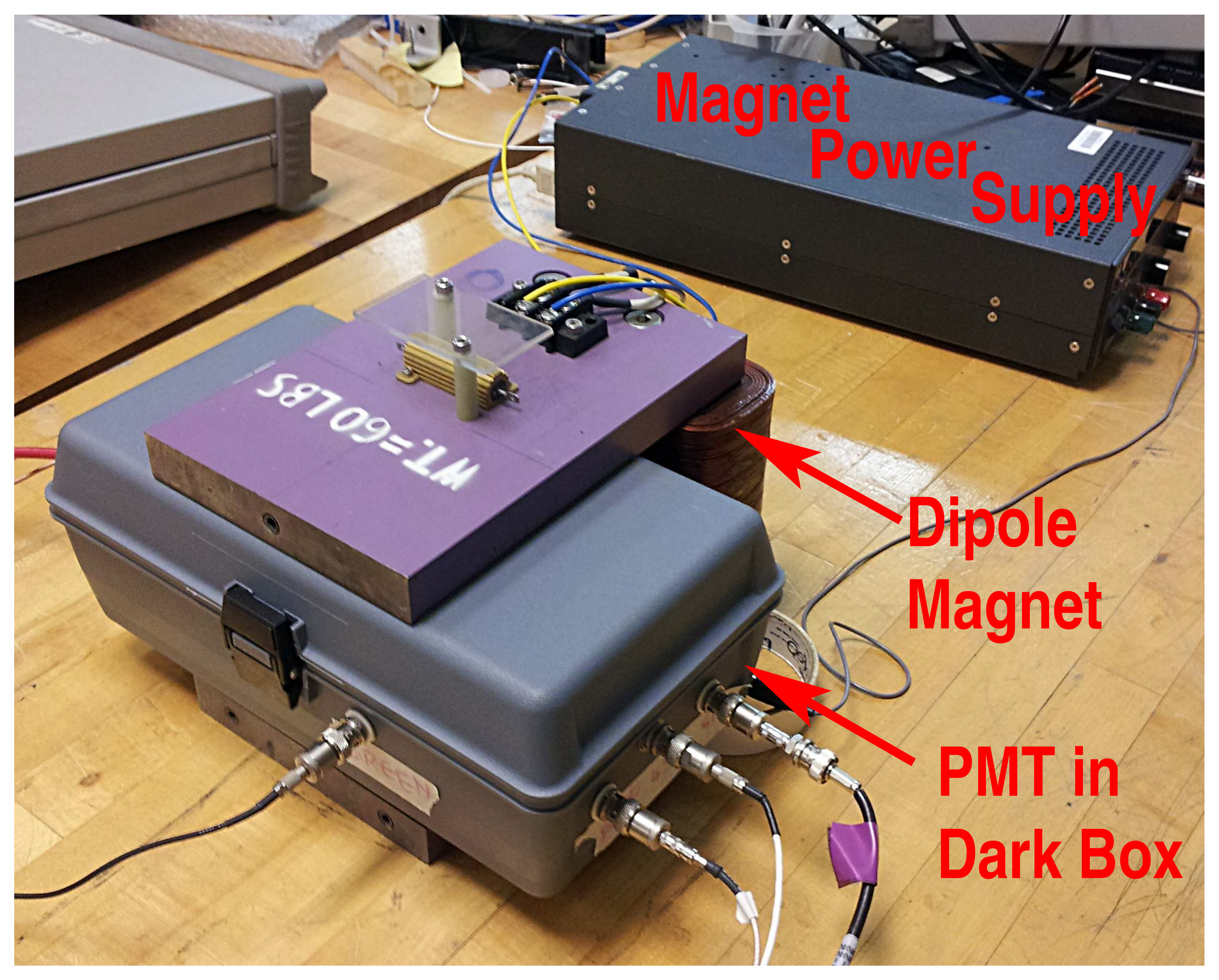} 
&
\hspace{-1.4in}
\includegraphics[clip,width=7.cm]{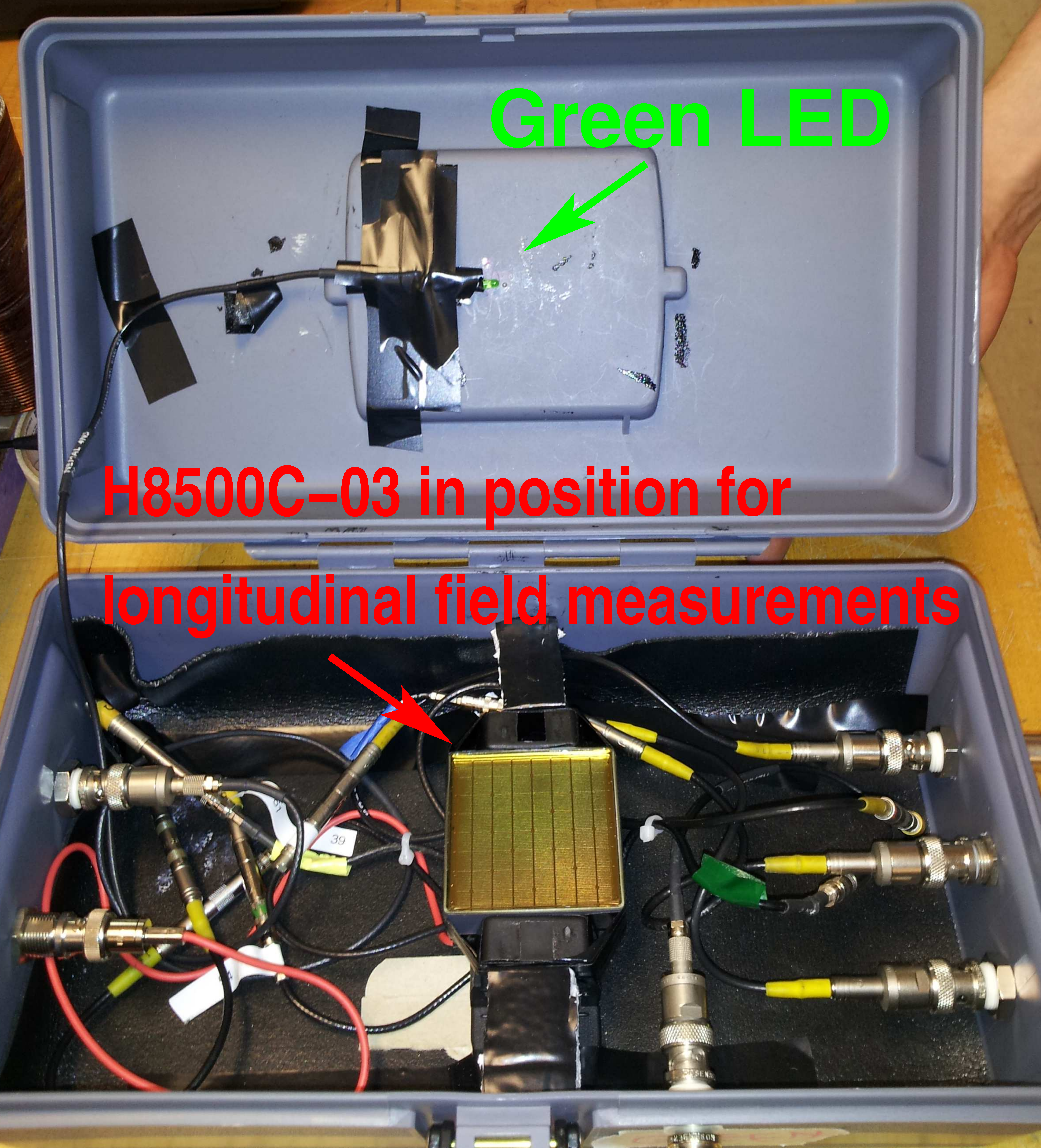} \\
\hspace{1.9in}
\includegraphics[clip,width=8.cm]{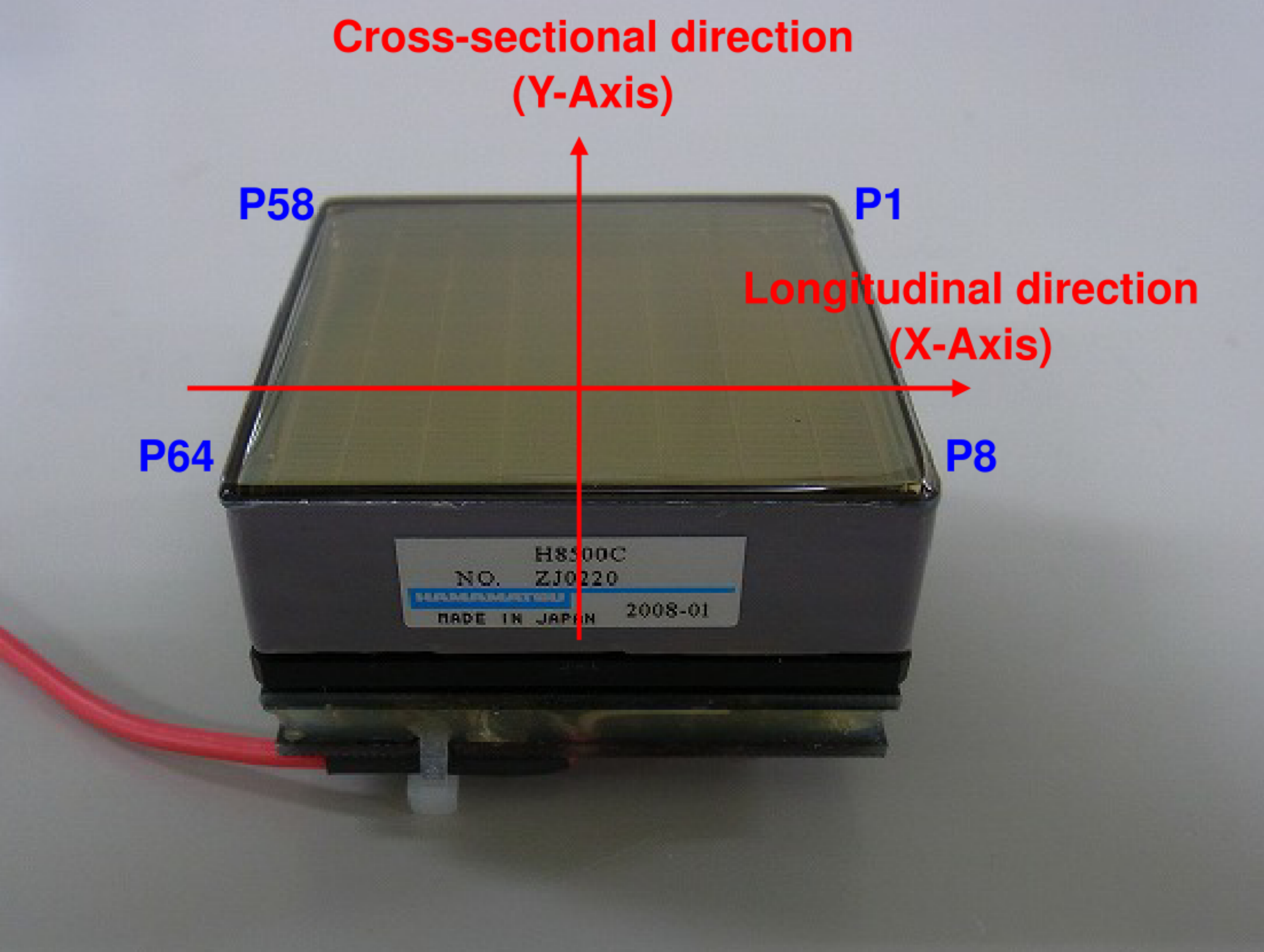}
\end{tabular}
\caption[]{
{Pictures of the experimental setup used for magnetic field measurements.} }
\label{exp_pics1}
\end{figure}

\begin{figure}[htbp]
\centering
\begin{tabular}{c}
\includegraphics[width=13.cm]{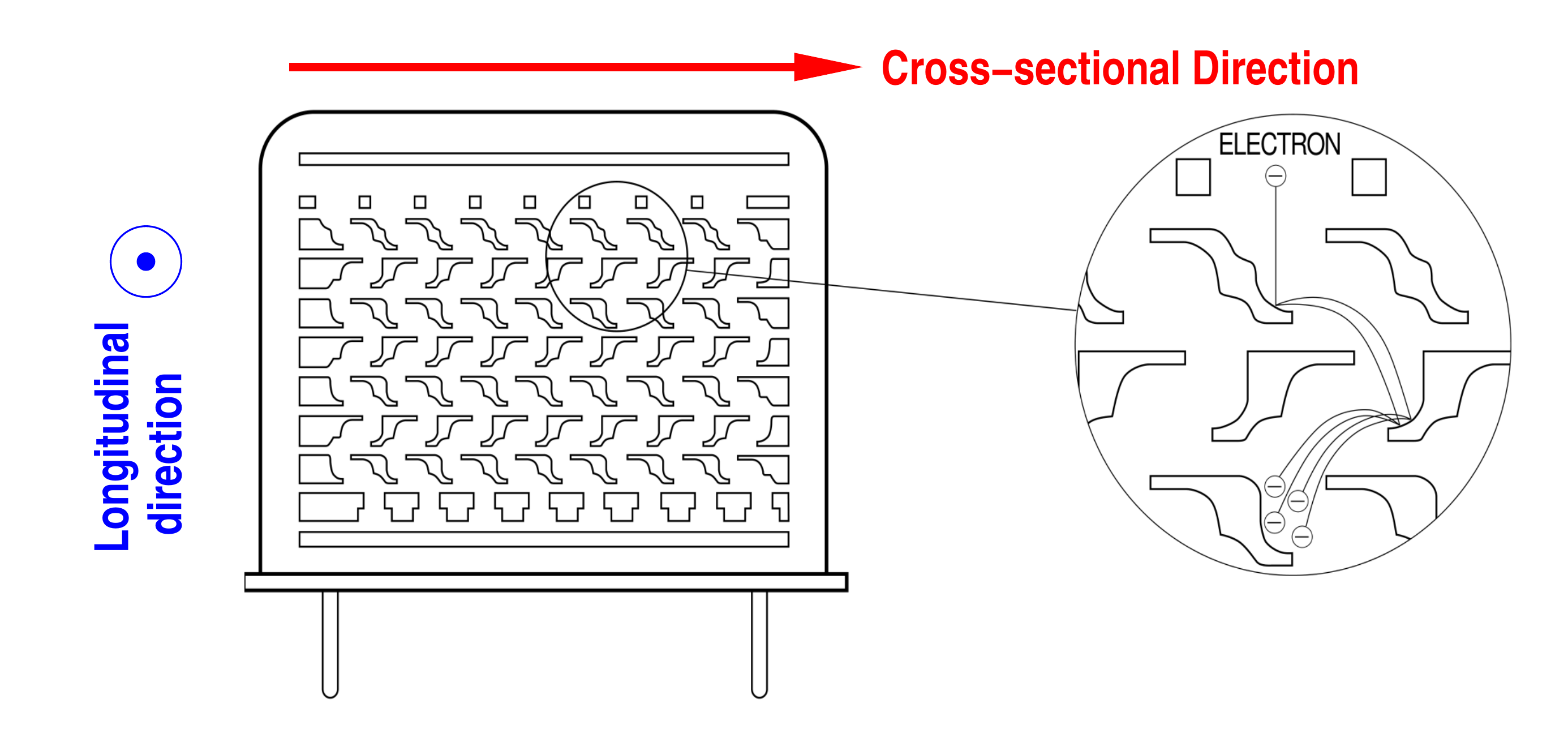} 
\end{tabular}
\caption[]{
{A schematic of the metal channel dynode structure. The transverse field orientations are 
also shown.} }
\label{exp_pics2}
\end{figure}

In Figs.~\ref{exp_pics1} and \ref{exp_pics2} we present few pictures of the experimental setup used 
to study the PMT behavior in magnetic field. In Fig.~\ref{exp_pics1}, top left panel, we show 
the PMT dark box (in grey) in between the dipole magnet plates (purple) with 
the coils visible at the back of the dark box. The magnet cable leads (in blue 
and yellow) connect the coils to the power supply. 
The four signal cables corresponding to outputs from 
pixel 45, 61, quad A and quad B can also be seen. In the panel on the top right, 
we show the PMT positioned inside the dark box for longitudinal magnetic 
field measurements, \emph{i.e.} the dipole magnetic field is perpendicular to 
the face of the PMT. In what follows, we will refer to this field orientation 
as $B_z$. The LED is attached to the box so that it can illuminate the 
face of the PMT. In the bottom panel the transverse field 
orientations are shown (i.e. field perpendicular to the sides of the PMT) with 
respect to the PMT and we labeled those as $B_x$ and $B_y$. The metal channel 
dynode structure of the PMT and the transverse field 
orientations are also depicted in Fig.~\ref{exp_pics2}. Note that the PMT 
response to magnetic field is symmetric with respect to a rotation around the 
longitudinal field $B_z$ axis. For a given alignment axis we can switch the field direction 
(plus or minus) by swapping the leads that power the coil. 
To expose the PMT to either purely longitudinal 
or transverse field components we monitored closely the PMT positioning 
relative to the magnetic field orientation and built rigid supports to hold 
the PMT in place and avoid possible position shifts during measurements. 

We performed two sets of measurements to study the PMT signal loss in magnetic 
fields. In one case we tuned the yield of (LED) photons to trigger large PMT 
signals, at the level of 30 to 50 photoelectrons, and in the zero field configuration 
this became one baseline. We then increased the magnetic field, in steps of 20 Gauss and 
followed the deviation of the PMT response from the baseline. For the second set 
of measurements the same procedure was followed but photon yields were tuned to
provide small PMT signals, at the single photoelectron level. 

\subsection{Effects of Magnetic Field on Large Signals}

\begin{figure}[htbp]
\vspace*{-0.1in}
\centering
\begin{tabular}{cc}
\vspace{-0.09in}
\includegraphics[width=8.8cm]{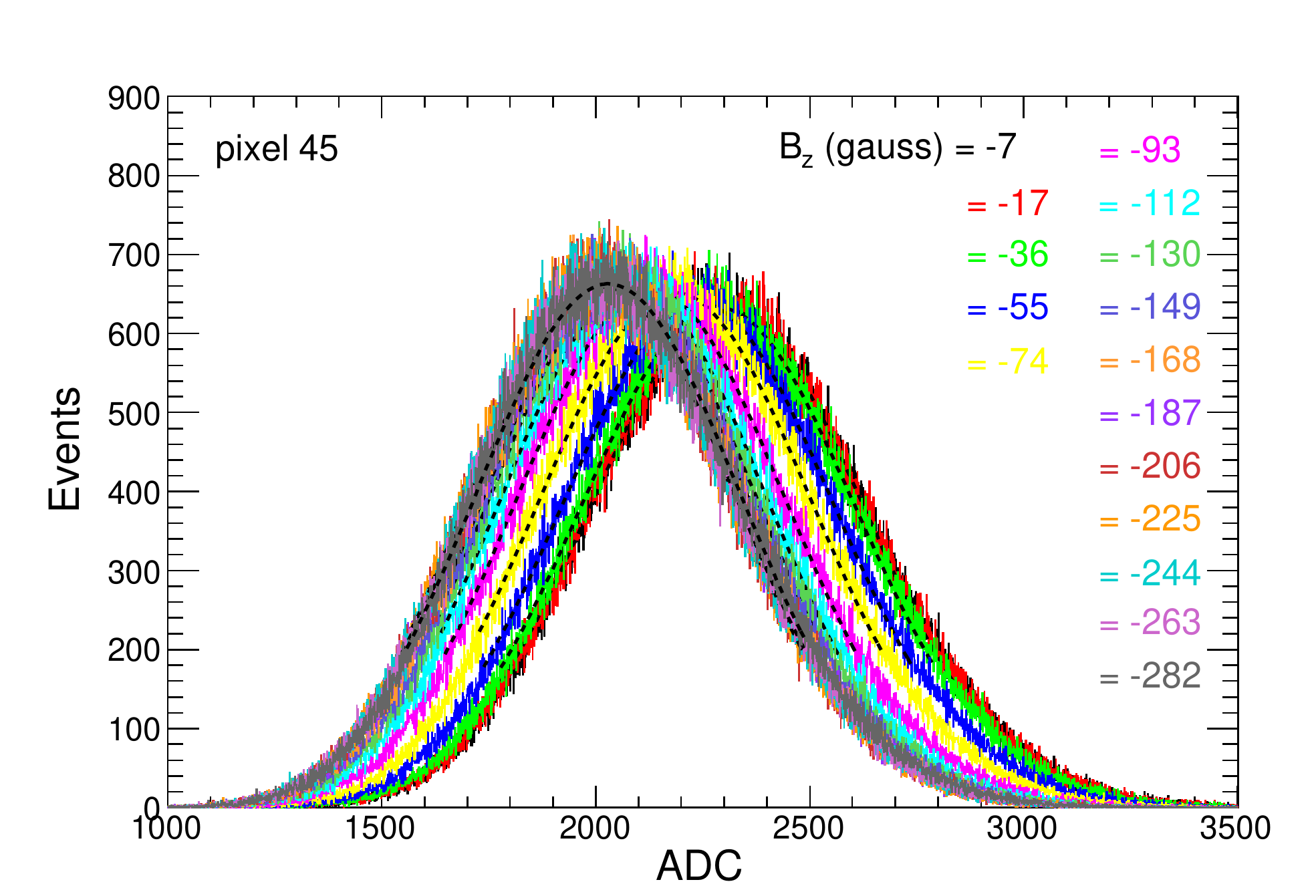}
&
\hspace{-0.4in}
\includegraphics[width=8.8cm]{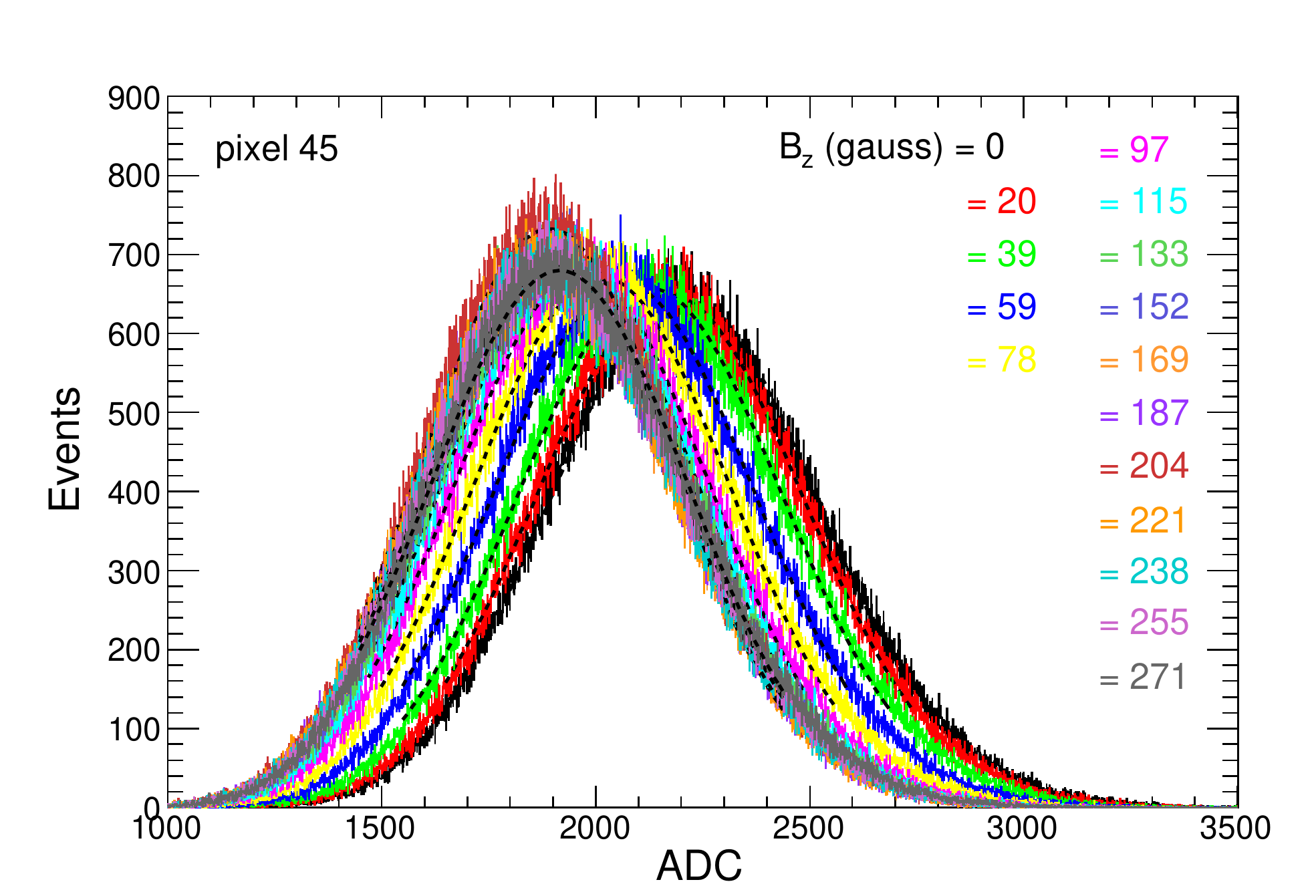} \\
\vspace{-0.09in}
\includegraphics[width=8.8cm]{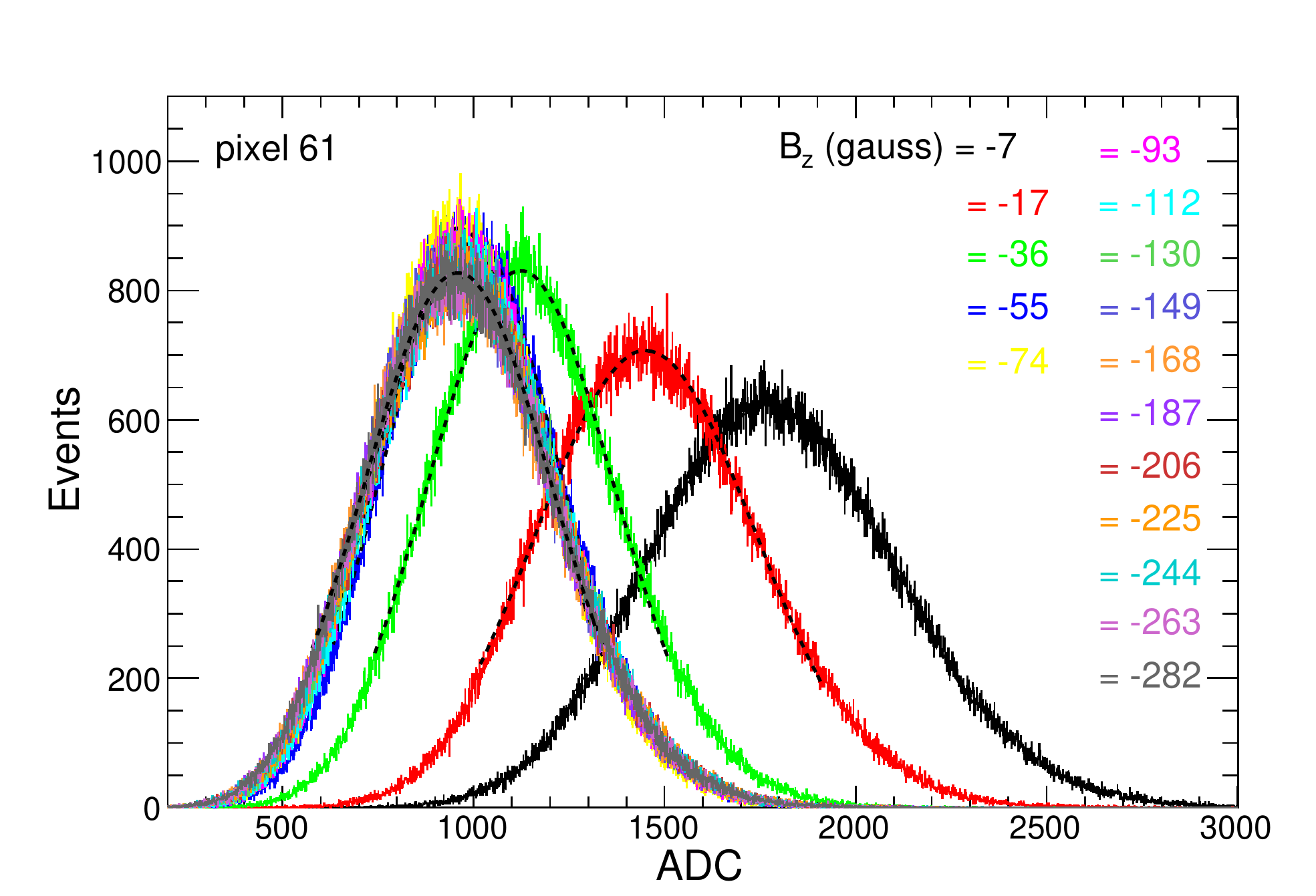}
&
\hspace{-0.4in}
\includegraphics[width=8.8cm]{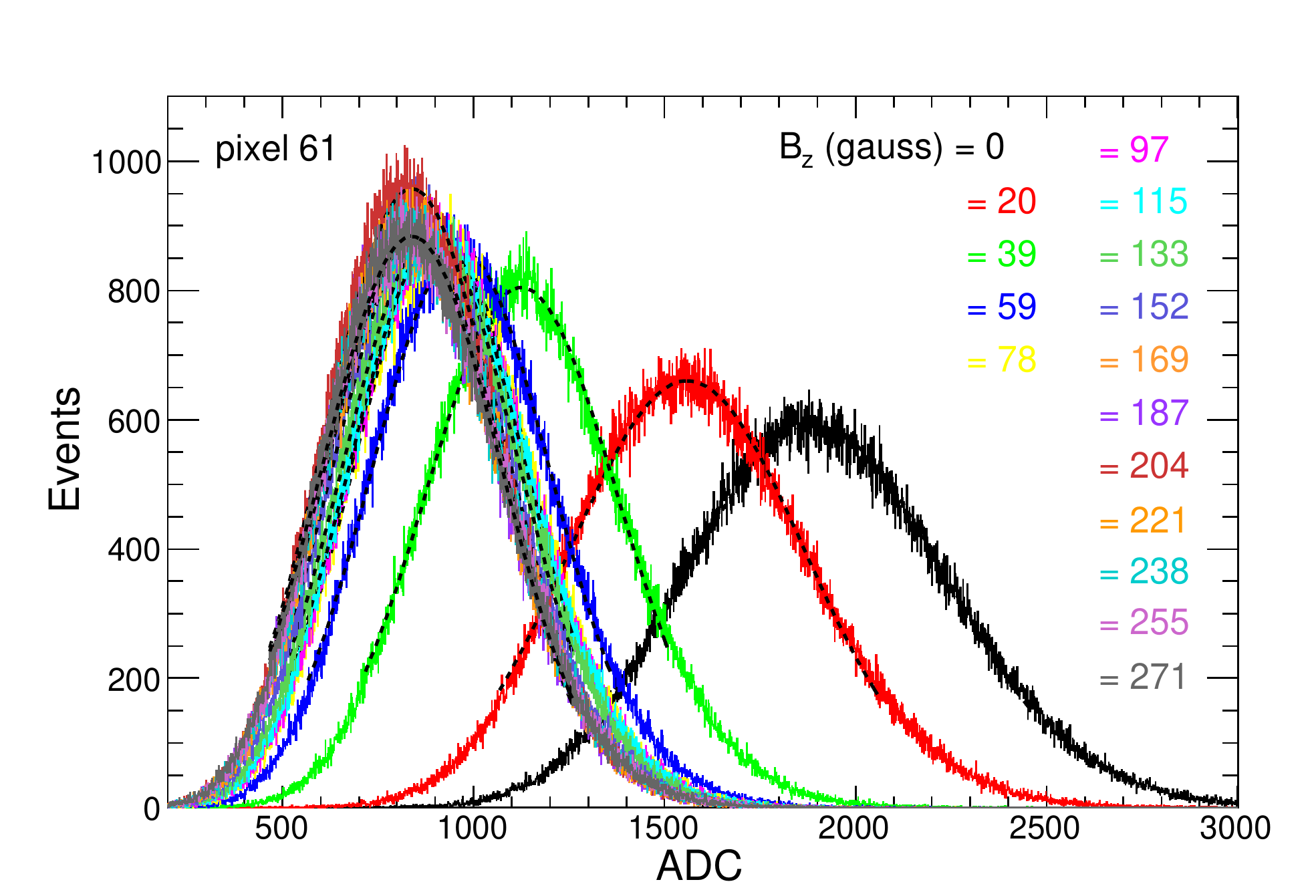} \\
\vspace{-0.09in}
\includegraphics[width=8.8cm]{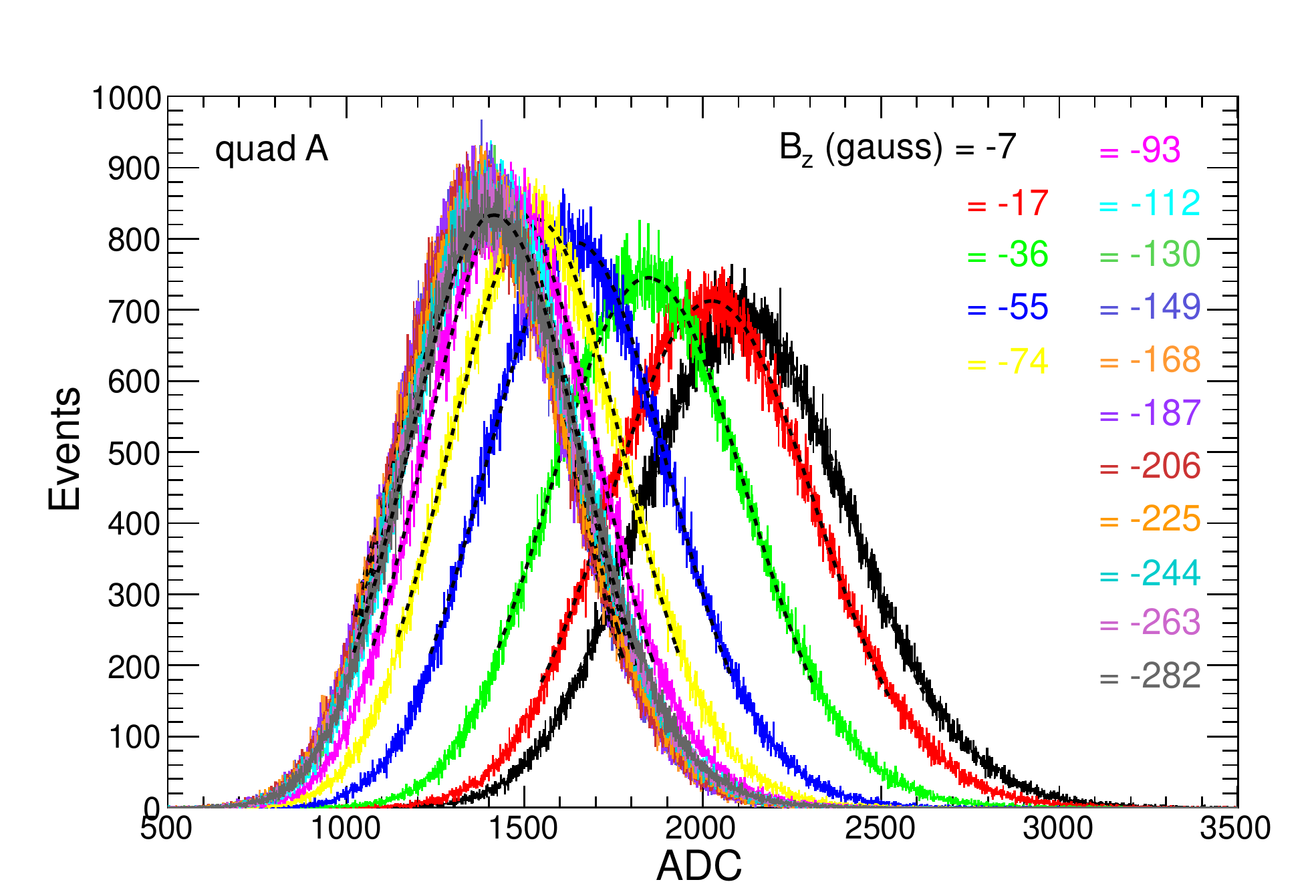}
&
\hspace{-0.4in}
\includegraphics[width=8.8cm]{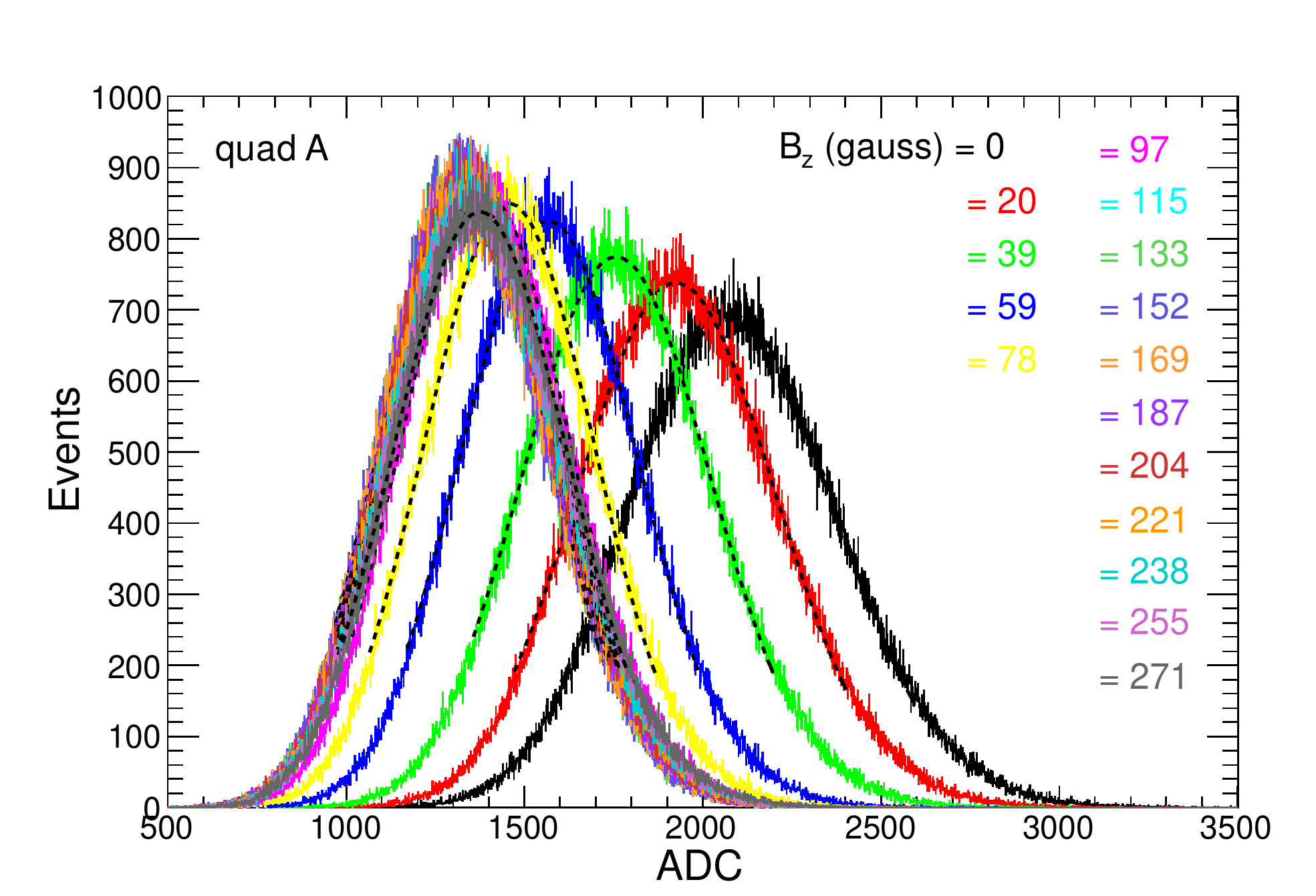} \\
\vspace{-0.09in}
\includegraphics[width=8.8cm]{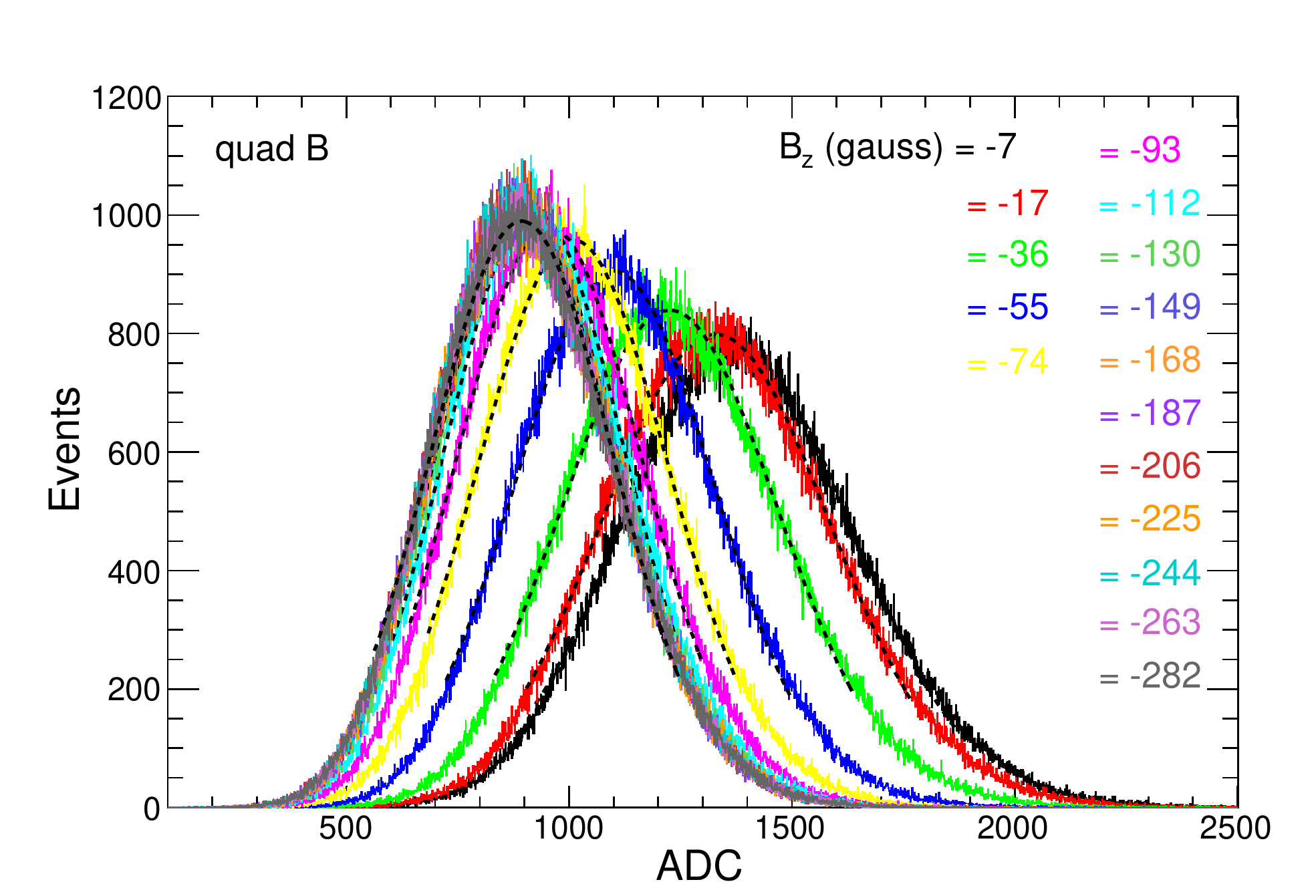}
&
\hspace{-0.4in}
\includegraphics[width=8.8cm]{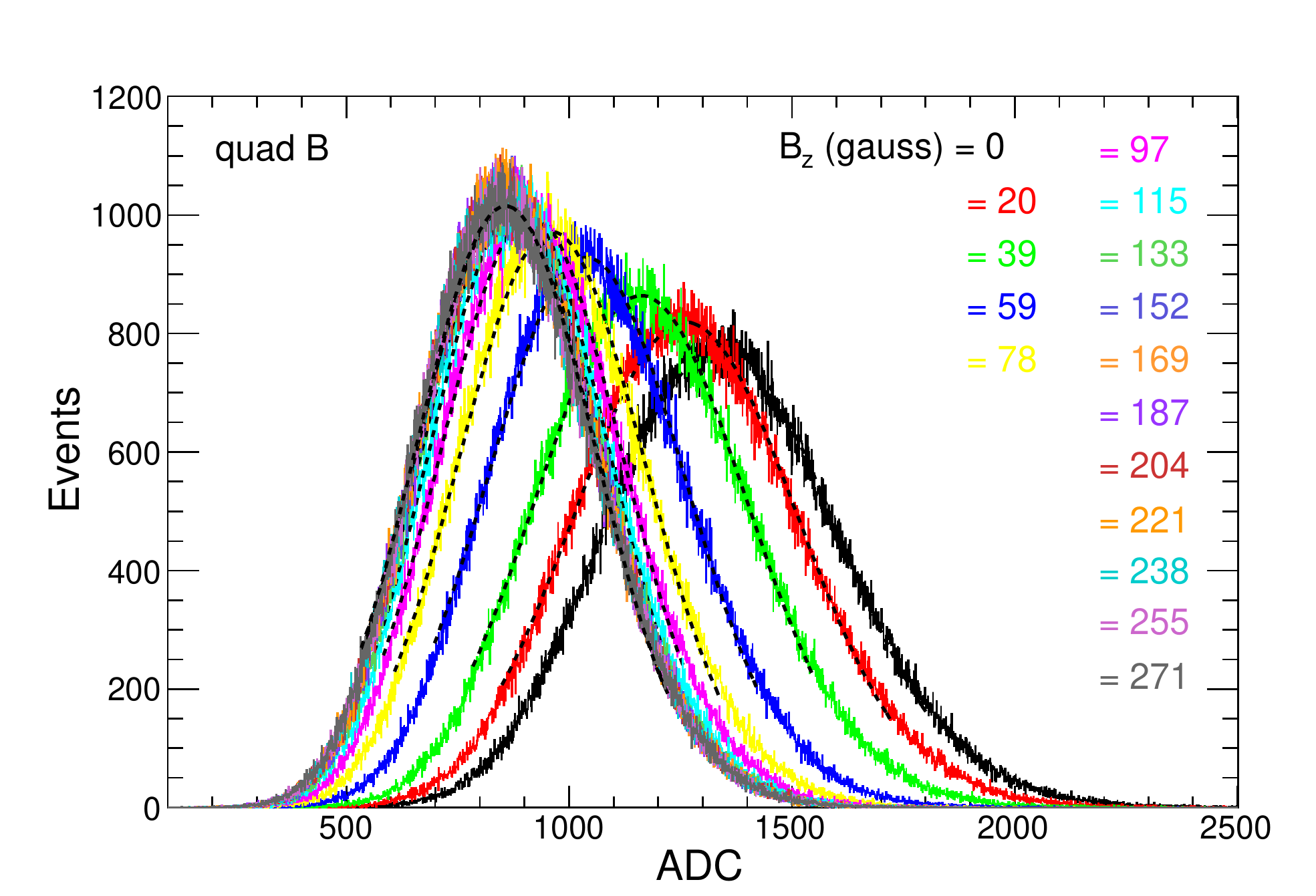}
\end{tabular}
\linespread{0.5}
\caption[]{
{Longitudinal $B_z$ magnetic field scan: ADC histograms for pixels 45 and 61 and for quads A and B.} }
\label{bz_many}
\end{figure}

\begin{figure}[htbp]
\vspace*{-0.1in}
\centering
\begin{tabular}{cc}
\vspace{-0.09in}
\includegraphics[width=8.8cm]{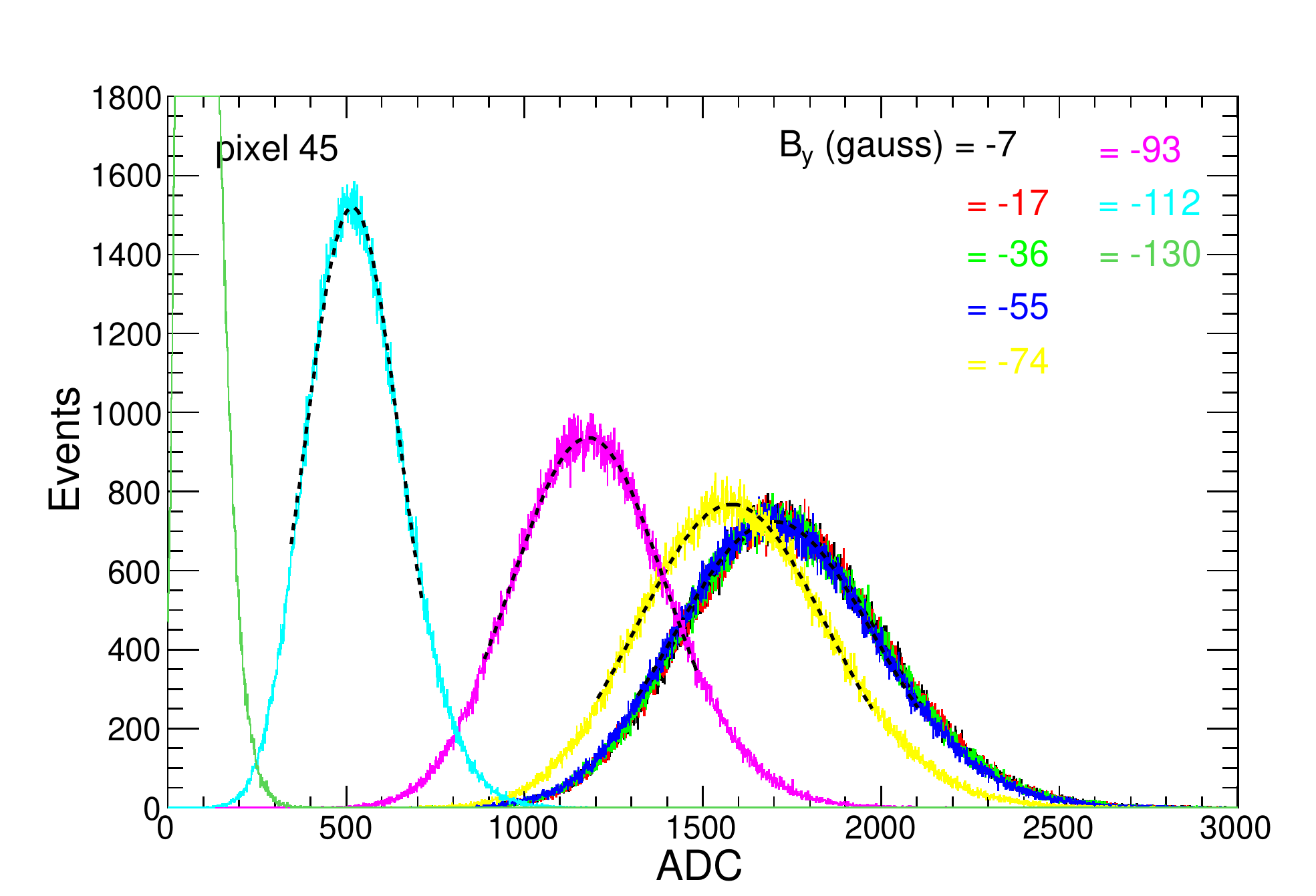}
&
\hspace{-0.4in}
\includegraphics[width=8.8cm]{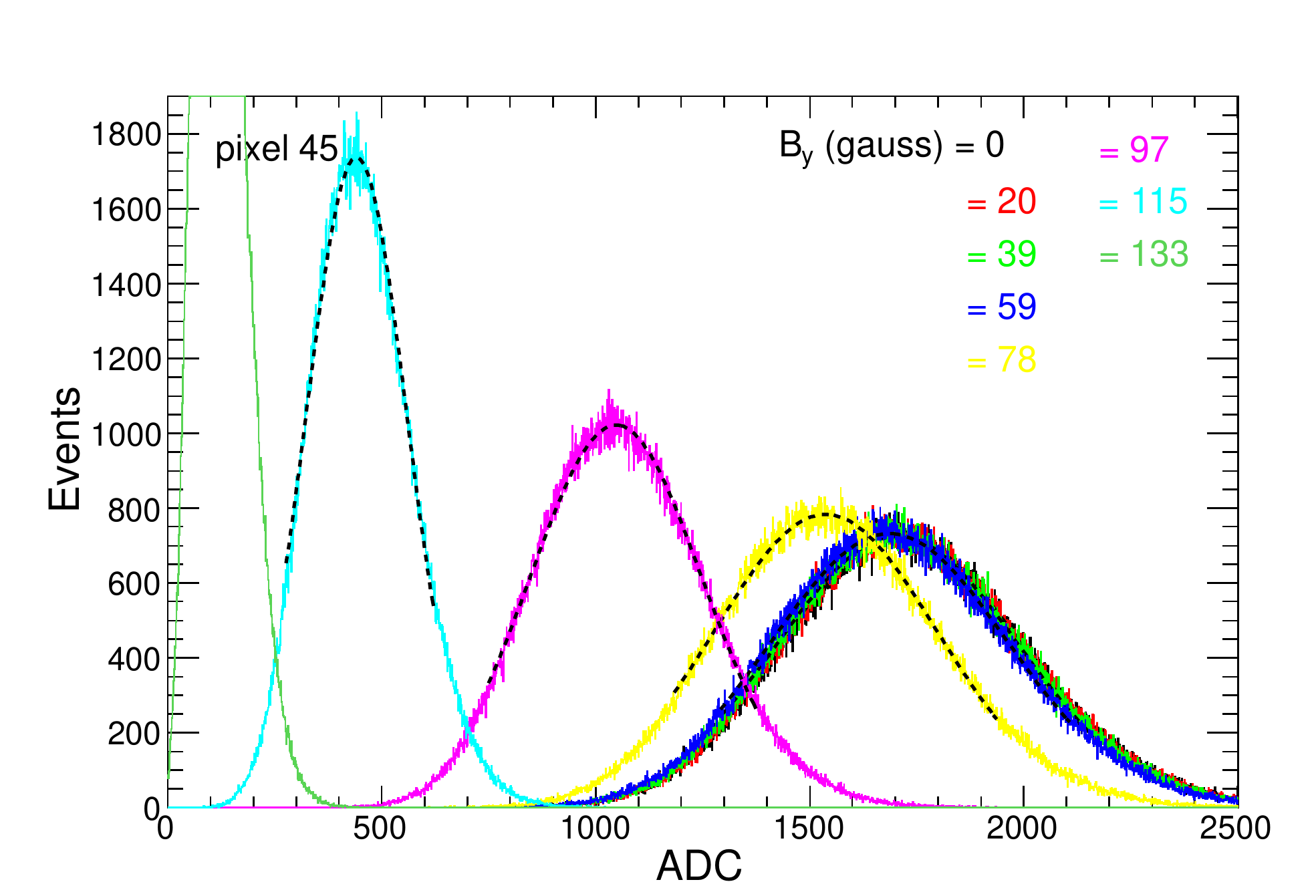} \\
\vspace{-0.09in}
\includegraphics[width=8.8cm]{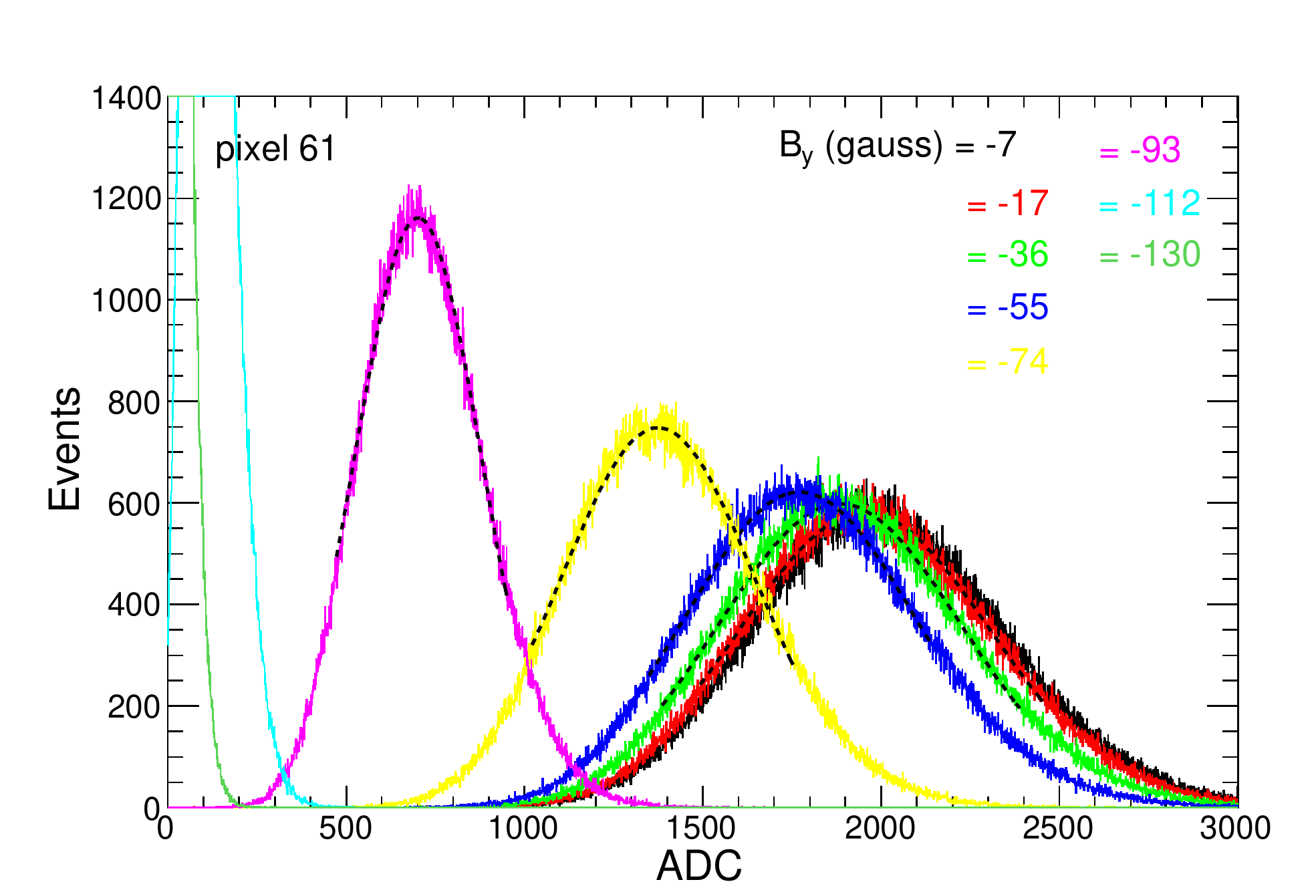}
&
\hspace{-0.4in}
\includegraphics[width=8.8cm]{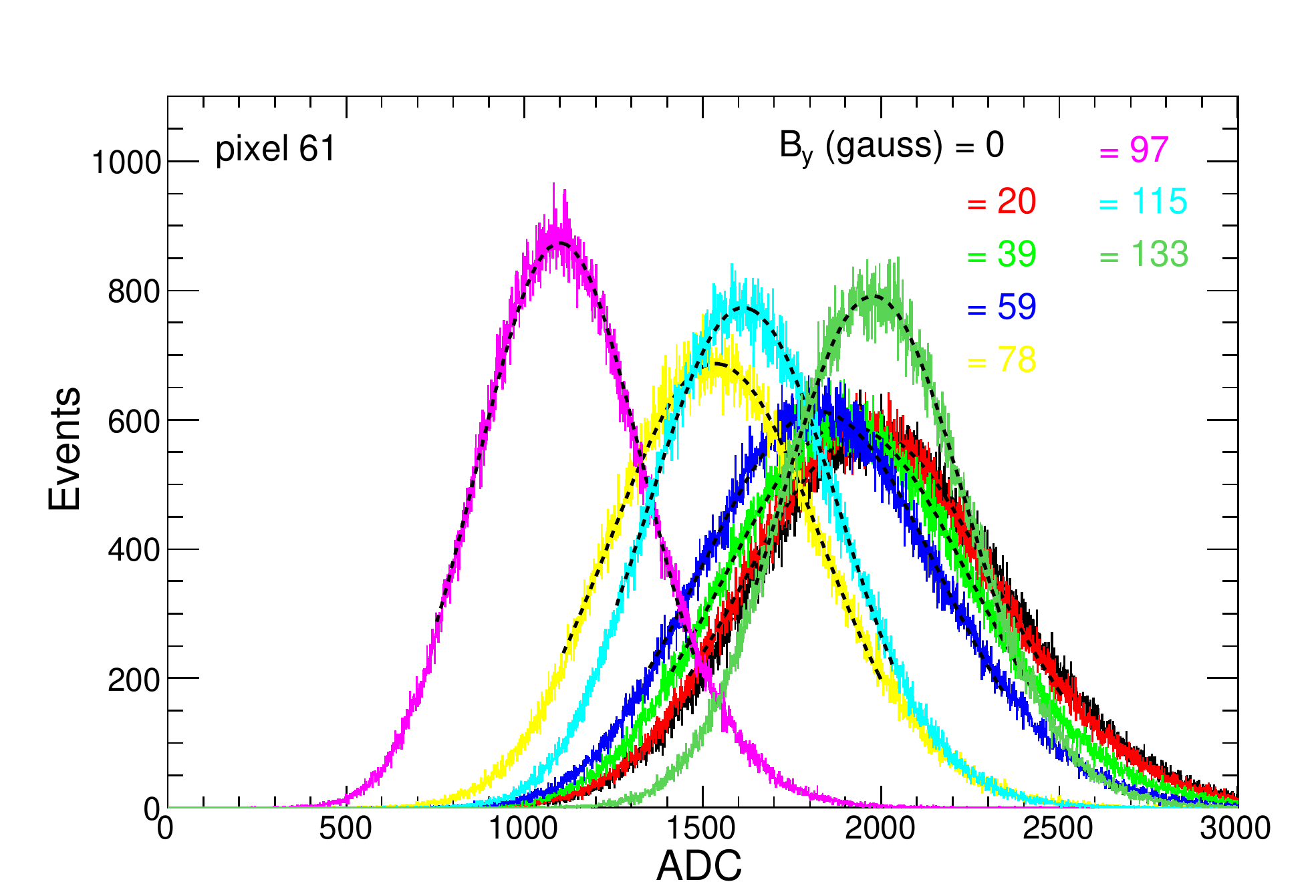} \\
\vspace{-0.09in}
\includegraphics[width=8.8cm]{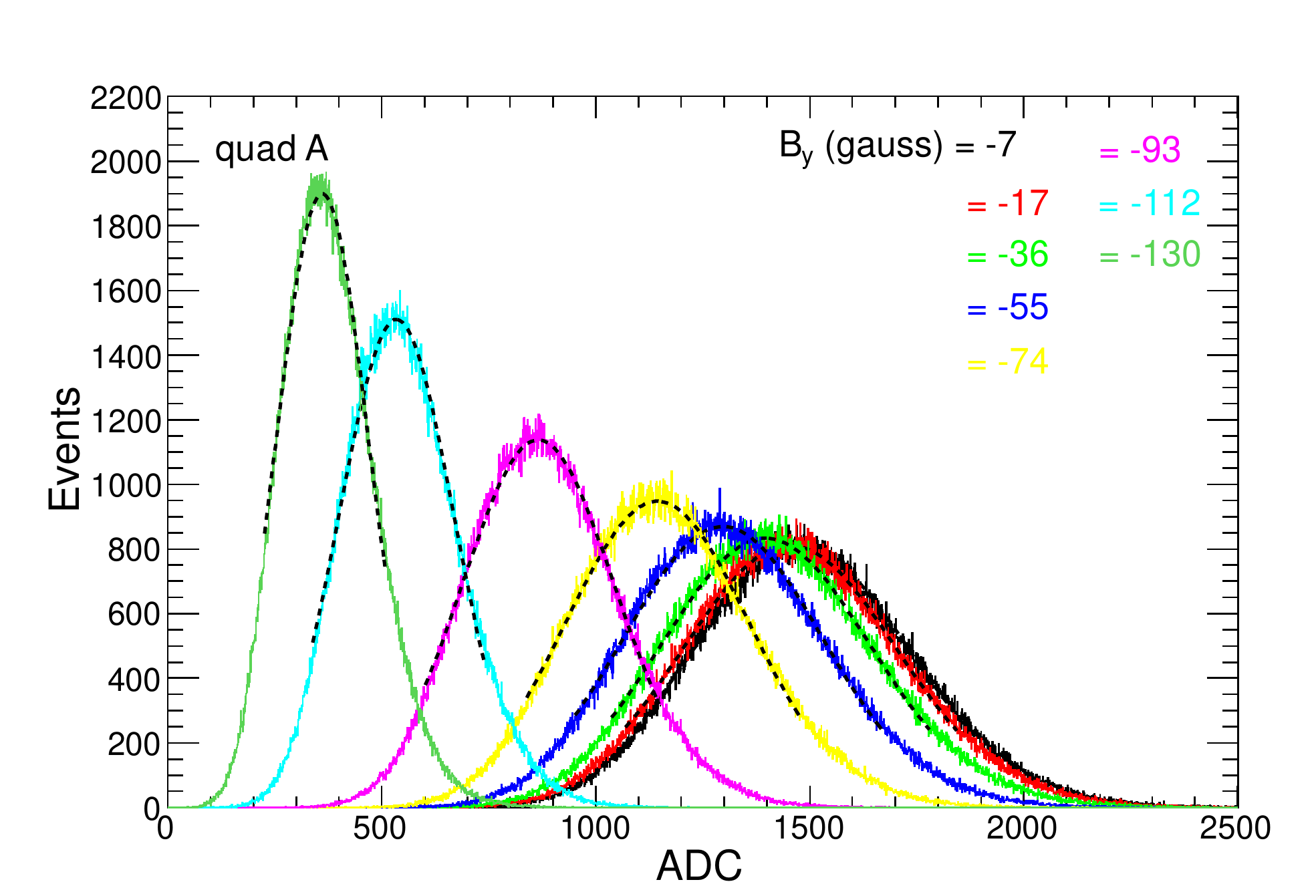}
&
\hspace{-0.4in}
\includegraphics[width=8.8cm]{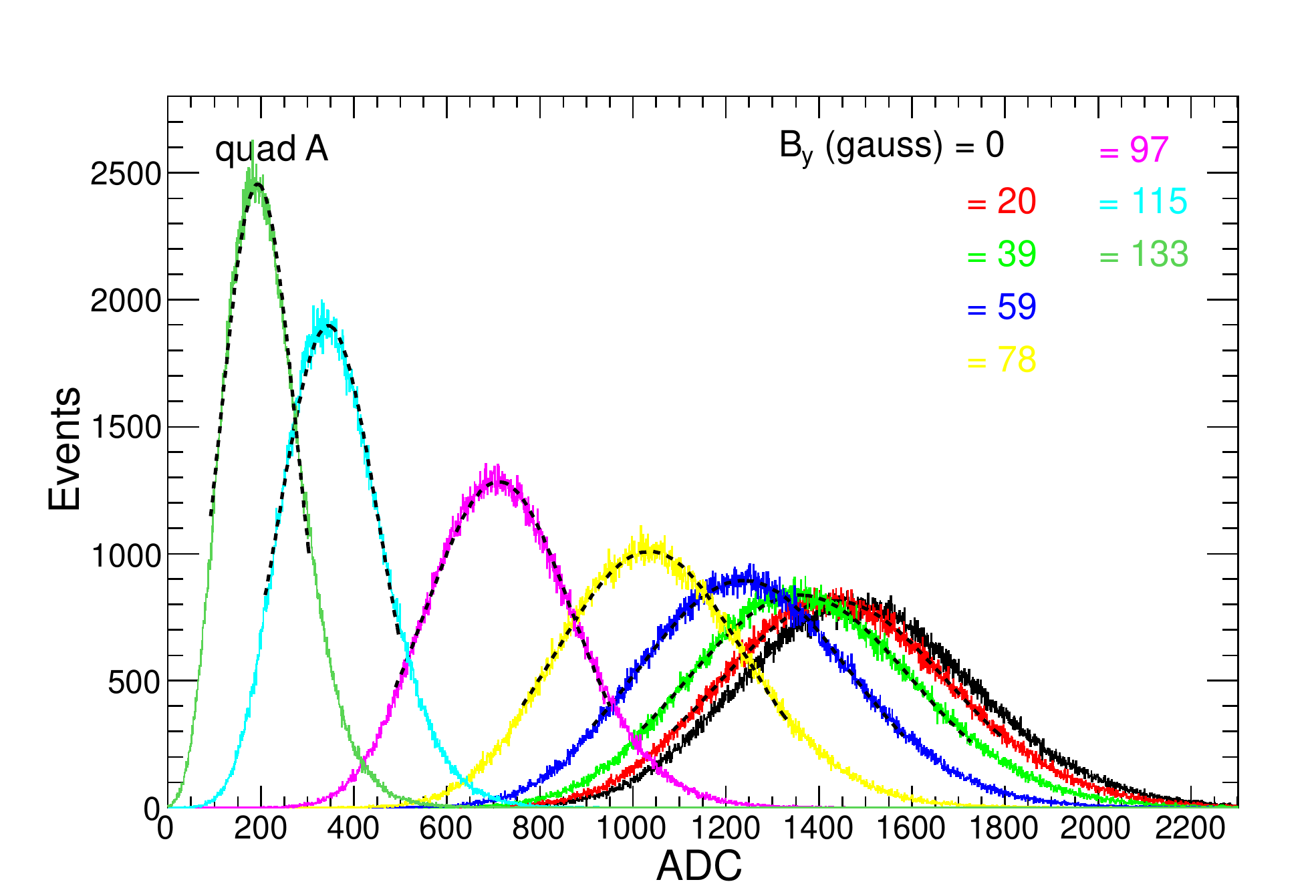} \\
\vspace{-0.09in}
\includegraphics[width=8.8cm]{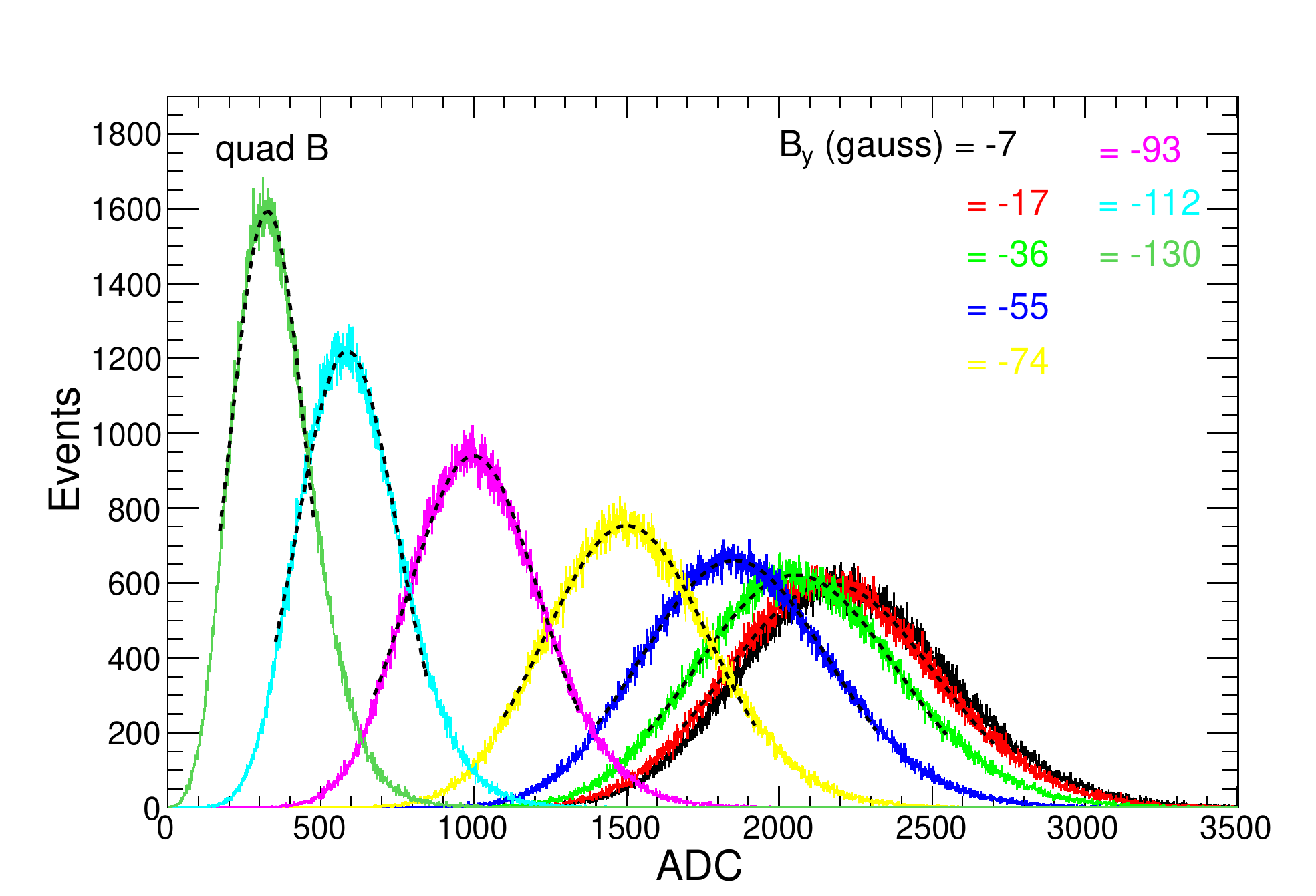}
&
\hspace{-0.4in}
\includegraphics[width=8.8cm]{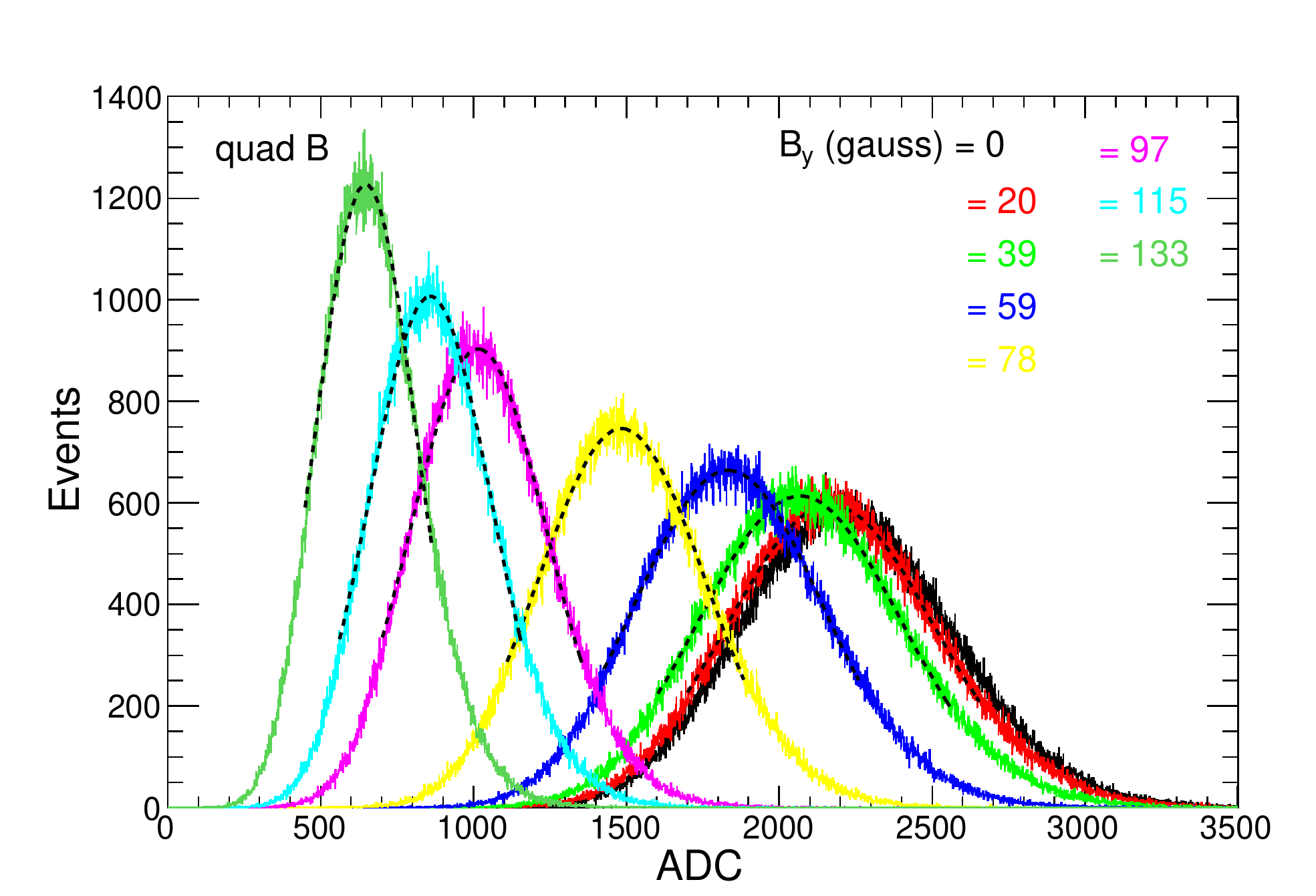}
\end{tabular}
\linespread{0.5}
\caption[]{
{Transverse $B_y$ magnetic field scan: ADC histograms for pixels 45 and 61 and for quads A and B.} }
\label{by_many}
\end{figure}

\begin{figure}[htbp]
\vspace*{-0.1in}
\centering
\begin{tabular}{cc}
\vspace{-0.09in}
\includegraphics[width=8.8cm]{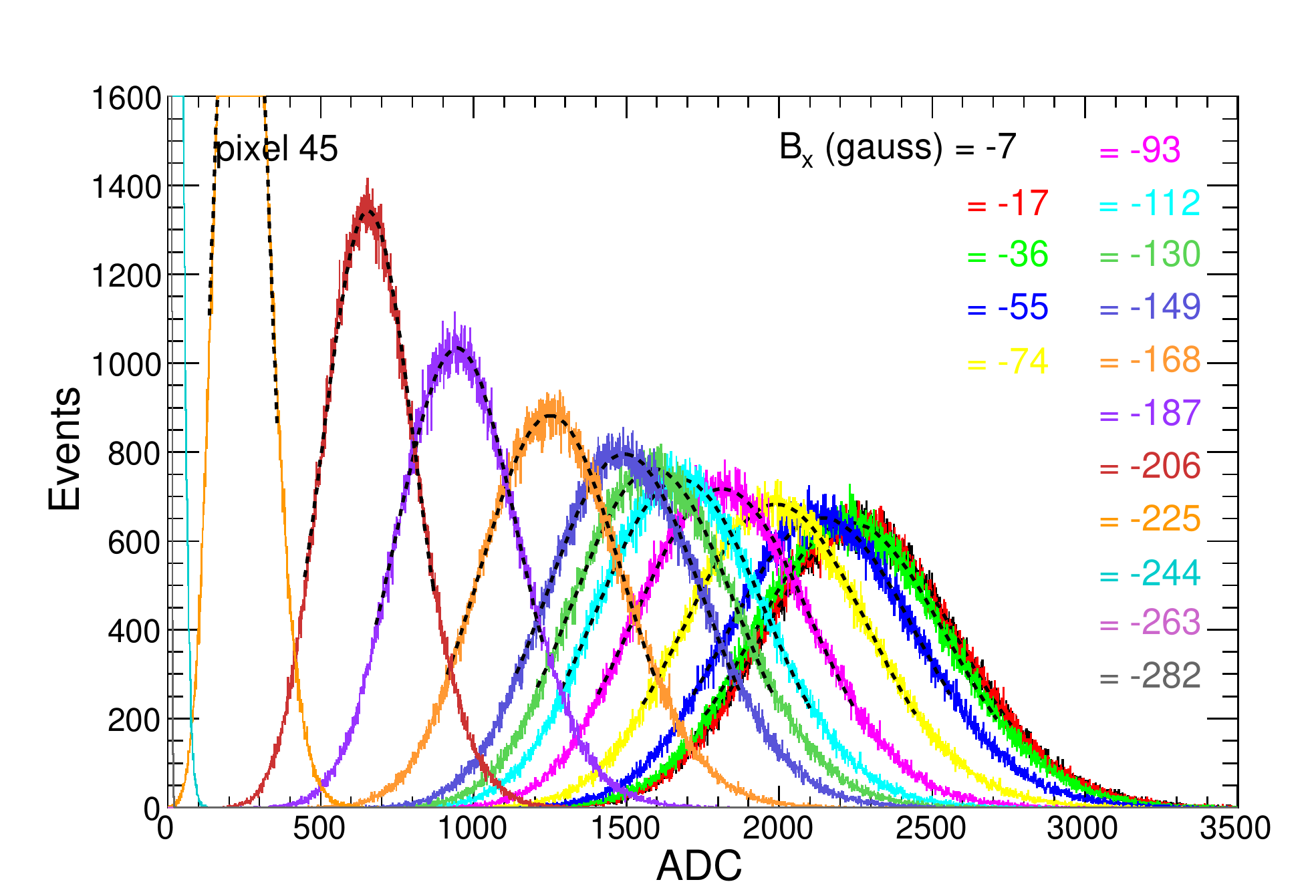}
&
\hspace{-0.4in}
\includegraphics[width=8.8cm]{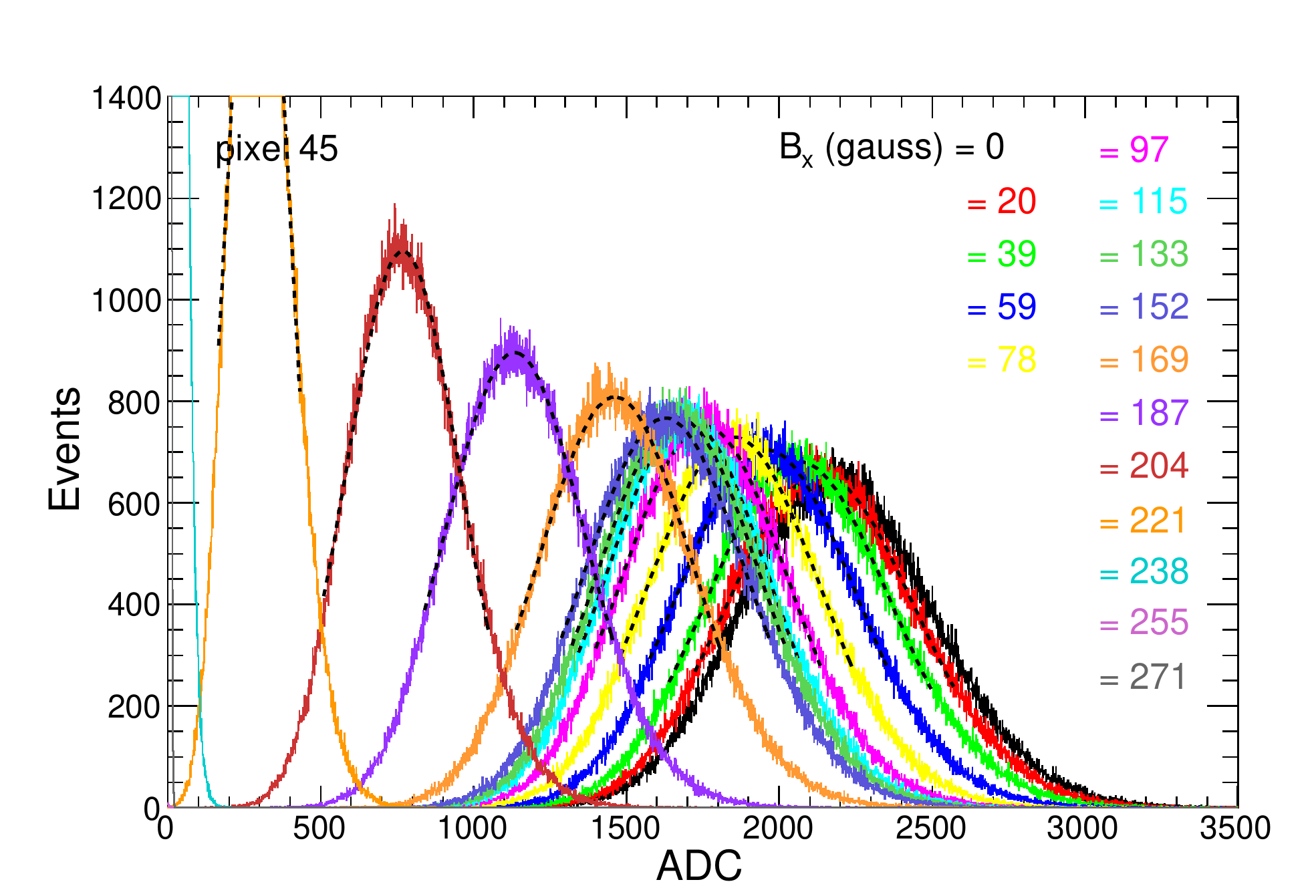} \\
\vspace{-0.09in}
\includegraphics[width=8.8cm]{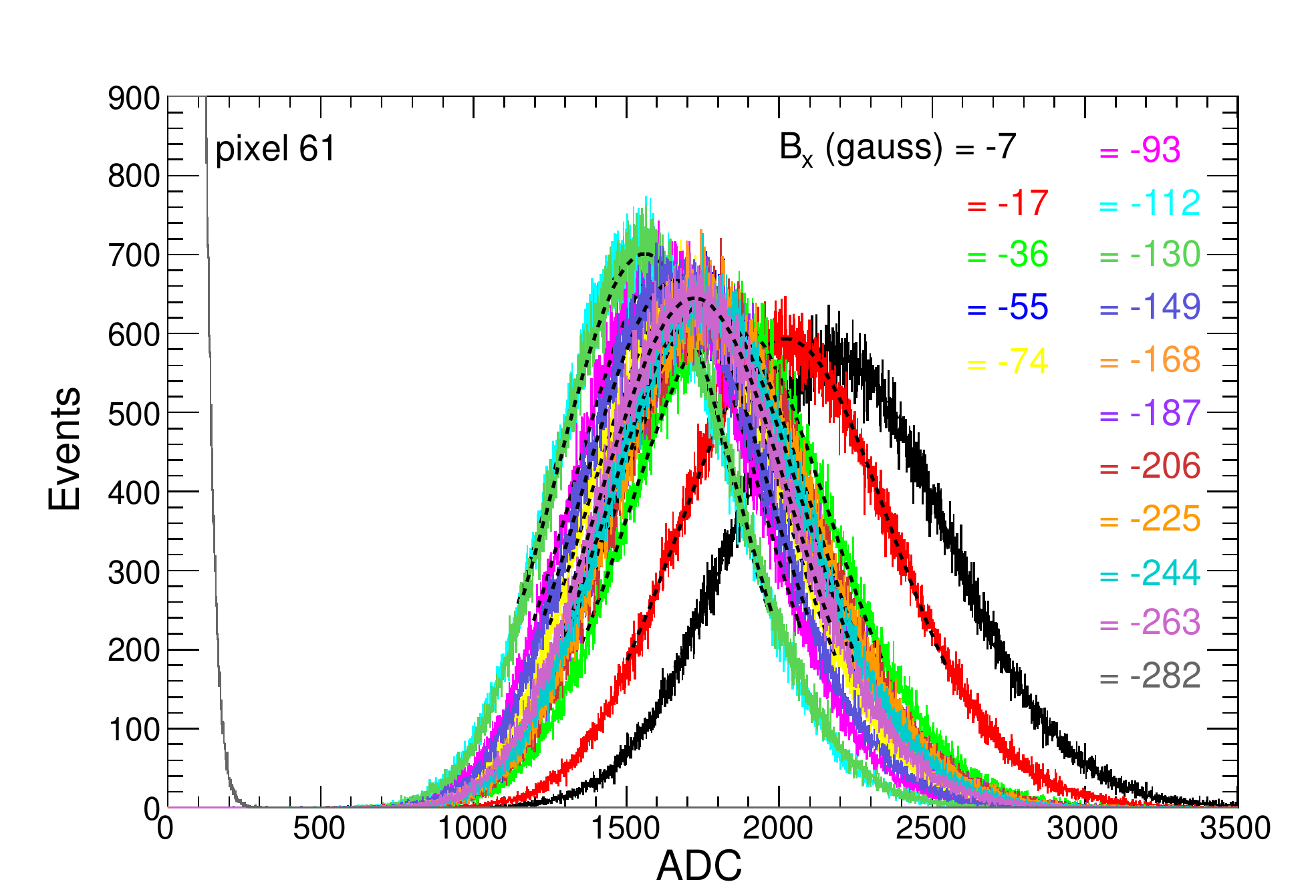}
&
\hspace{-0.4in}
\includegraphics[width=8.8cm]{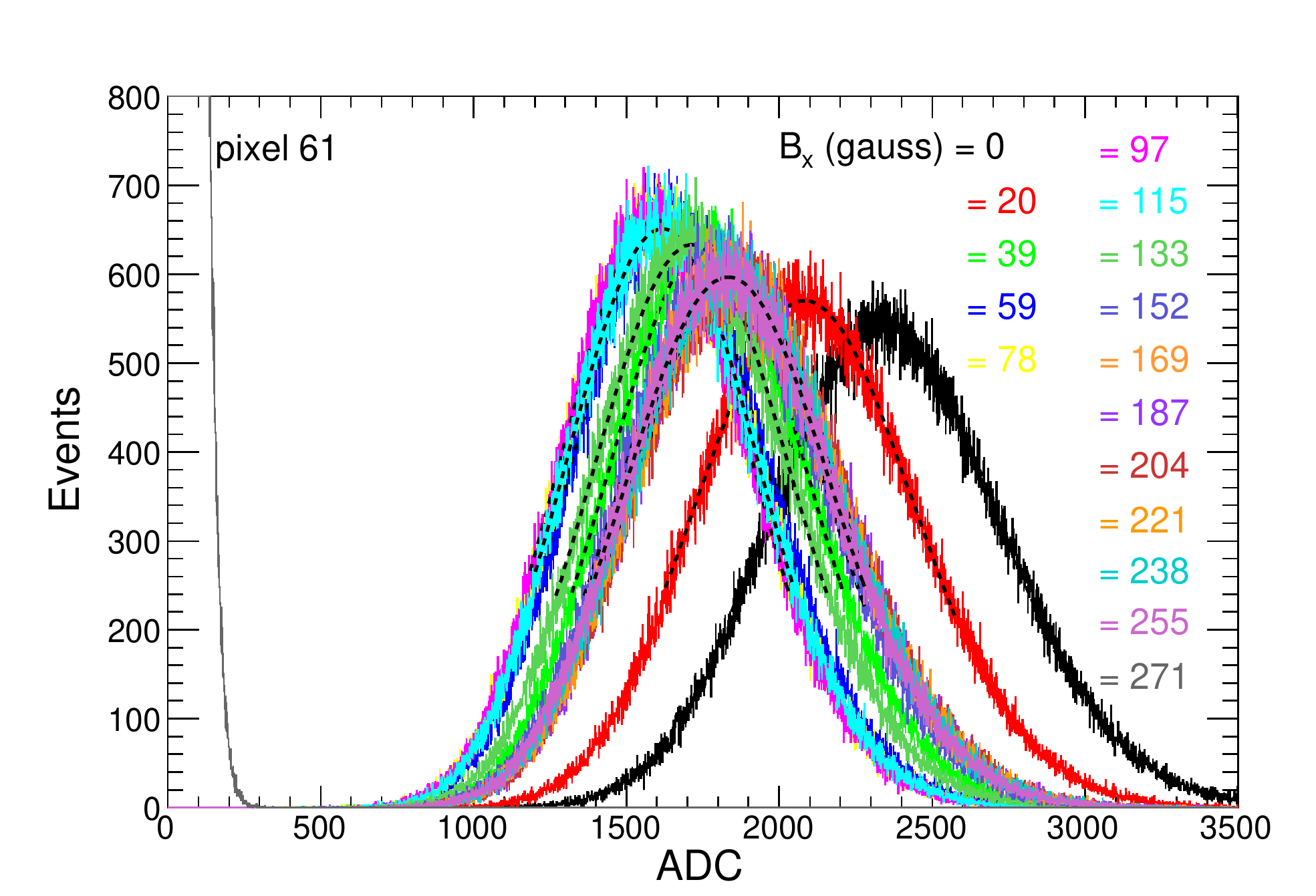} \\
\vspace{-0.09in}
\includegraphics[width=8.8cm]{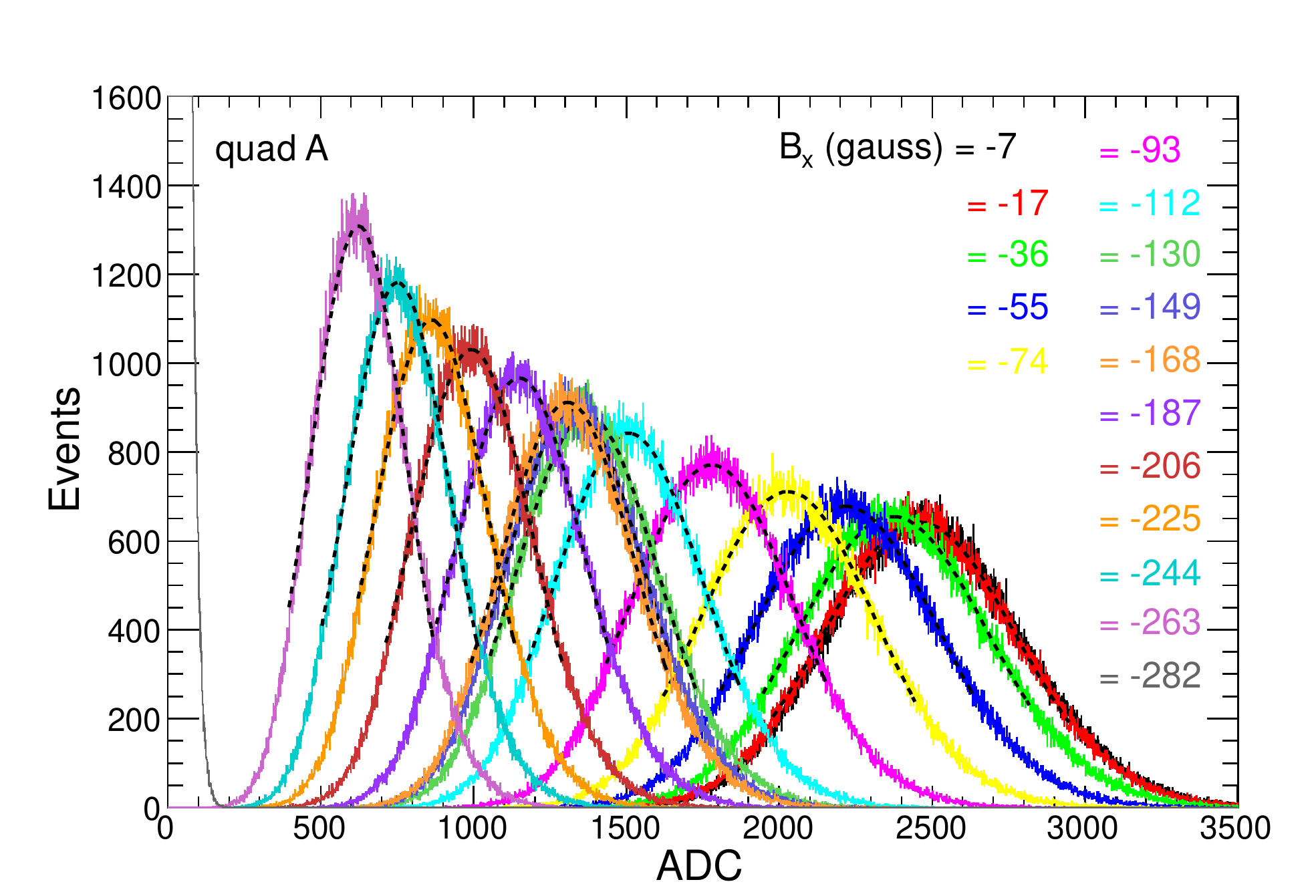}
&
\hspace{-0.4in}
\includegraphics[width=8.8cm]{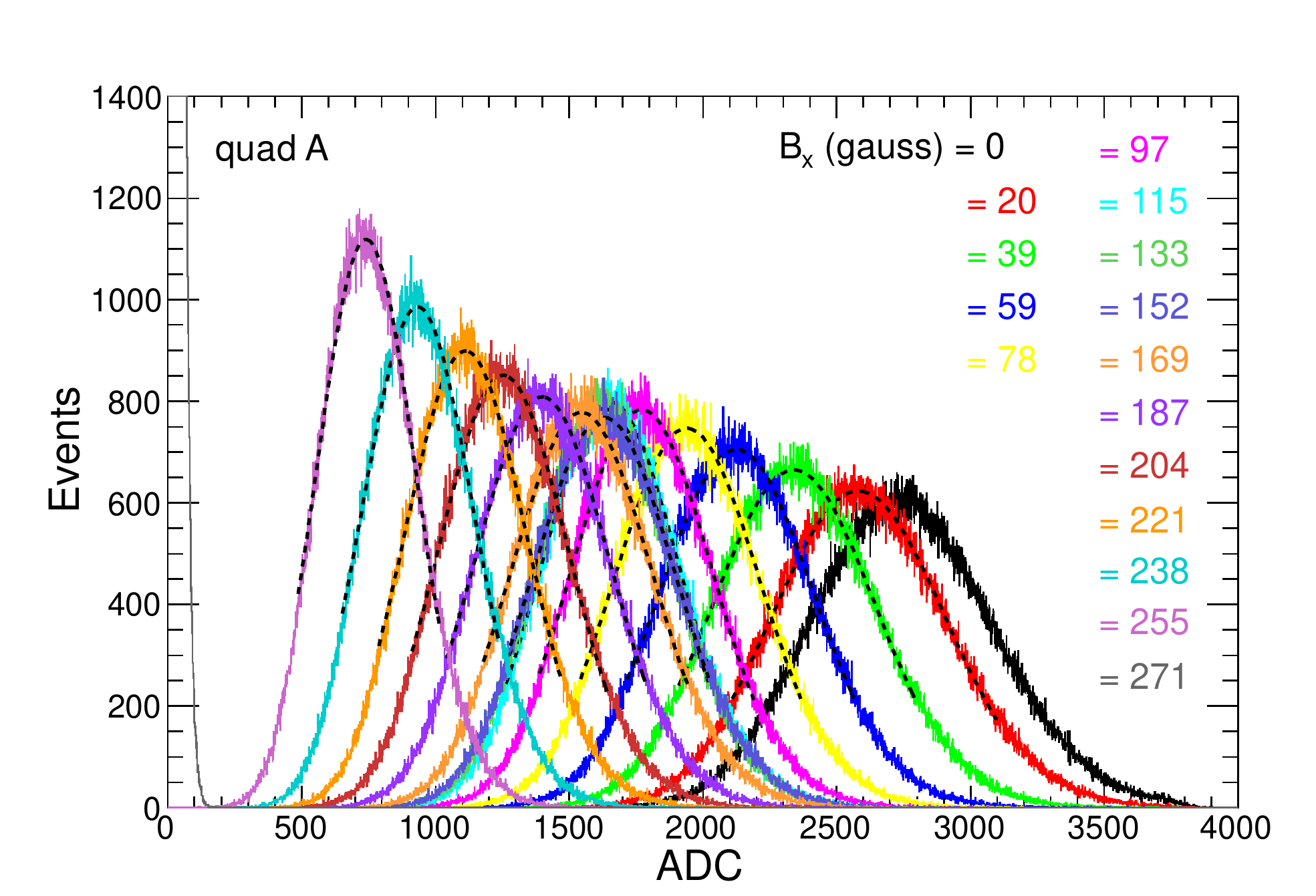} \\
\vspace{-0.09in}
\includegraphics[width=8.8cm]{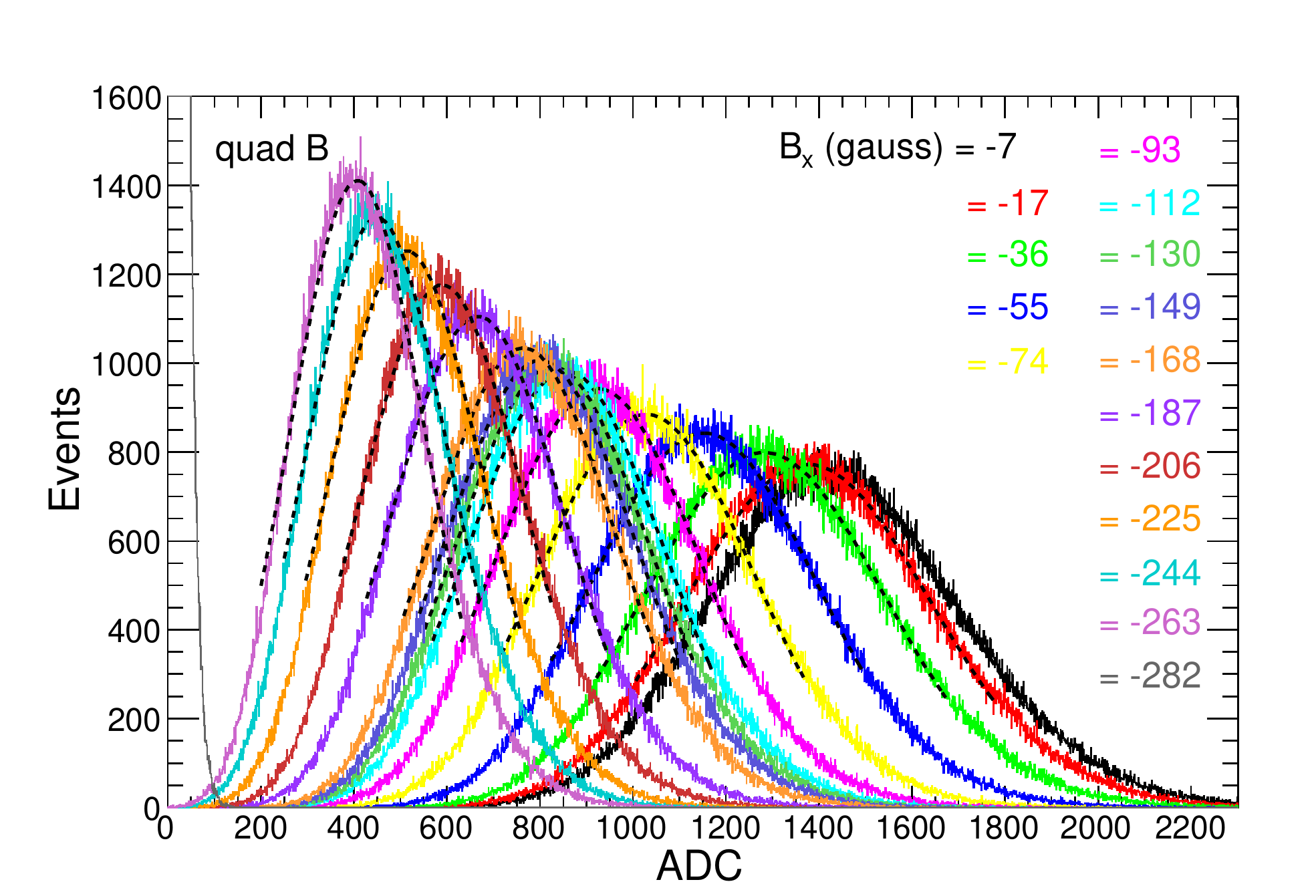}
&
\hspace{-0.4in}
\includegraphics[width=8.8cm]{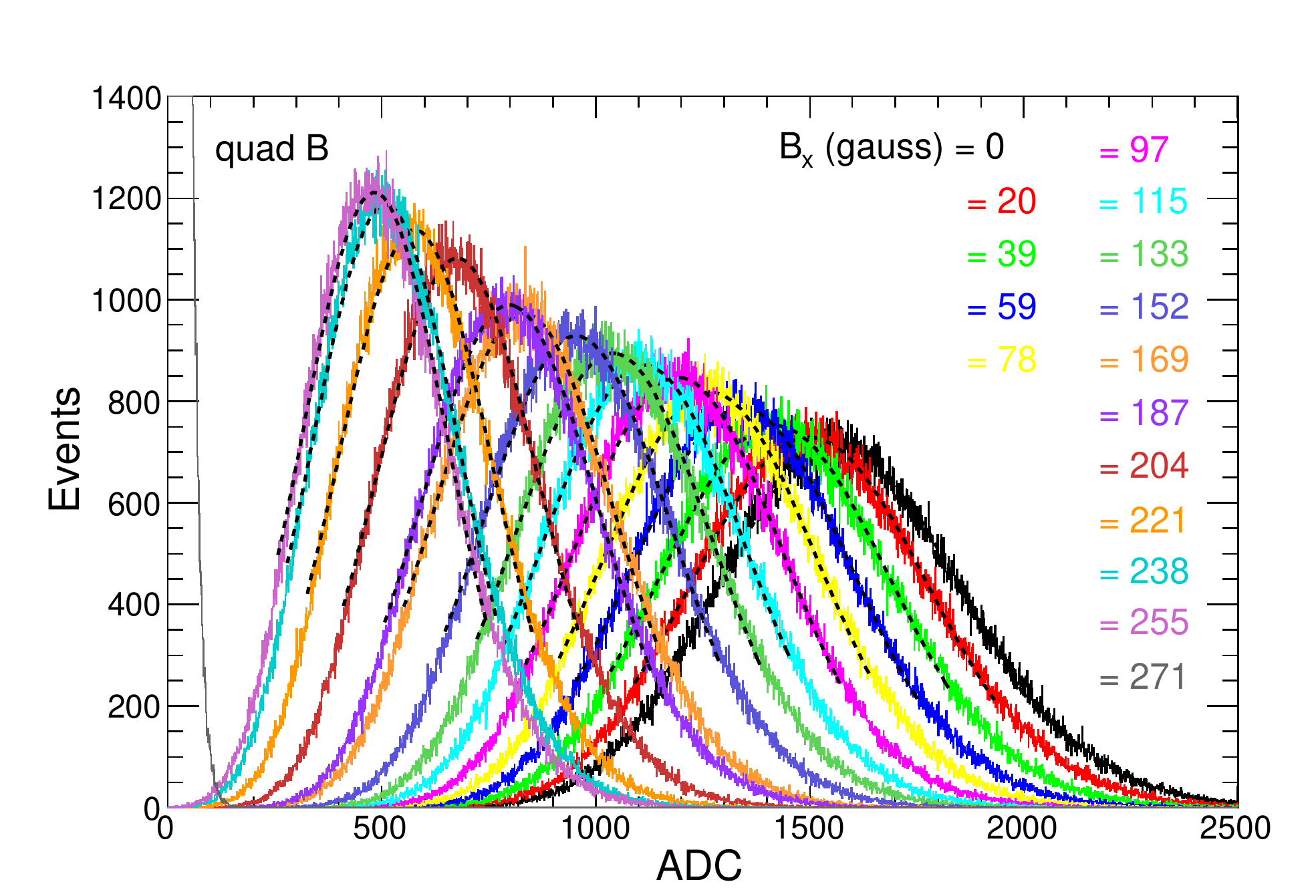}
\end{tabular}
\linespread{0.5}
\caption[]{
{Transverse $B_x$ magnetic field scan: ADC histograms for pixels 45 and 61 and for quads A and B.} }
\label{bx_many}
\end{figure}

In Figs.~\ref{bz_many}, \ref{by_many} and \ref{bx_many} we present the change in 
the output with magnetic field as 
recorded by the ADC from pixels 45 and 61 and from the two groups of 16 pixels, 
quad A and quad B. Each of the ADC distributions shown were recorded at a 
given field setting and have been pedestal subtracted. While there is a 
small signal degradation for 
the central pixel (45) with an increasing longitudinal $Bz$ field, as seen 
in Fig.~\ref{bz_many}, top left and right panels, the edge pixel (61) experiences 
greater losses as indicated by the more pronounced shift of the mean of the ADC 
distributions when going from the no field configuration (black curve) to a 
magnetic field close to 300 Gauss (gray curve) (second left and right panels). 
The quads' response to an increasing field can be viewed as an average over the 
central and edge pixels behavior. It is interesting to note that the signal 
progressively degrades with an increasing longitudinal magnetic field up to 
about 100 Gauss 
but then the degradation saturates and the output stays constant up to 
field magnitudes of $\approx$300 Gauss. This pattern is observed for both 
central and edge pixels as well as for quads.

The loss of signal in a transverse magnetic field $B_x$ and $B_y$ is more 
dramatic as seen in Figs.~\ref{bx_many} and \ref{by_many}. In particular, 
the PMT output degrades rapidly under a $B_y$ transverse field, 
being reduced to zero at $\approx$130 Gauss. A transverse $B_x$ field 
has a similar effect at $\approx$244 Gauss for the central pixel and 
at $\approx$282 Gauss for the edge pixel and quads. The variation of 
the edge pixel output with a transverse magnetic field follows a different 
pattern as seen in Figs.~\ref{bx_many} and \ref{by_many}, second left and right 
panels.

In Fig.~\ref{summary_many} we summarize our studies of a longitudinal 
and transverse magnetic field effect on large outputs from central and 
edge pixels and groups of pixels. To quantify the effects 
we fitted each of the pedestal subtracted ADC distributions 
with a Gaussian function as shown in 
Figs.~\ref{bz_many}, \ref{by_many} and \ref{bx_many}. For each magnetic 
field setting the mean of the ADC distribution thus obtained has then 
been normalized to the no-field value. The results for a longitudinal 
magnetic field are shown in Fig.~\ref{summary_many}, first panel. The same 
pattern emerges when looking at the variation of the output with field from 
central and edge pixels and from quads. There is only a difference in 
the magnitude of the field induced losses, the edge pixel 
experiencing the largest reduction, at most 55\%, and the central pixel 
the smallest, at most 15\%. If the quad configuration were to be used in production, 
then a 10\% signal loss would be expected at 30 Gauss. For applications like 
SoLID where the PMT is placed in a high magnetic field environment, 
up to 200 Gauss, shielding that would reduce the longitudinal field from 
hundreds of Gauss to 30 Gauss should be sufficient to ensure a 
proper functionality of the system. In practice, the longitudinal component of the field 
is typically the most difficult to shield, so a PMT like H8500C-03 which experiences only 
small output losses up to longitudinal fields of the order of tens of Gauss 
makes more practical and cost-effective shielding designs possible. 
The bottom panels summarize the signal 
reduction in transverse magnetic field $B_x$ and $B_y$, respectively. Here 
the losses can be very large even at modest field values and very good shielding  
against transverse fields would be necessary for a good functionality of the PMT.

\subsection{Effects of Magnetic Field on Single Photoelectron Signals}

\begin{figure}[htbp]
\vspace*{-0.1in}
\centering
\begin{tabular}{cc}
\includegraphics[width=9.3cm]{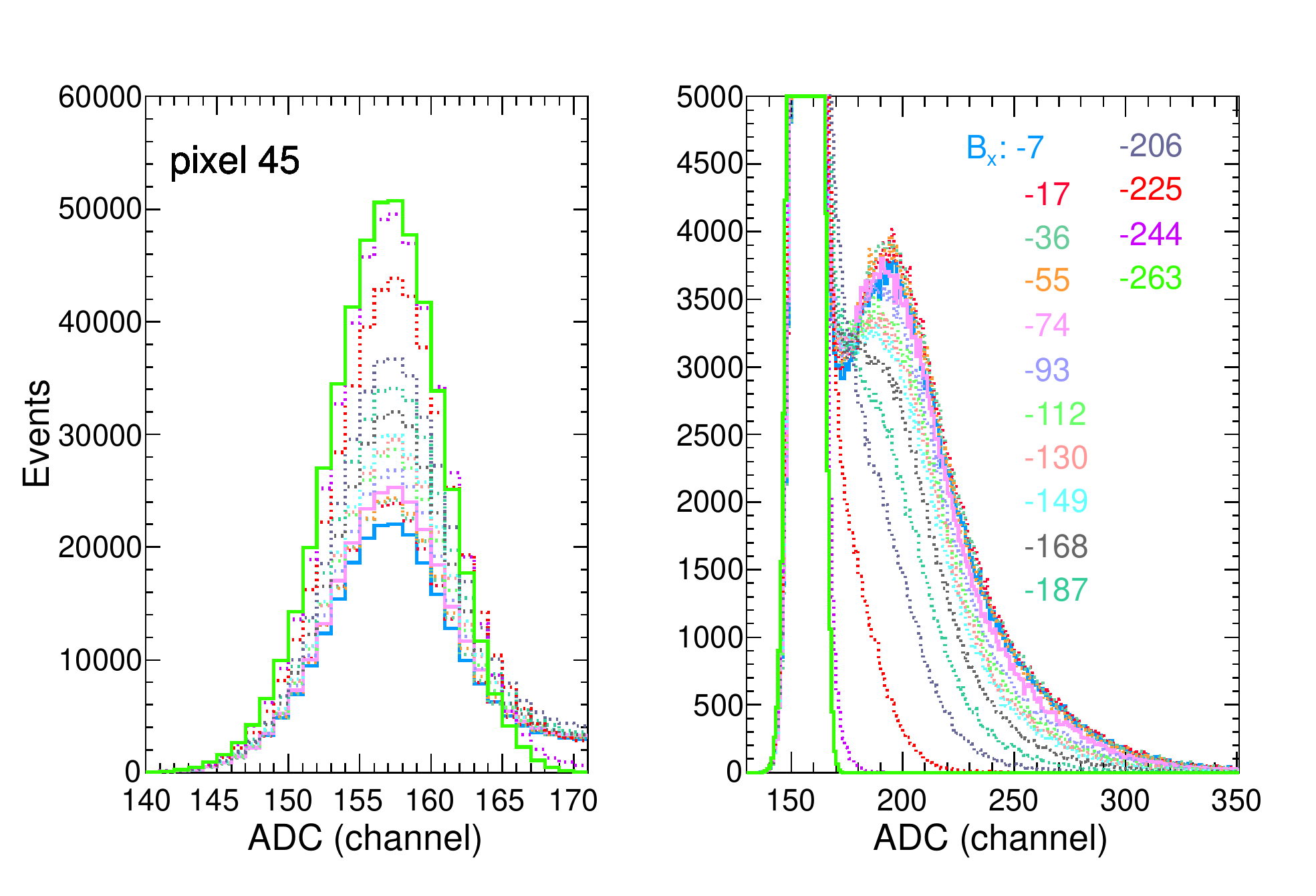}
&
\hspace{-0.4in}
\includegraphics[width=9.3cm]{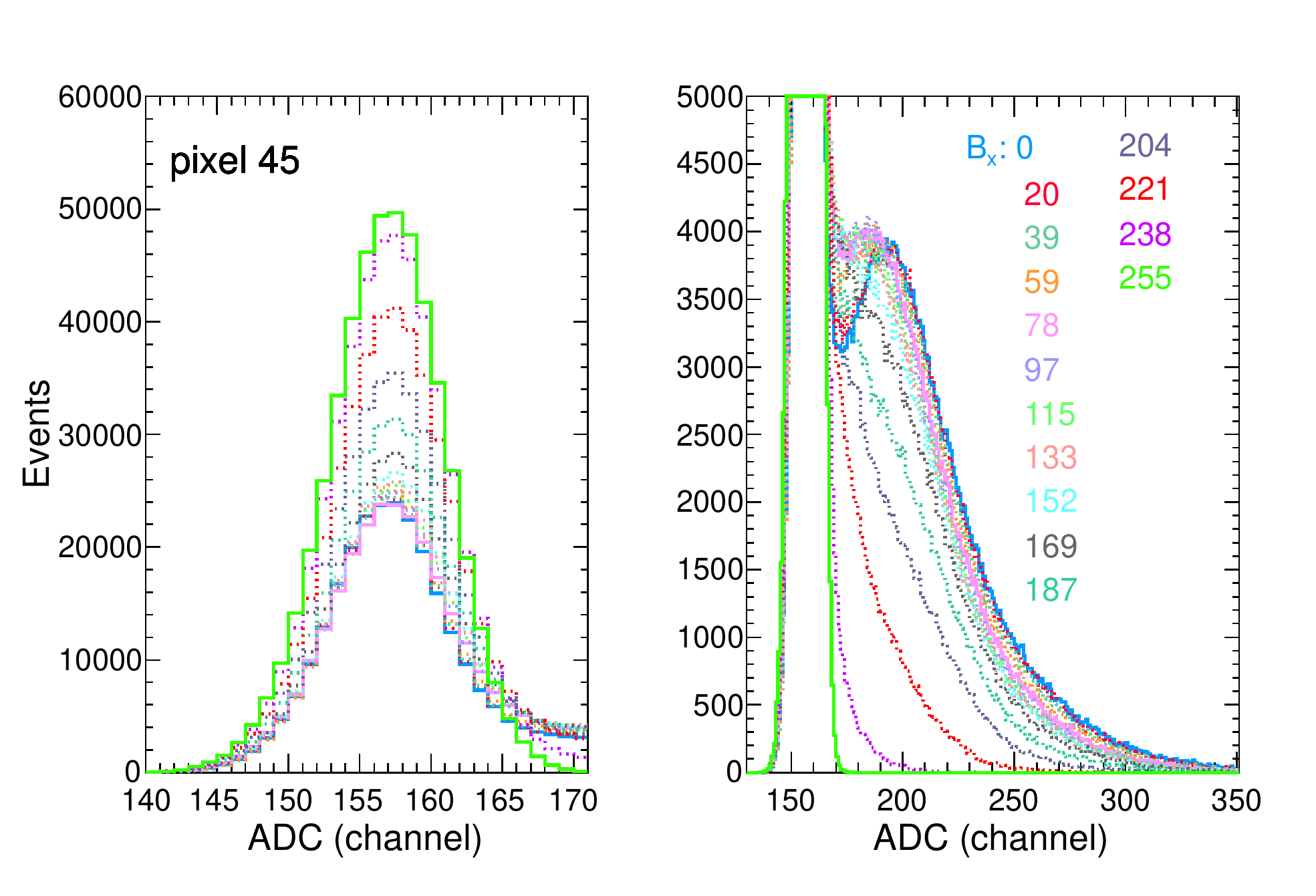} \\
\includegraphics[width=9.3cm]{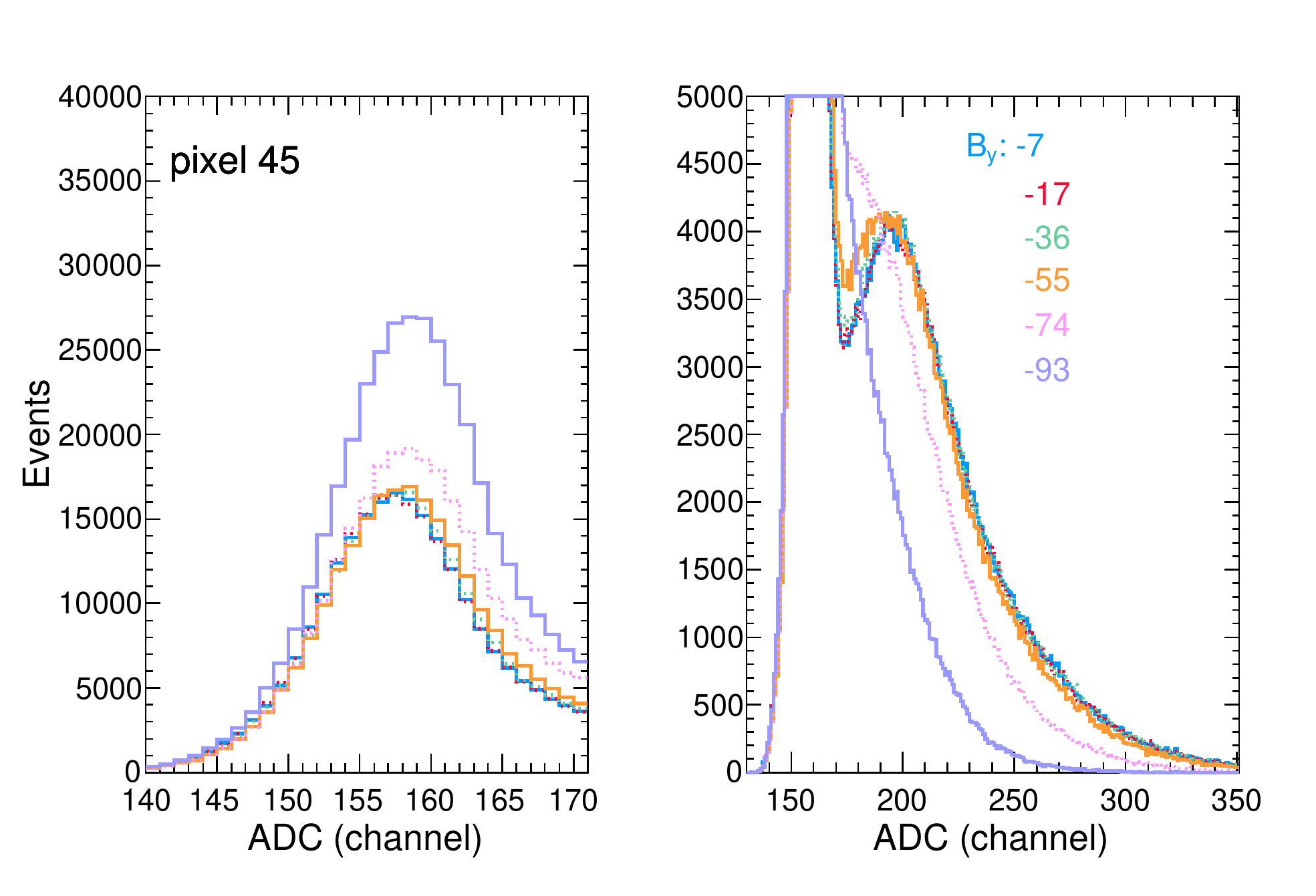}
&
\hspace{-0.4in}
\includegraphics[width=9.3cm]{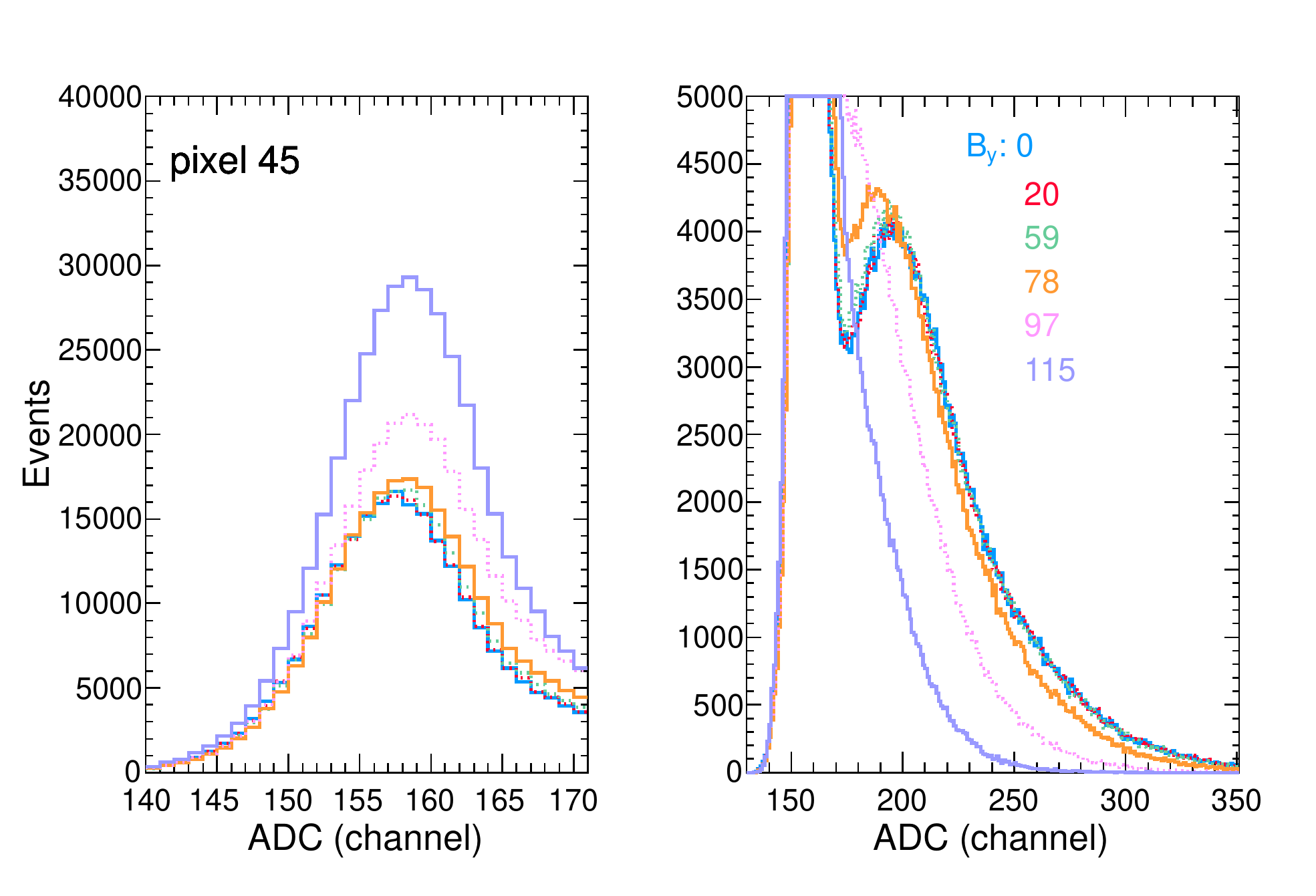} \\
\includegraphics[width=9.3cm]{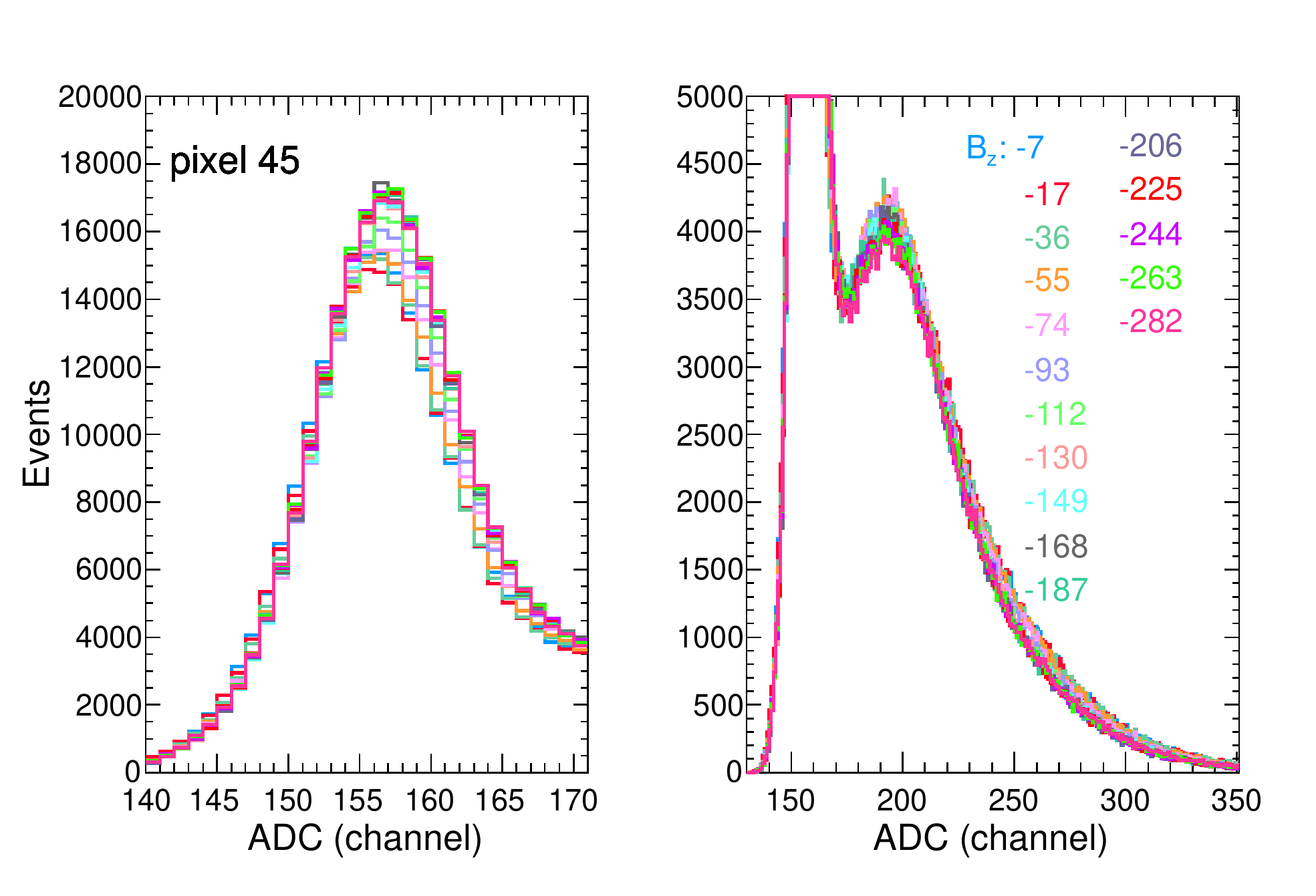}
&
\hspace{-0.4in}
\includegraphics[width=9.3cm]{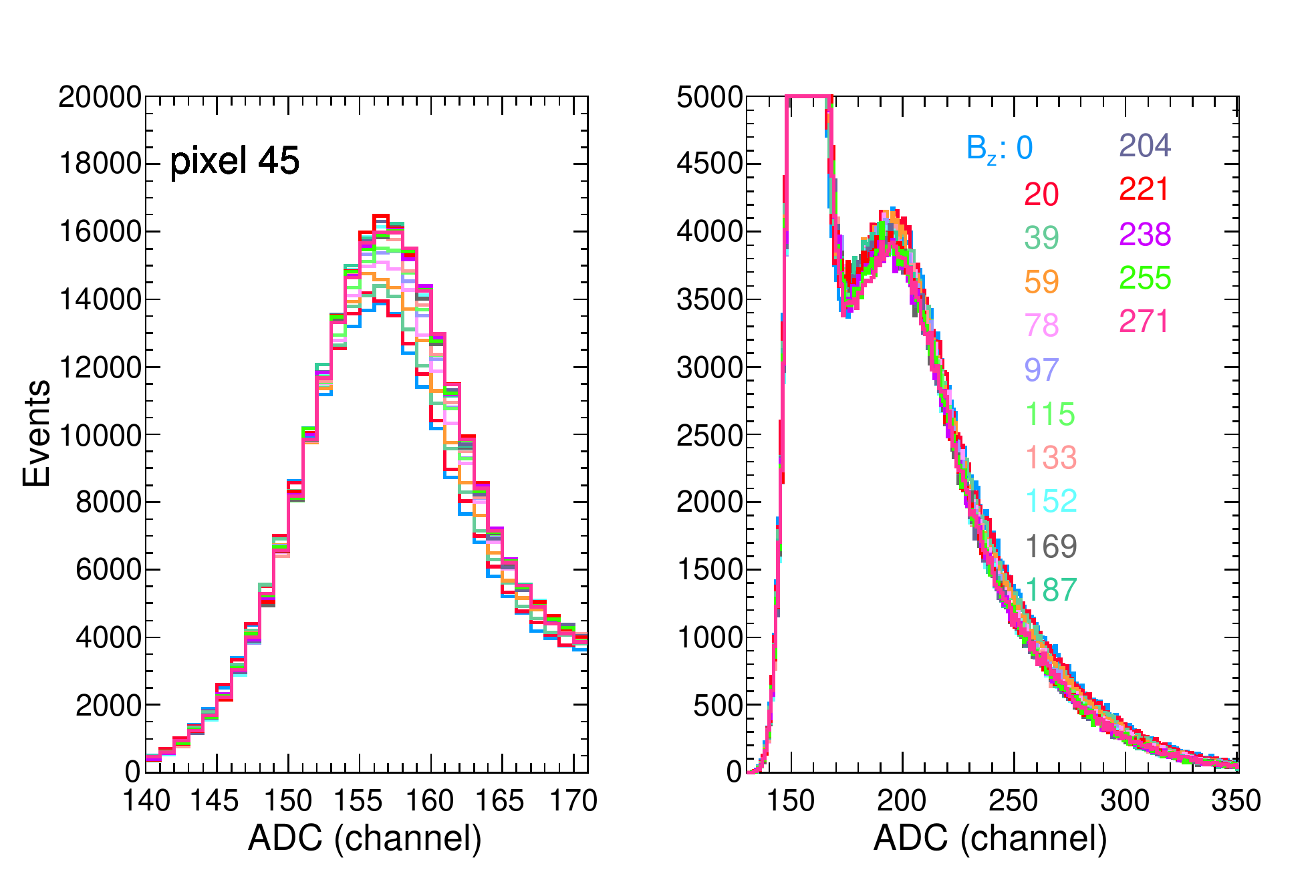} \\
\end{tabular}
\linespread{0.5}
\caption[]{
{Impact of a longitudinal and transverse magnetic field on single photoelectron outputs from 
pixel 45.} }
\label{45_one}
\end{figure}

\begin{figure}[htbp]
\vspace*{-0.1in}
\centering
\begin{tabular}{cc}
\includegraphics[width=9.3cm]{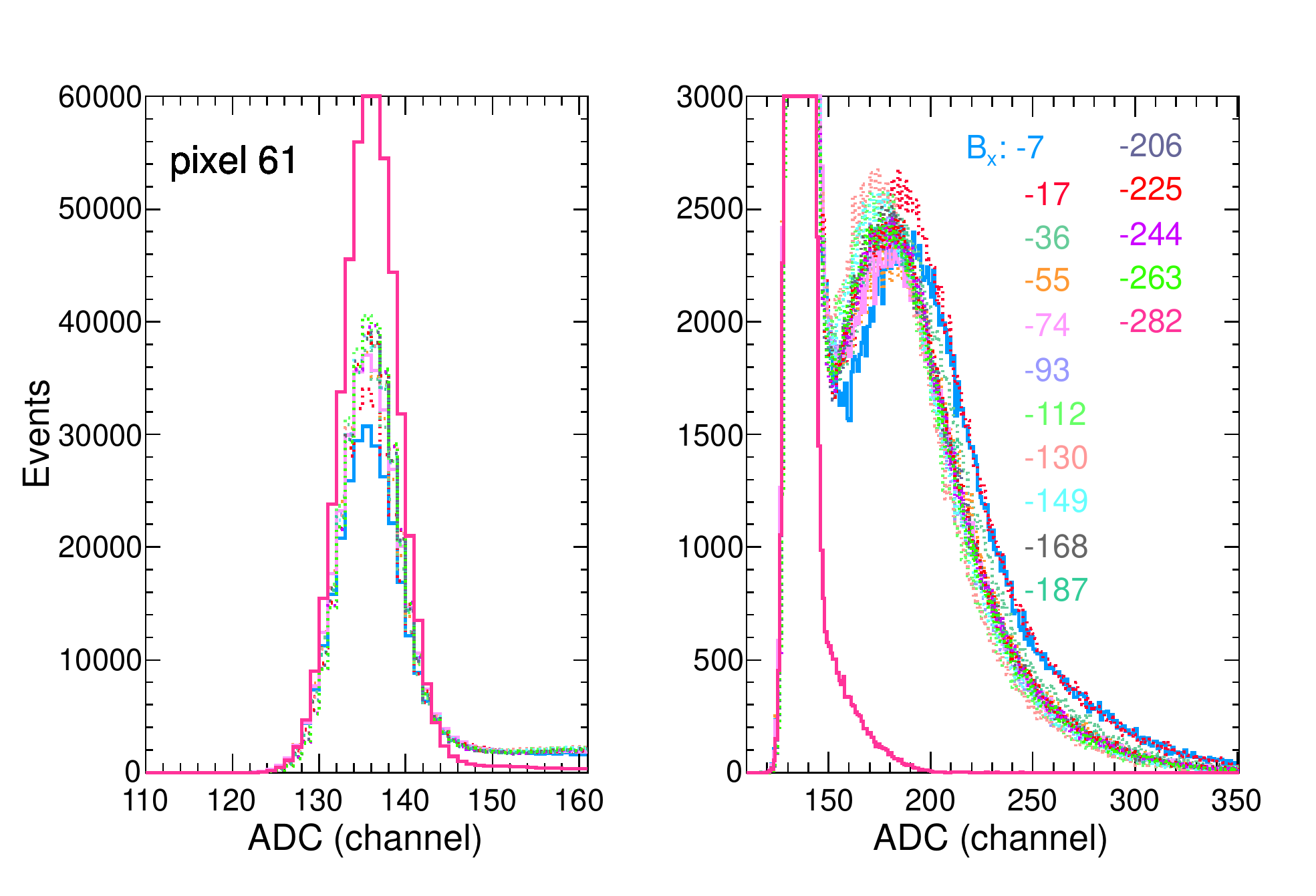}
&
\hspace{-0.4in}
\includegraphics[width=9.3cm]{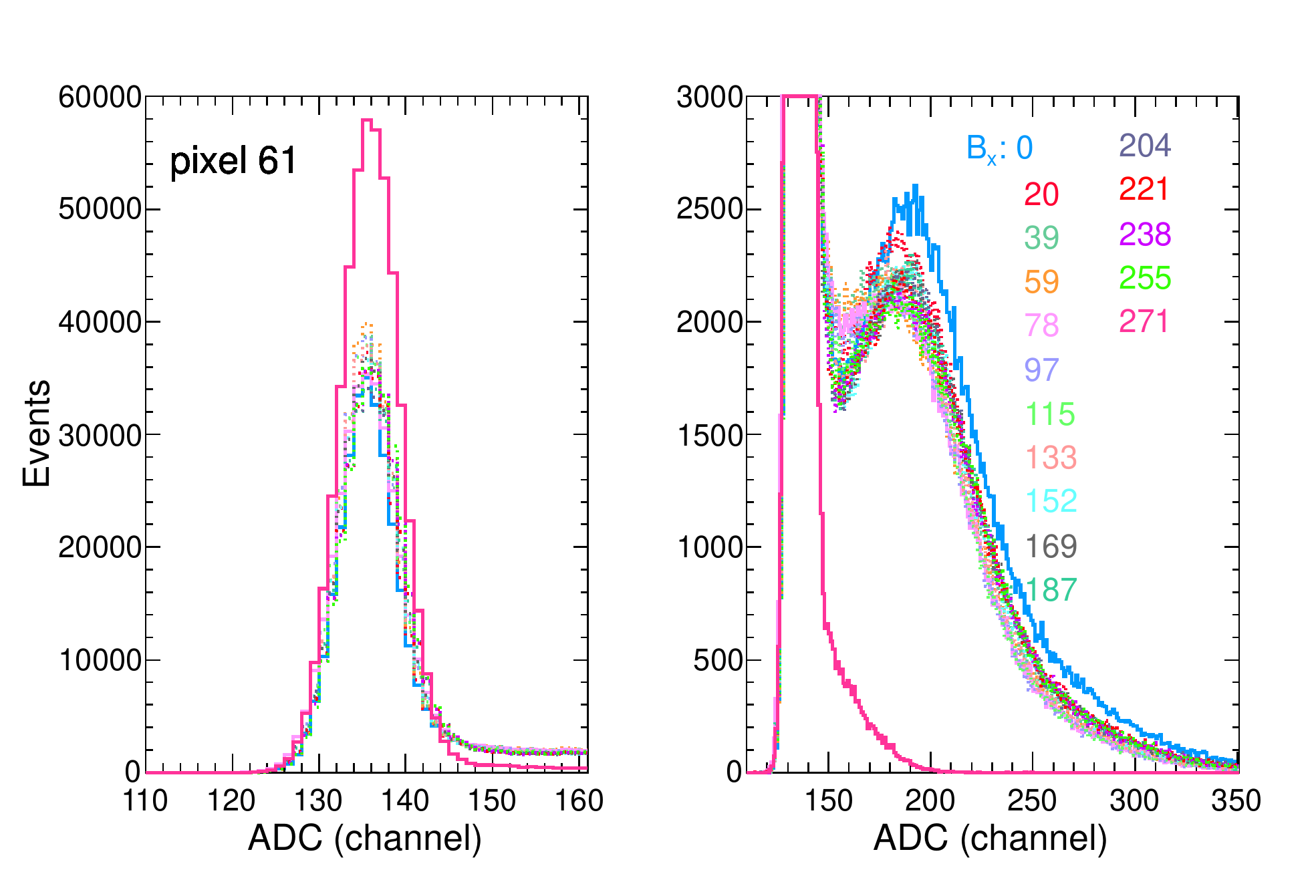} \\
\includegraphics[width=9.3cm]{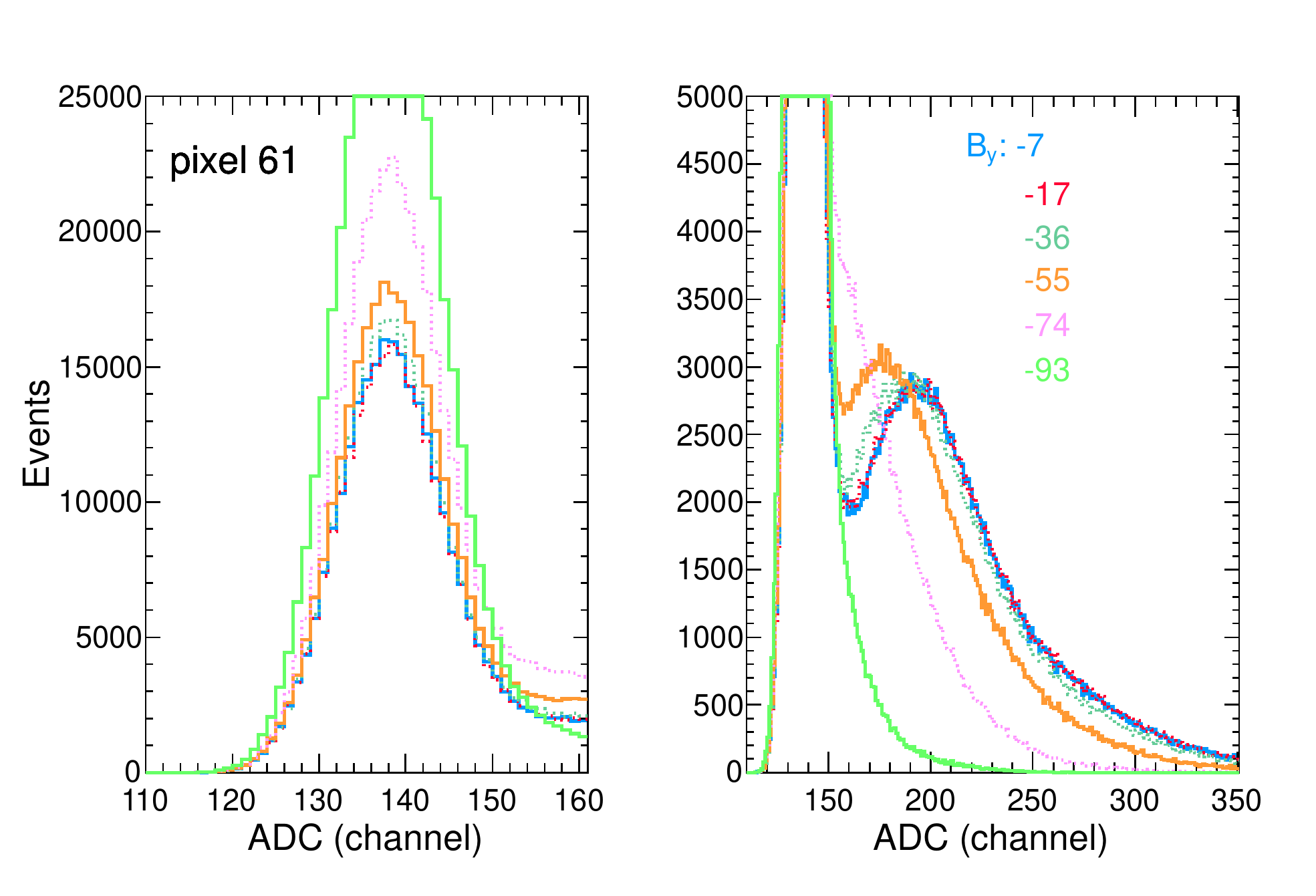}
&
\hspace{-0.4in}
\includegraphics[width=9.3cm]{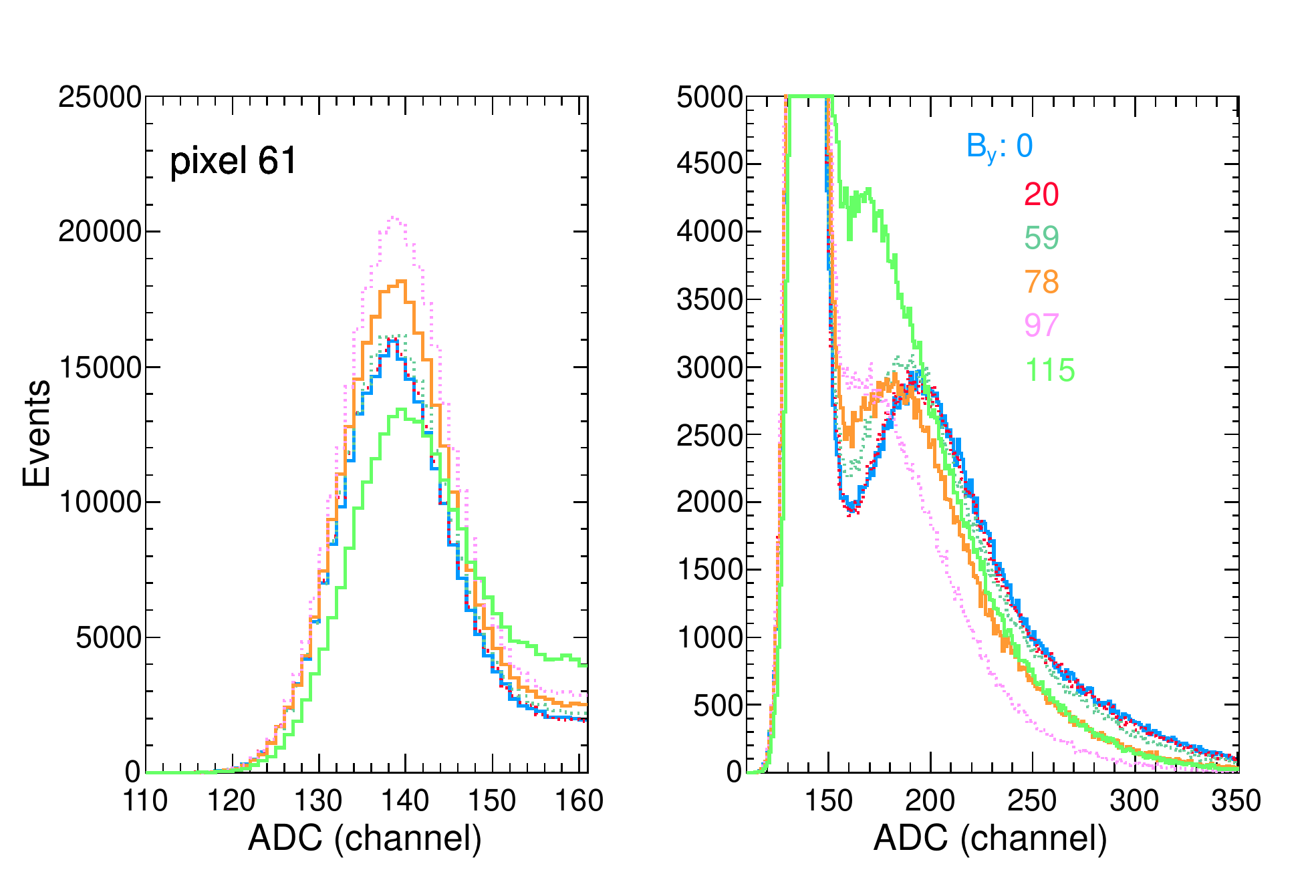} \\
\includegraphics[width=9.3cm]{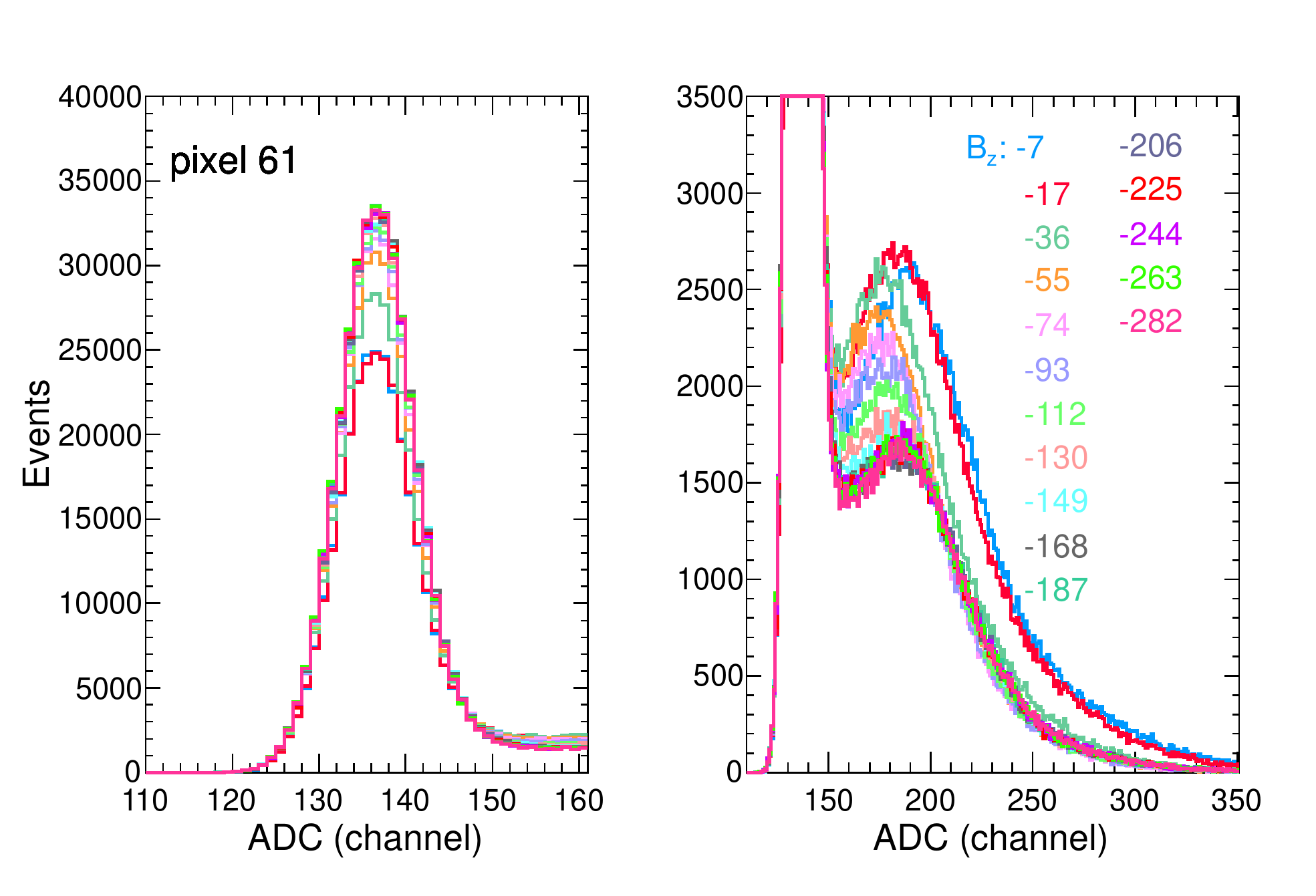}
&
\hspace{-0.4in}
\includegraphics[width=9.3cm]{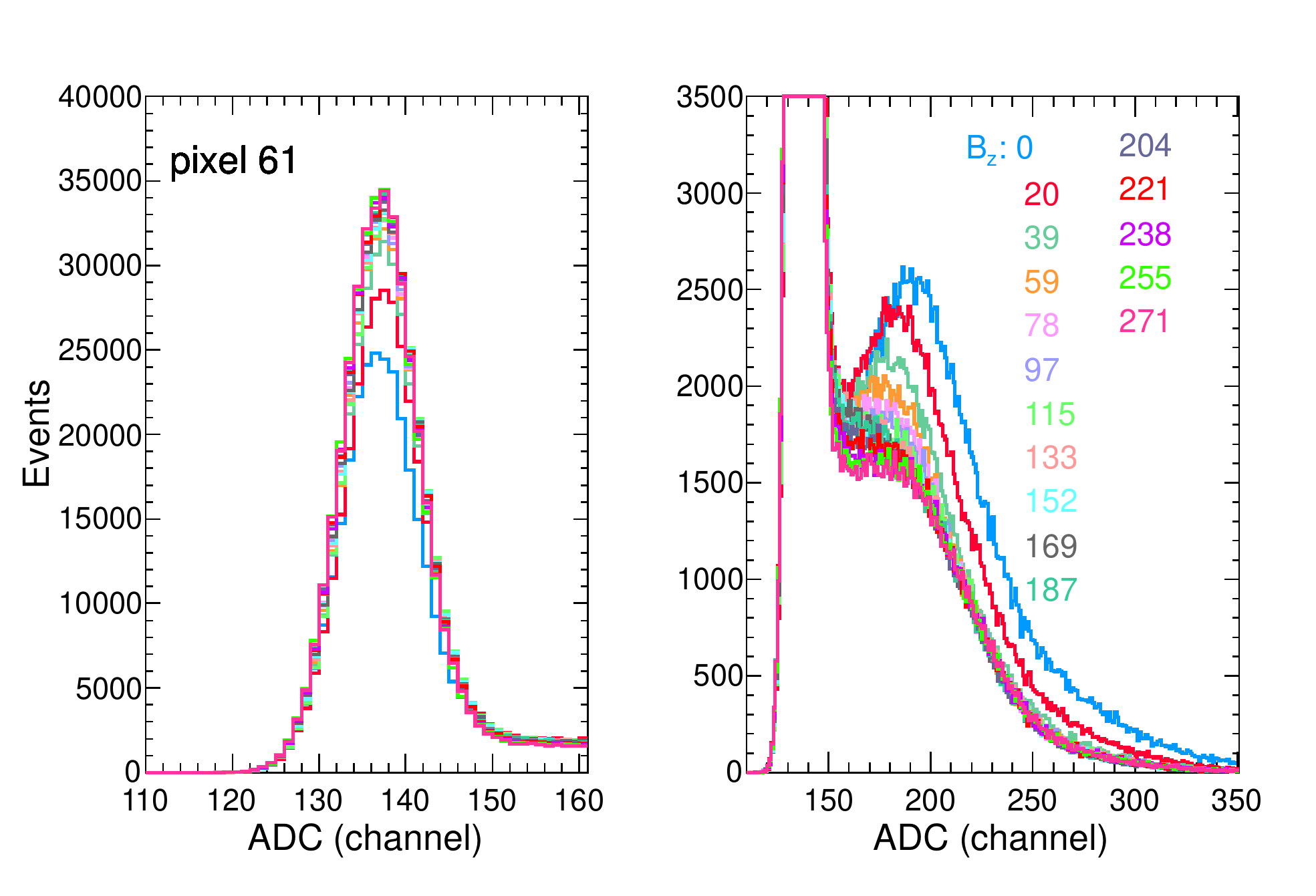} \\
\end{tabular}
\linespread{0.5}
\caption[]{
{Impact of a longitudinal and transverse magnetic field on single photoelectron outputs from 
pixel 61.} }
\label{61_one}
\end{figure}

\begin{figure}[htbp]
\vspace*{-0.1in}
\centering
\begin{tabular}{cc}
\includegraphics[width=9.3cm]{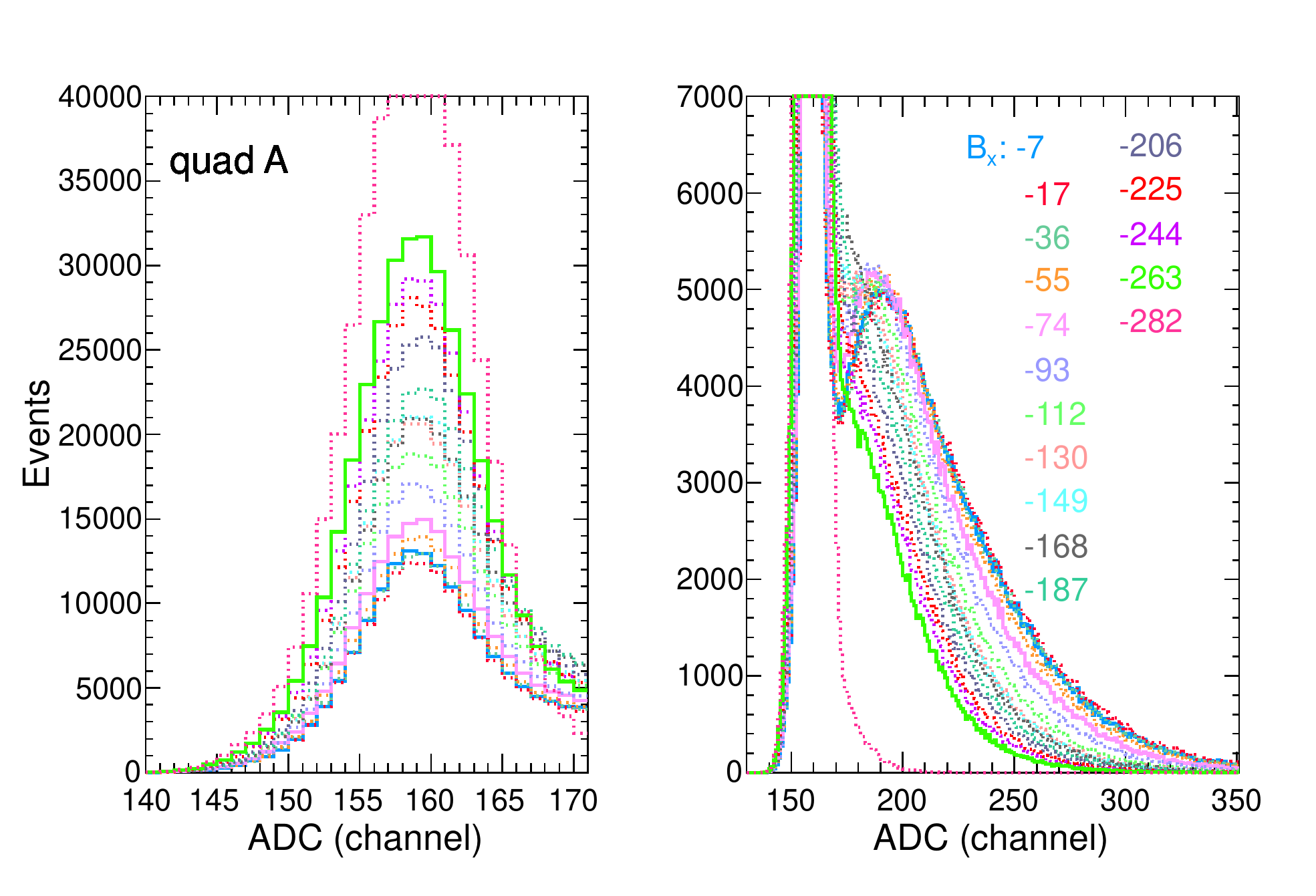}
&
\hspace{-0.4in}
\includegraphics[width=9.3cm]{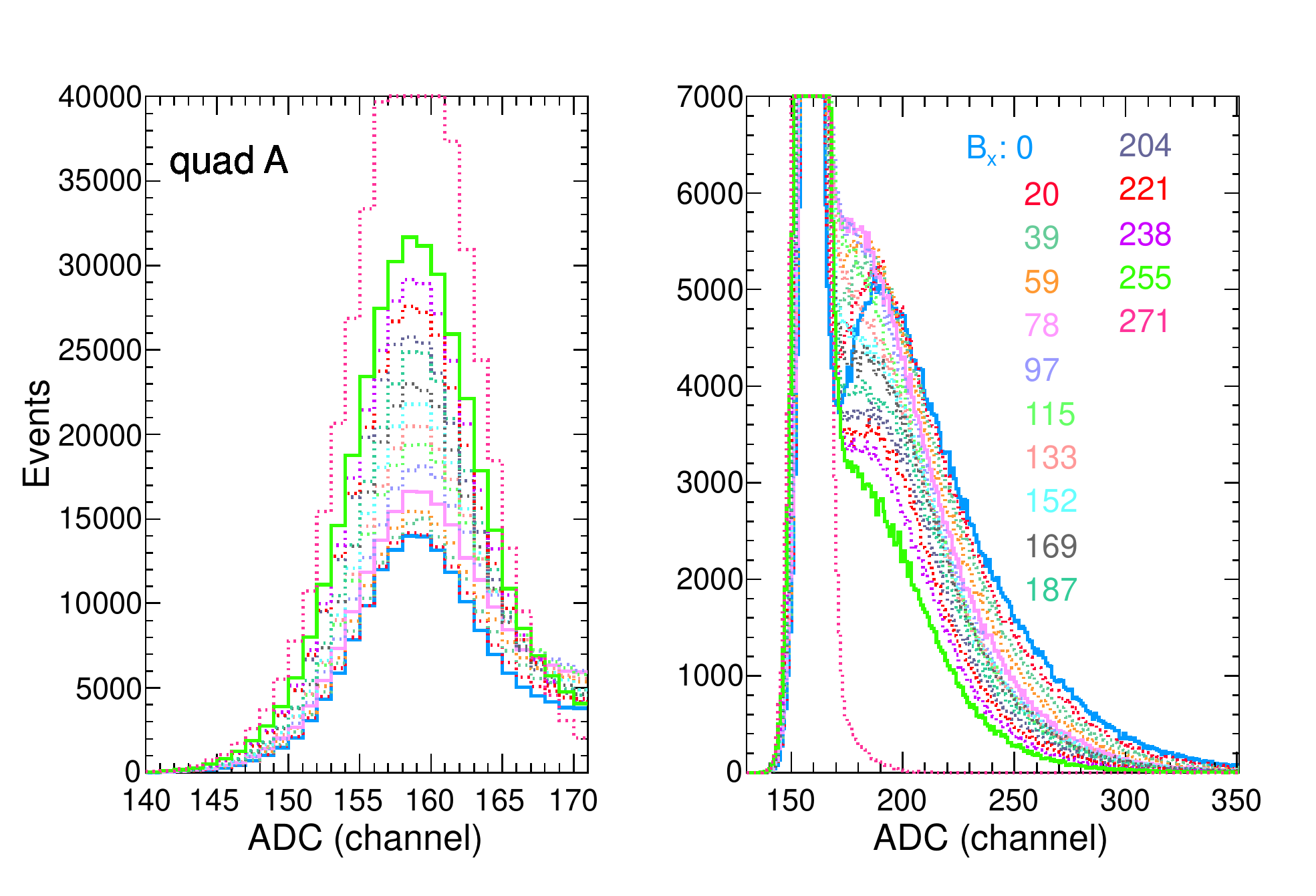} \\
\includegraphics[width=9.3cm]{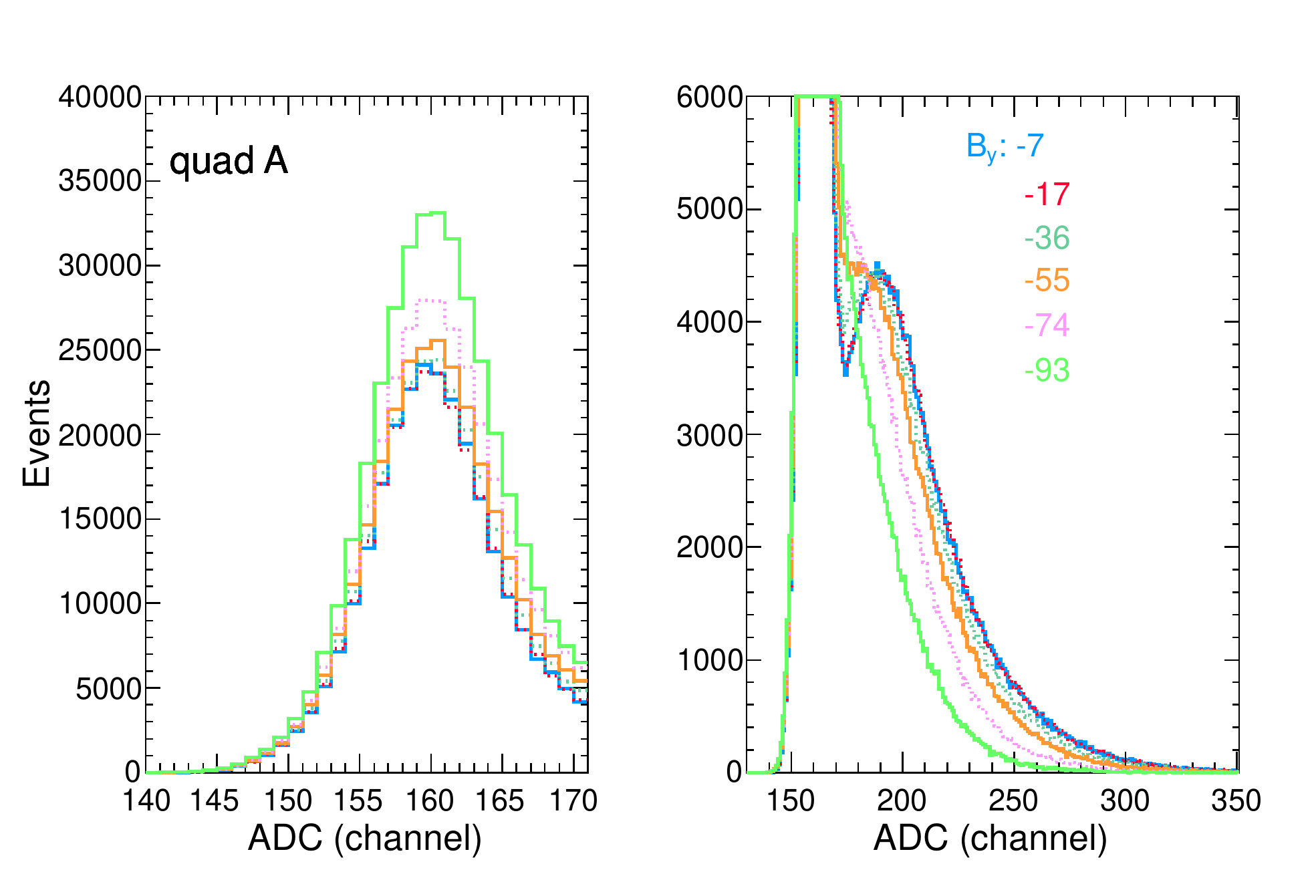}
&
\hspace{-0.4in}
\includegraphics[width=9.3cm]{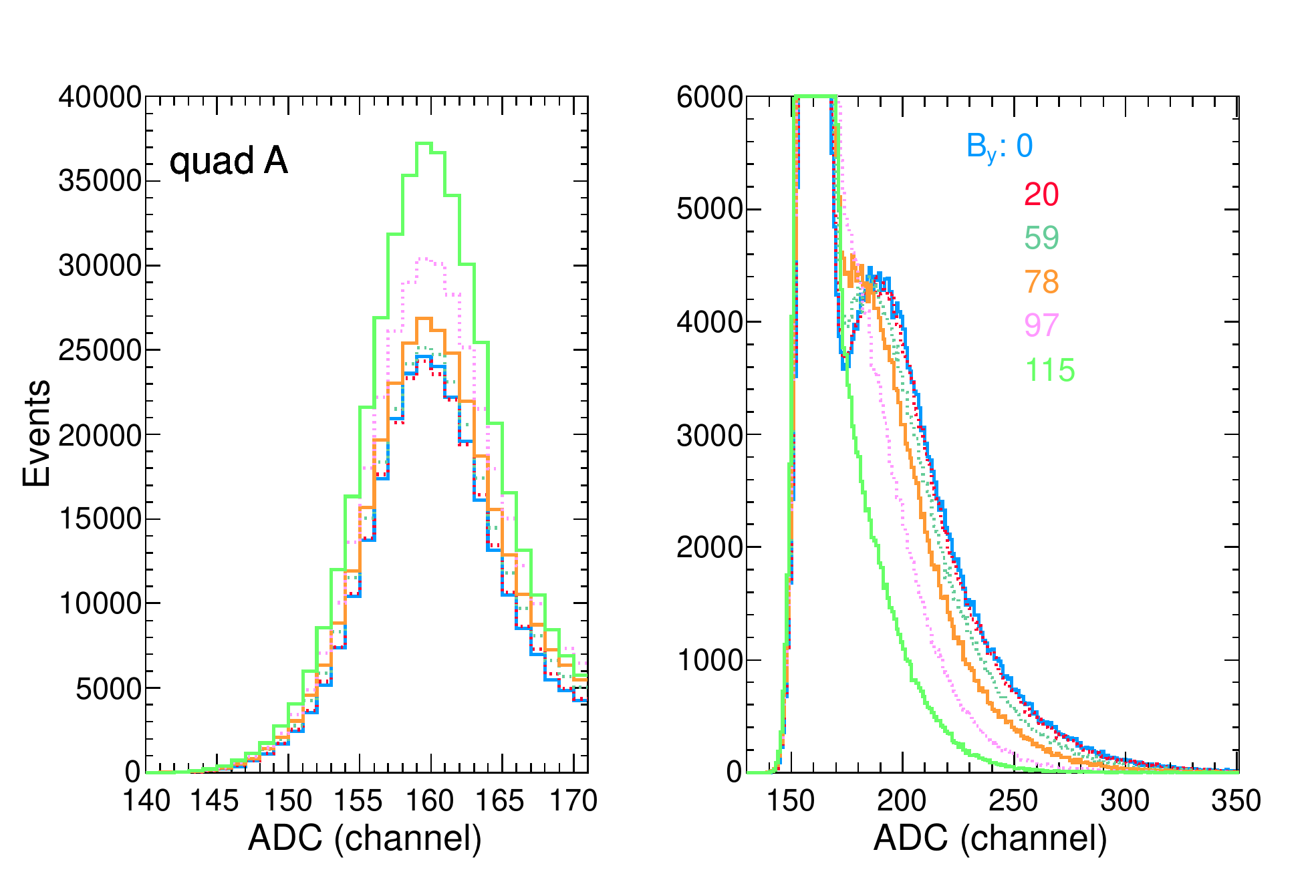} \\
\includegraphics[width=9.3cm]{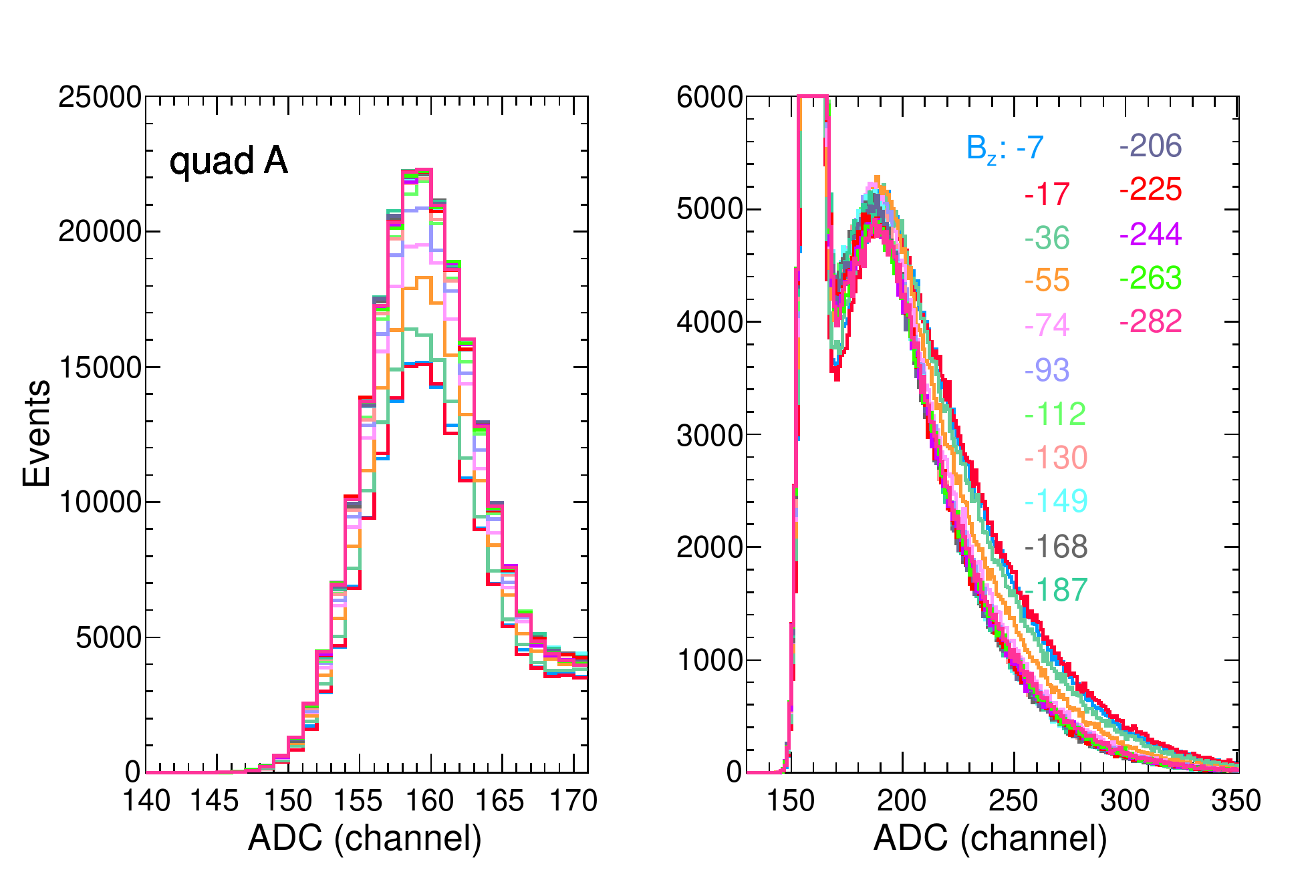}
&
\hspace{-0.4in}
\includegraphics[width=9.3cm]{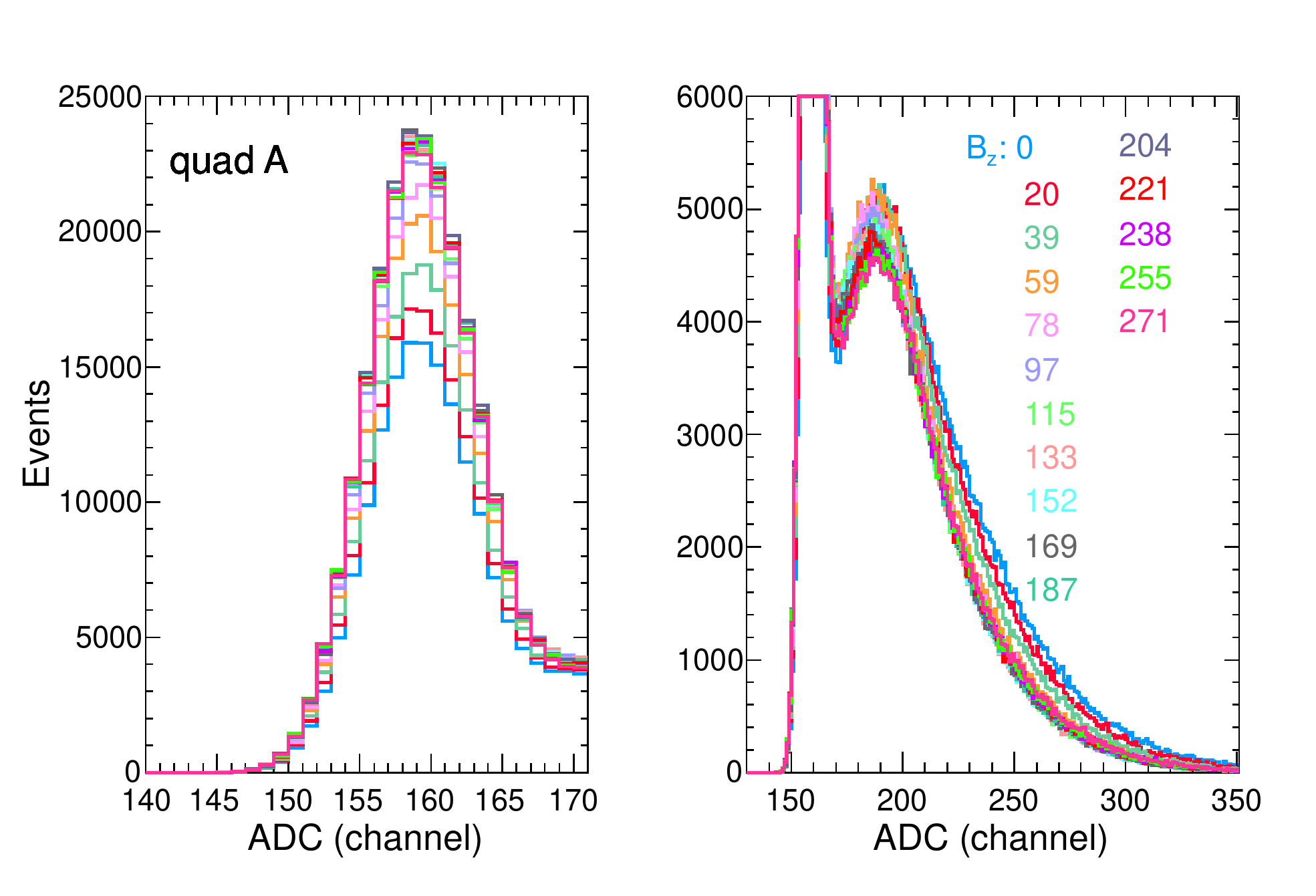} \\
\end{tabular}
\linespread{0.5}
\caption[]{
{Impact of a longitudinal and transverse magnetic field on single photoelectron outputs from 
quad A.} }
\label{quad_a_one}
\end{figure}

\begin{figure}[htbp]
\vspace*{-0.1in}
\centering
\begin{tabular}{cc}
\includegraphics[width=9.3cm]{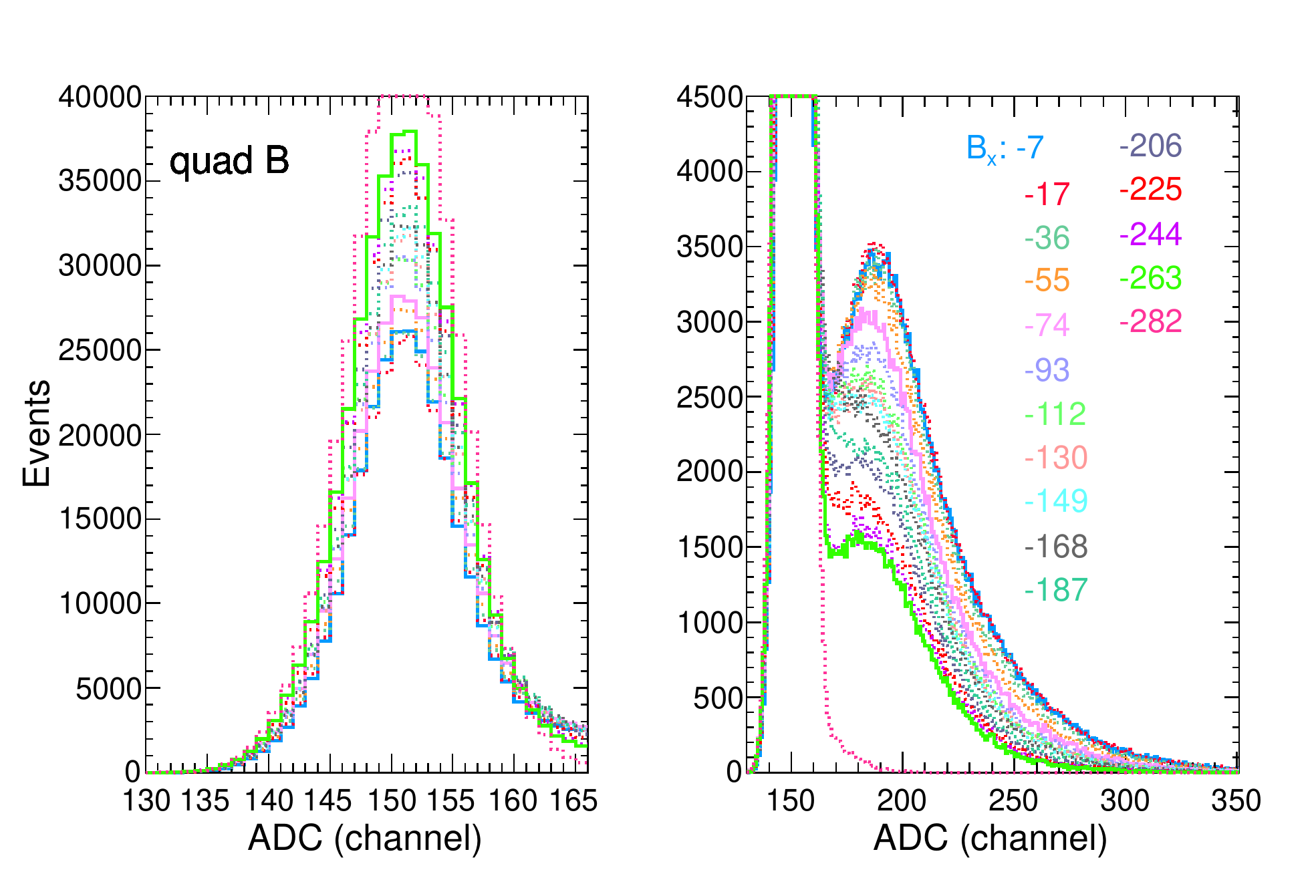}
&
\hspace{-0.4in}
\includegraphics[width=9.3cm]{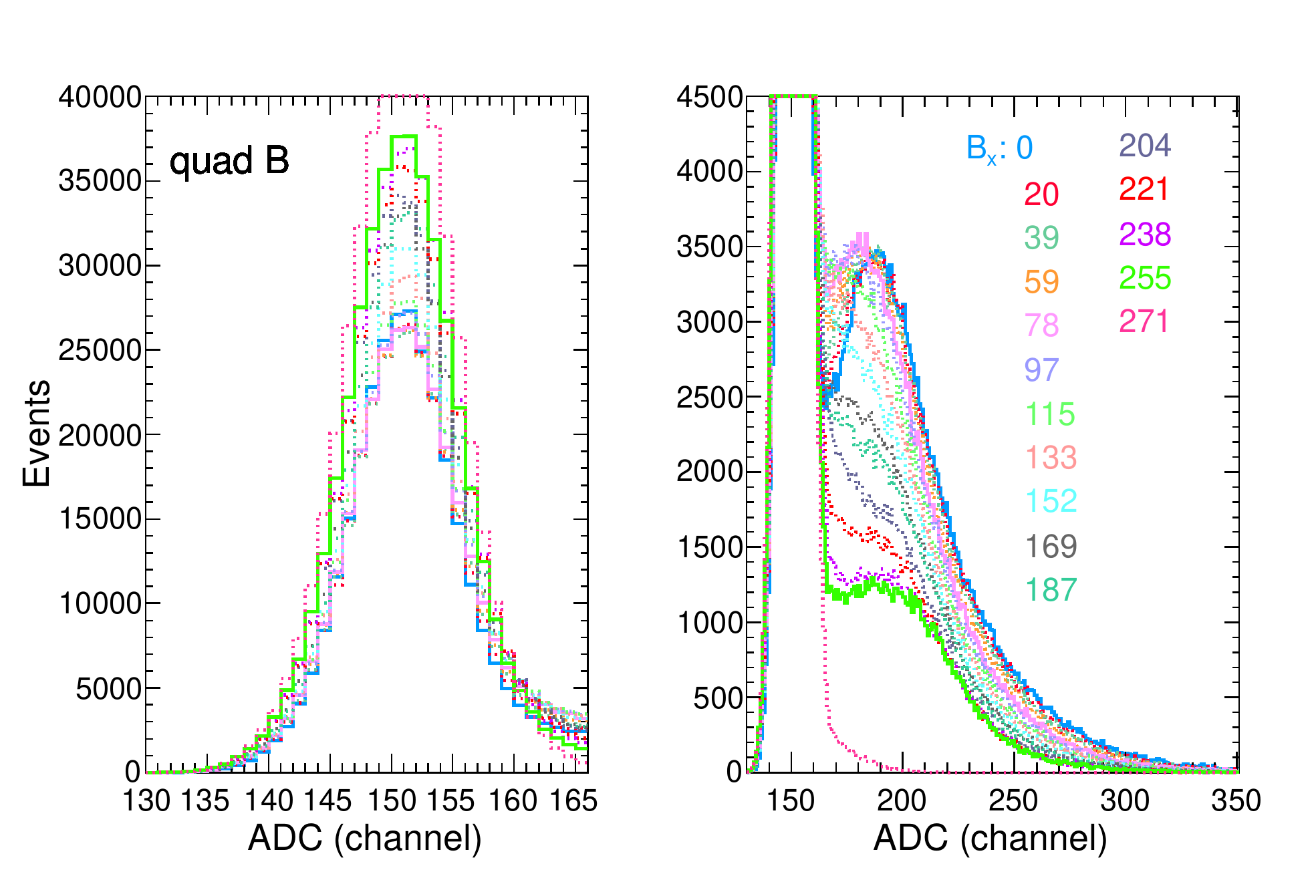} \\
\includegraphics[width=9.3cm]{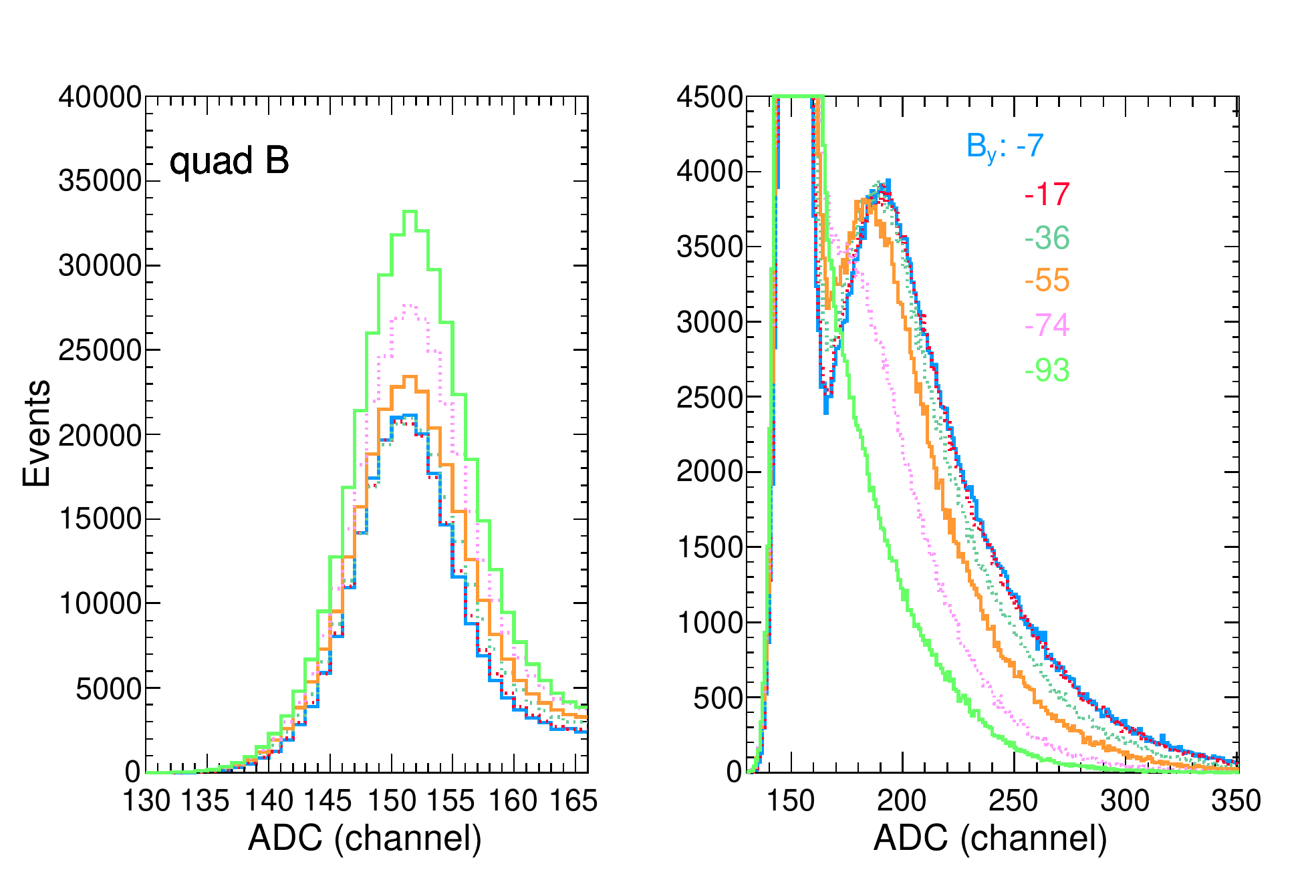}
&
\hspace{-0.4in}
\includegraphics[width=9.3cm]{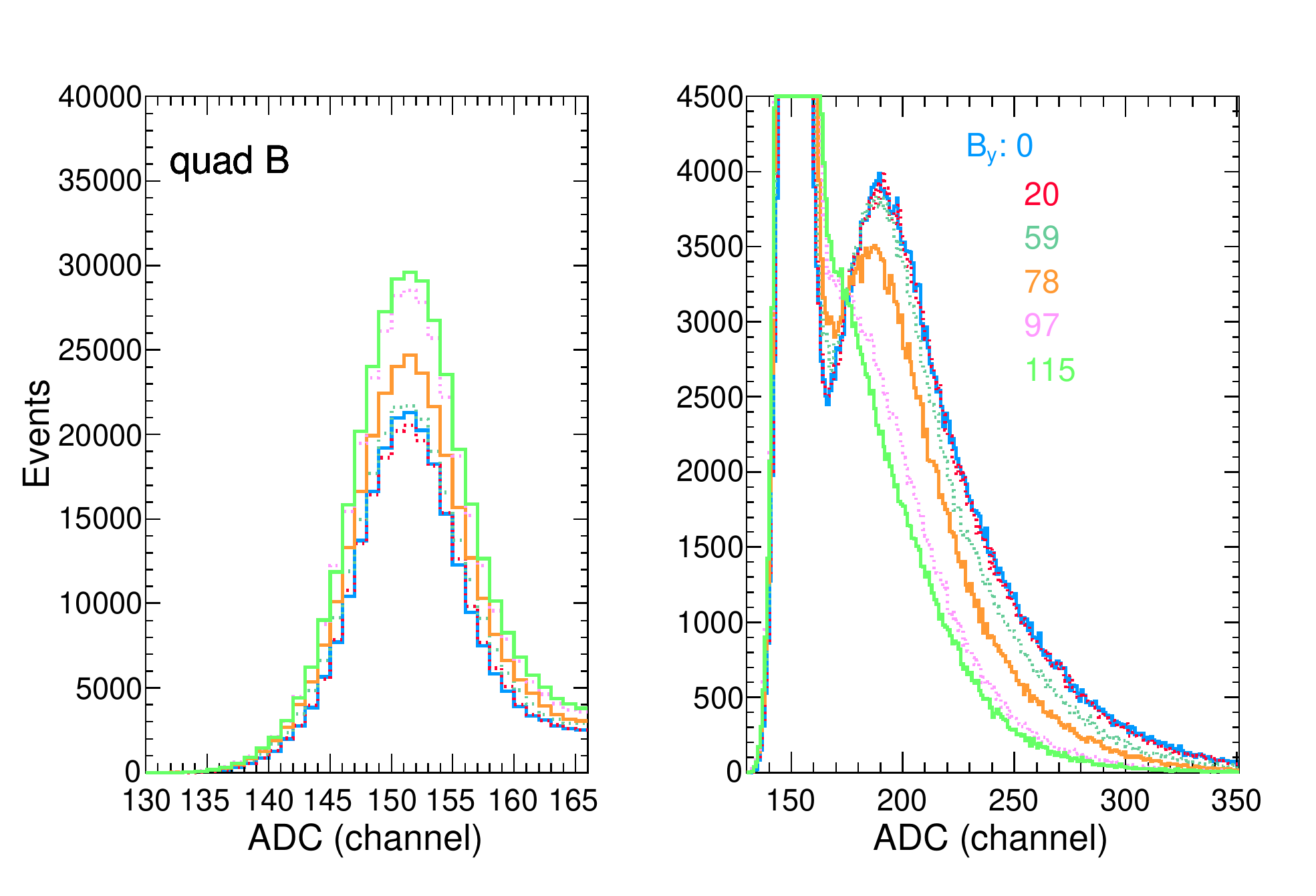} \\
\includegraphics[width=9.3cm]{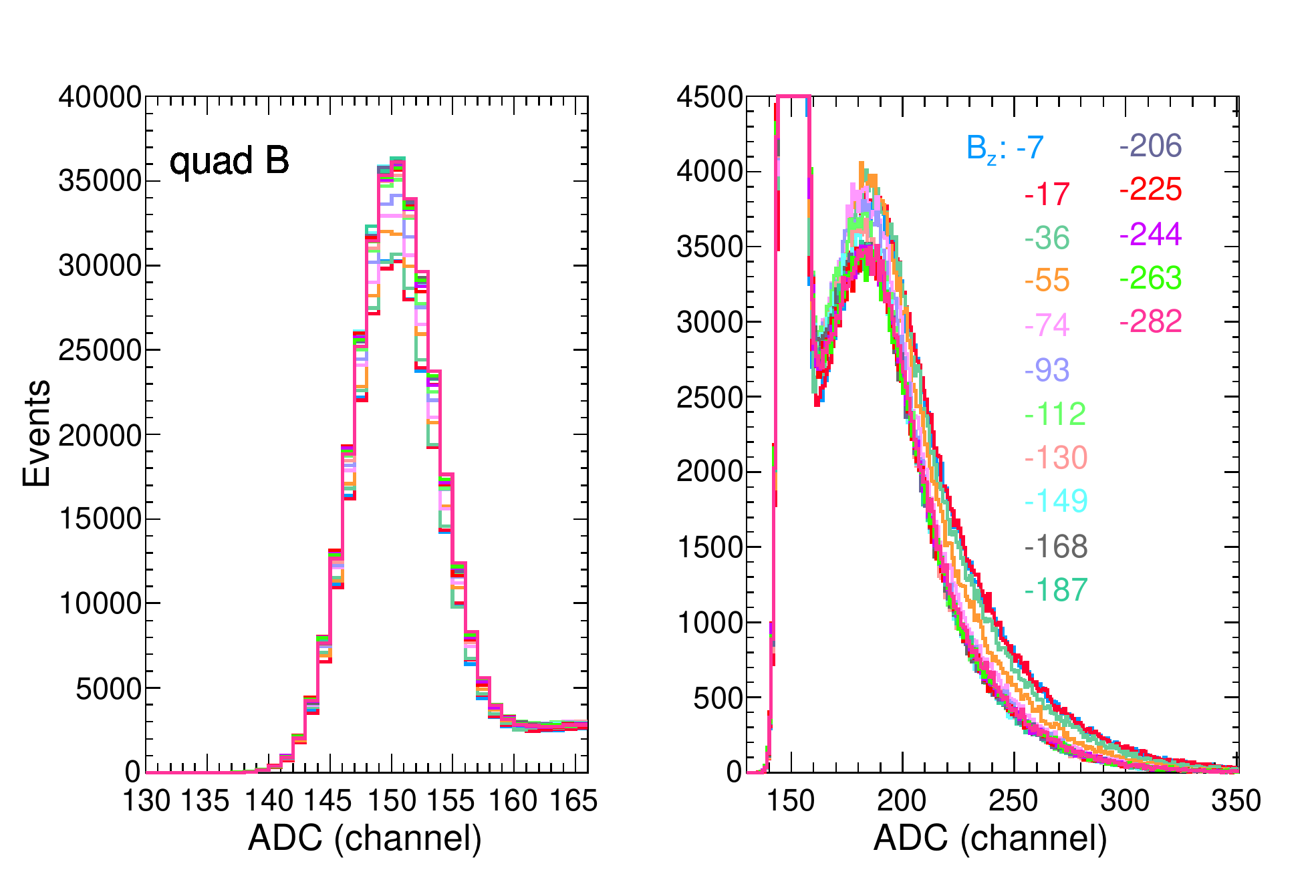}
&
\hspace{-0.4in}
\includegraphics[width=9.3cm]{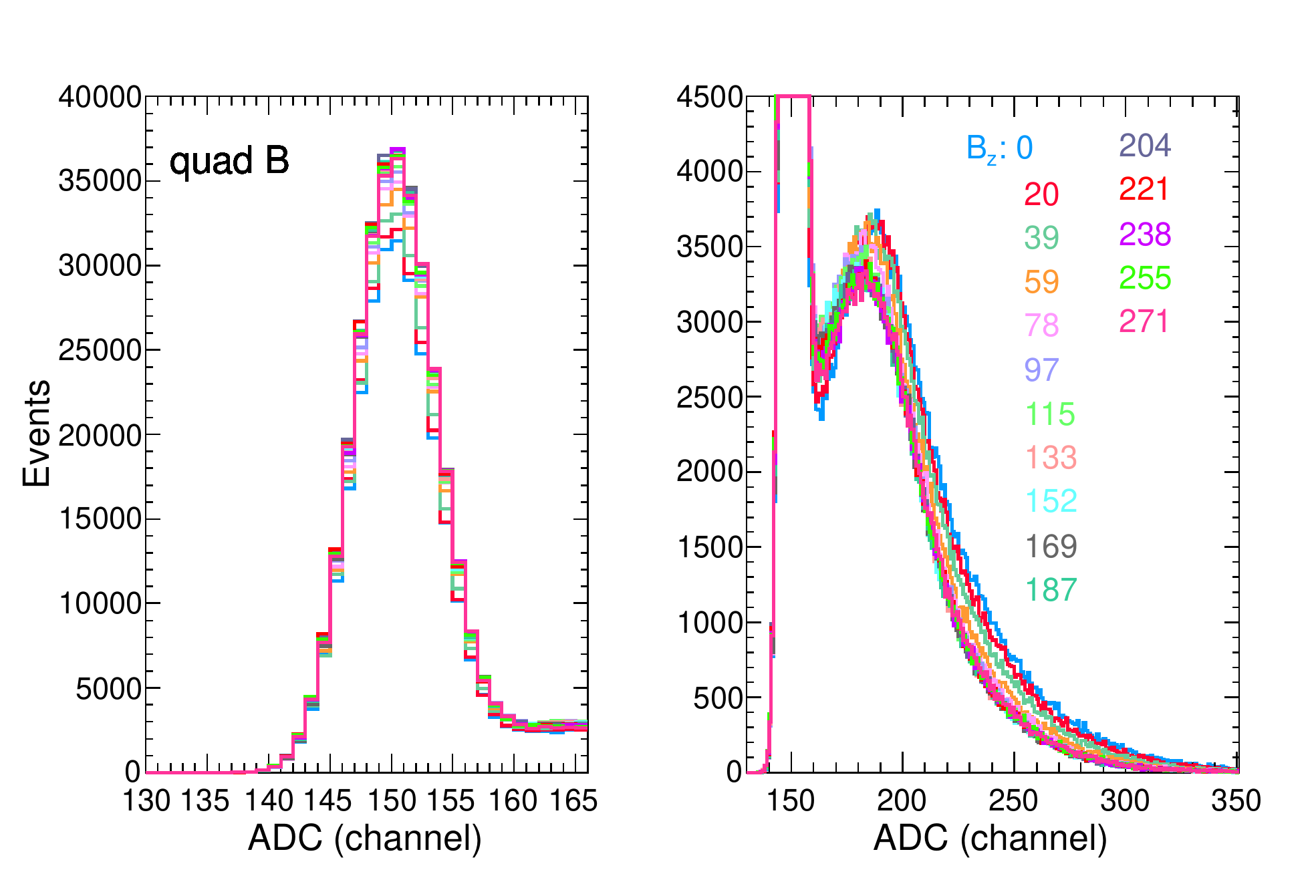} \\
\end{tabular}
\linespread{0.5}
\caption[]{
{Impact of a longitudinal and transverse magnetic field on single photoelectron outputs from 
quad B.} }
\label{quad_b_one}
\end{figure}

We studied the effect of a longitudinal and transverse magnetic field 
on signals at the single photoelectron level as well. ADC distributions 
from pixel 45, 61 and quads A and B are shown in Figs.~\ref{45_one}, 
\ref{61_one}, \ref{quad_a_one} and \ref{quad_b_one} for various field settings. 
We highlight the pedestal and signal regions separately for all settings. 
After tunning the LED to maximize the output of single photoelectrons, we 
applied a magnetic field and varied its magnitude in small steps. The purpose of 
the study was to provide insight into the mechanism by which the pixel and quad 
outputs get diminished with increasing magnetic field. In particular there could 
be two dominant mechanisms by which the signal is lost. One hypothesis is 
that the applied magnetic field would disrupt the electrical field that focuses 
the photoelectron extracted from the photocathode onto the first dynode. In our 
case this type of event would ``transfer'' hits from the signal region to the pedestal 
region and the position of the single photoelectron peak would not be affected in the 
ADC distribution. Another possible effect could be the loss of gain 
down the dynode chain as the magnetic field applied disrupts the amplification. 
This would result in a diminishing gain with field and would be visible as a 
smaller separation between the single photoelectron peak and the pedestal 
on the ADC distribution. The latter effect can be compensated for by increasing 
the high-voltage or otherwise amplifying the signal. The former scenario, however, 
cannot be corrected as the initial single photoelectron is lost.

\begin{figure}[htbp]
\vspace*{-0.2in}
\centering
\begin{tabular}{ccc}
\includegraphics[width=6.5cm]{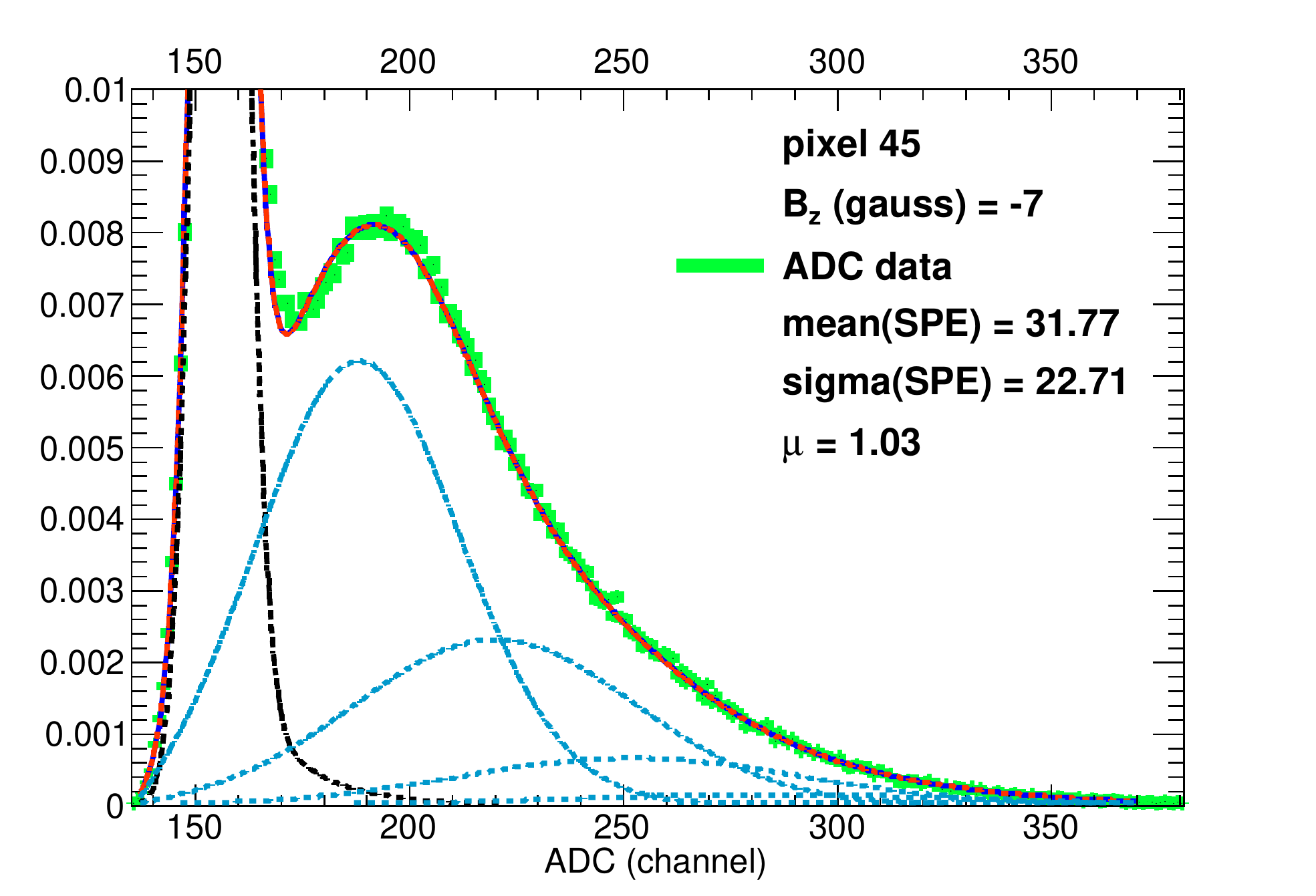}
&
\hspace{-0.4in}
\includegraphics[width=6.5cm]{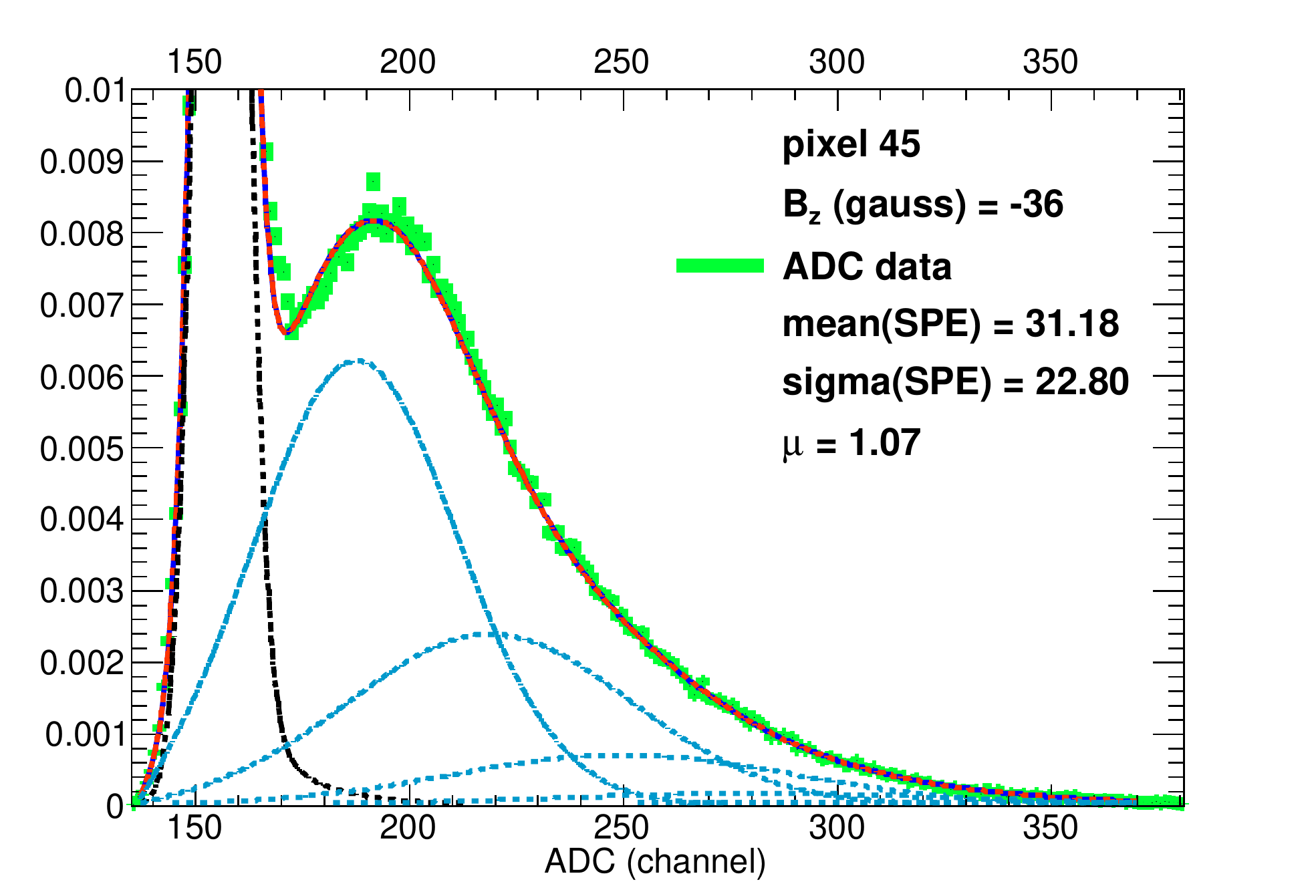}
&
\hspace{-0.4in}
\includegraphics[width=6.5cm]{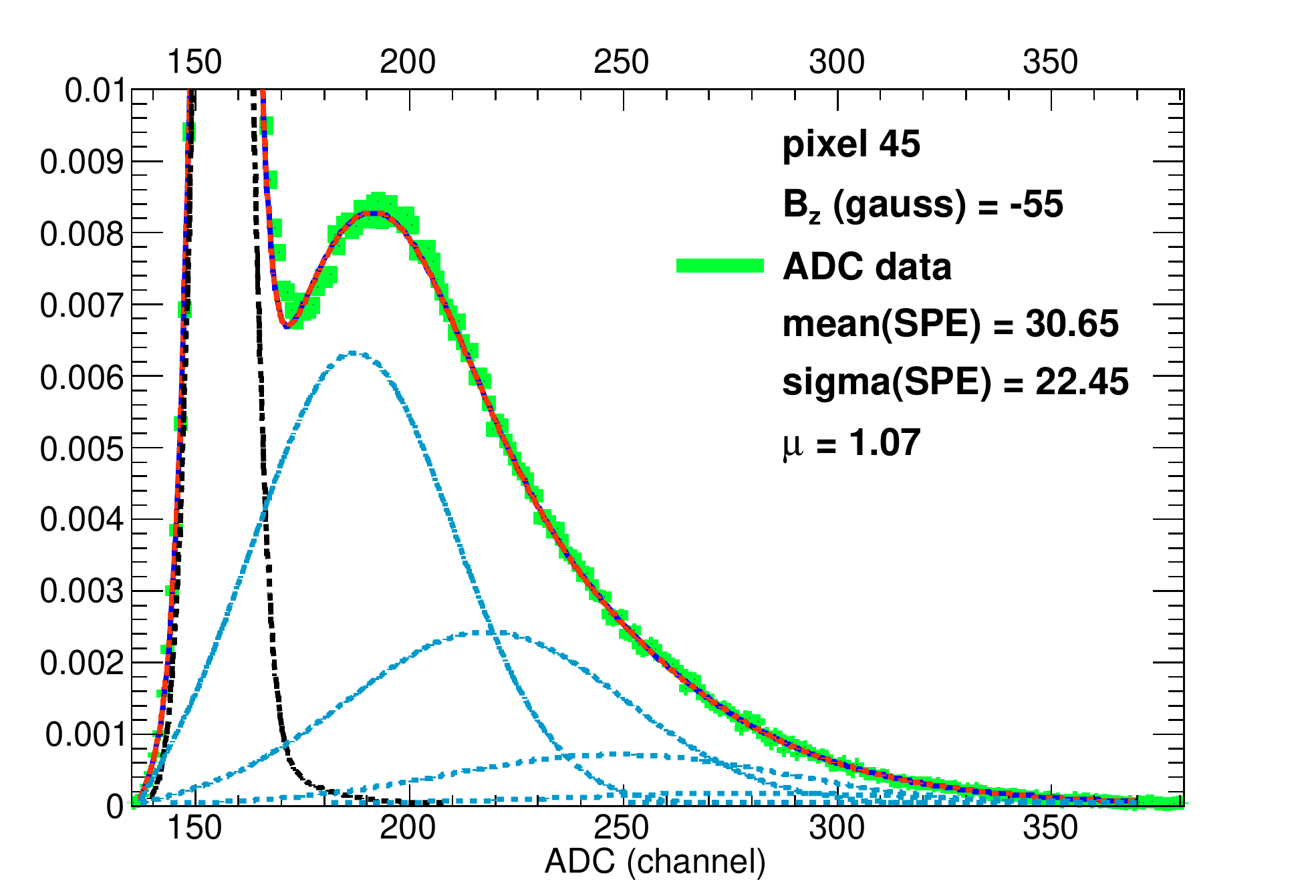} \\
\includegraphics[width=6.5cm]{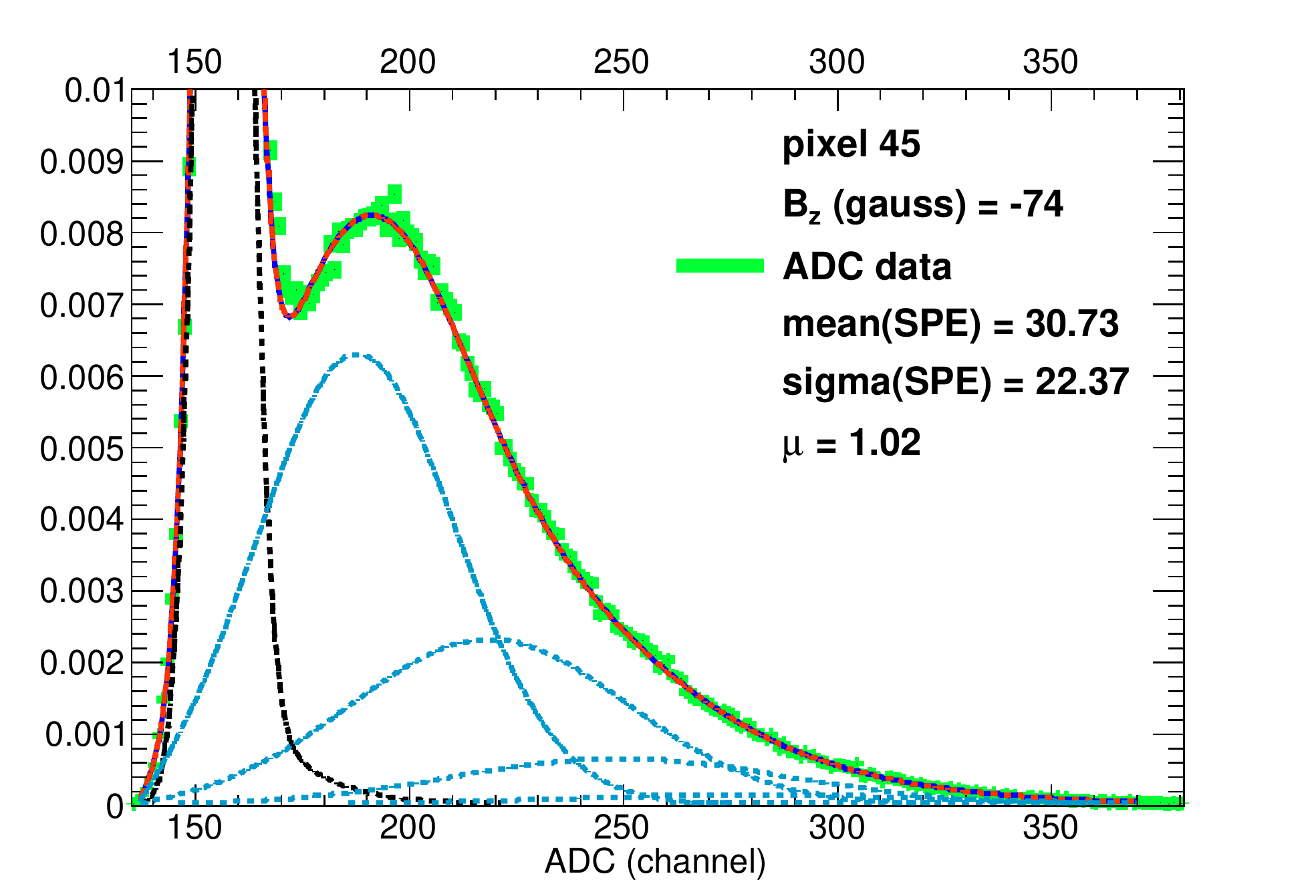}
&
\hspace{-0.4in}
\includegraphics[width=6.5cm]{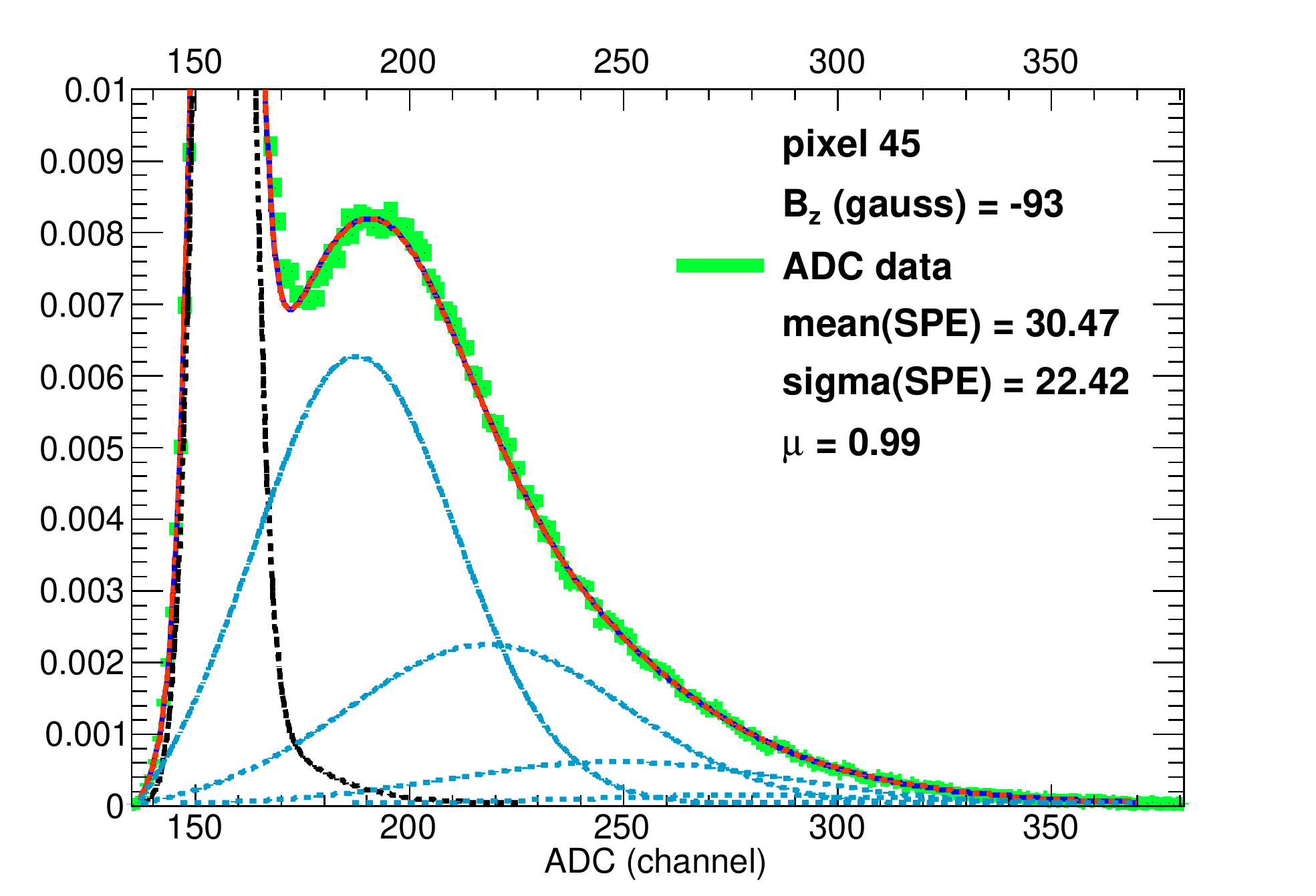}
&
\hspace{-0.4in}
\includegraphics[width=6.5cm]{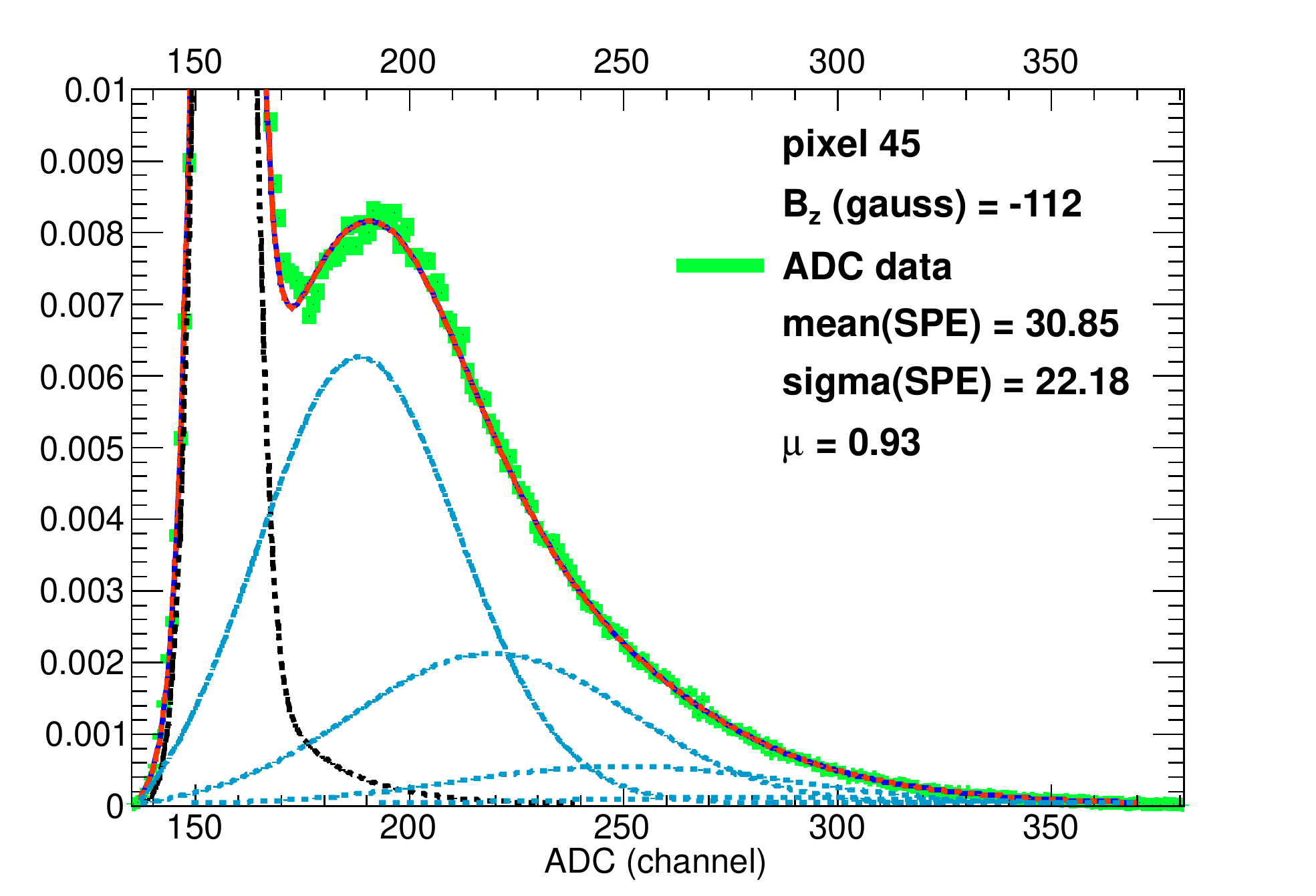} \\
\includegraphics[width=6.5cm]{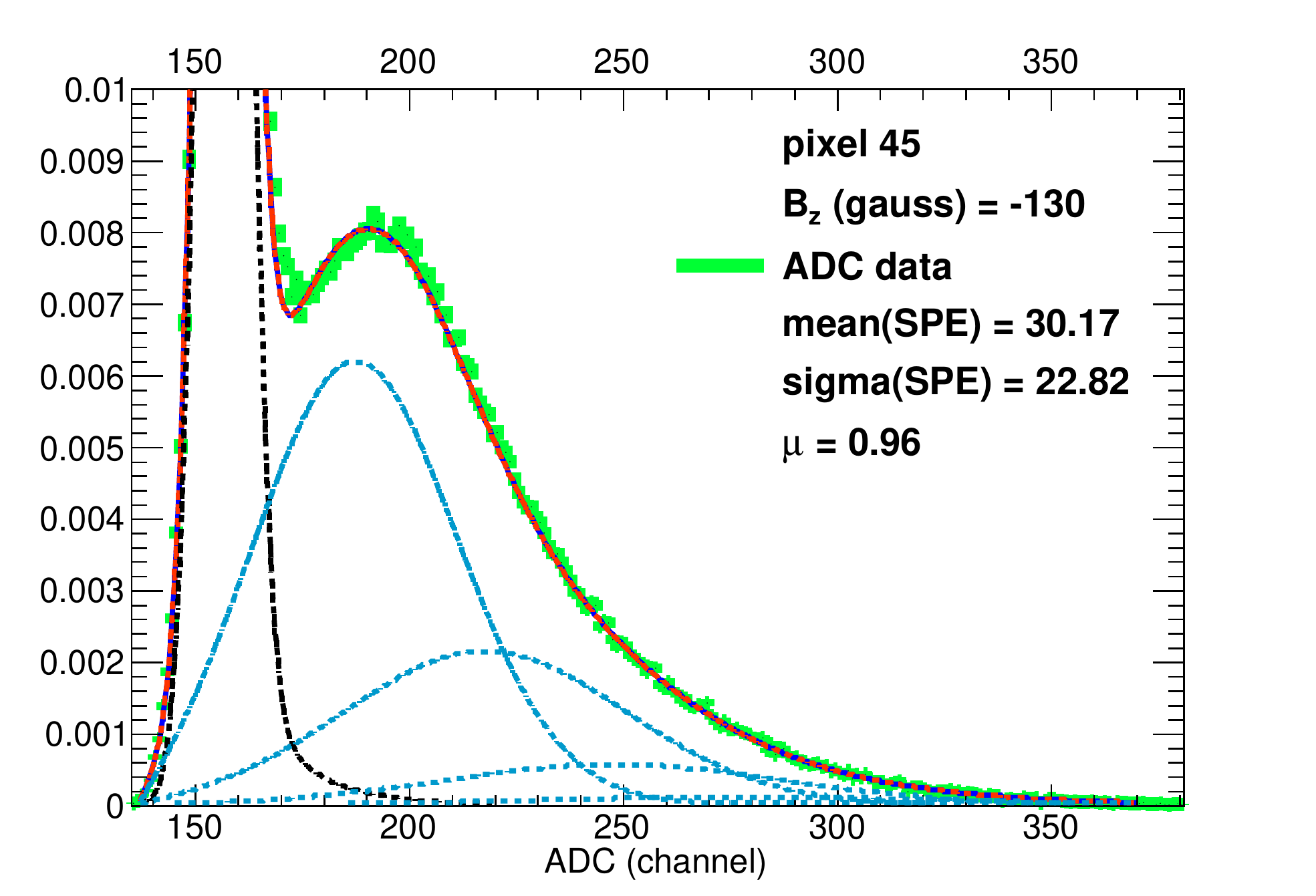} 
&
\hspace{-0.4in}
\includegraphics[width=6.5cm]{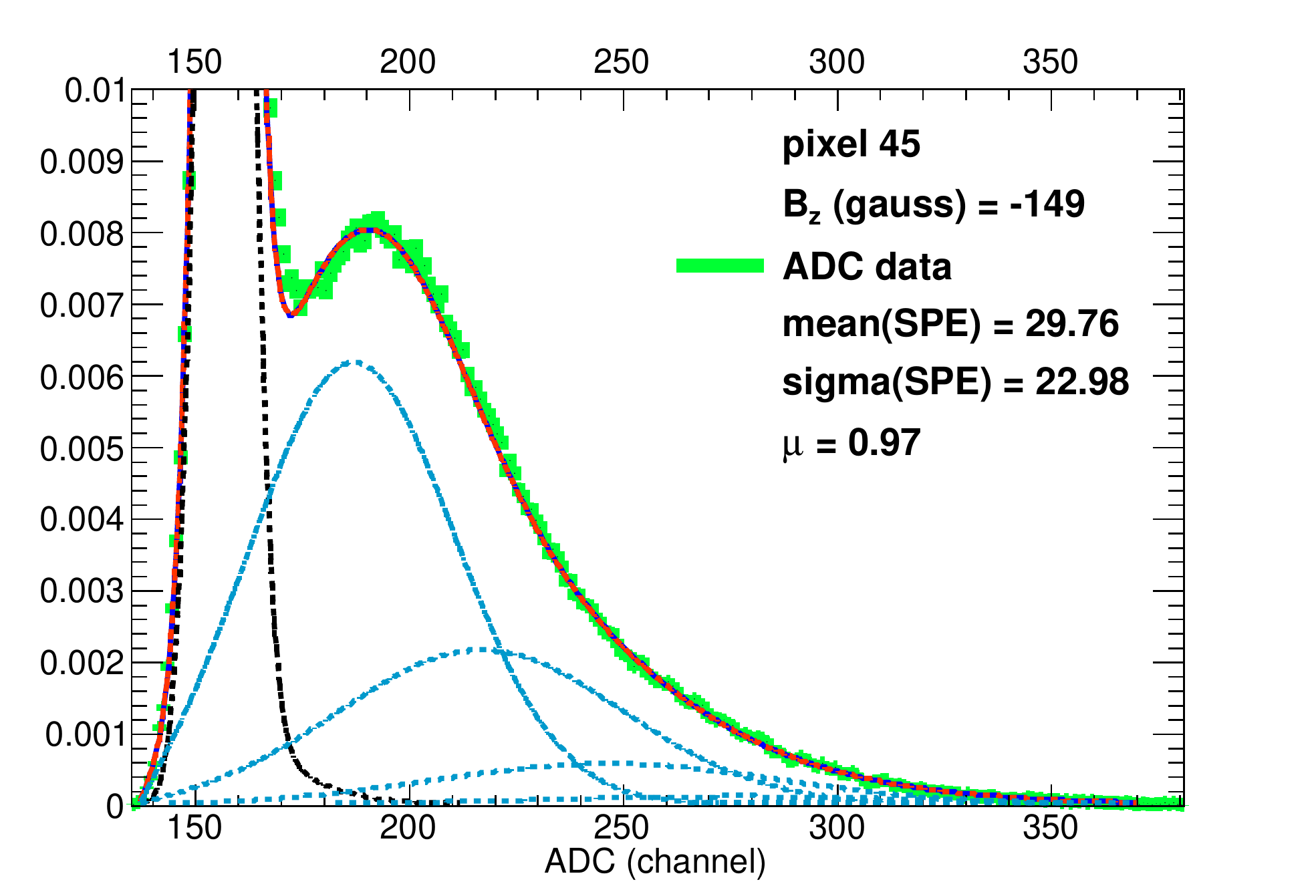}
&
\hspace{-0.4in}
\includegraphics[width=6.5cm]{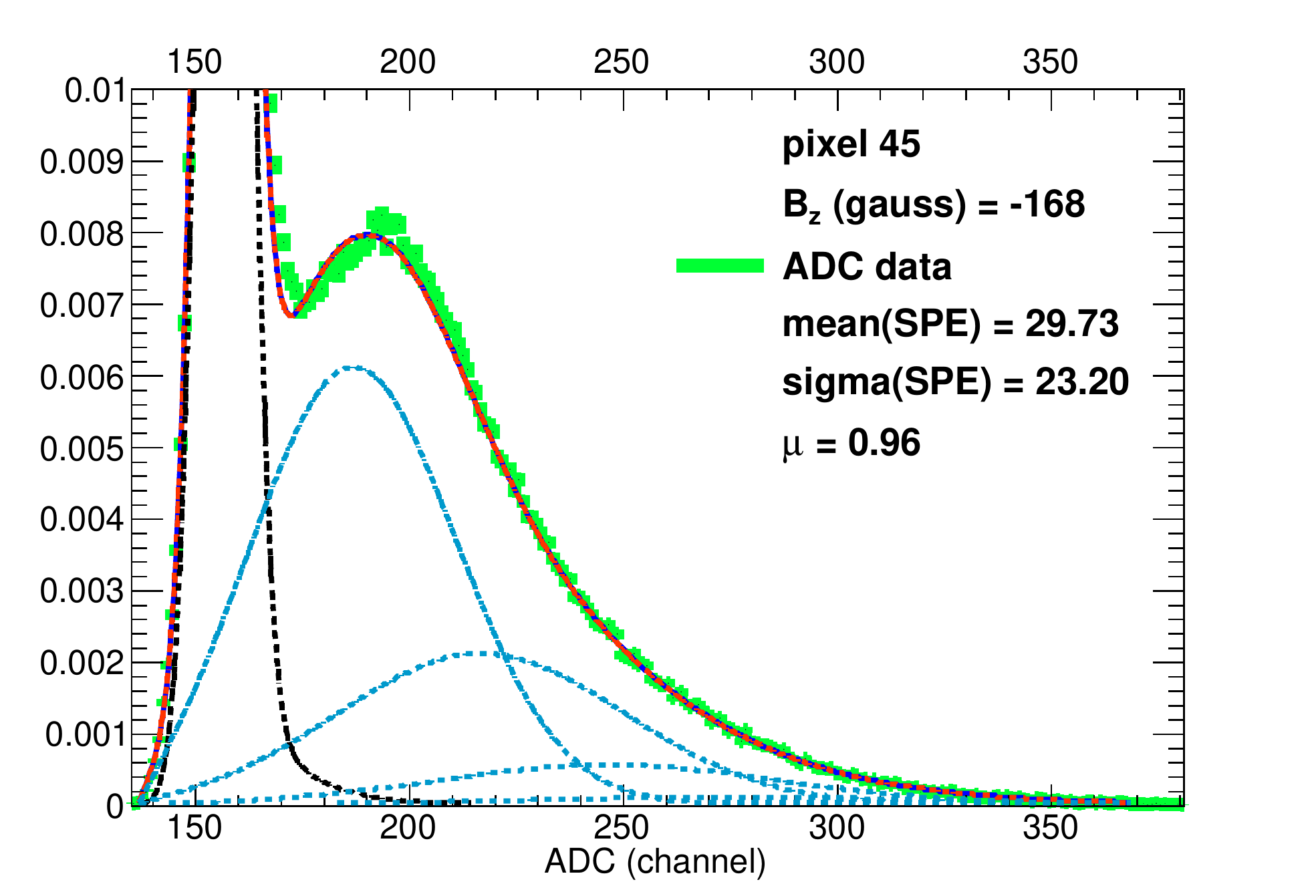} \\
\includegraphics[width=6.5cm]{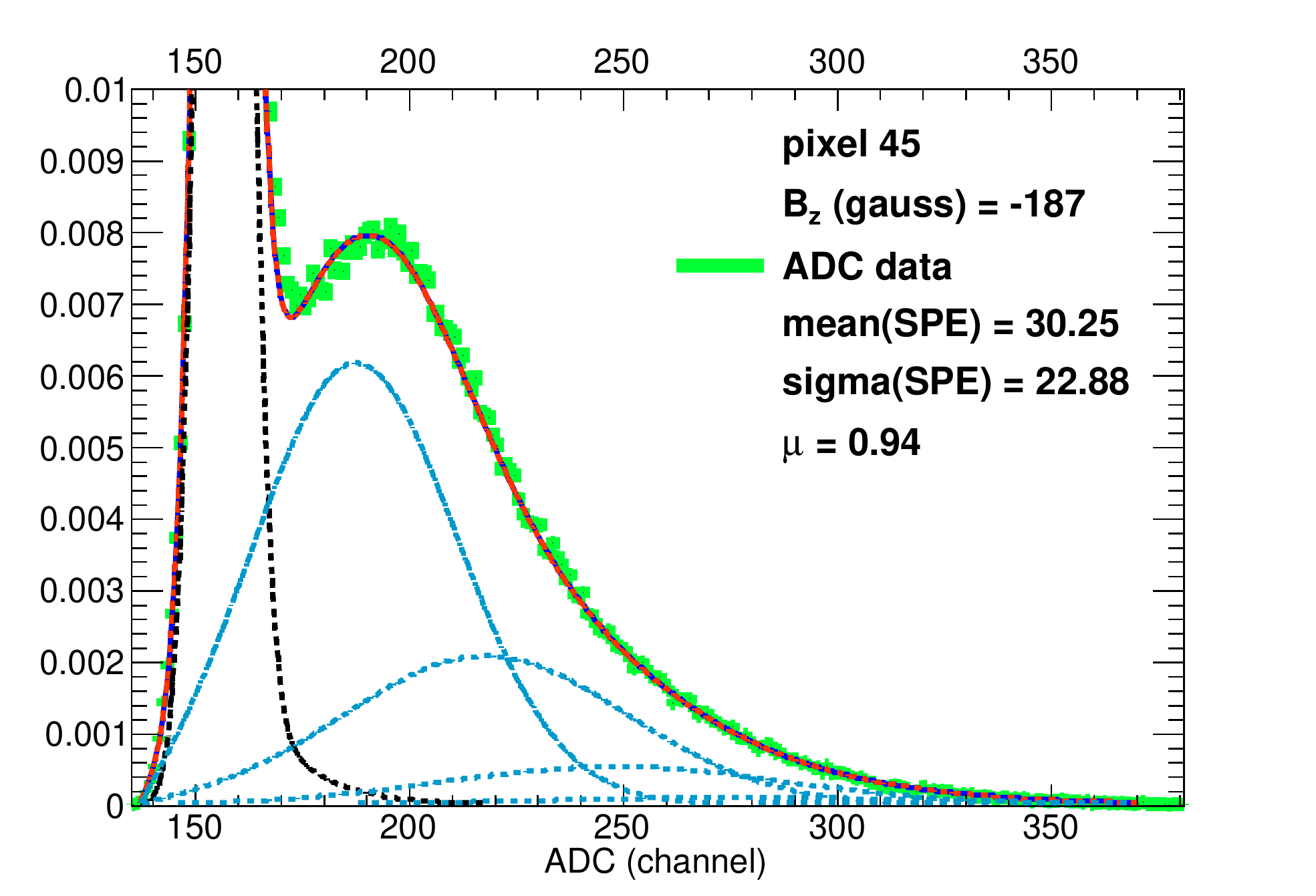}
&
\hspace{-0.4in}
\includegraphics[width=6.5cm]{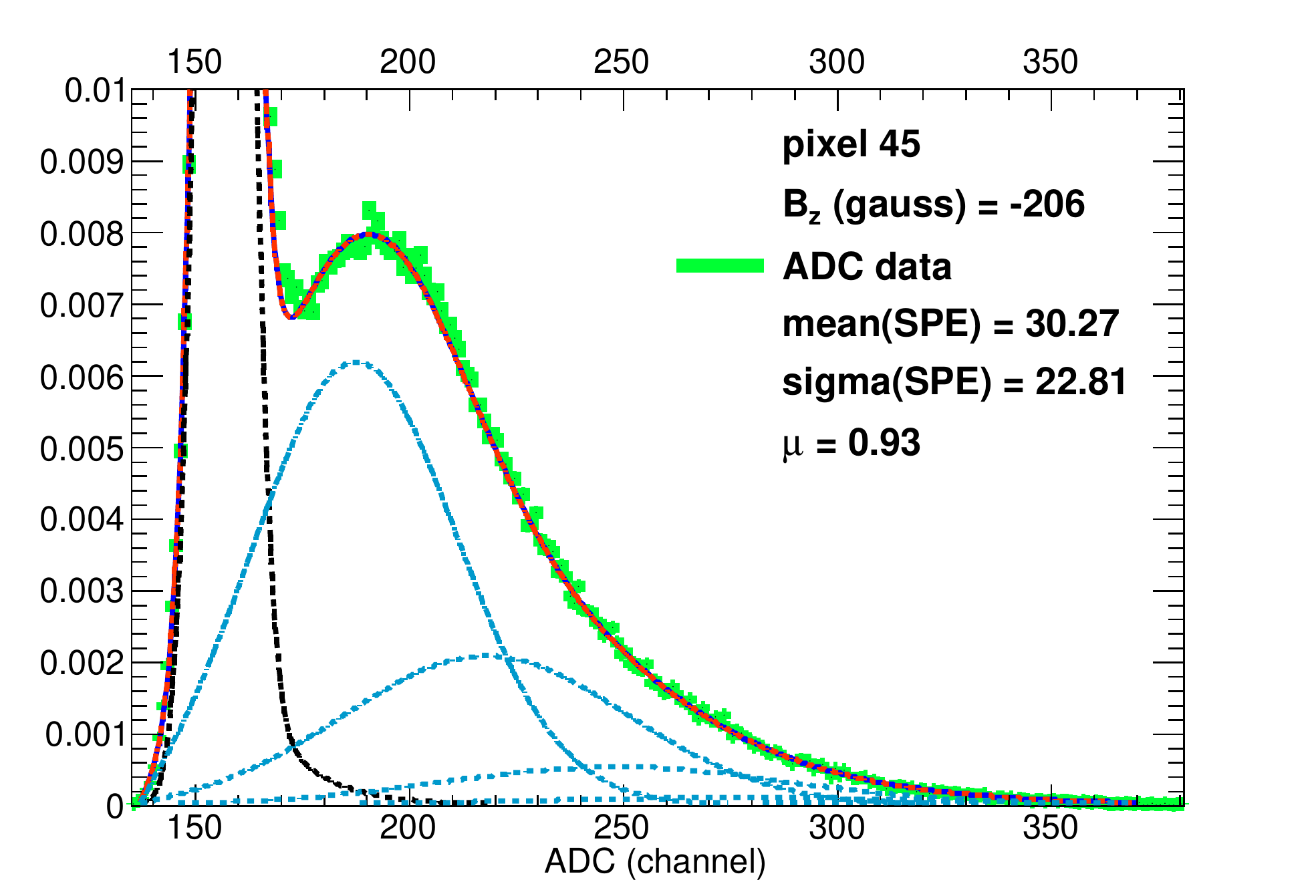} 
&
\hspace{-0.4in}
\includegraphics[width=6.5cm]{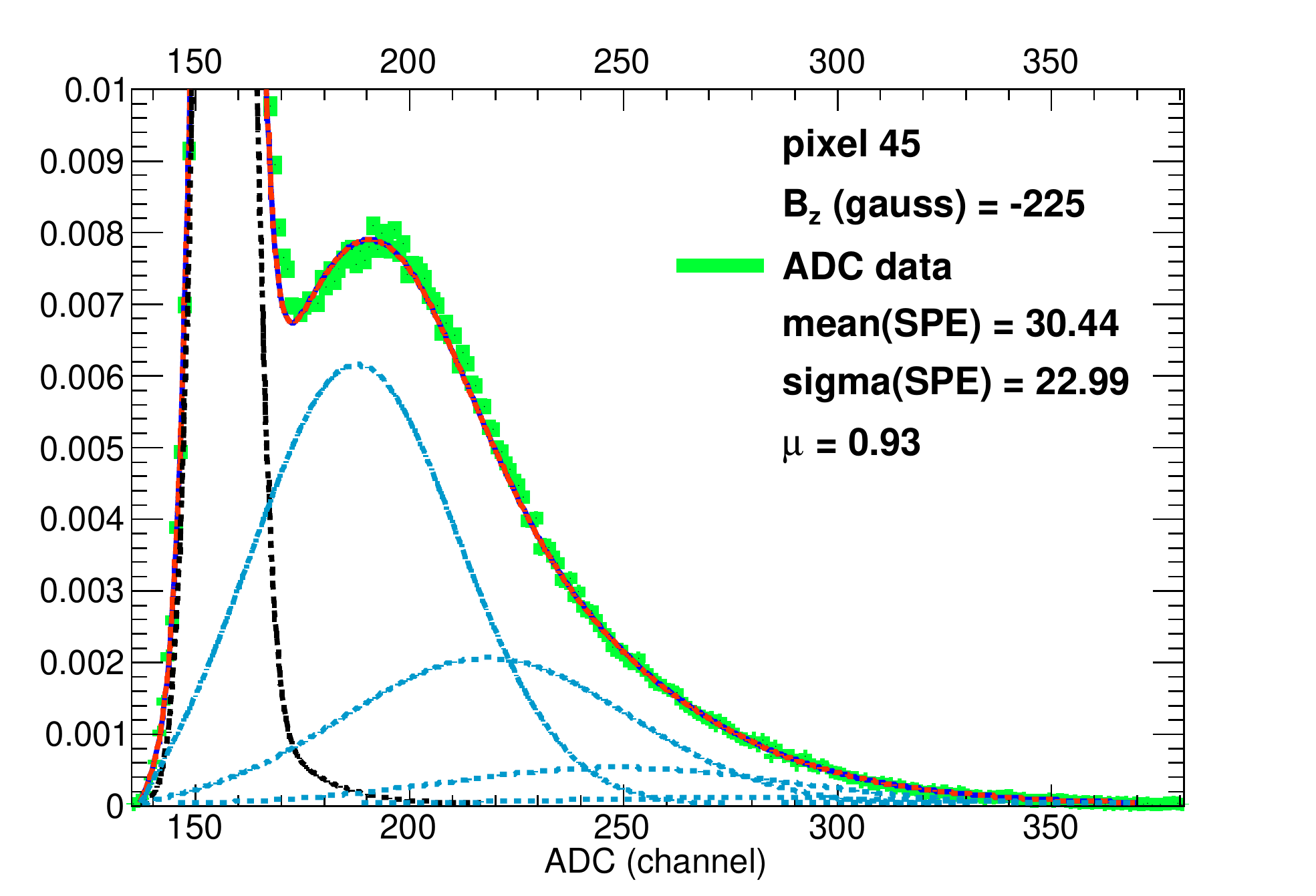} \\
\includegraphics[width=6.5cm]{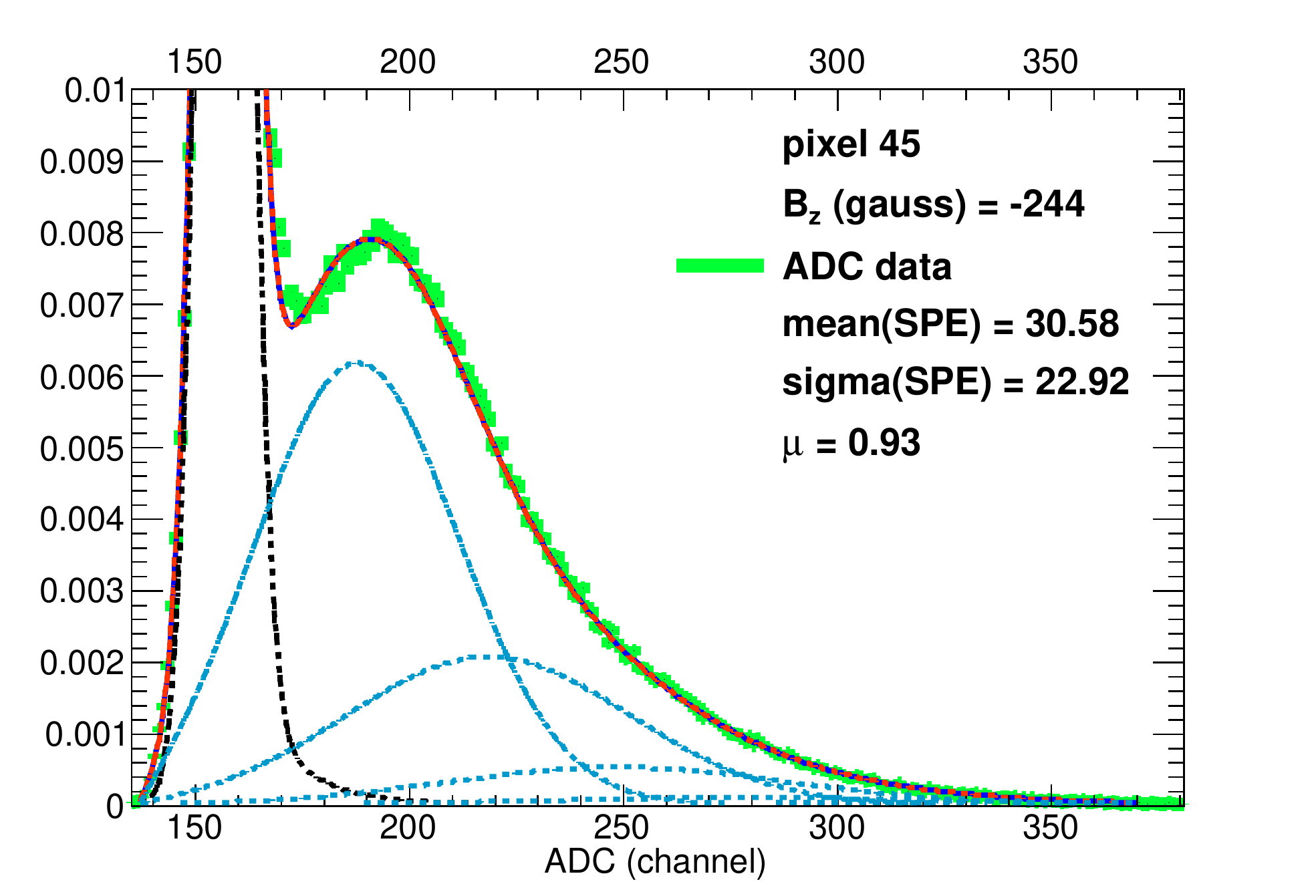}
&
\hspace{-0.4in}
\includegraphics[width=6.5cm]{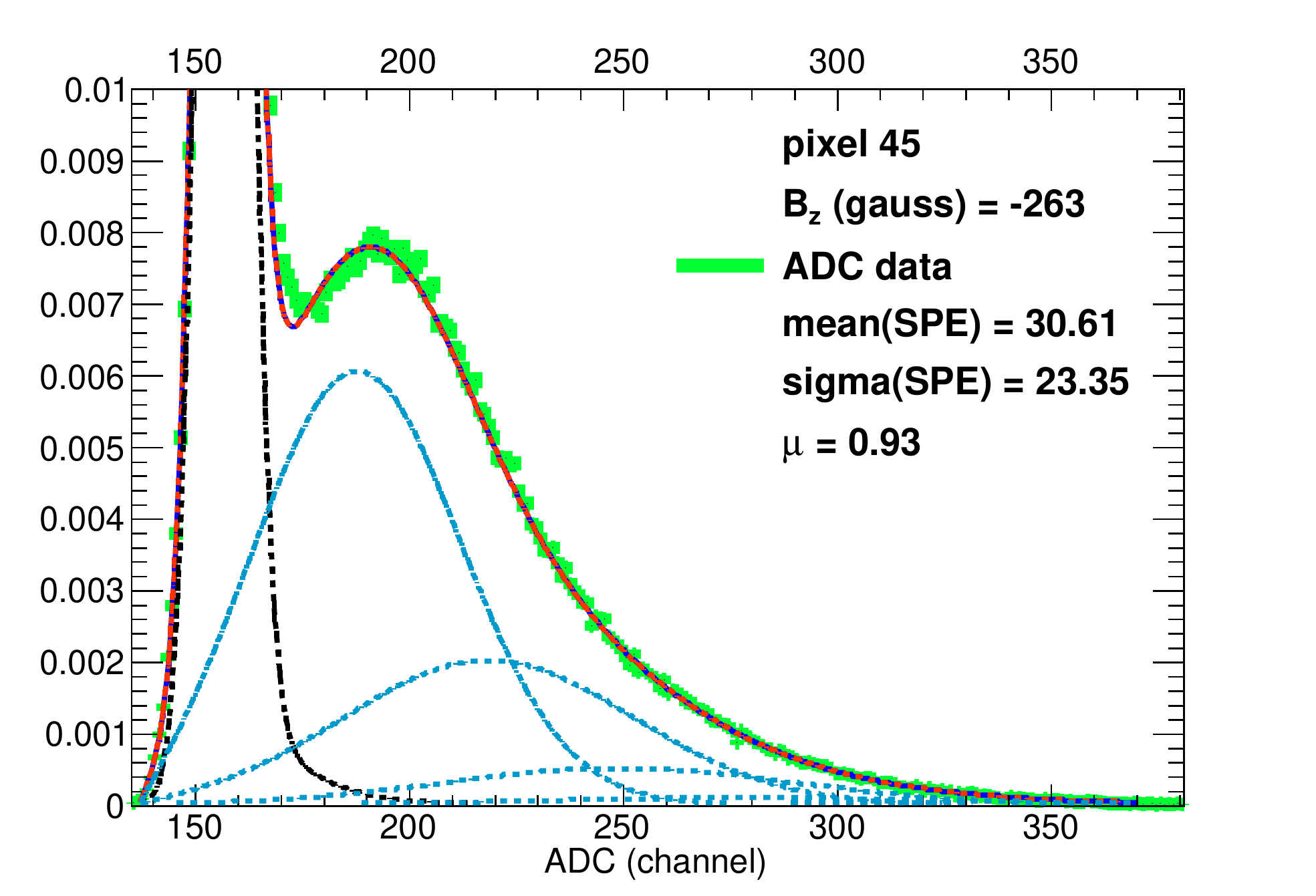}
&
\hspace{-0.4in}
\includegraphics[width=6.5cm]{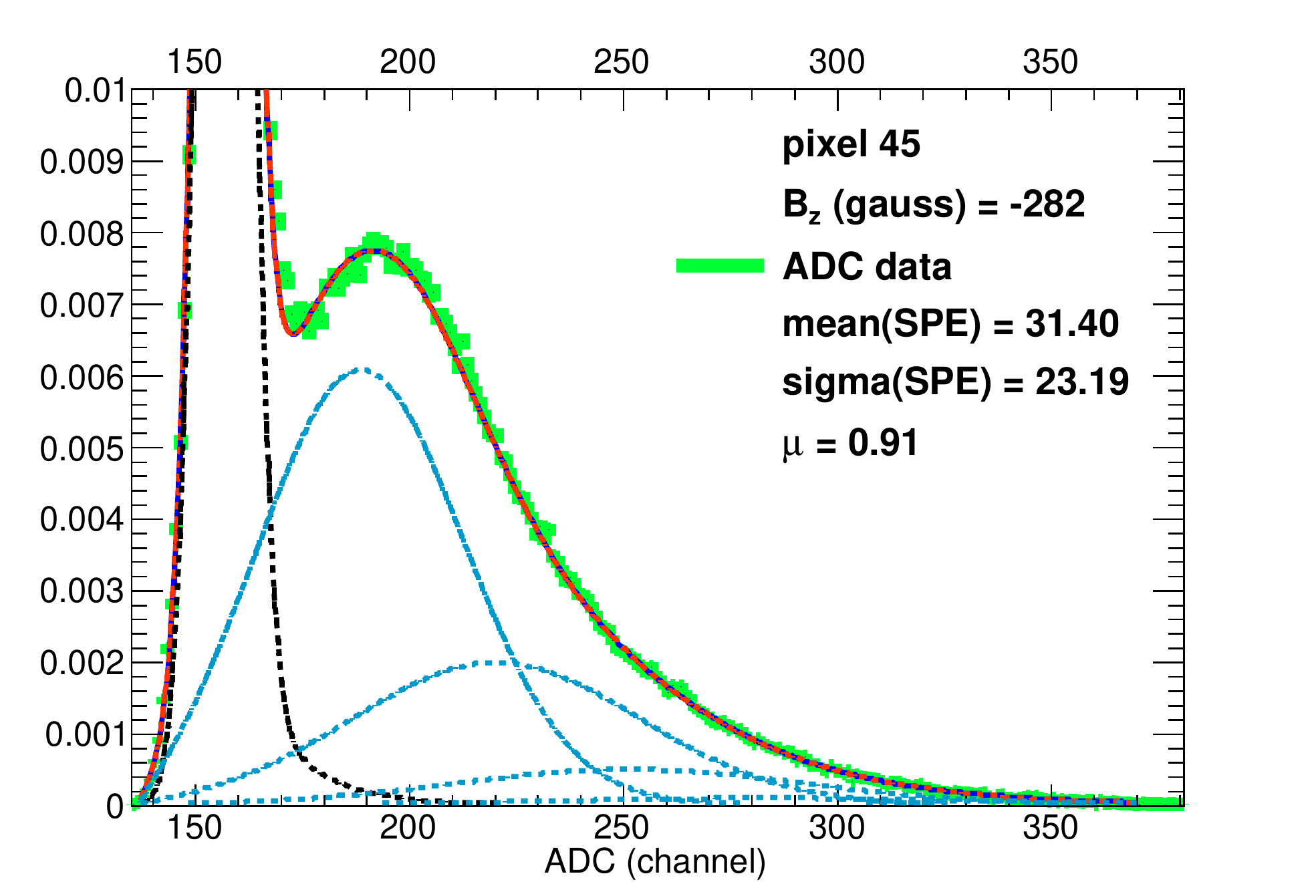} \\

\end{tabular}
\linespread{0.5}
\caption[]{
{Fits of the ADC distributions from pixel 45 for various longitudinal magnetic field settings.} }
\label{45_fit}
\end{figure}

\begin{figure}[htbp]
\vspace*{-0.2in}
\centering
\begin{tabular}{ccc}
\includegraphics[width=6.5cm]{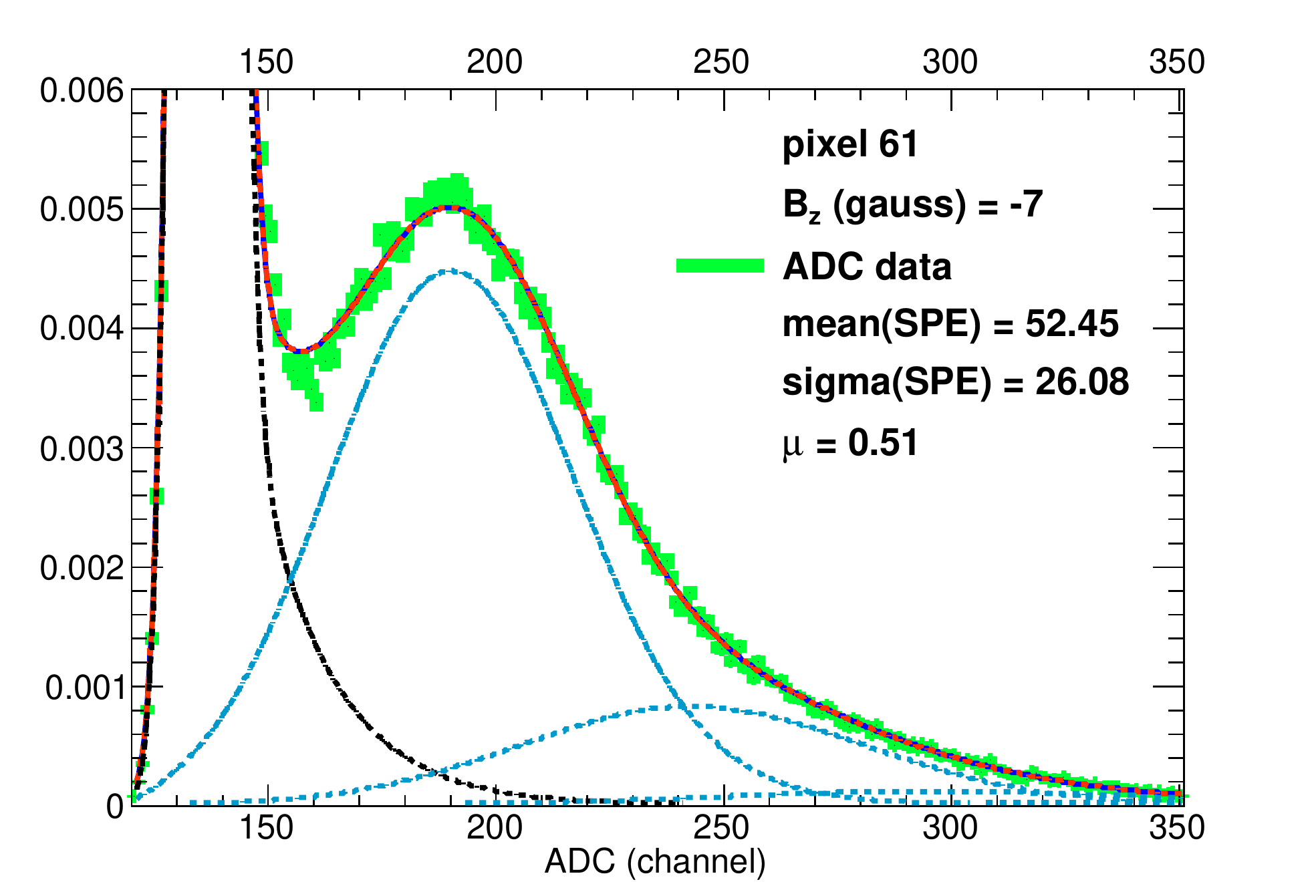}
&
\hspace{-0.4in}
\includegraphics[width=6.5cm]{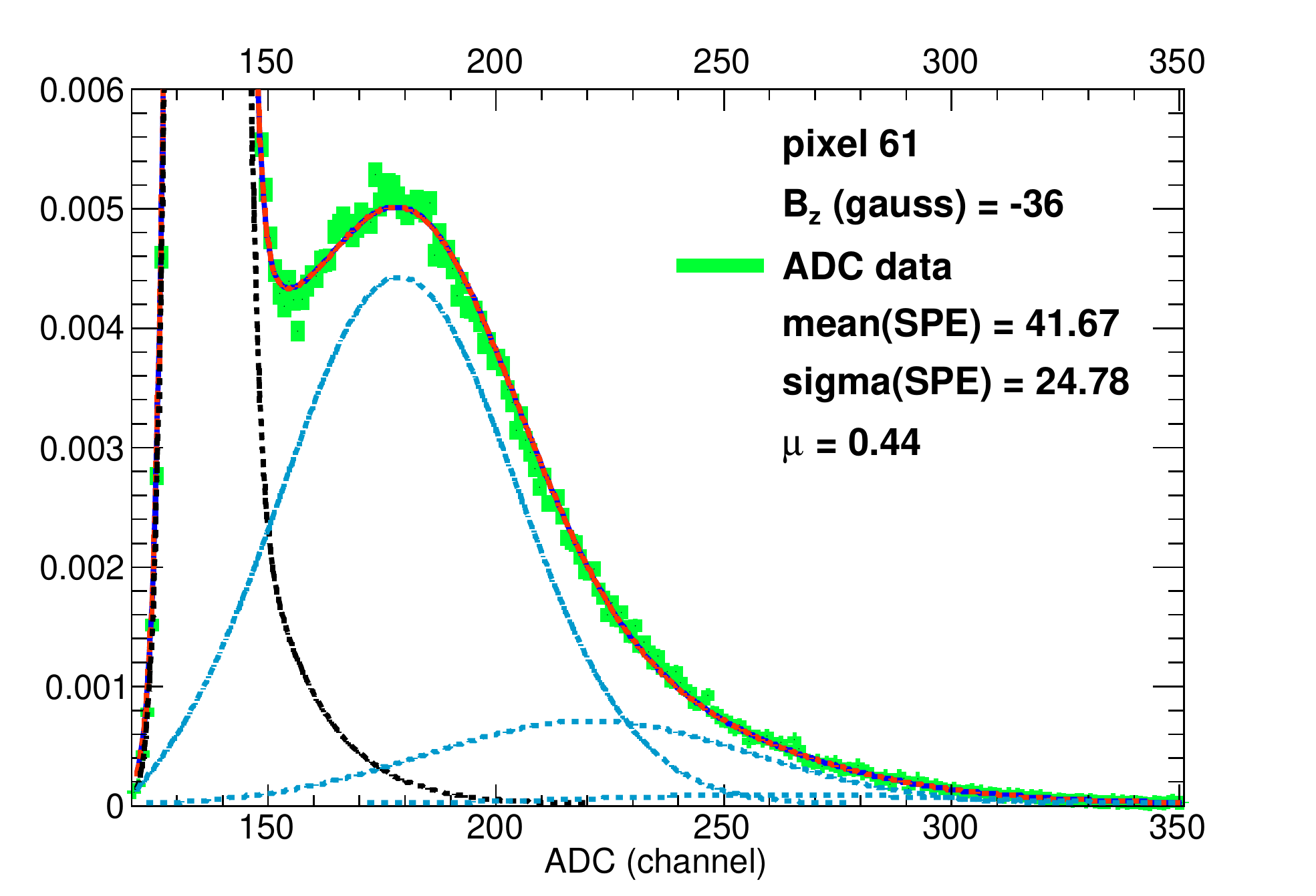}
&
\hspace{-0.4in}
\includegraphics[width=6.5cm]{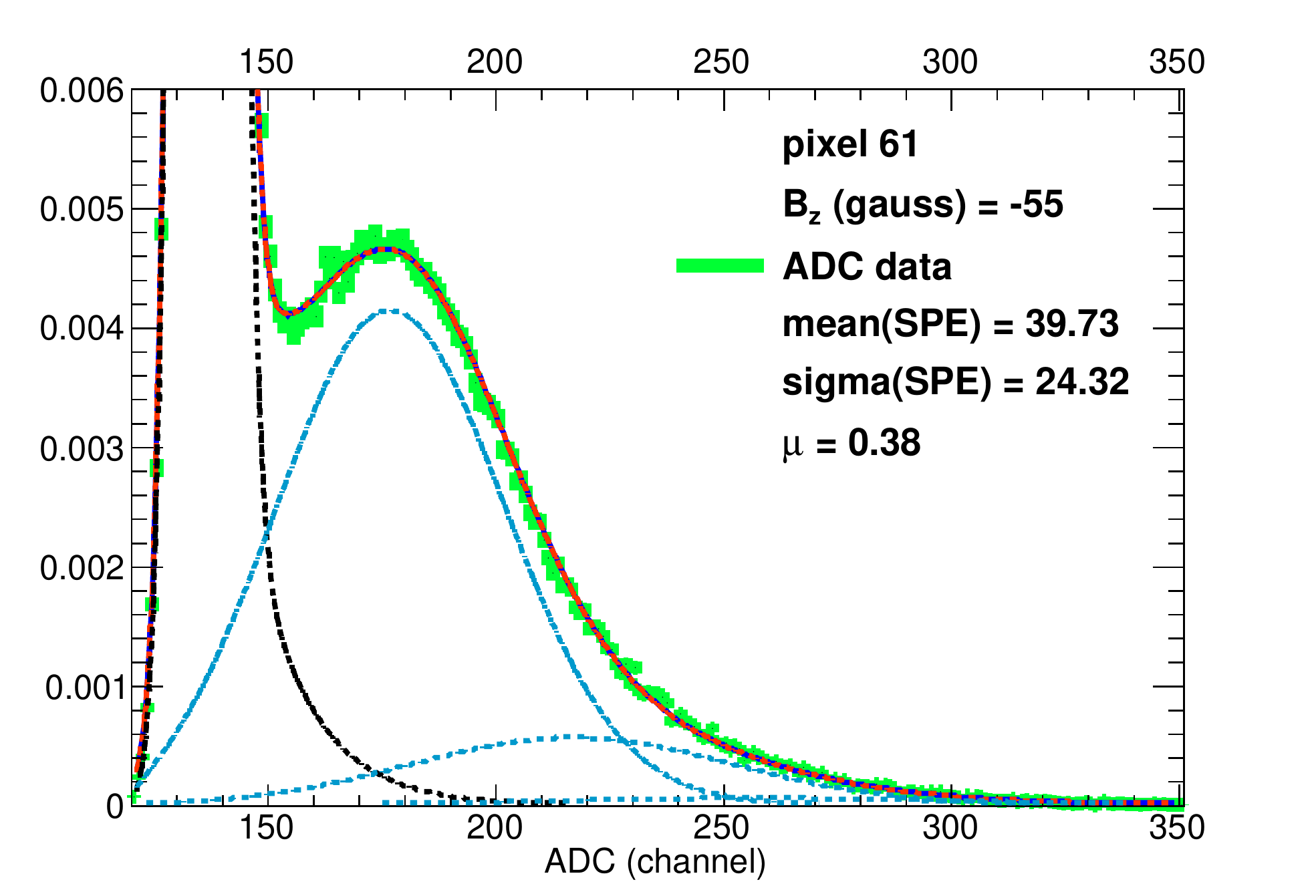} \\
\includegraphics[width=6.5cm]{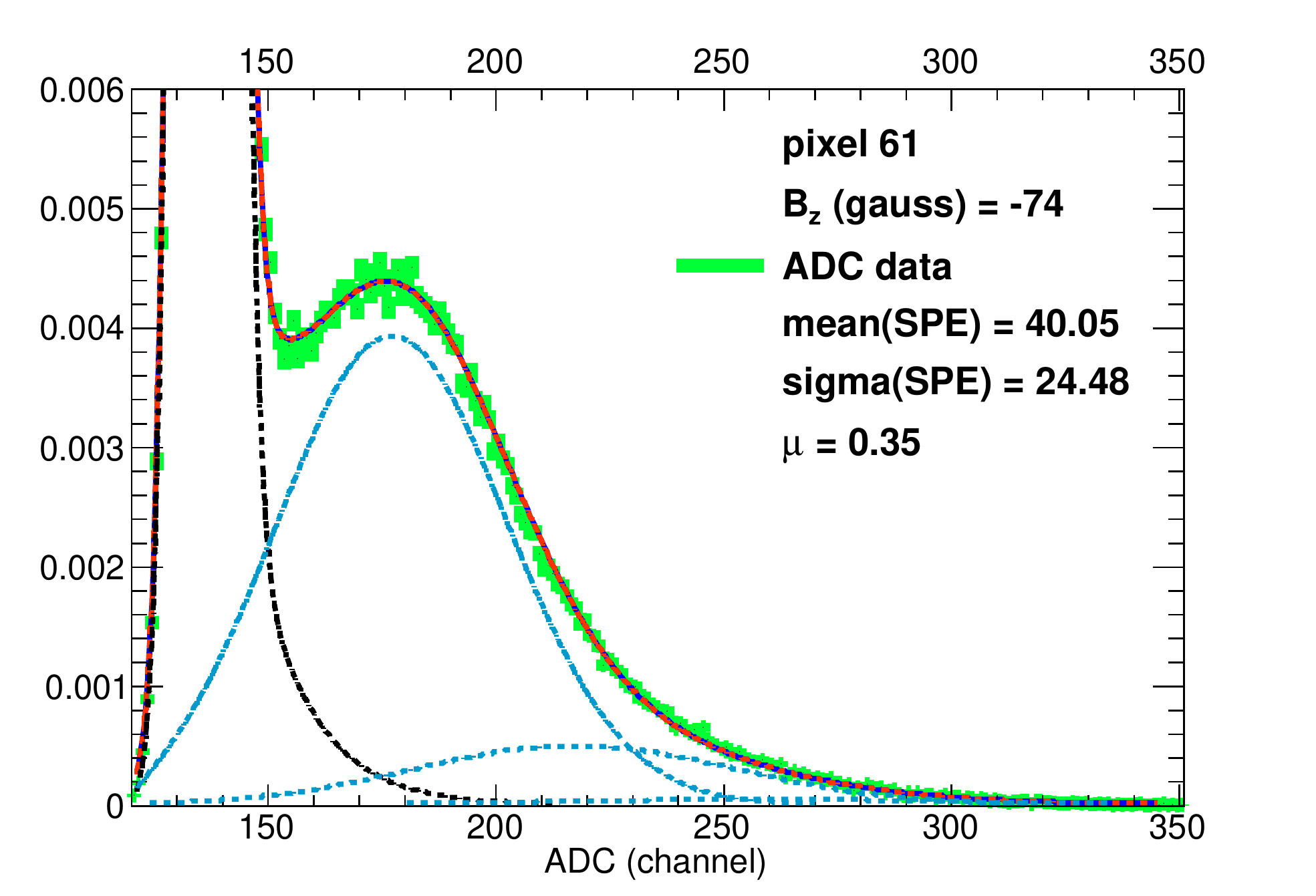}
&
\hspace{-0.4in}
\includegraphics[width=6.5cm]{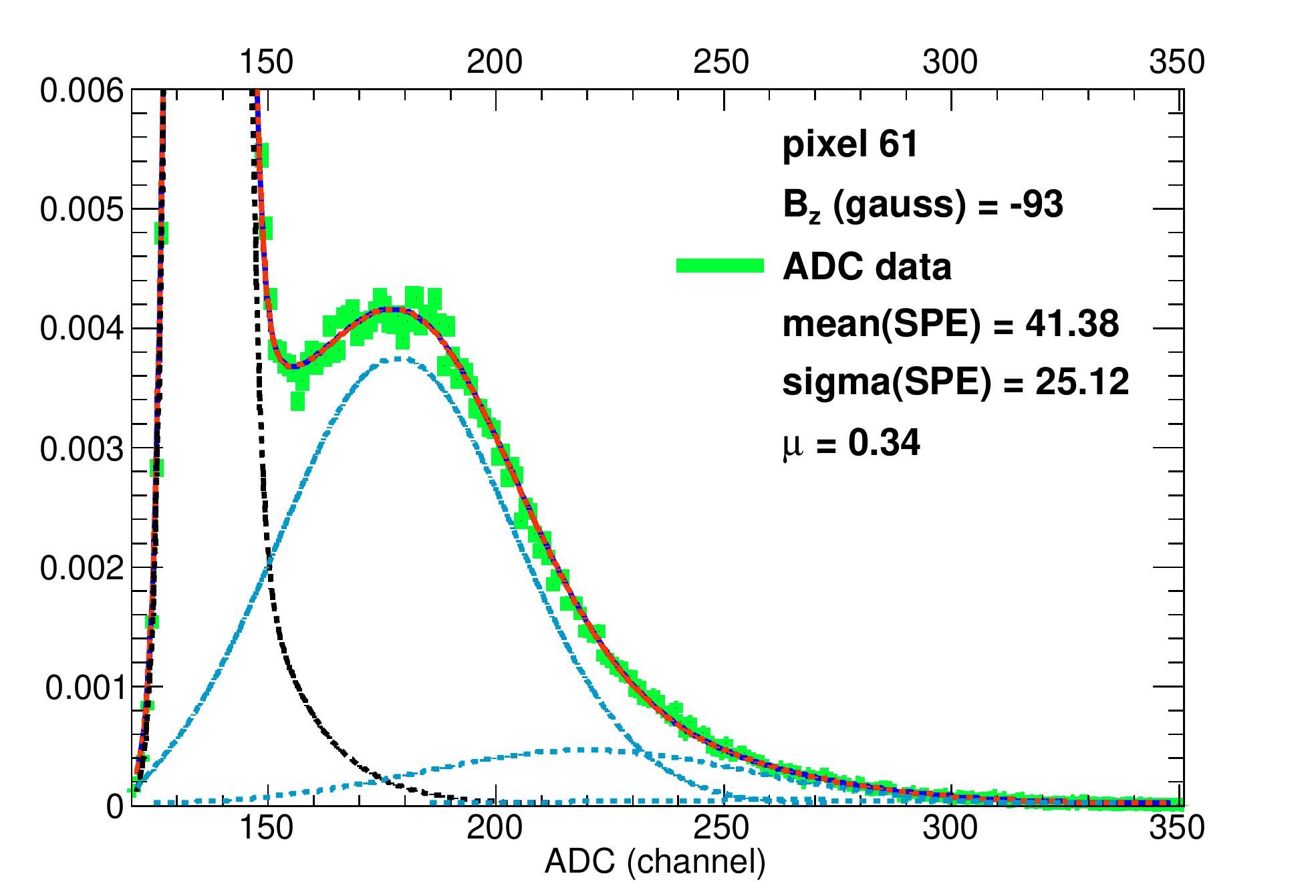}
&
\hspace{-0.4in}
\includegraphics[width=6.5cm]{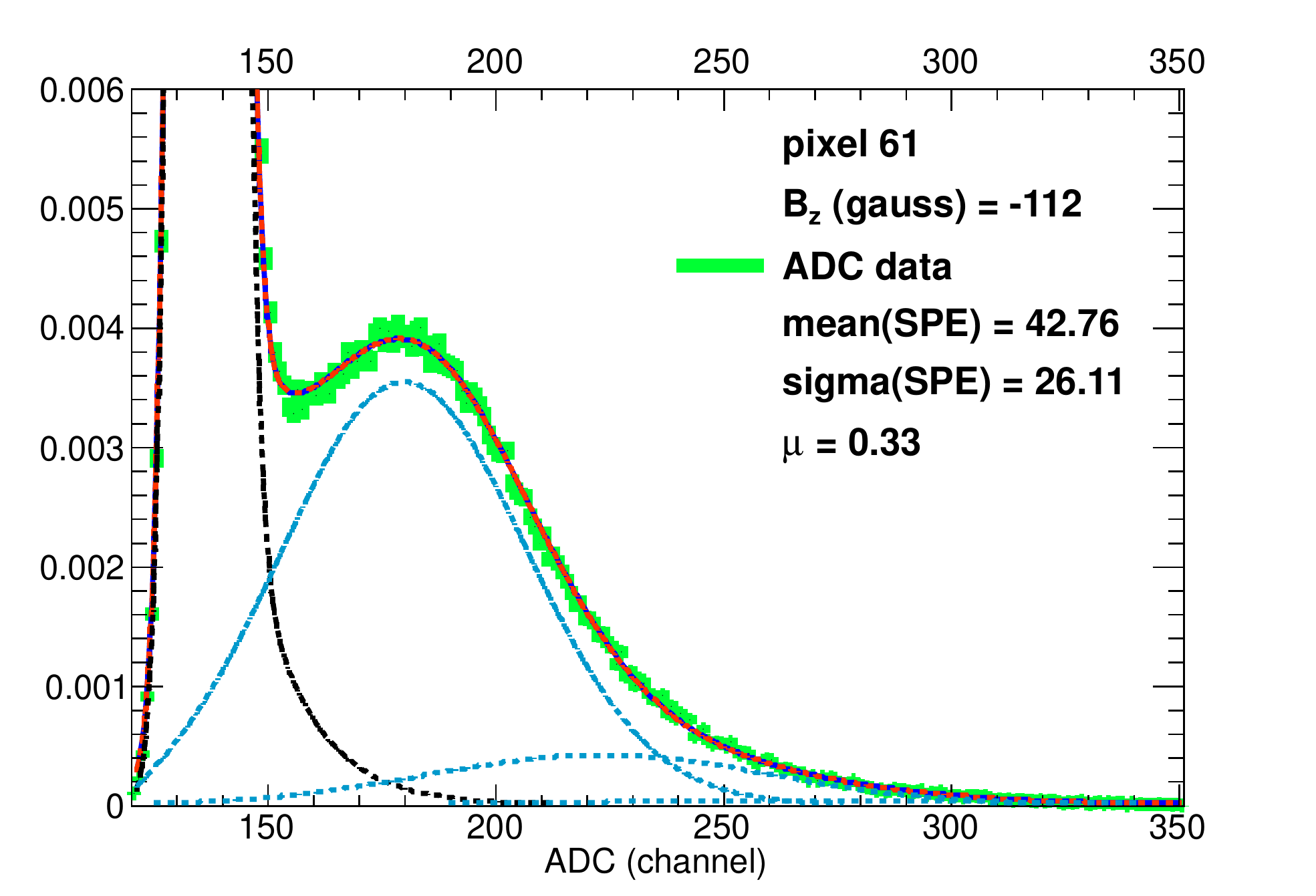} \\
\includegraphics[width=6.5cm]{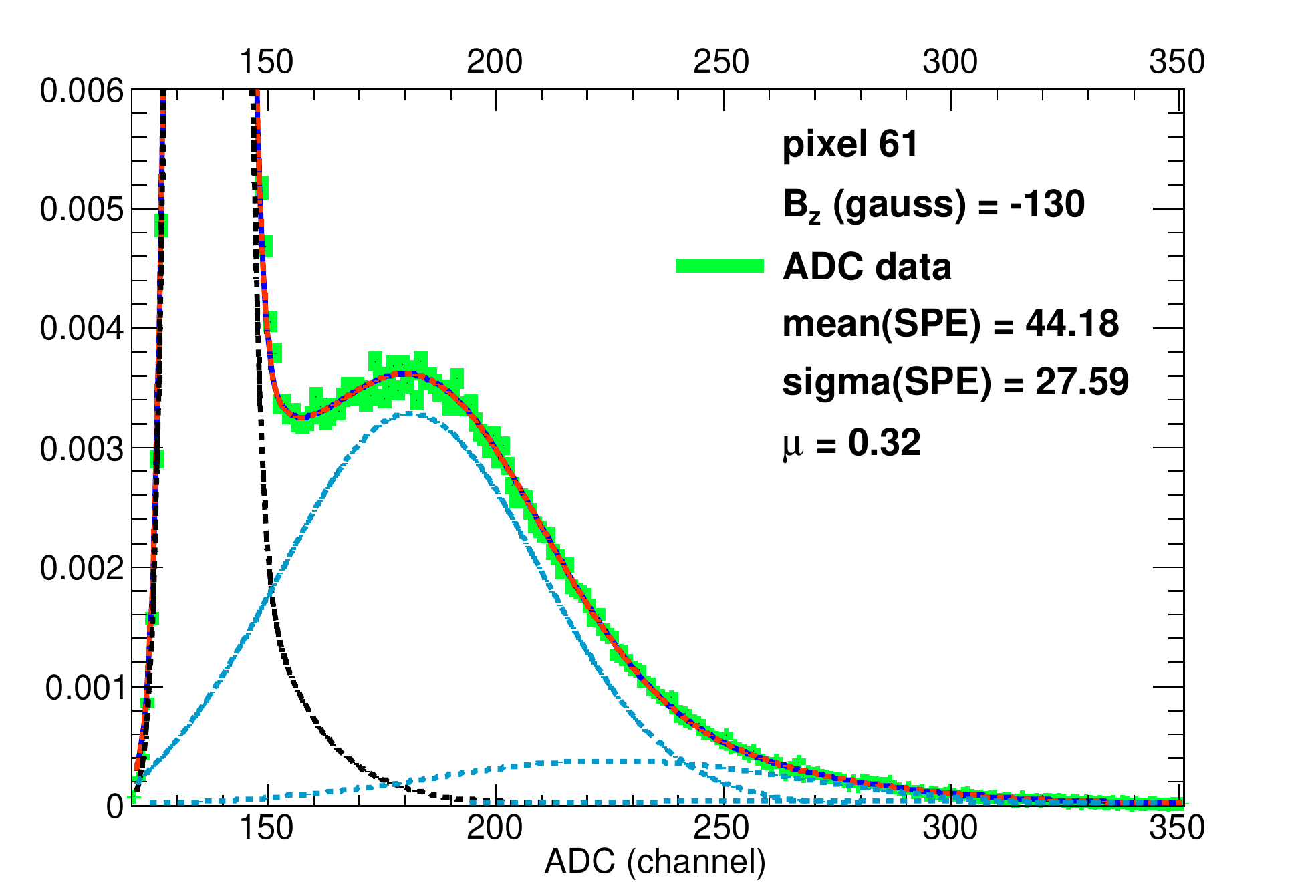} 
&
\hspace{-0.4in}
\includegraphics[width=6.5cm]{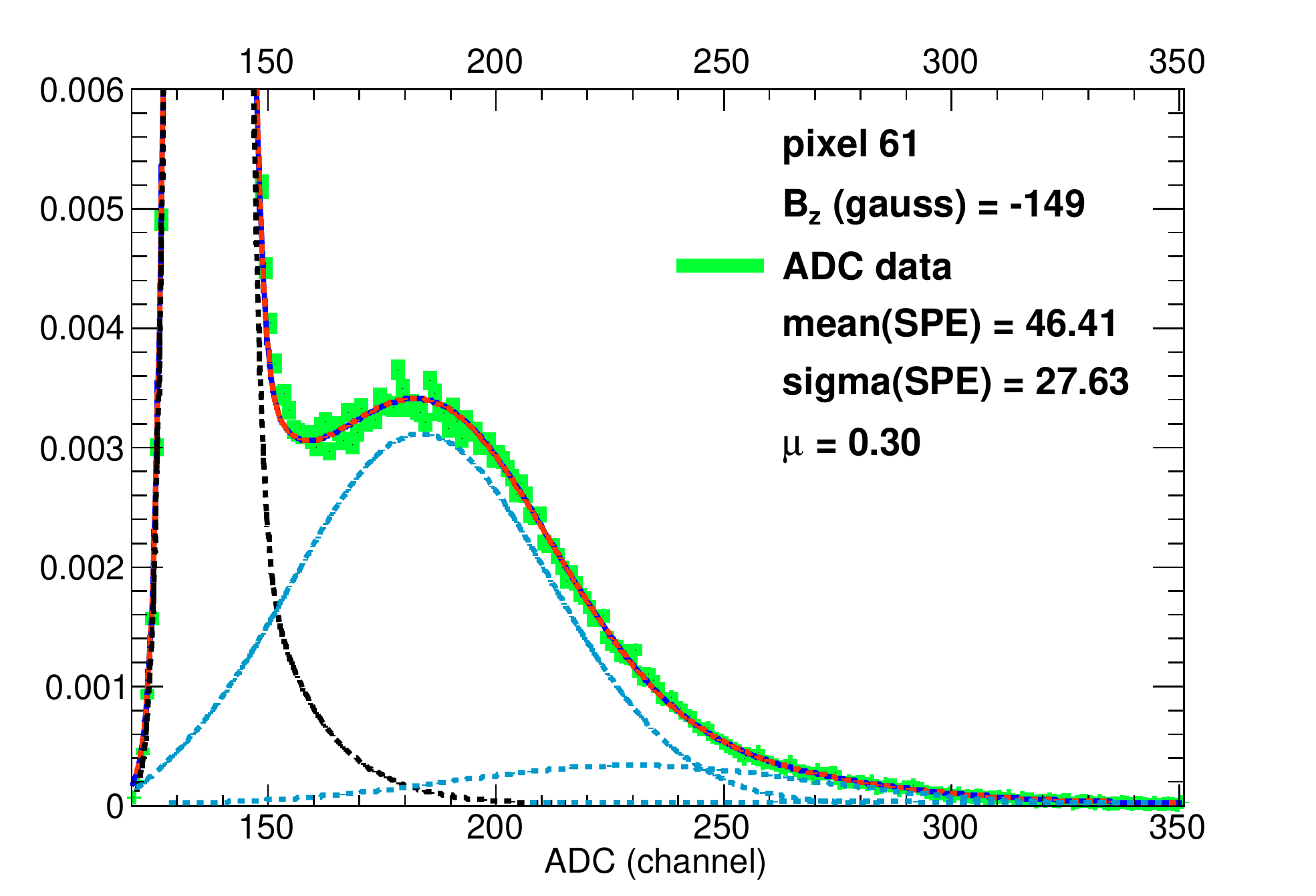}
&
\hspace{-0.4in}
\includegraphics[width=6.5cm]{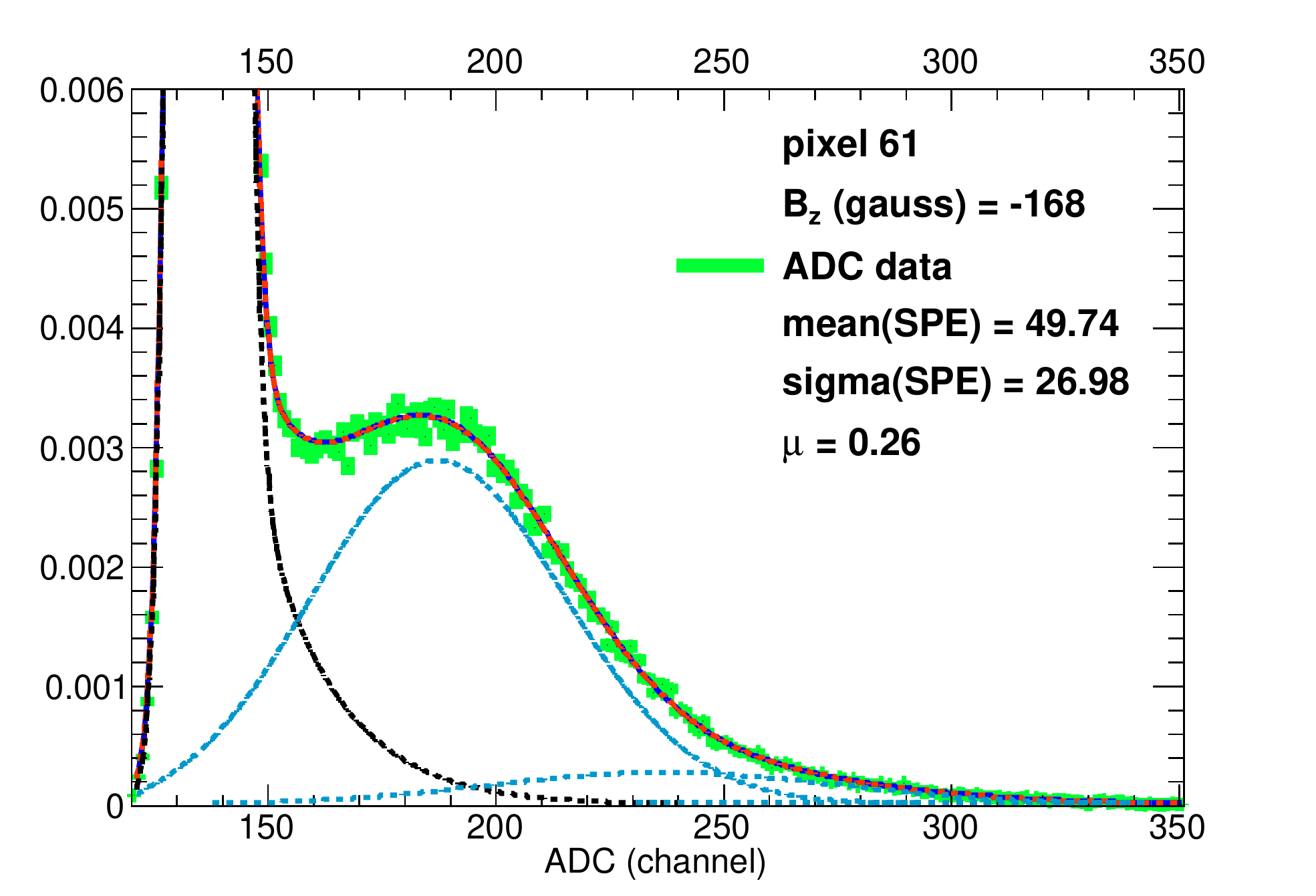} \\
\includegraphics[width=6.5cm]{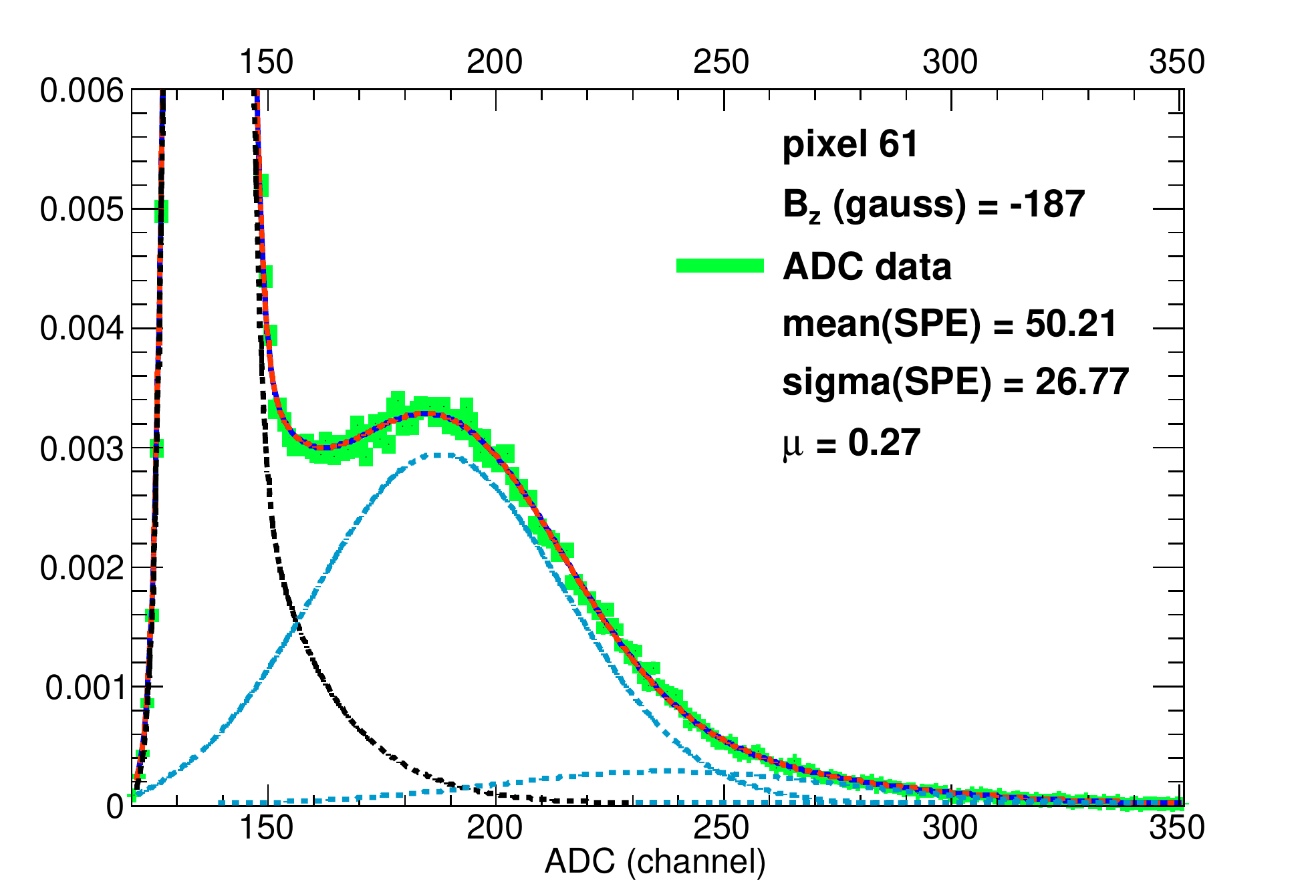}
&
\hspace{-0.4in}
\includegraphics[width=6.5cm]{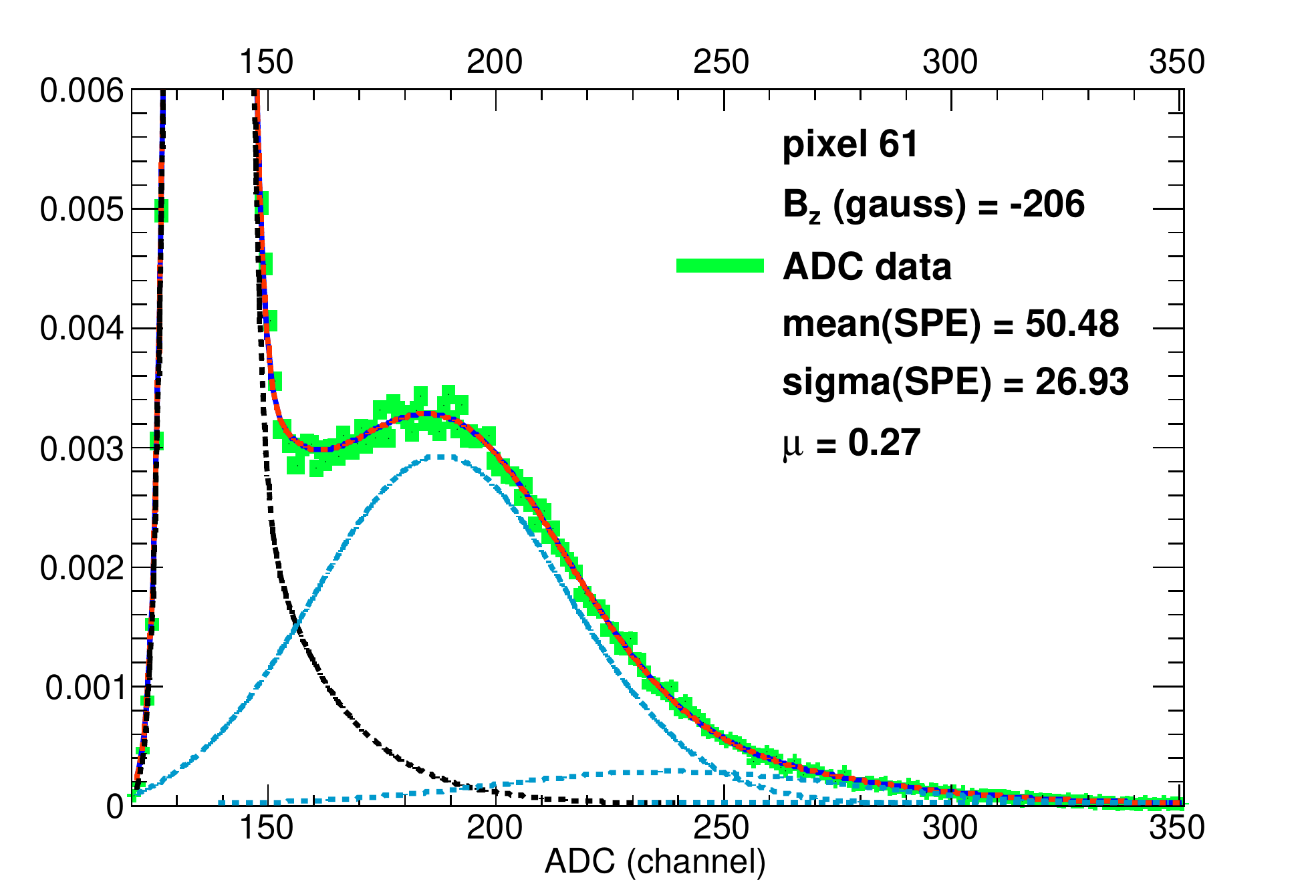} 
&
\hspace{-0.4in}
\includegraphics[width=6.5cm]{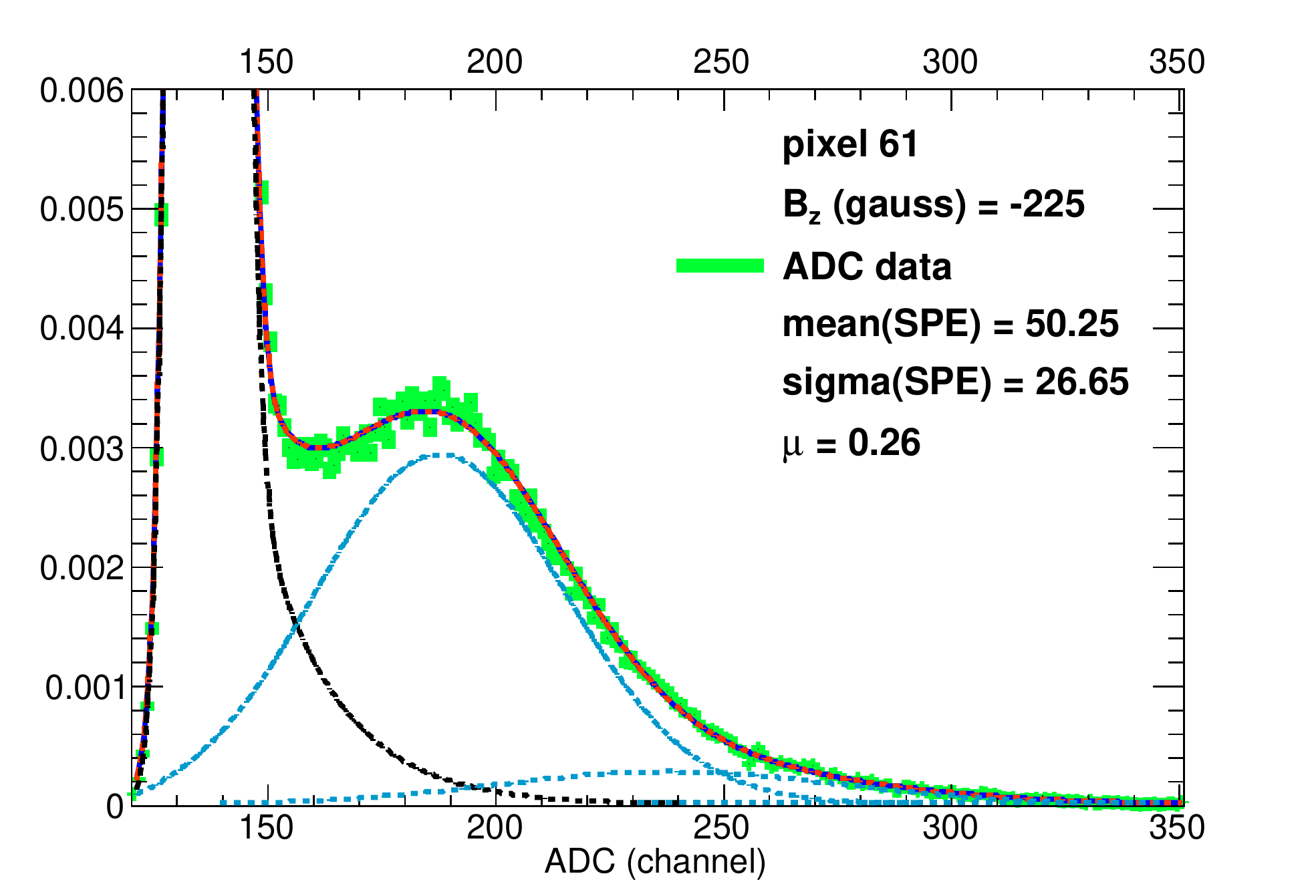} \\
\includegraphics[width=6.5cm]{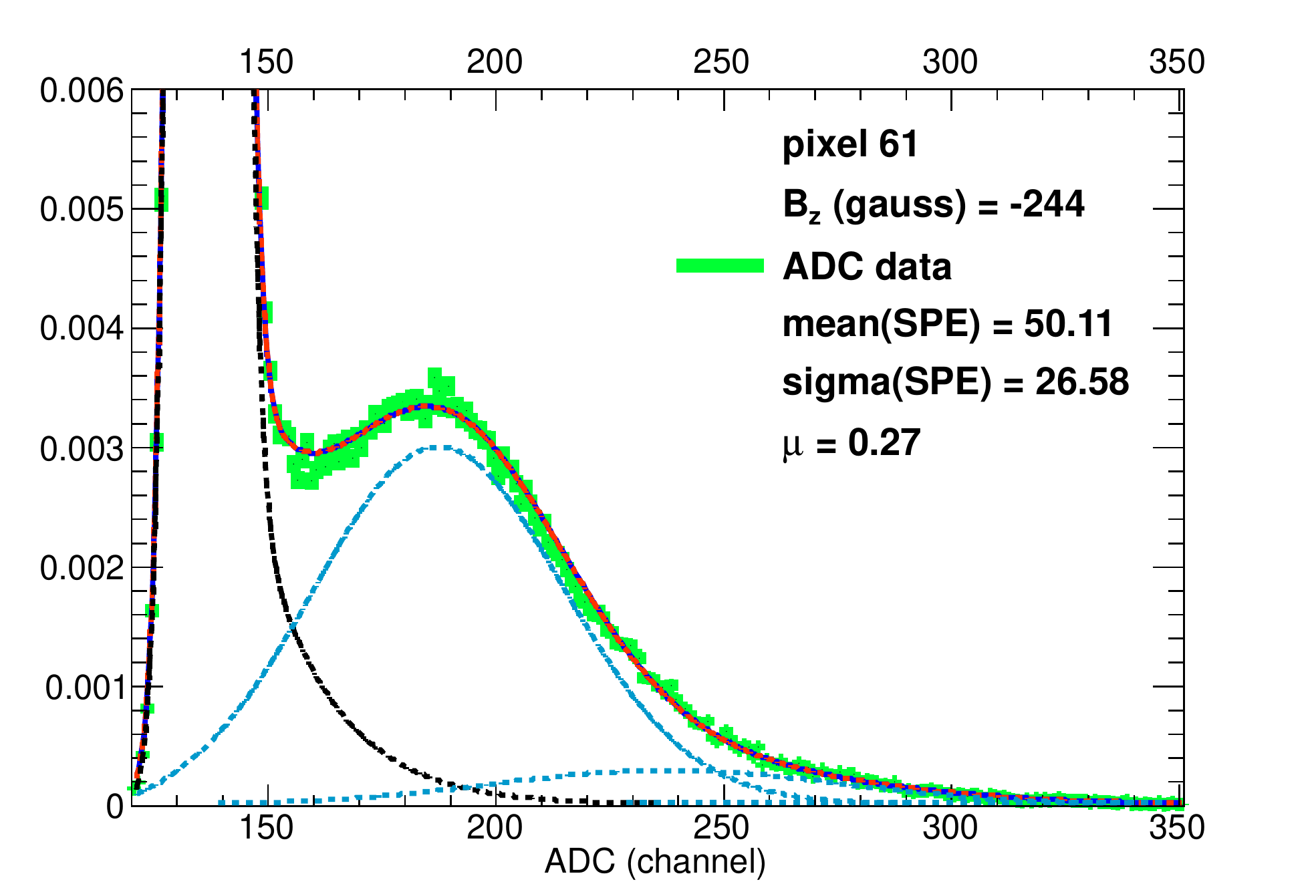}
&
\hspace{-0.4in}
\includegraphics[width=6.5cm]{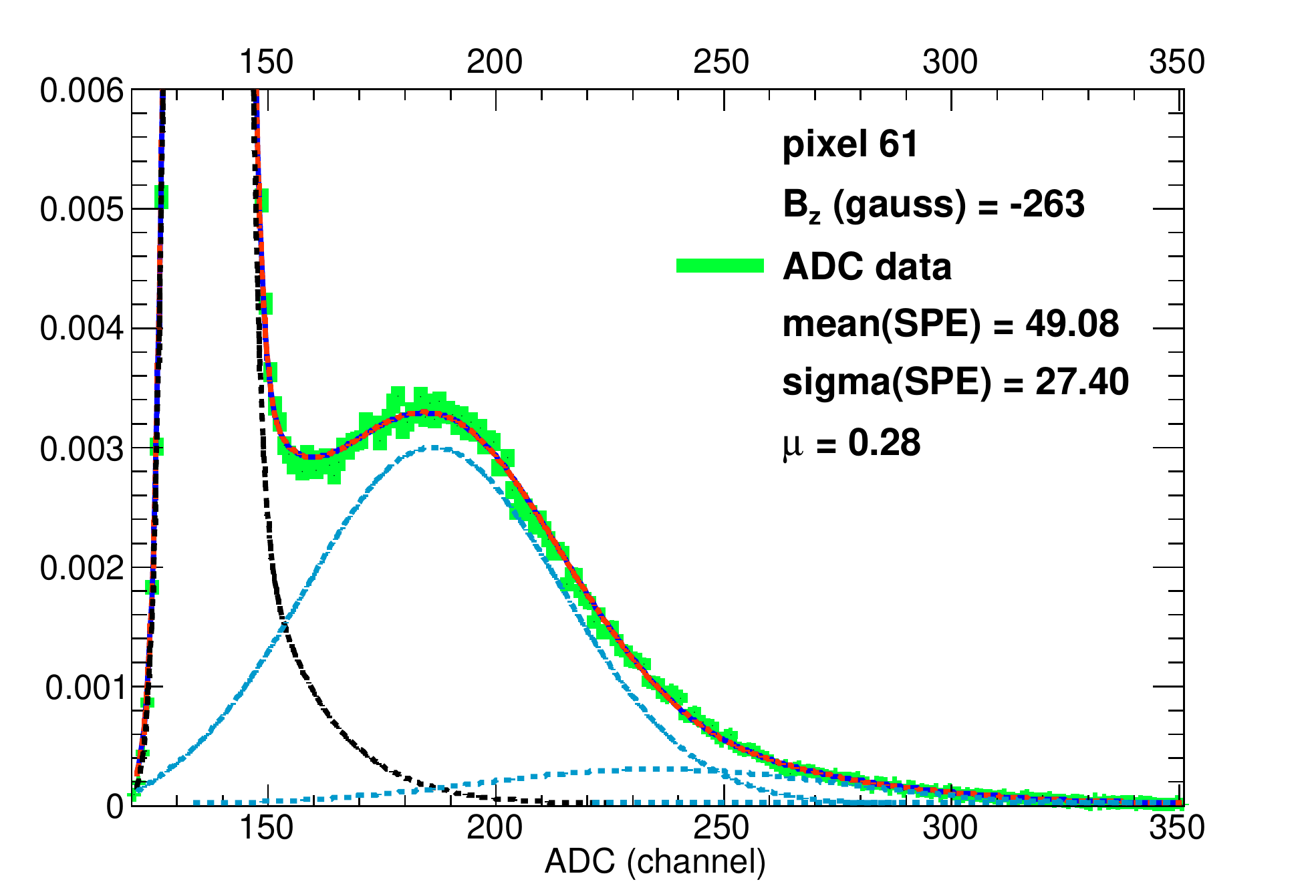}
&
\hspace{-0.4in}
\includegraphics[width=6.5cm]{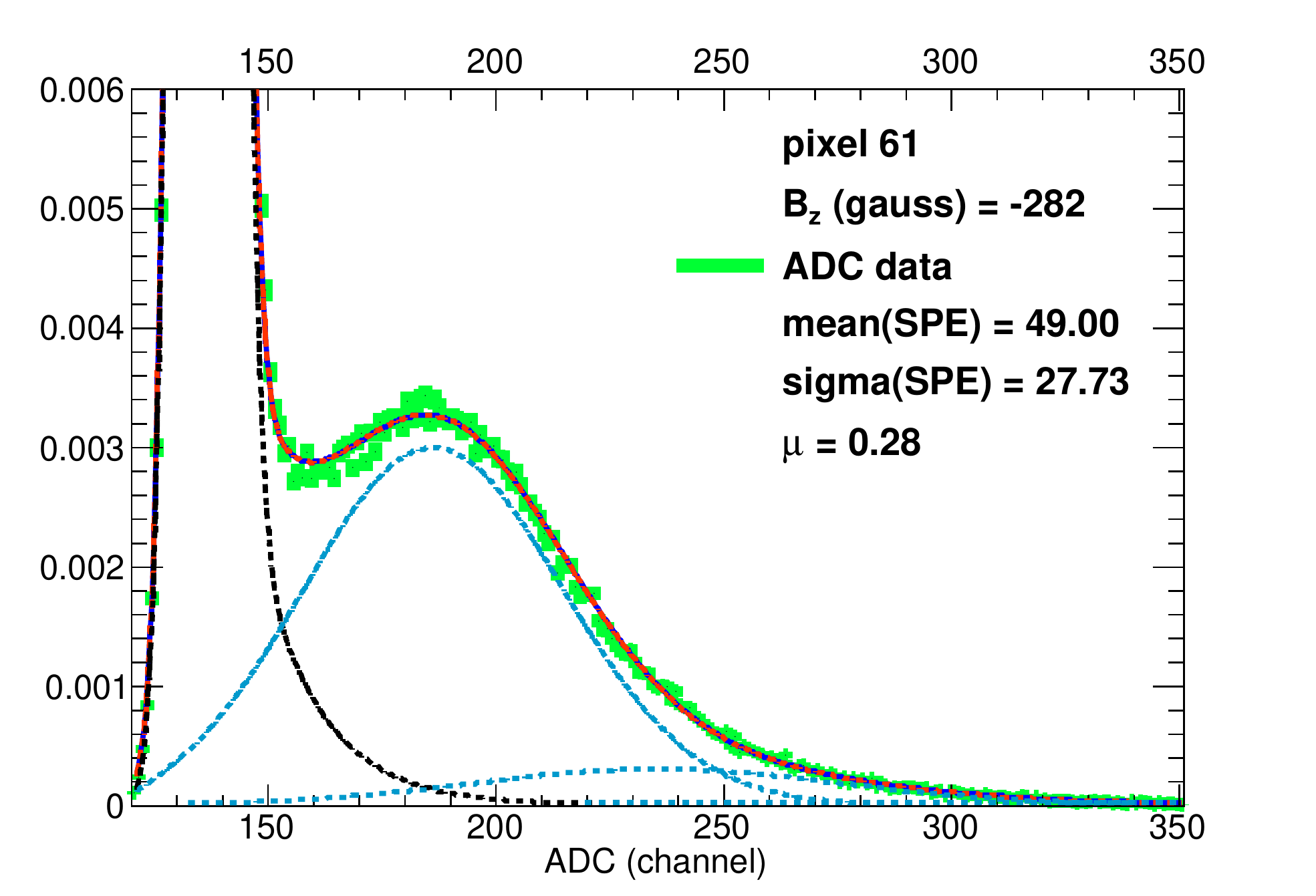} \\

\end{tabular}
\linespread{0.5}
\caption[]{
{Fits of the ADC distributions from pixel 61 for various longitudinal magnetic field settings.} }
\label{61_fit}
\end{figure}

\begin{figure}[htbp]
\vspace*{-0.2in}
\centering
\begin{tabular}{ccc}
\includegraphics[width=6.5cm]{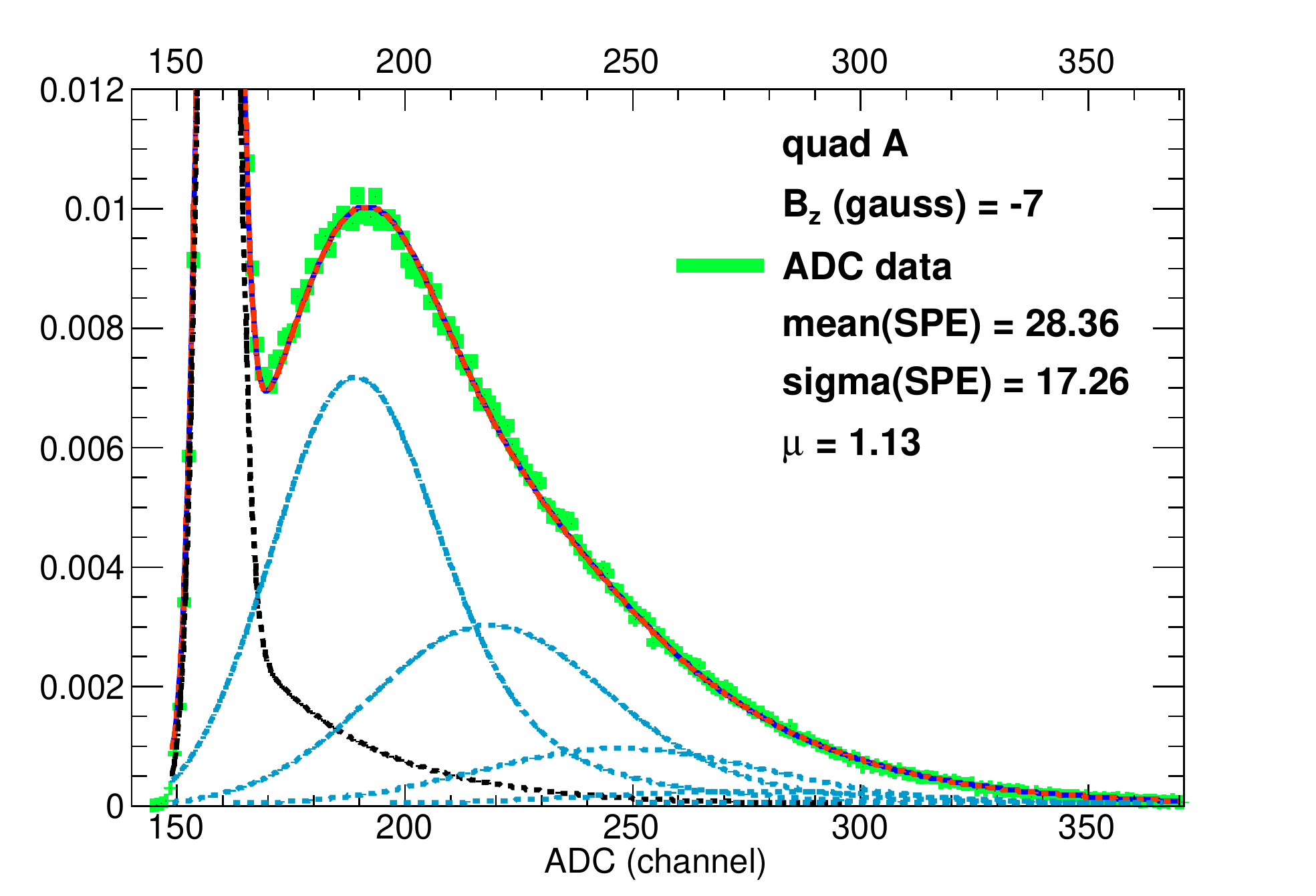}
&
\hspace{-0.4in}
\includegraphics[width=6.5cm]{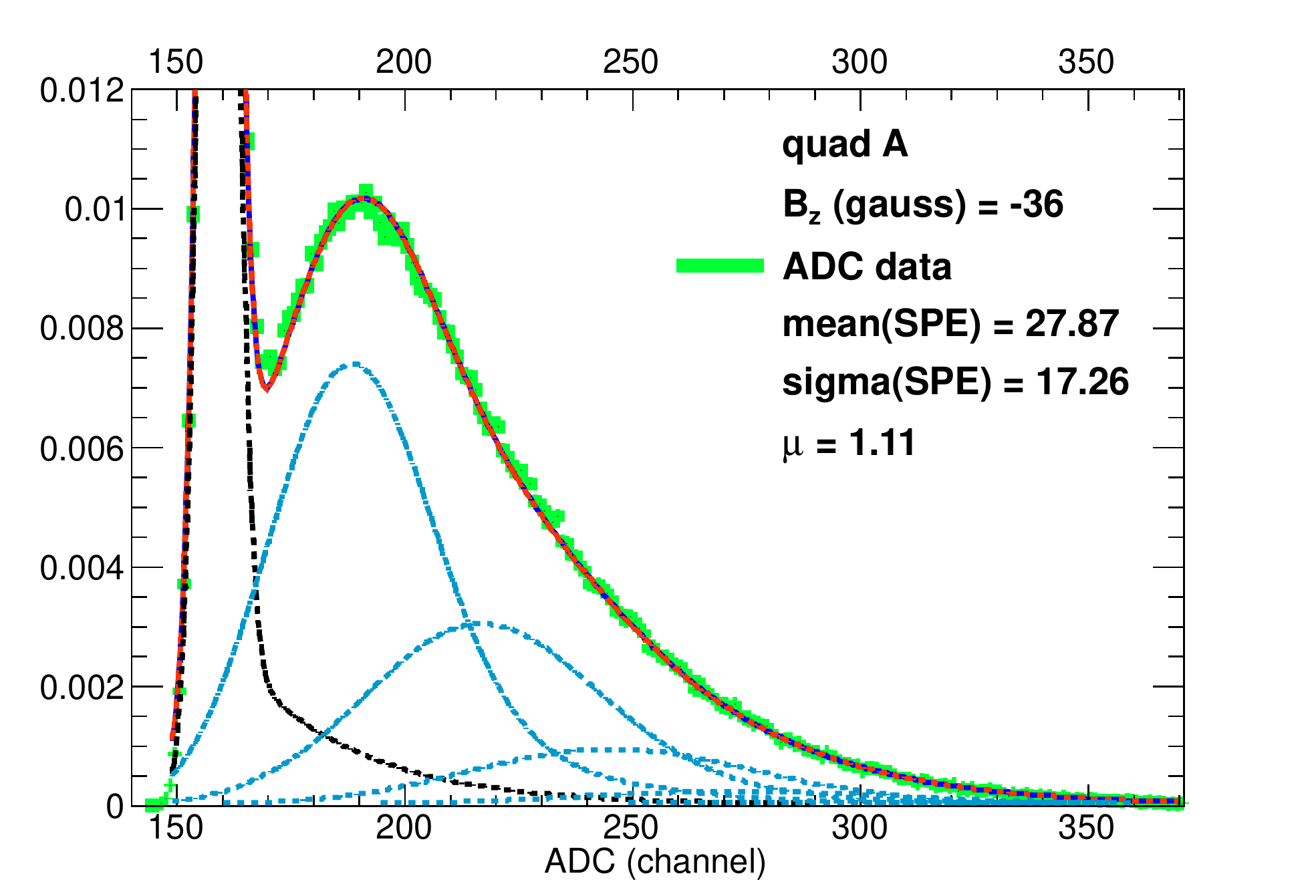}
&
\hspace{-0.4in}
\includegraphics[width=6.5cm]{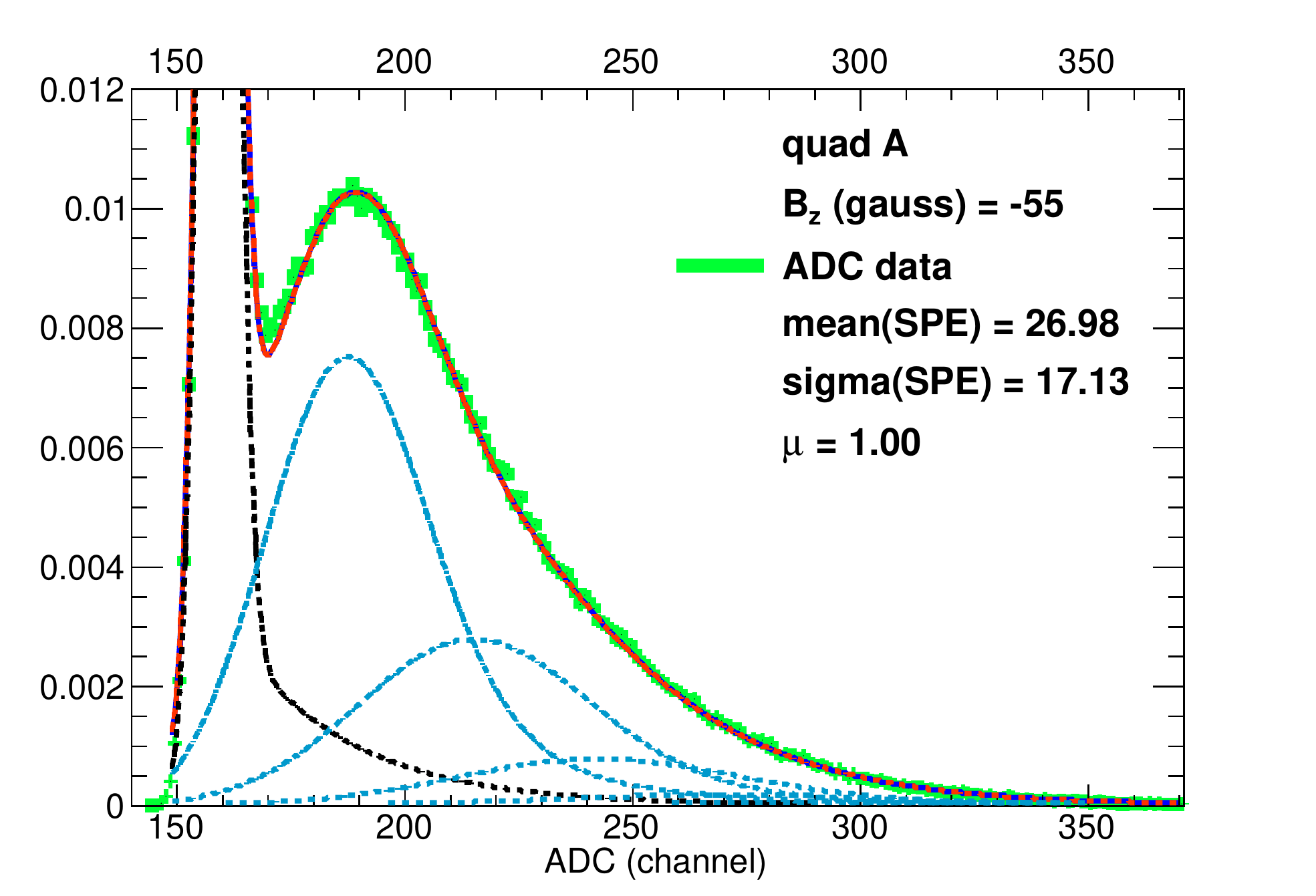} \\
\includegraphics[width=6.5cm]{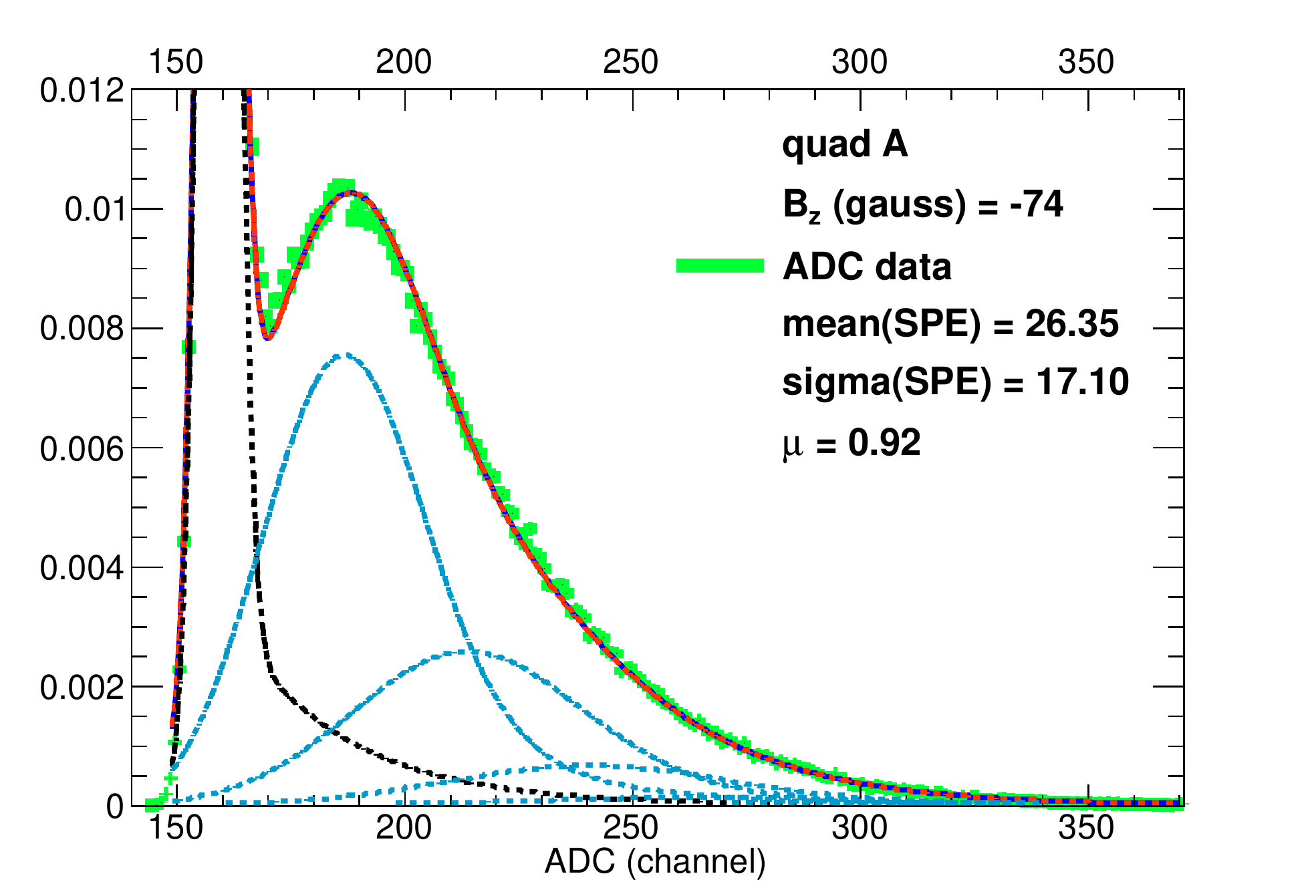}
&
\hspace{-0.4in}
\includegraphics[width=6.5cm]{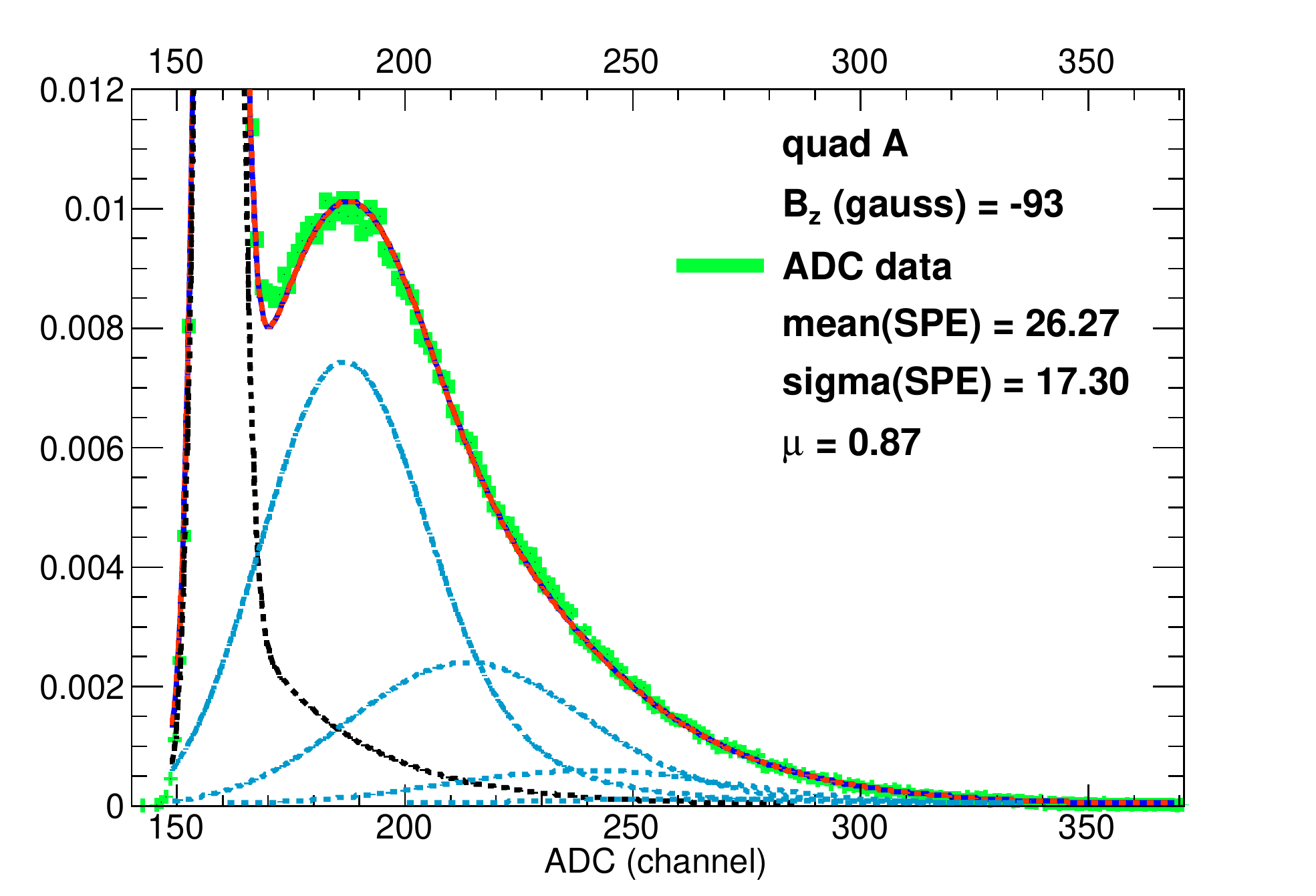}
&
\hspace{-0.4in}
\includegraphics[width=6.5cm]{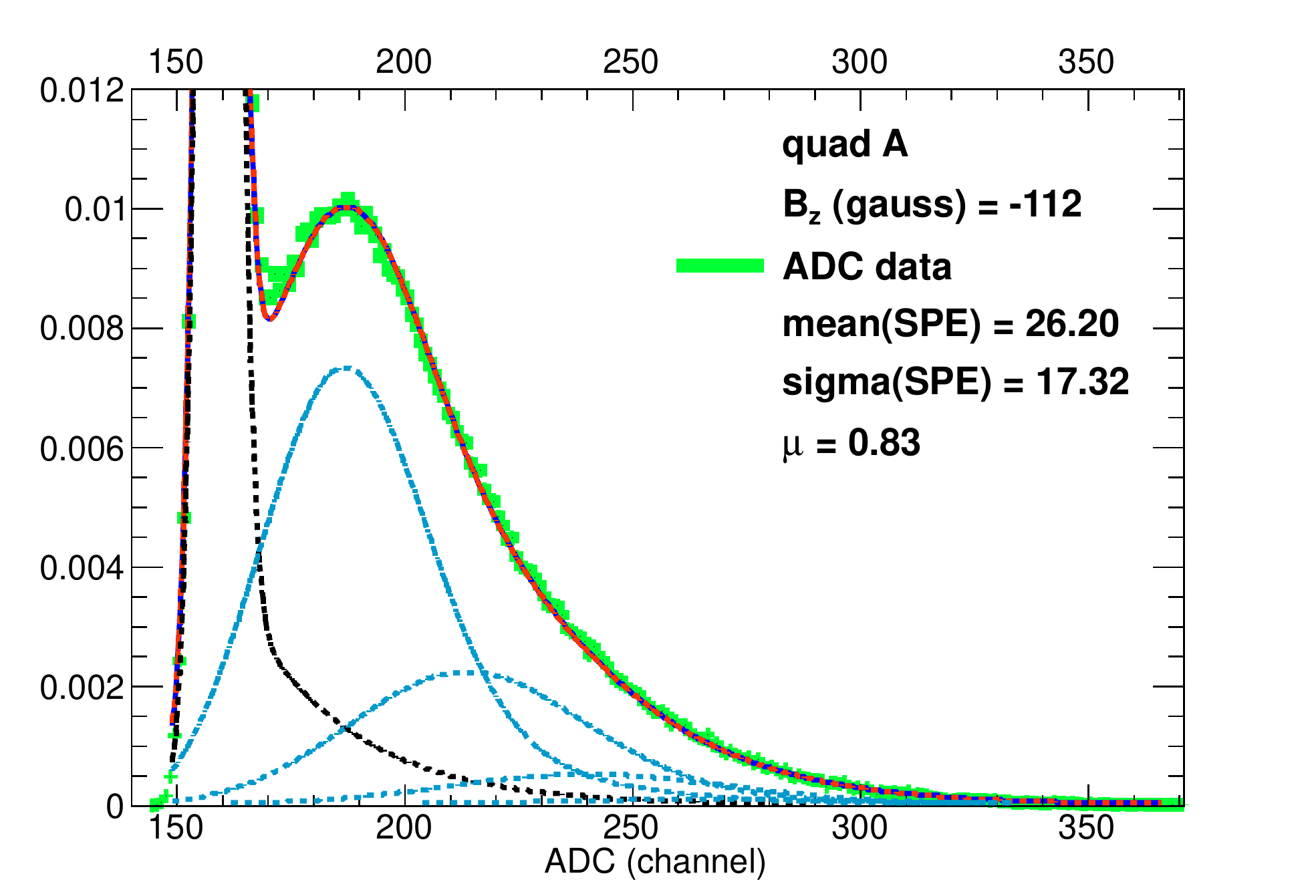} \\
\includegraphics[width=6.5cm]{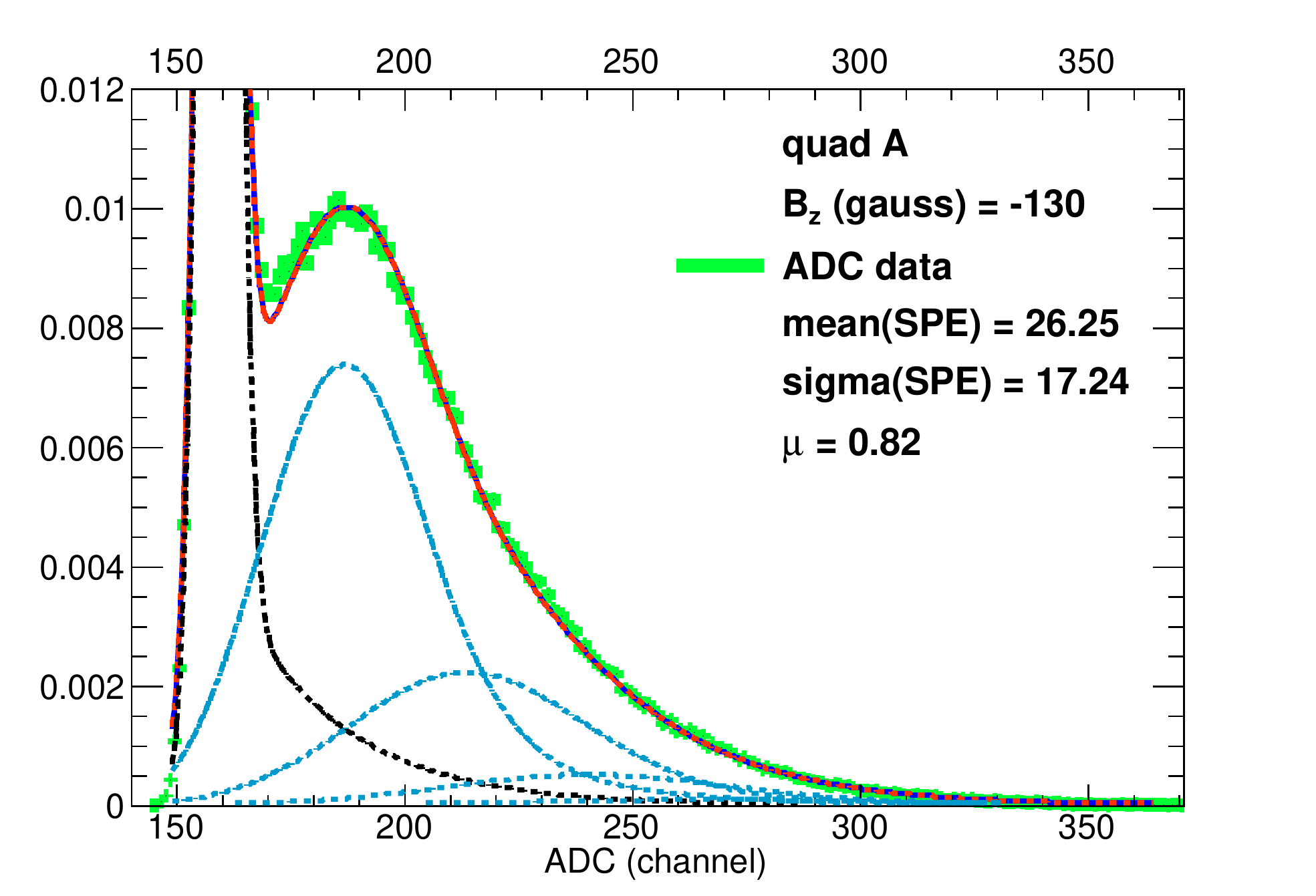} 
&
\hspace{-0.4in}
\includegraphics[width=6.5cm]{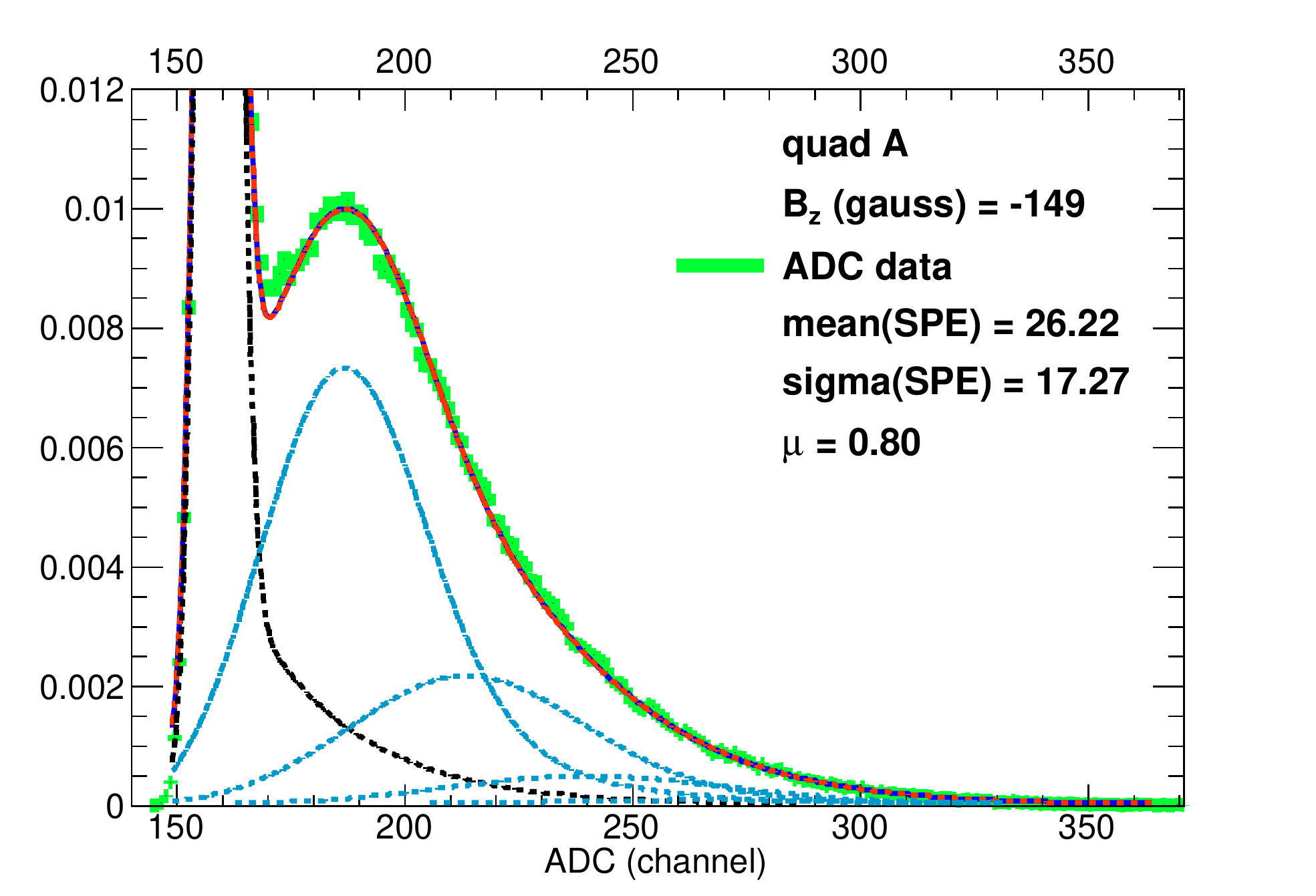}
&
\hspace{-0.4in}
\includegraphics[width=6.5cm]{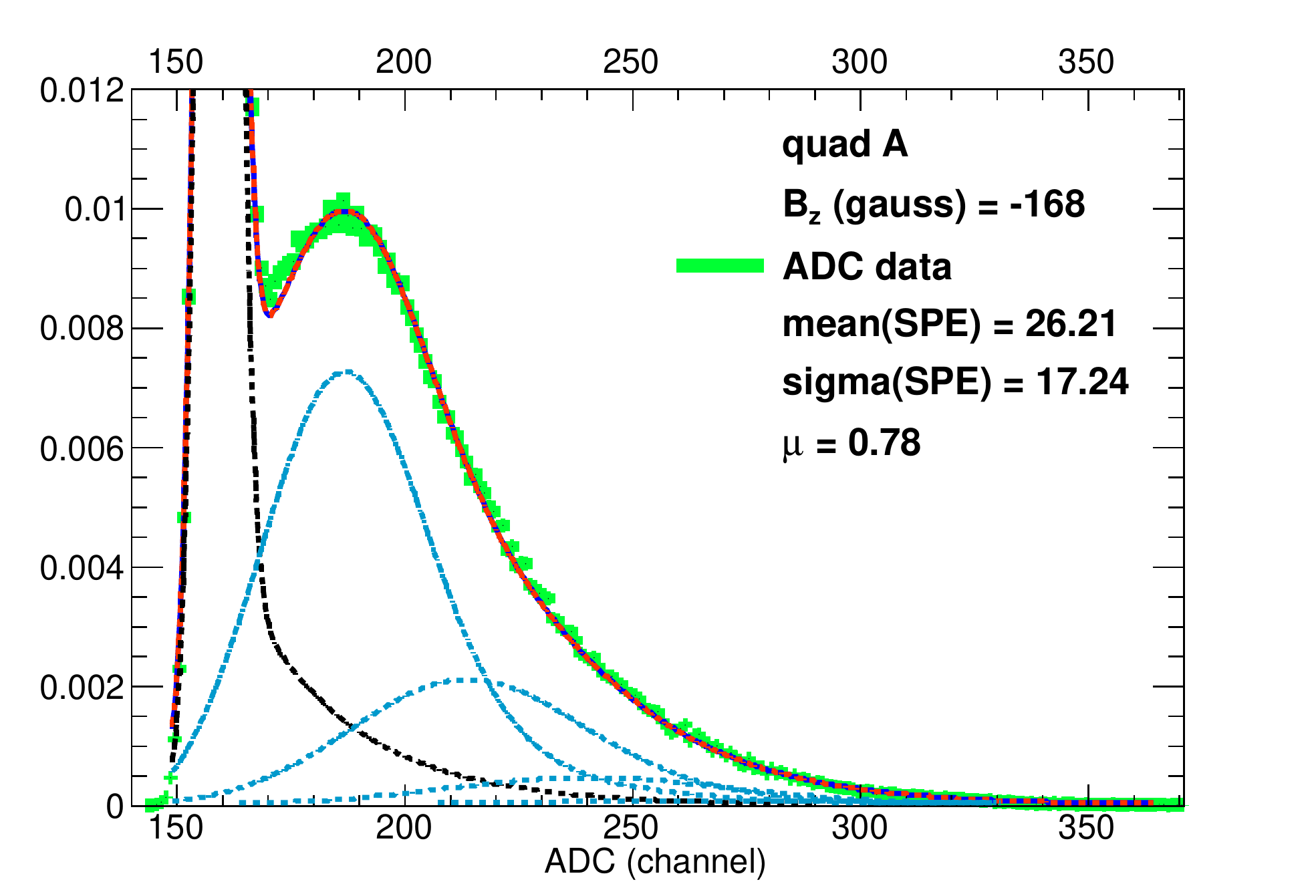} \\
\includegraphics[width=6.5cm]{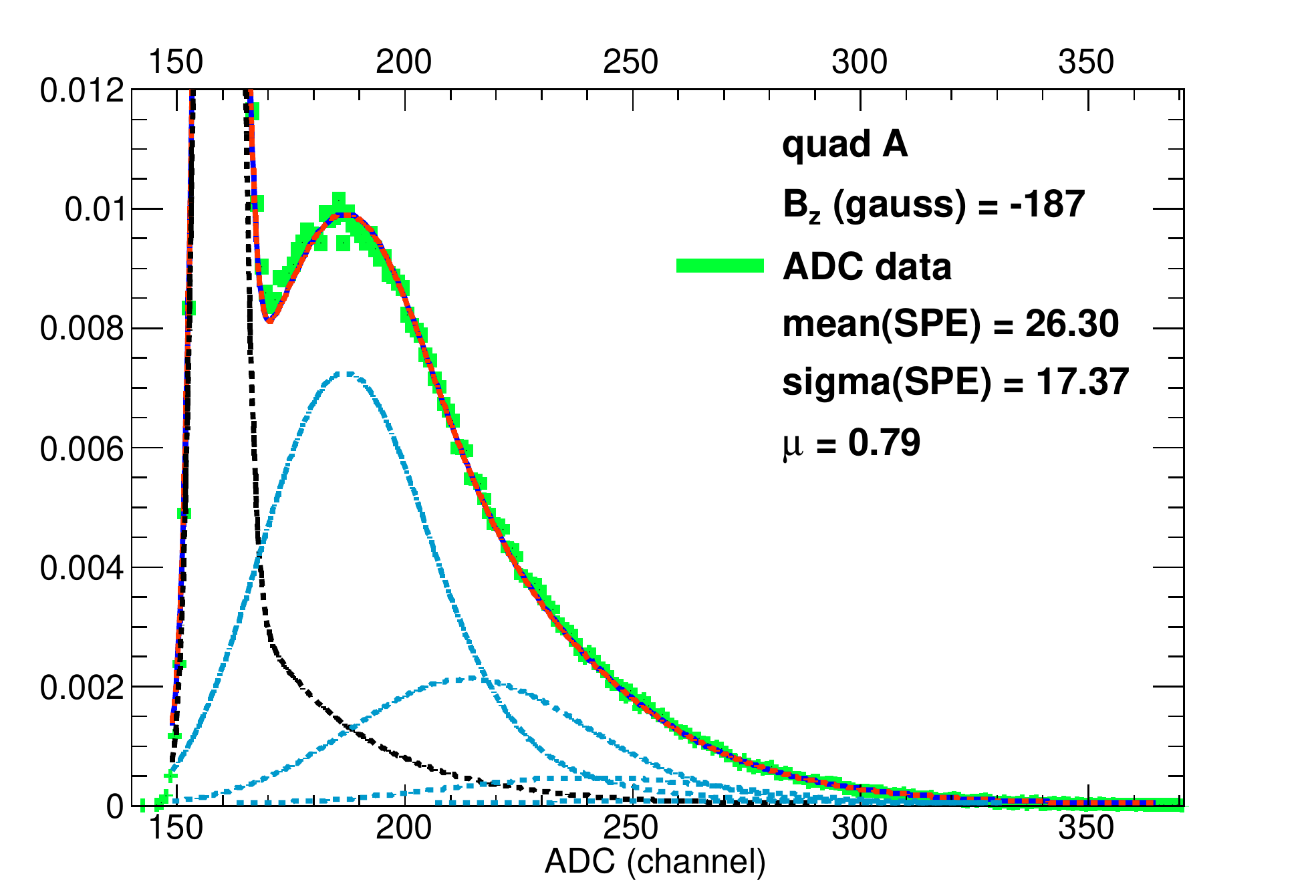}
&
\hspace{-0.4in}
\includegraphics[width=6.5cm]{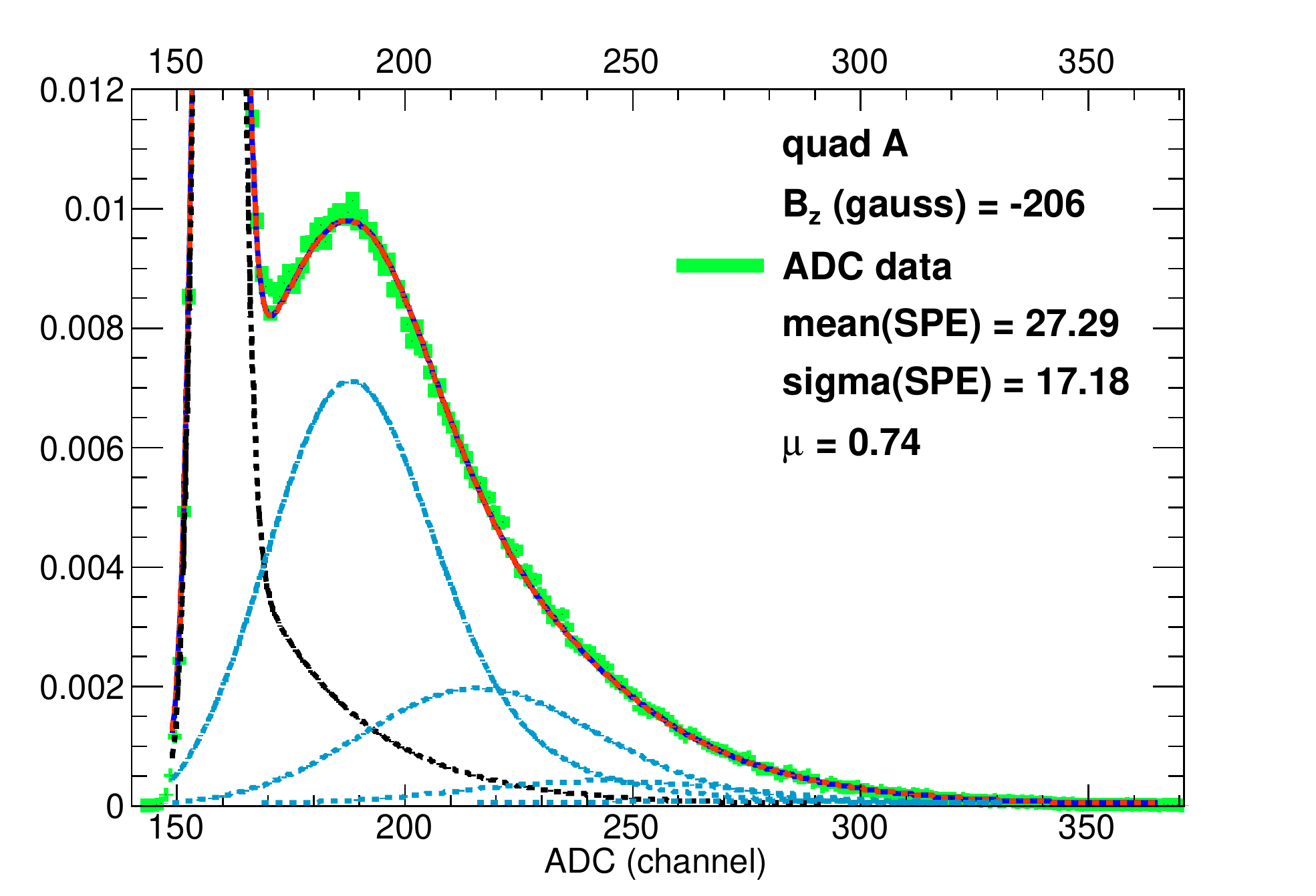} 
&
\hspace{-0.4in}
\includegraphics[width=6.5cm]{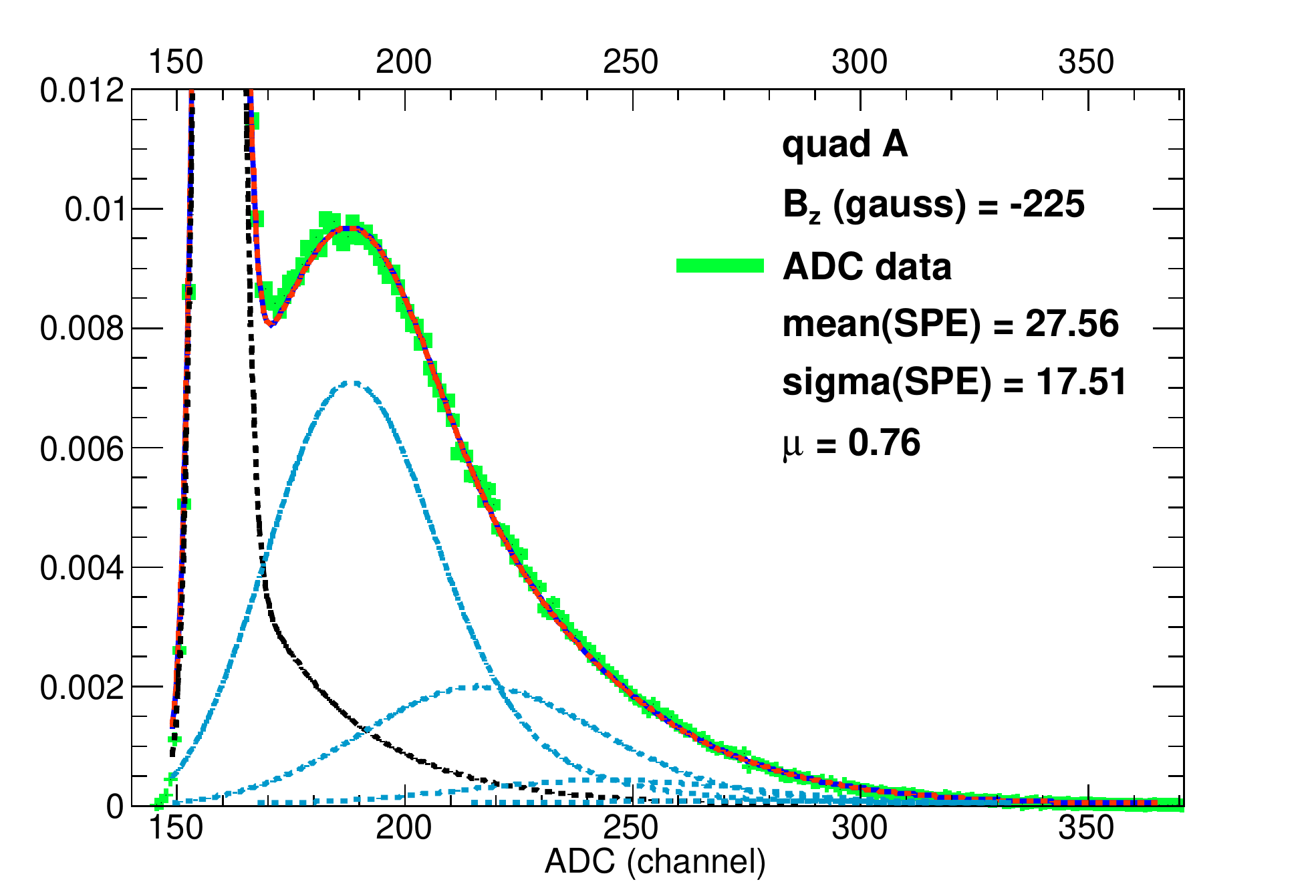} \\
\includegraphics[width=6.5cm]{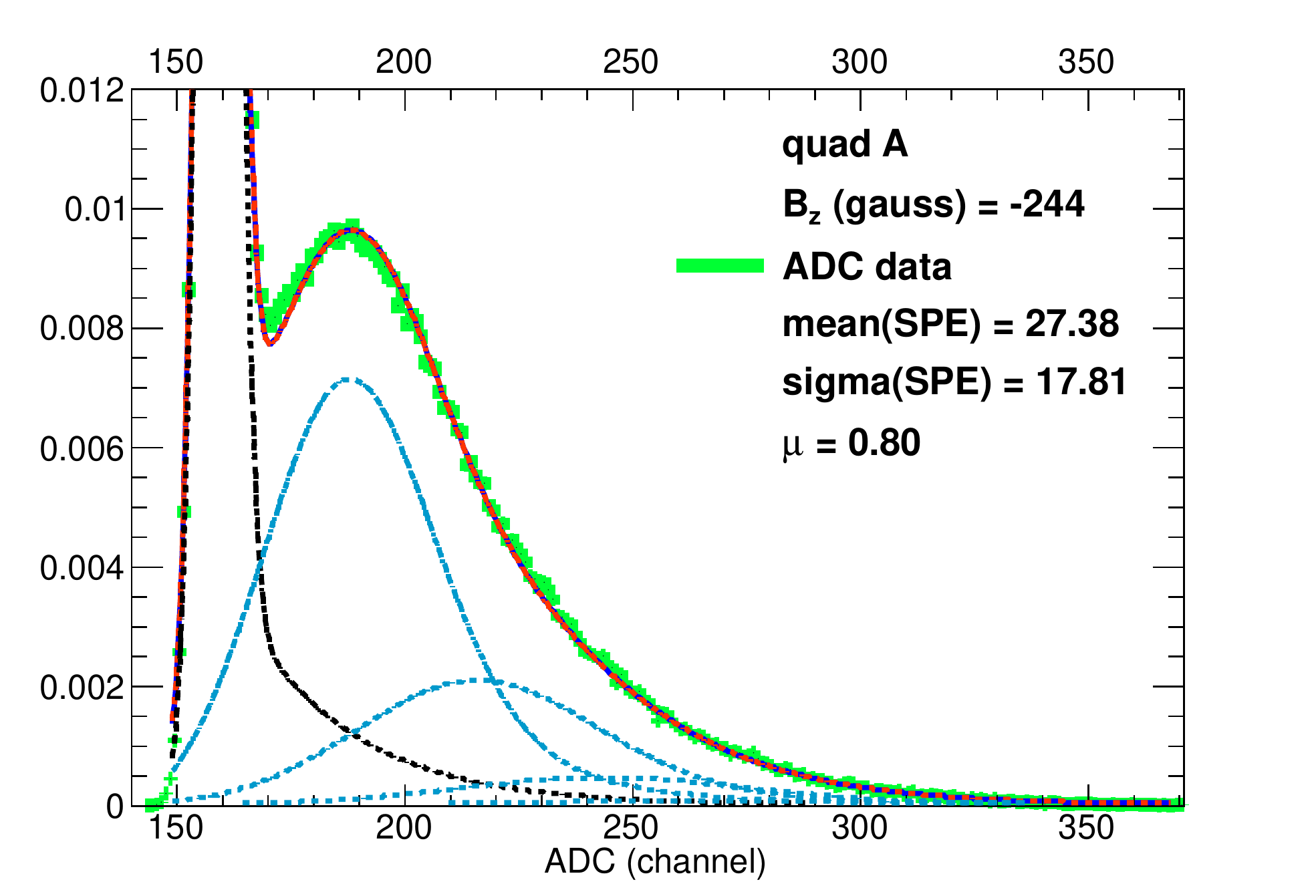}
&
\hspace{-0.4in}
\includegraphics[width=6.5cm]{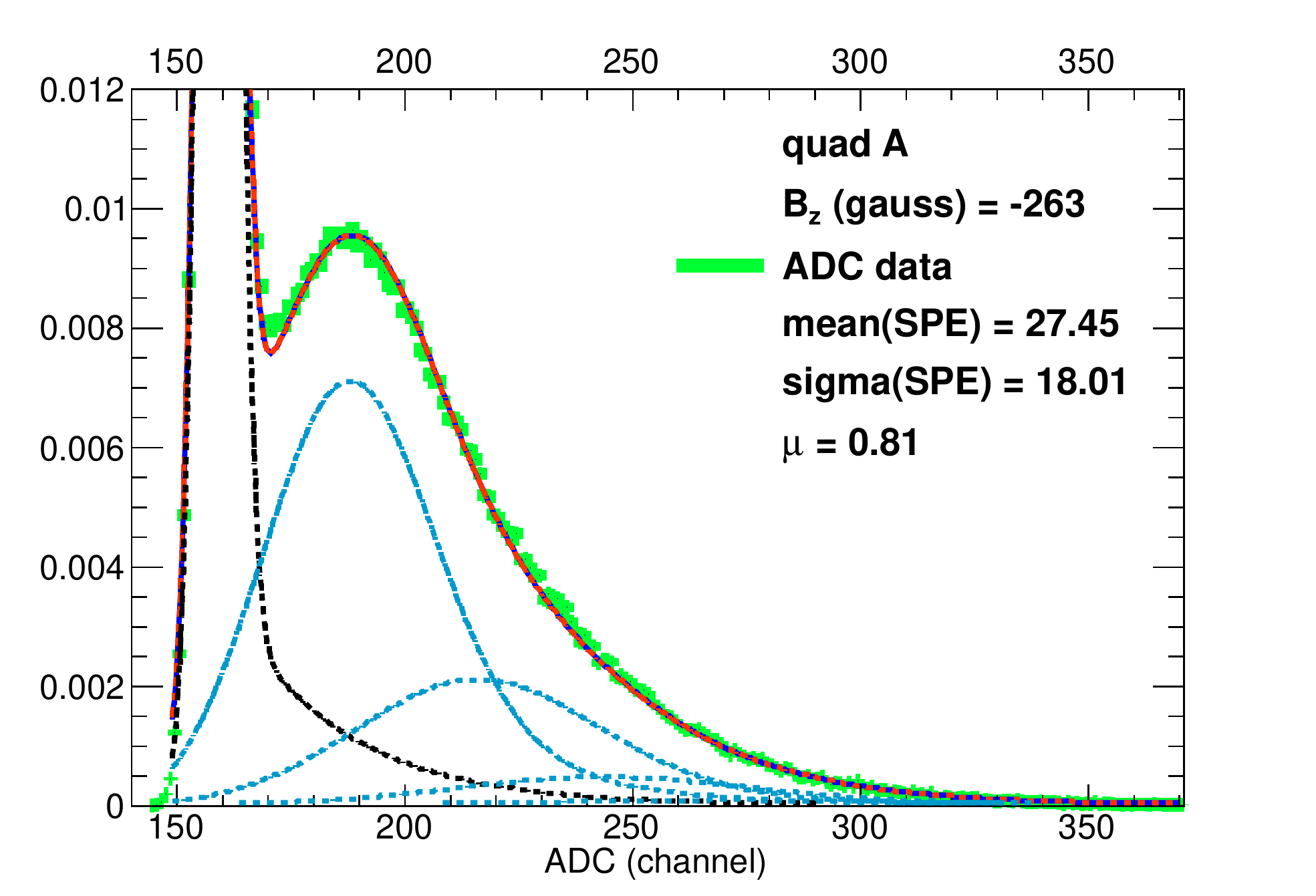}
&
\hspace{-0.4in}
\includegraphics[width=6.5cm]{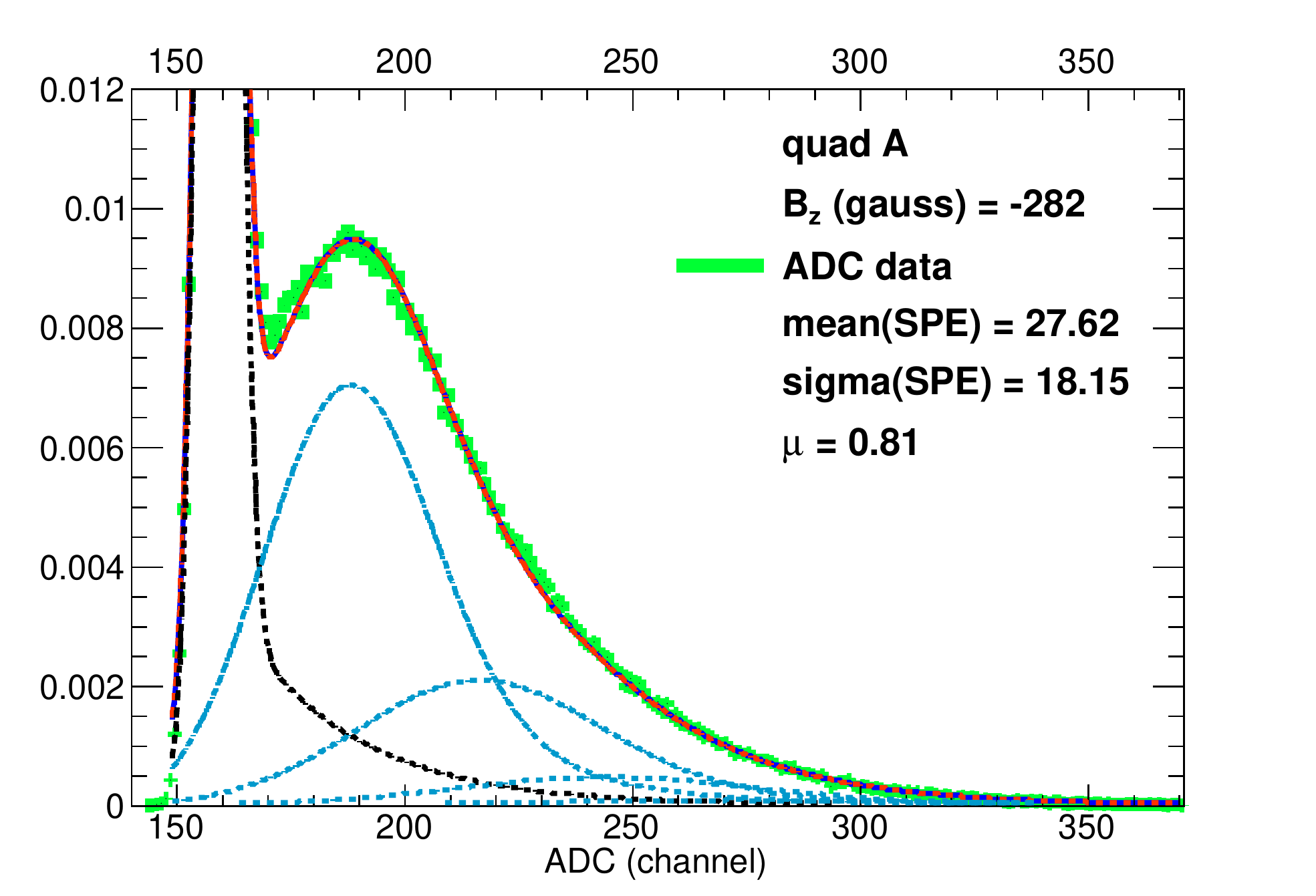} \\

\end{tabular}
\linespread{0.5}
\caption[]{
{Fits of the ADC distributions from quad A for various longitudinal magnetic field settings.} }
\label{quad_a_fit}
\end{figure}

\begin{figure}[htbp]
\vspace*{-0.2in}
\centering
\begin{tabular}{ccc}
\includegraphics[width=6.5cm]{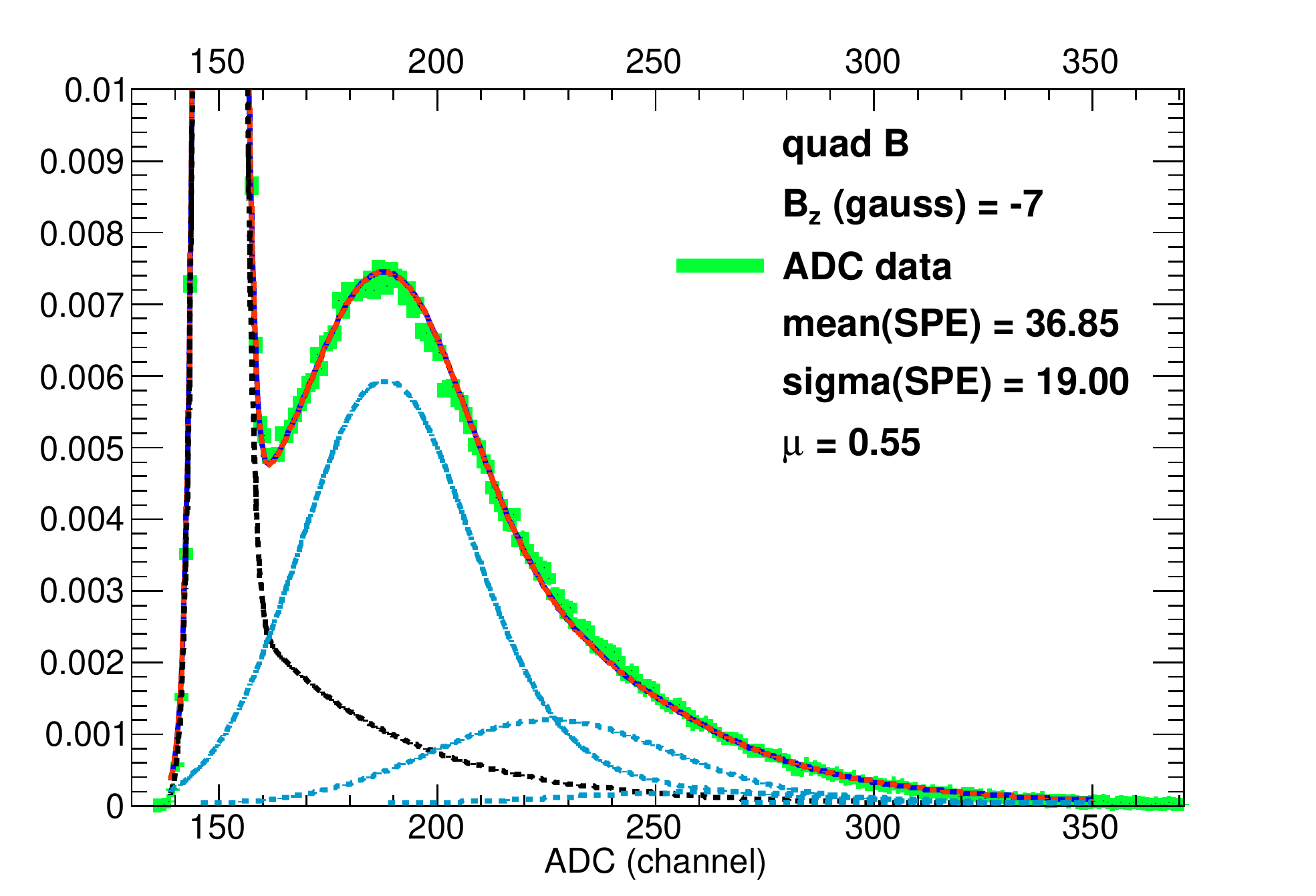}
&
\hspace{-0.4in}
\includegraphics[width=6.5cm]{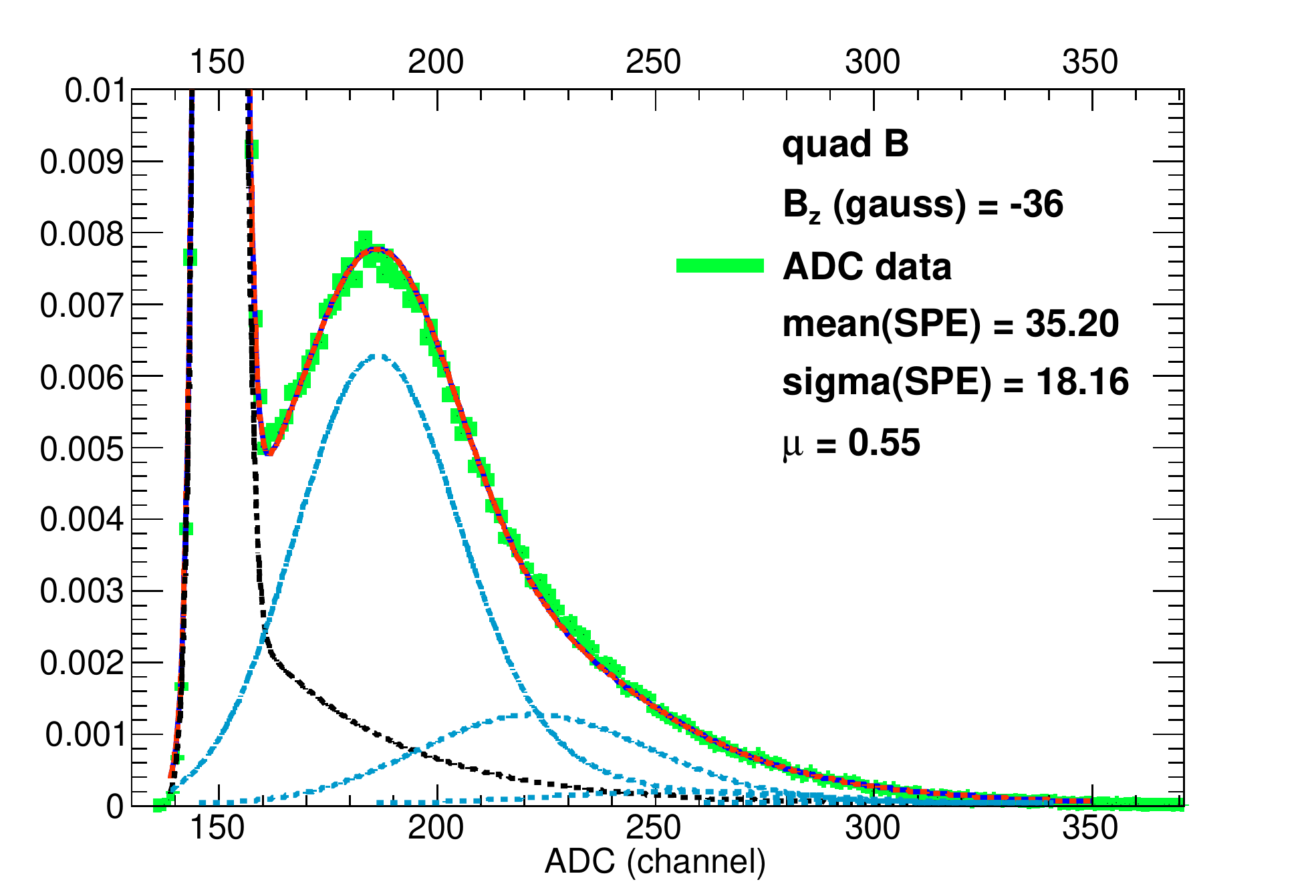}
&
\hspace{-0.4in}
\includegraphics[width=6.5cm]{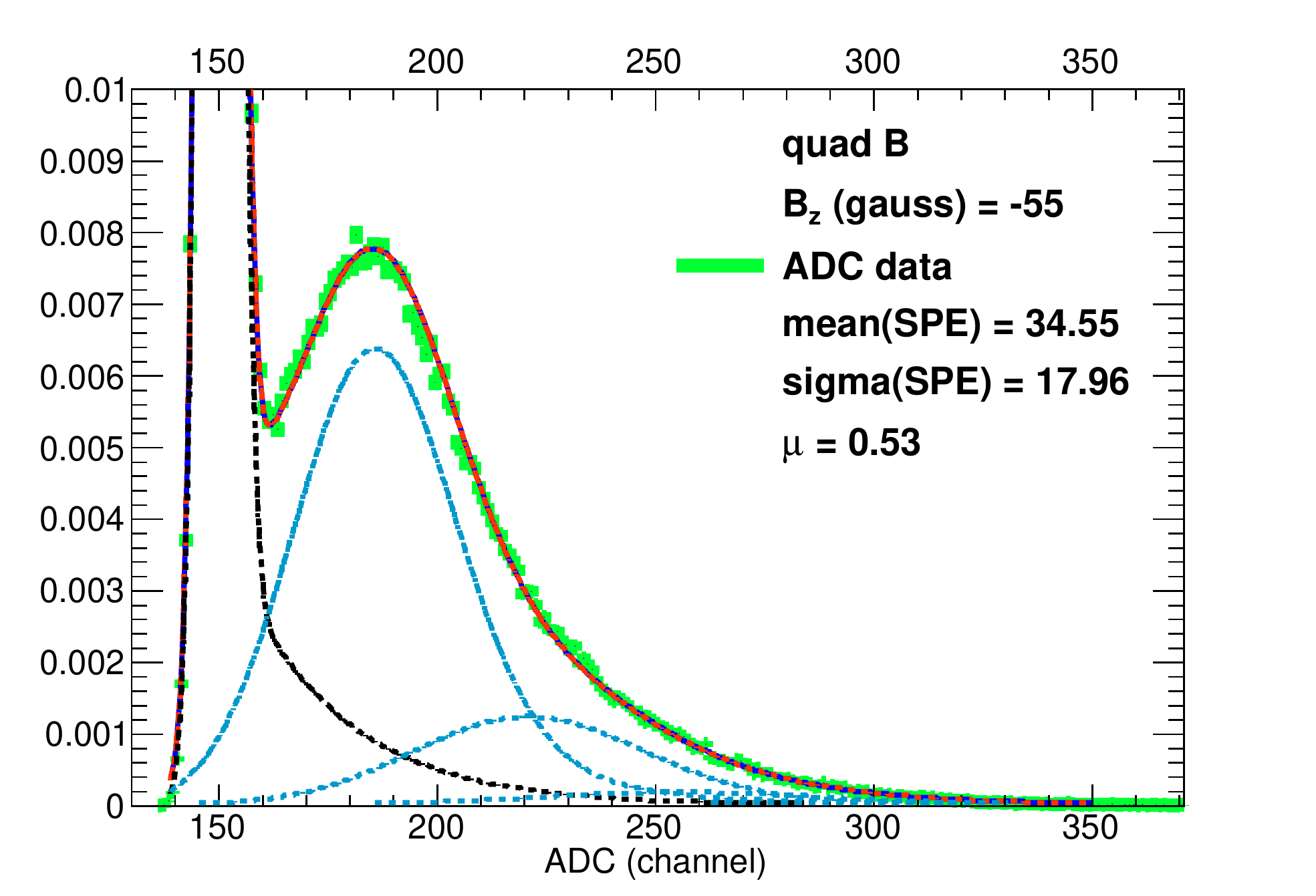} \\
\includegraphics[width=6.5cm]{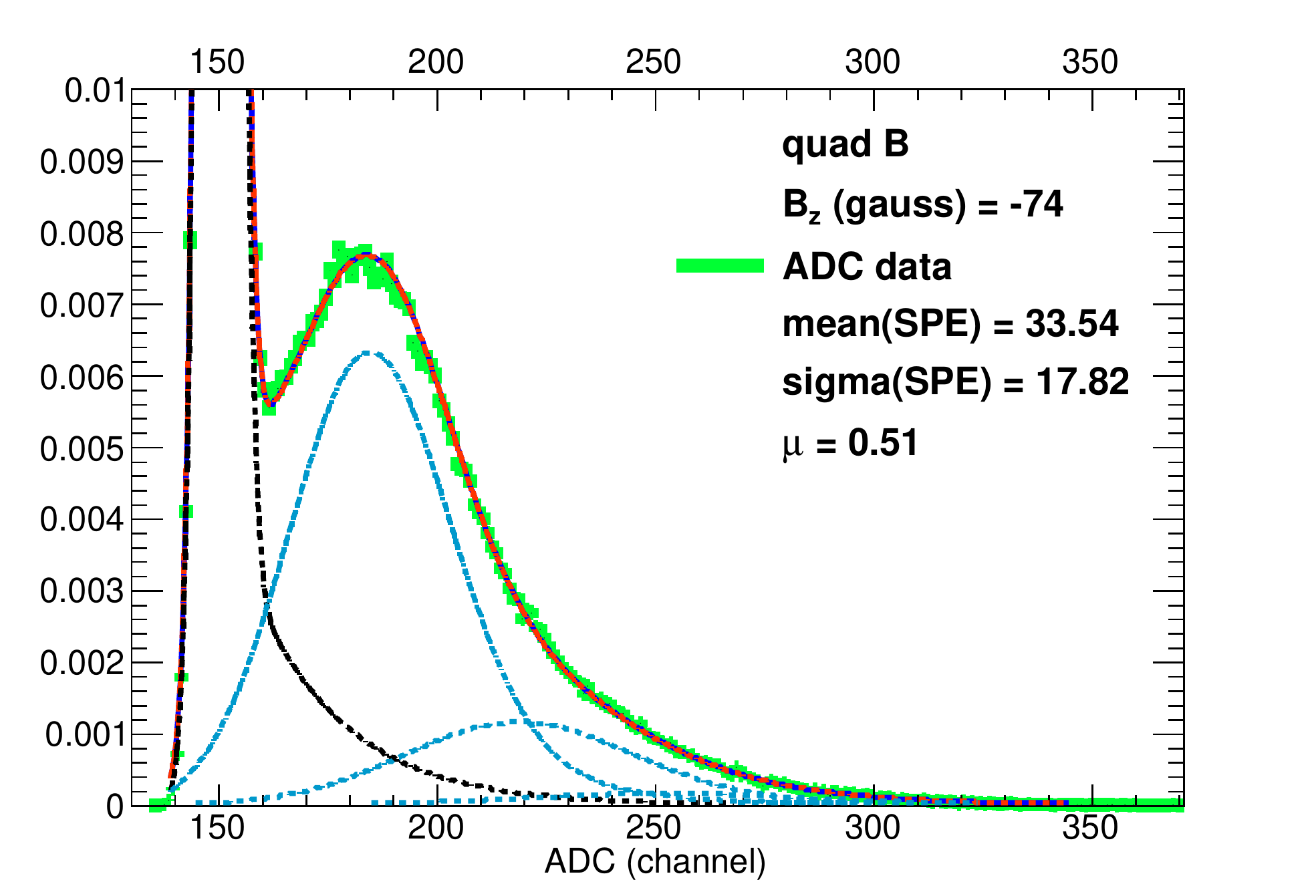}
&
\hspace{-0.4in}
\includegraphics[width=6.5cm]{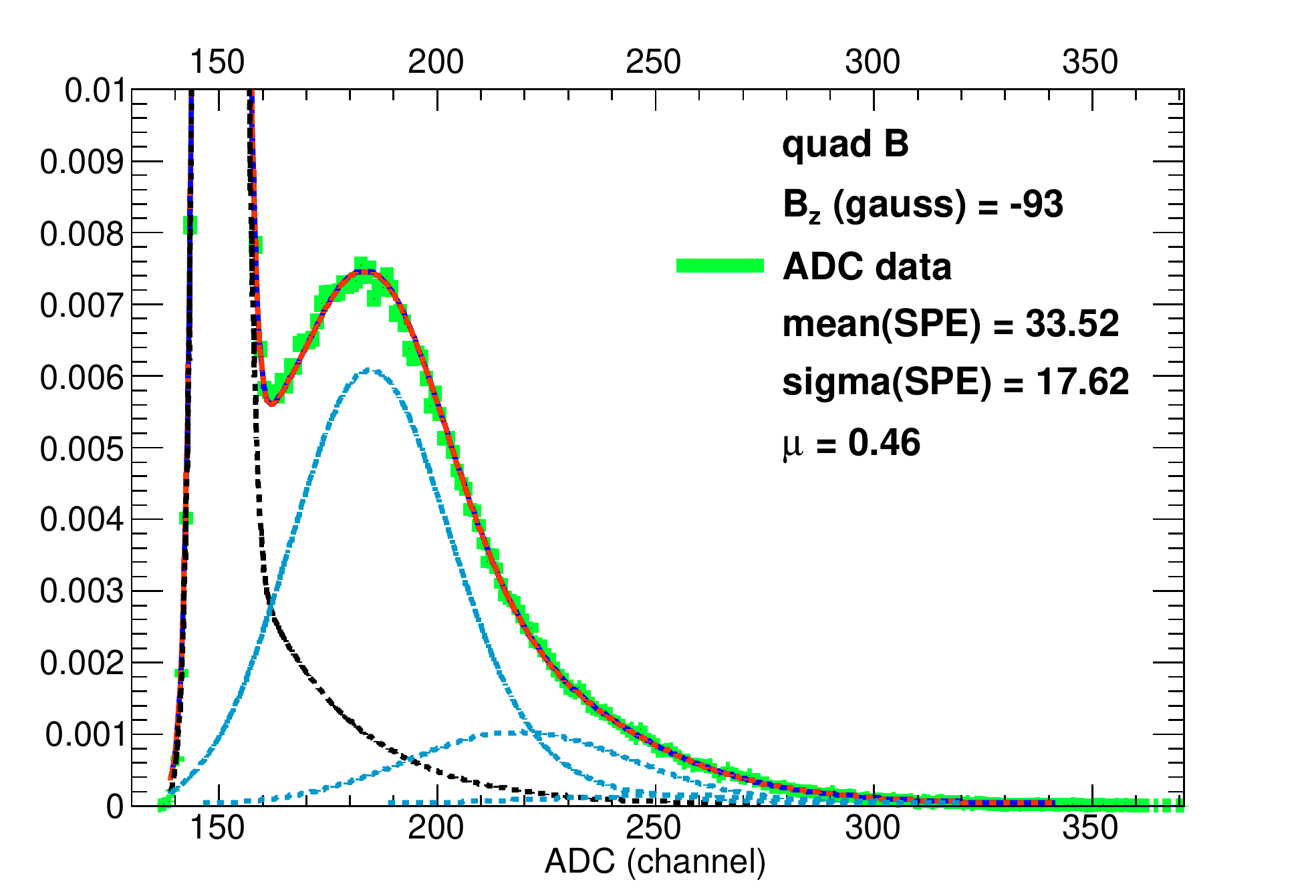}
&
\hspace{-0.4in}
\includegraphics[width=6.5cm]{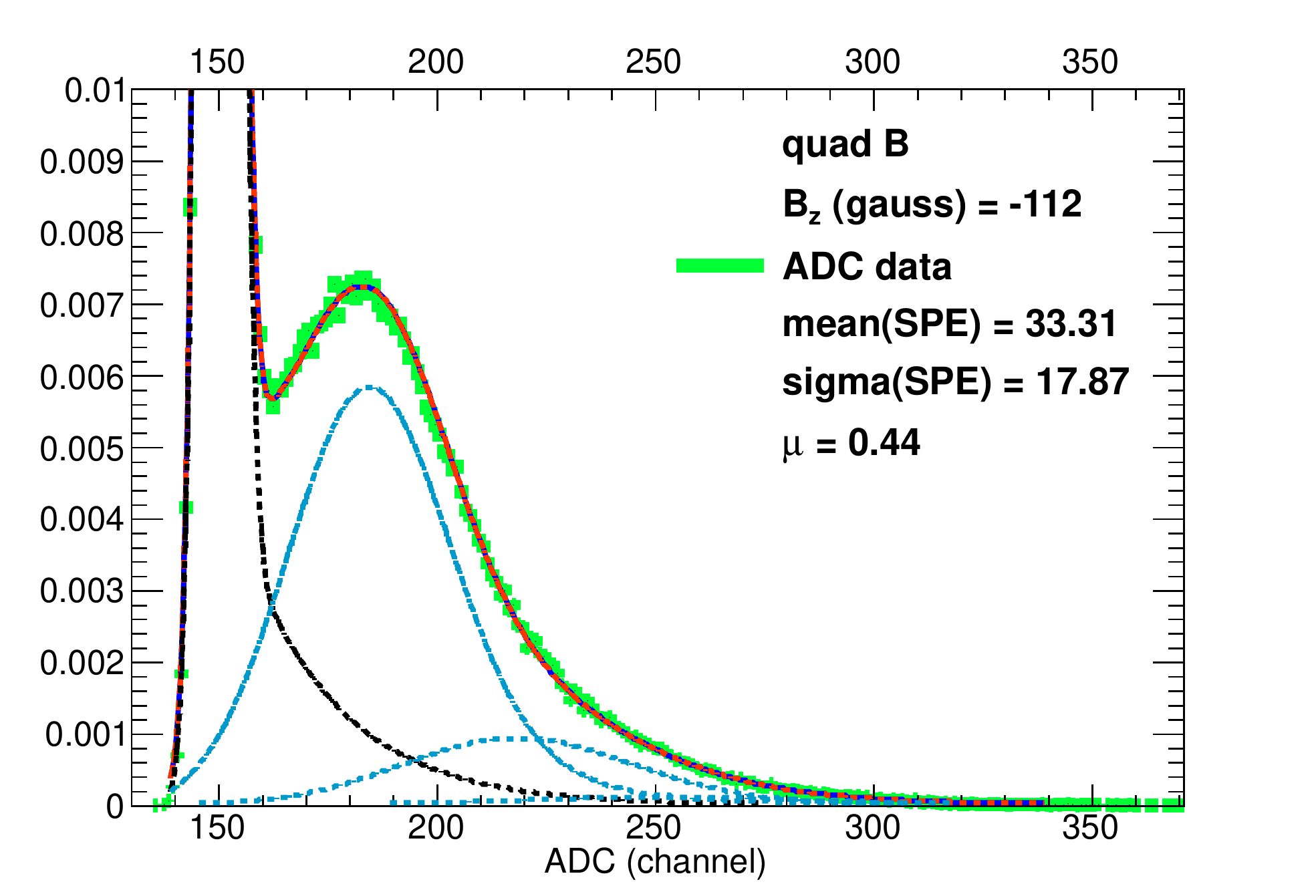} \\
\includegraphics[width=6.5cm]{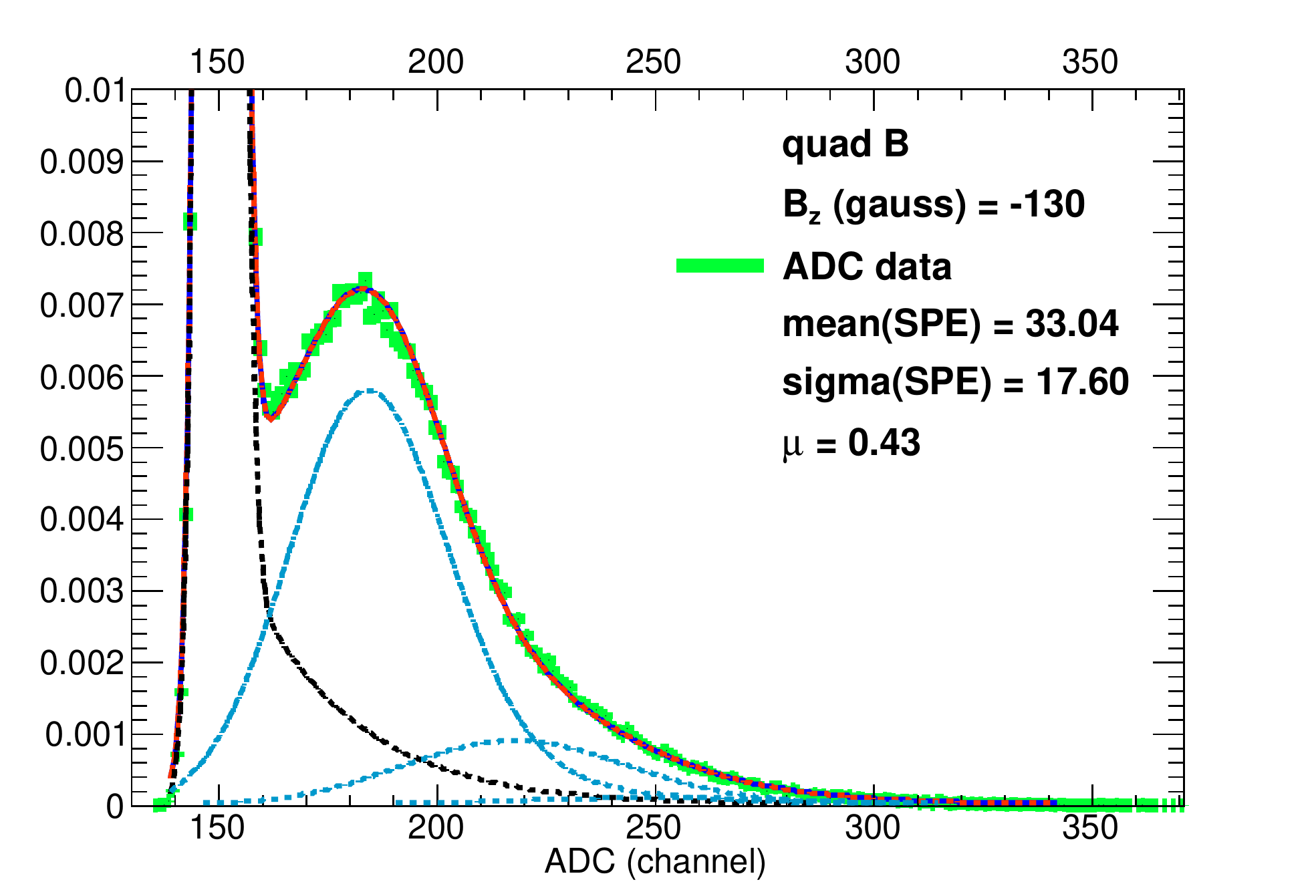} 
&
\hspace{-0.4in}
\includegraphics[width=6.5cm]{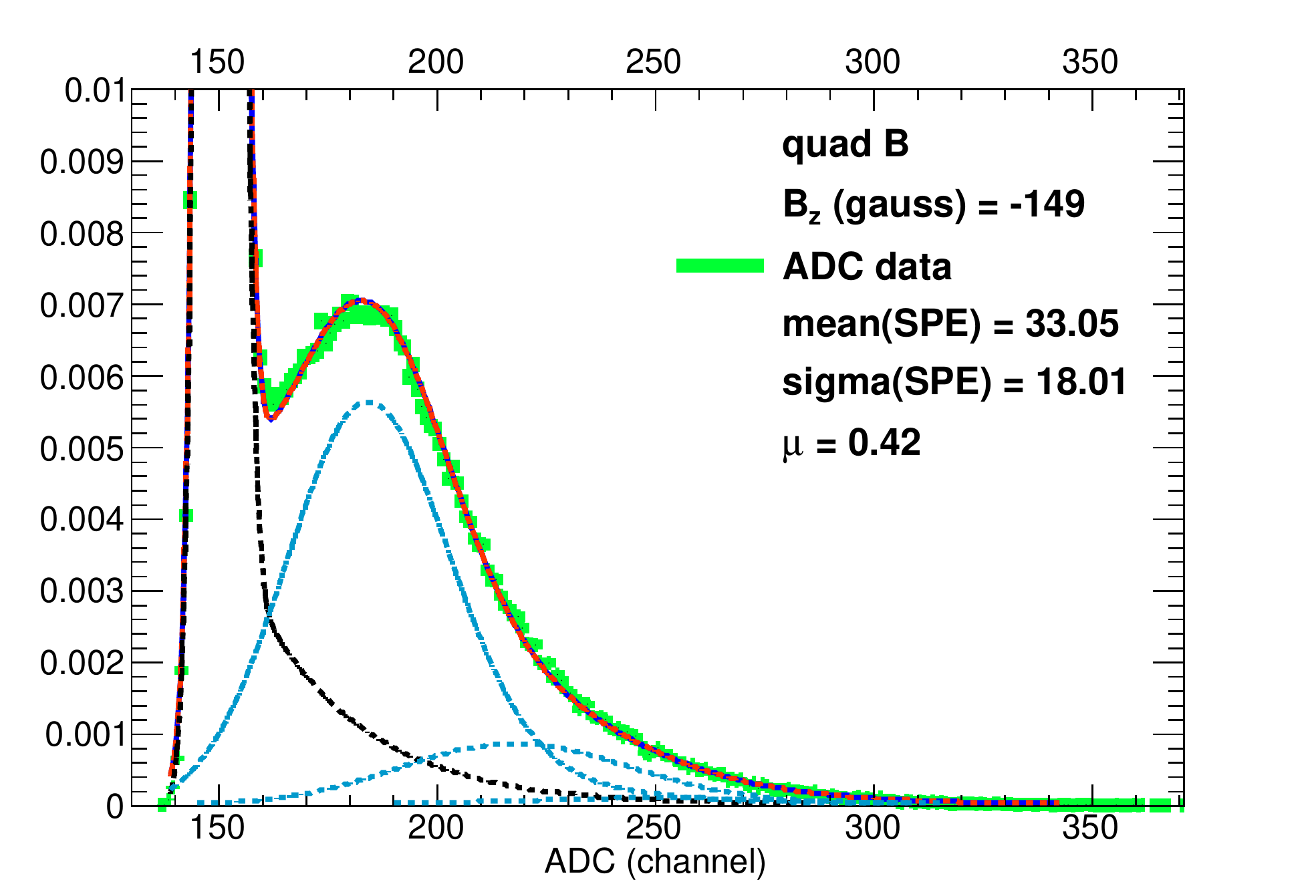}
&
\hspace{-0.4in}
\includegraphics[width=6.5cm]{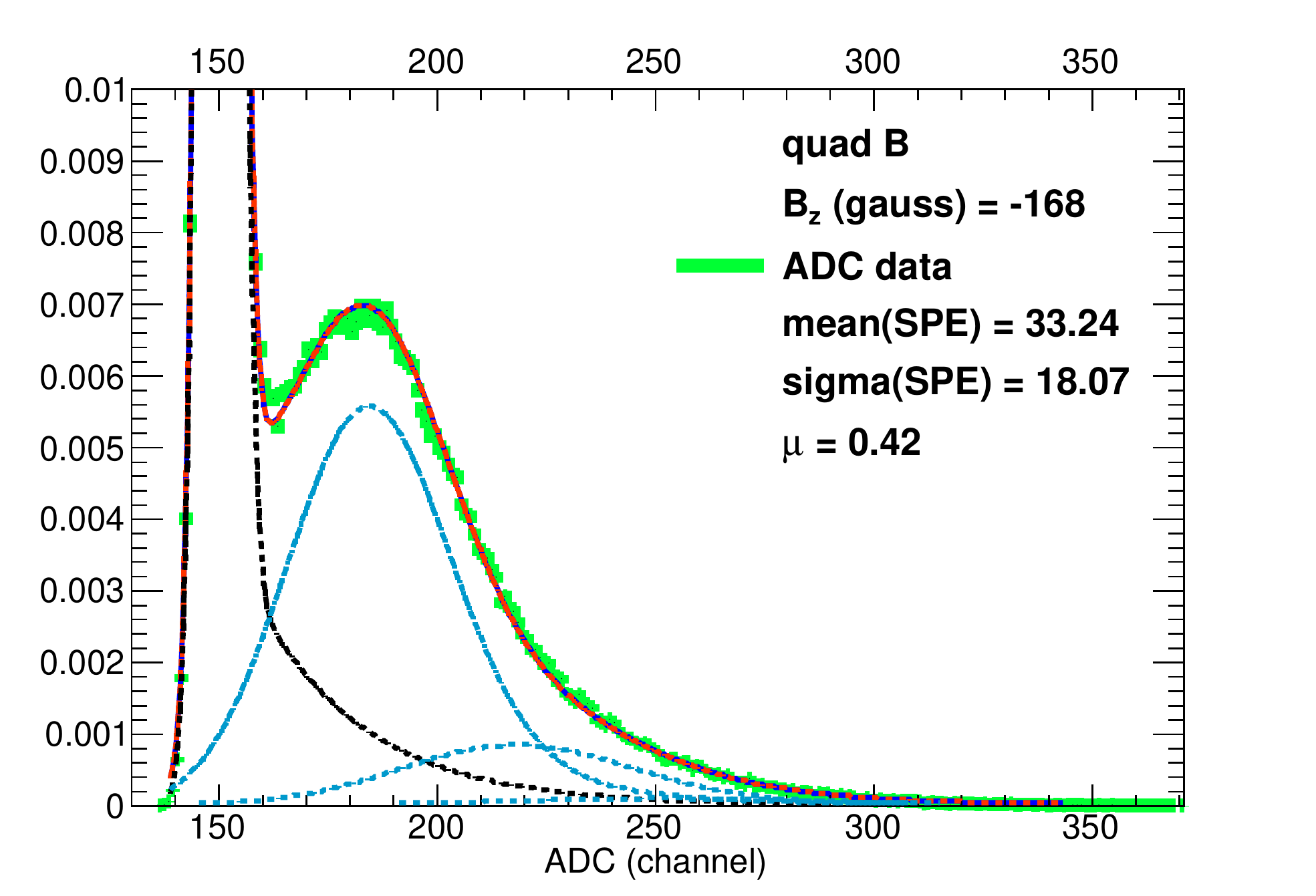} \\
\includegraphics[width=6.5cm]{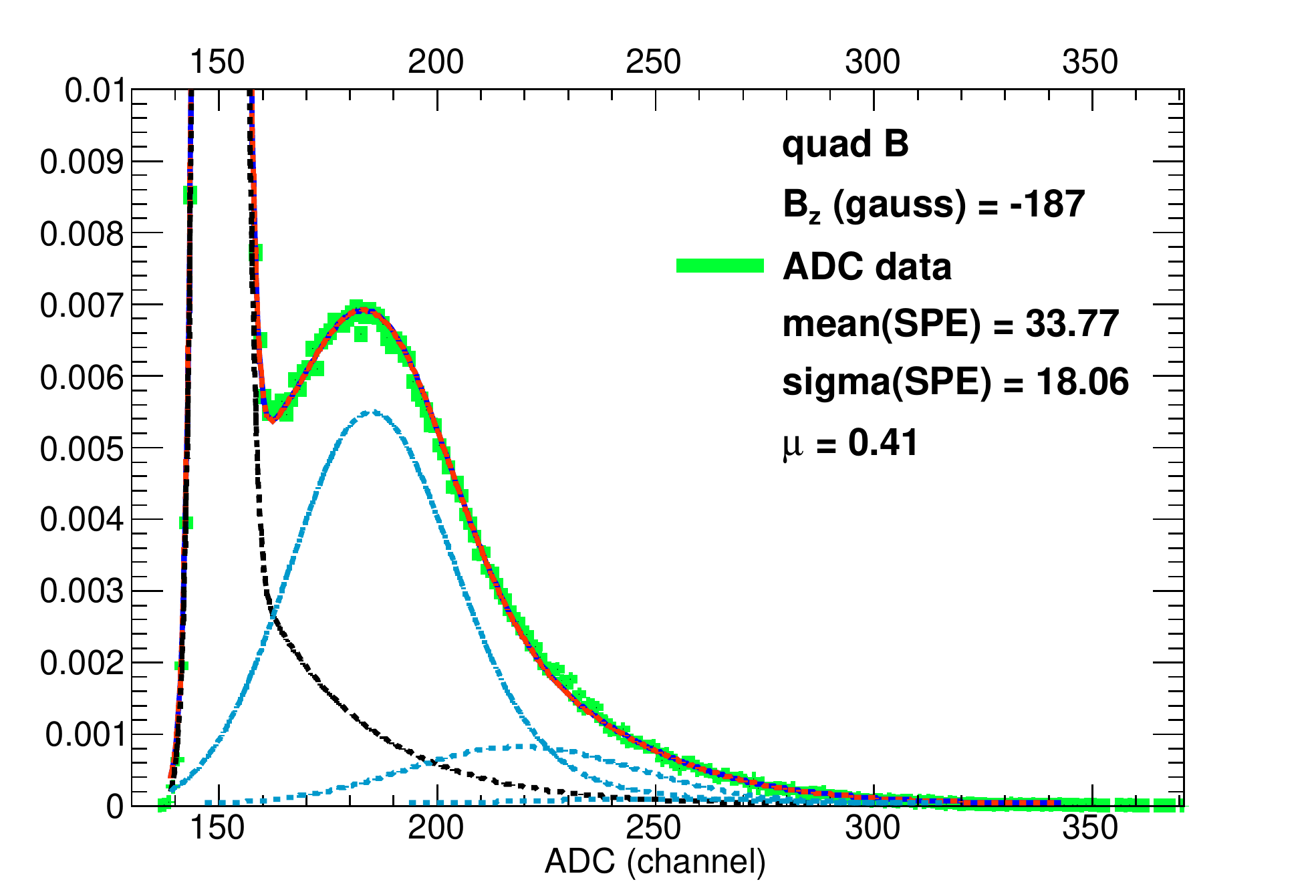}
&
\hspace{-0.4in}
\includegraphics[width=6.5cm]{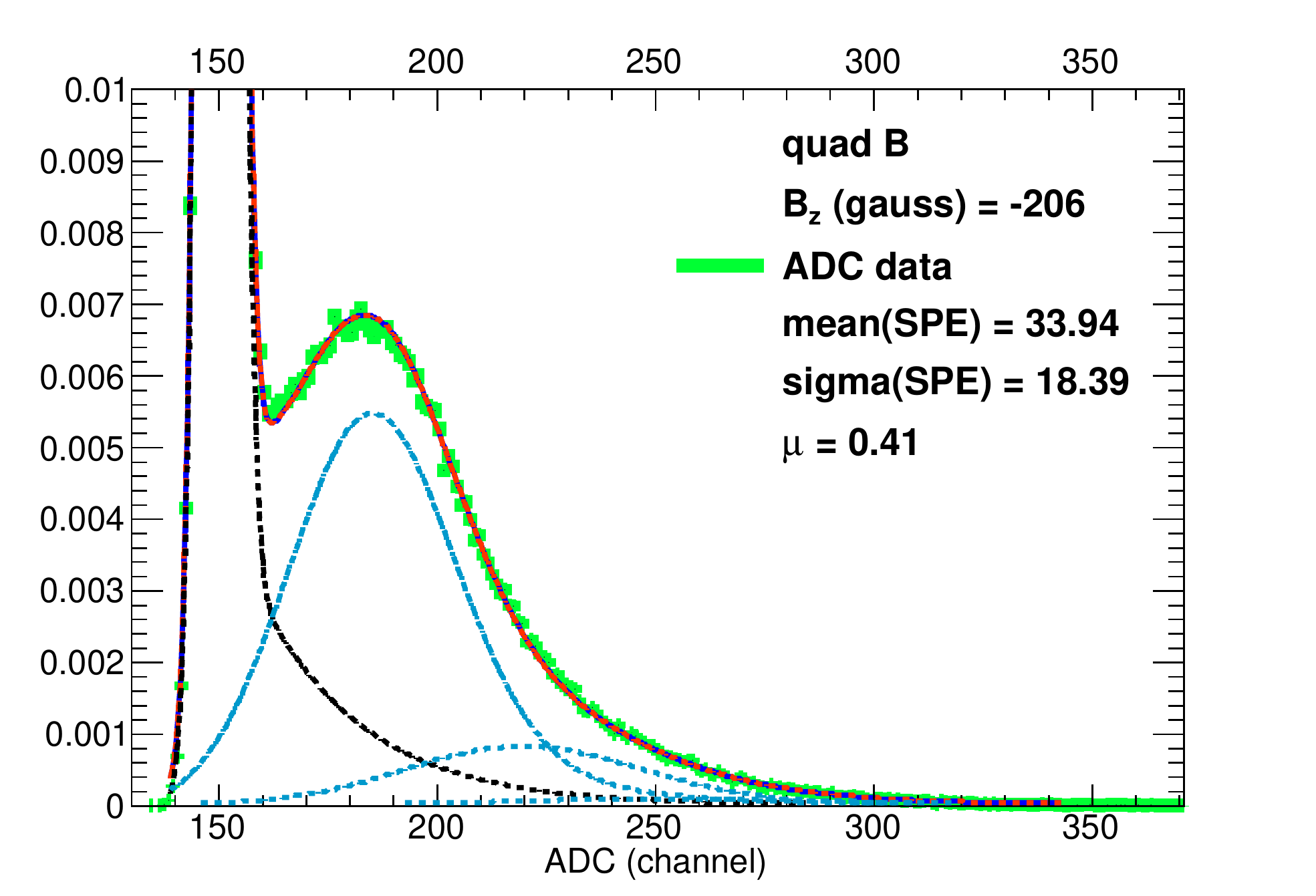} 
&
\hspace{-0.4in}
\includegraphics[width=6.5cm]{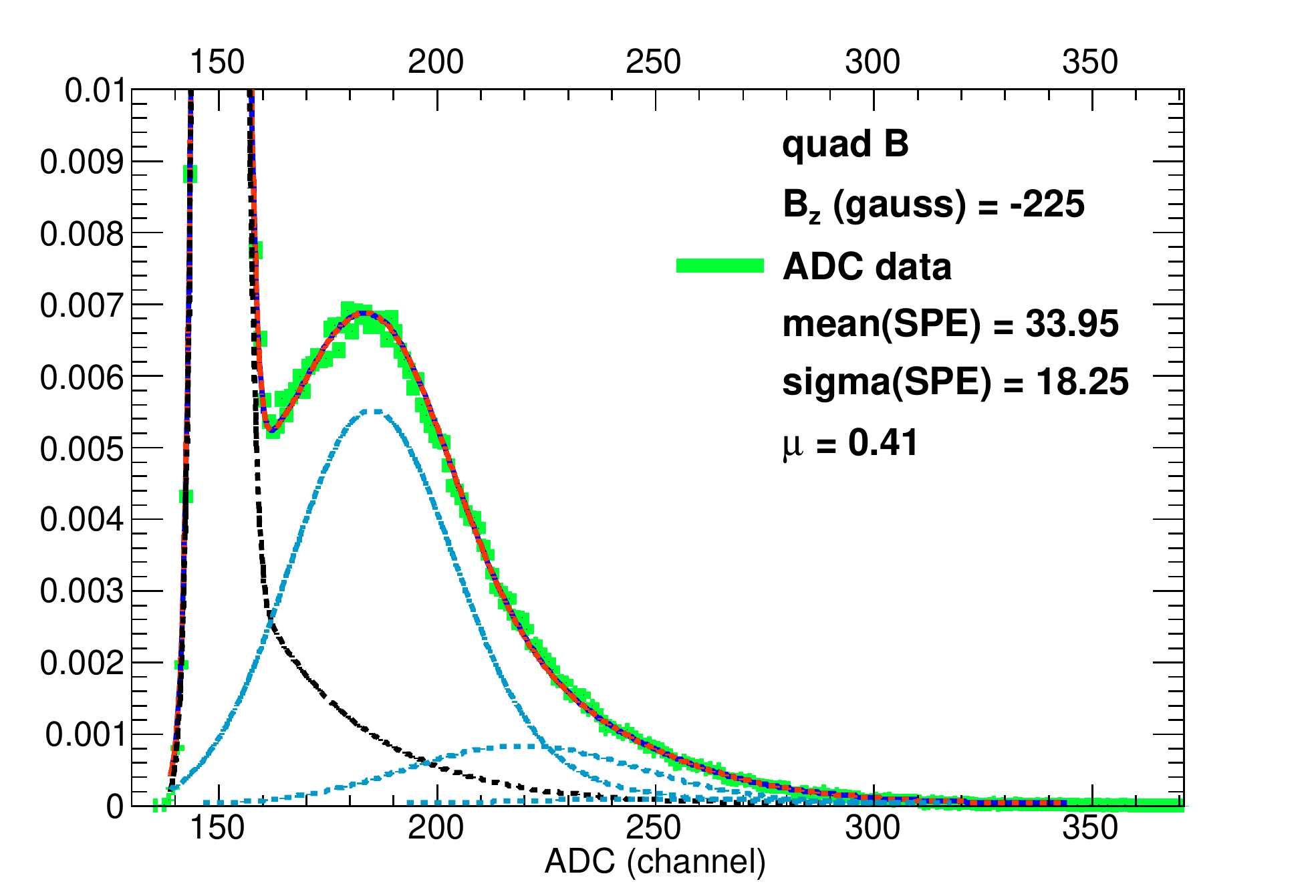} \\
\includegraphics[width=6.5cm]{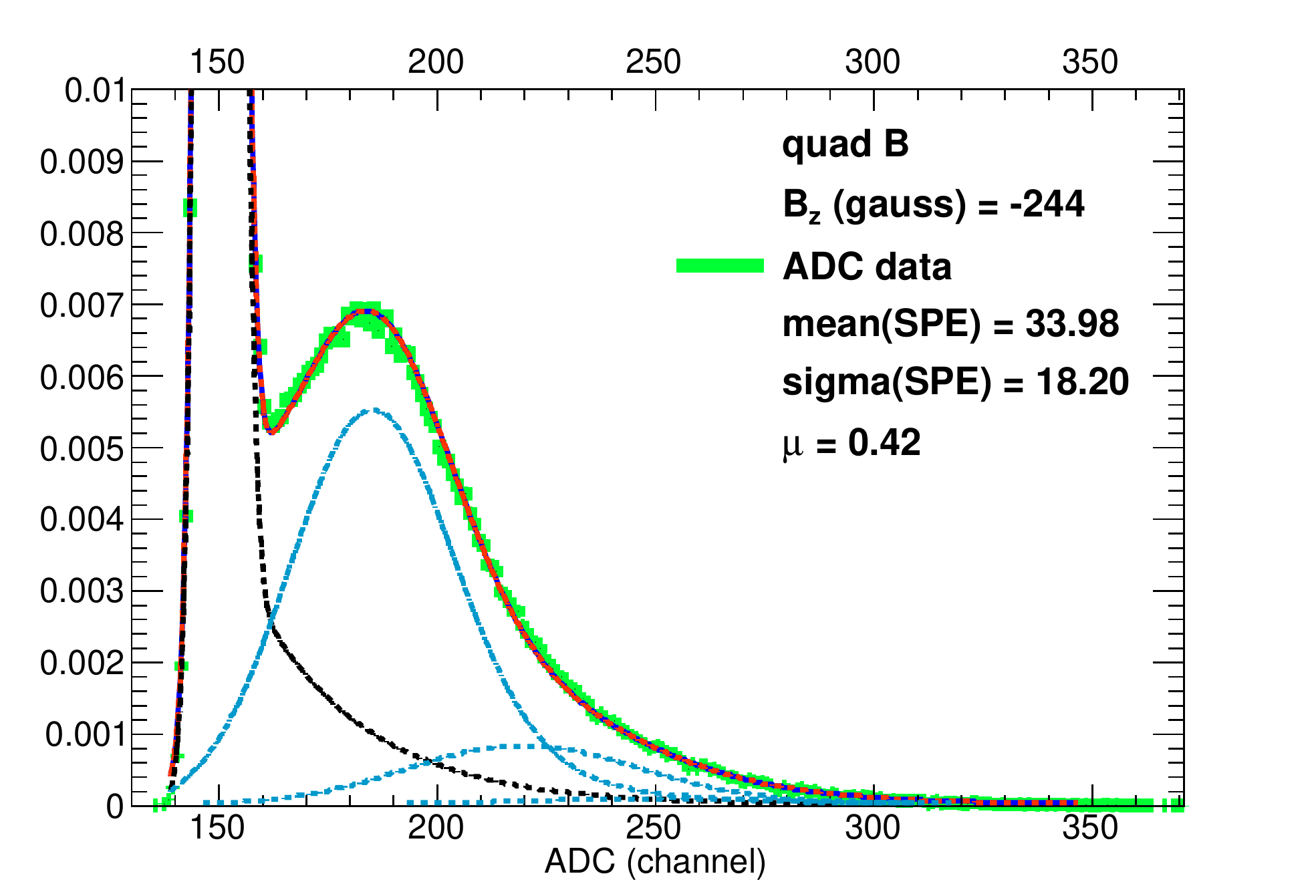}
&
\hspace{-0.4in}
\includegraphics[width=6.5cm]{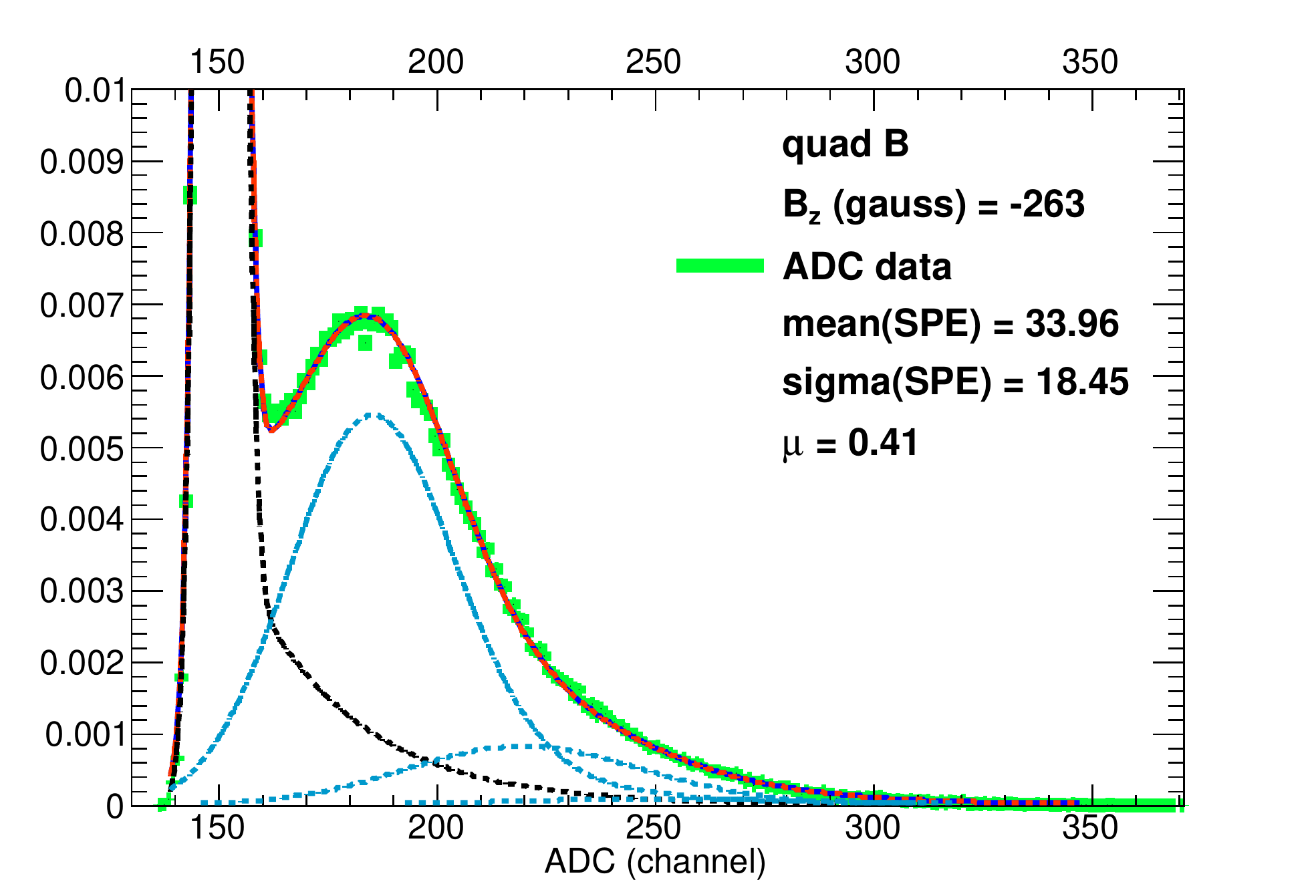}
&
\hspace{-0.4in}
\includegraphics[width=6.5cm]{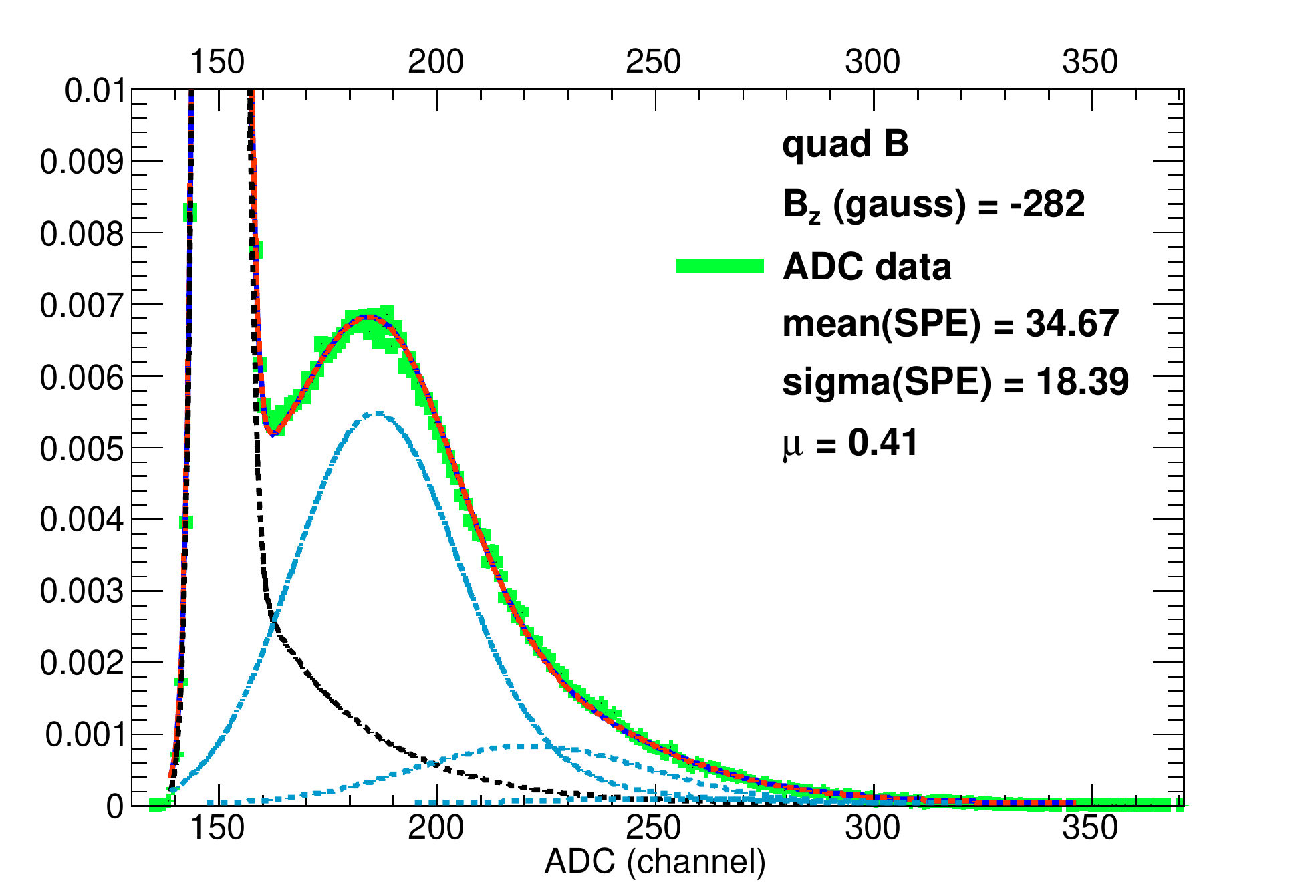} \\

\end{tabular}
\linespread{0.5}
\caption[]{
{Fits of the ADC distributions from quad B for various longitudinal magnetic field settings.} }
\label{quad_b_fit}
\end{figure}

\begin{figure}[htbp]
\vspace*{-0.1in}
\centering
\begin{tabular}{cc}
\includegraphics[width=9.4cm]{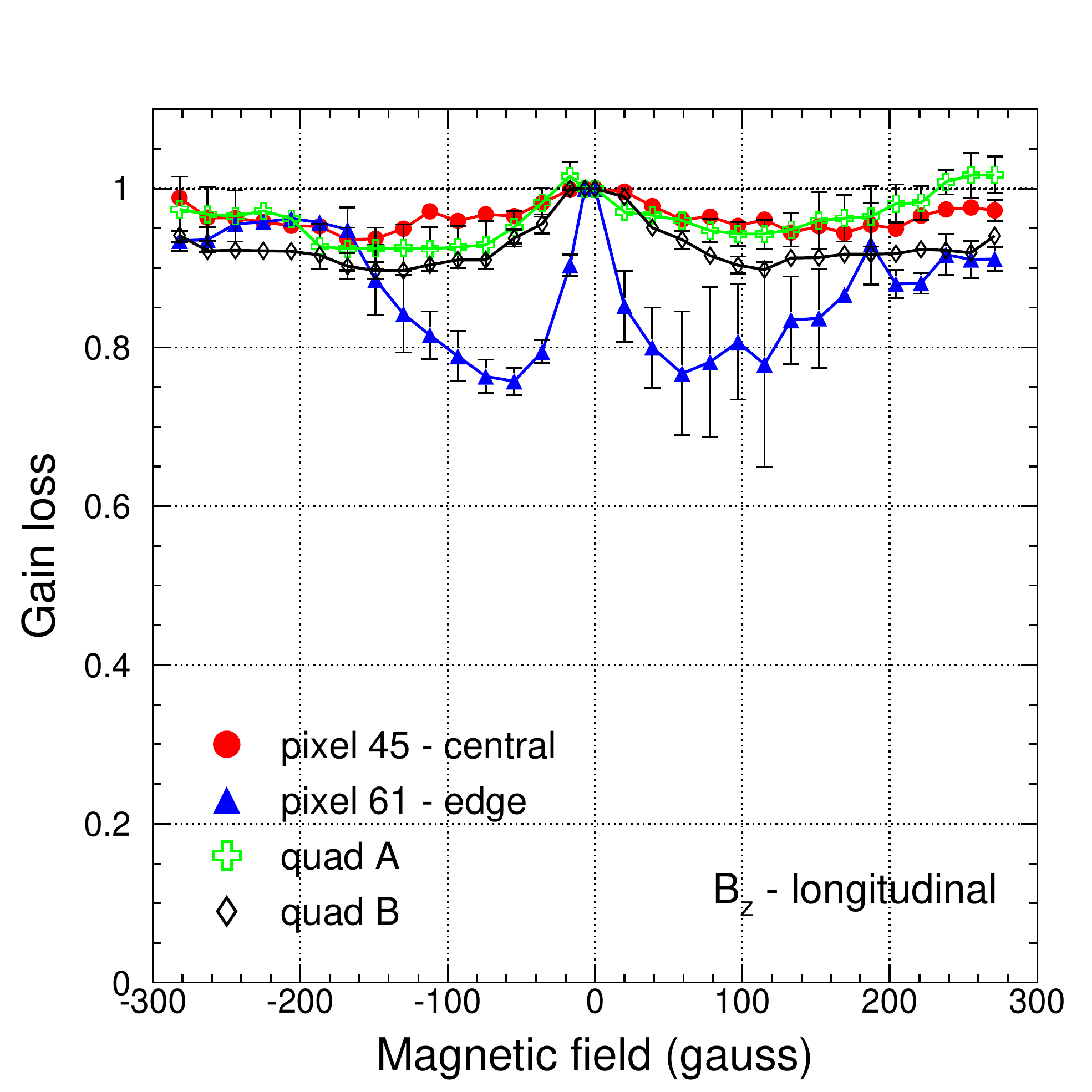}
&
\hspace{-0.35in}
\includegraphics[width=9.4cm]{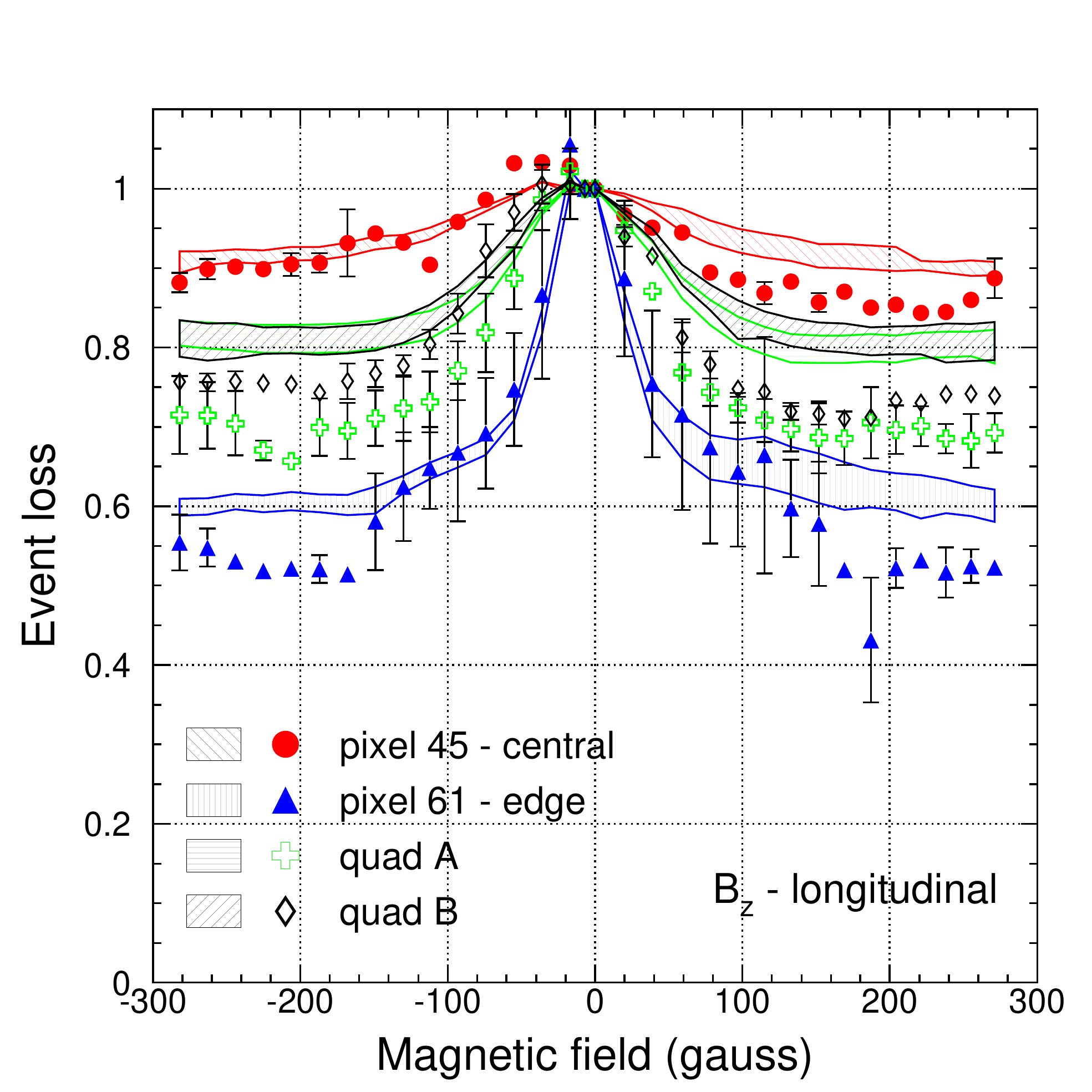} 
\end{tabular}
\linespread{0.5}
\caption[]{
{Gain (left) and single photoelectron (right) losses in a longitudinal magnetic field (see text 
for details).} }
\label{losses}
\end{figure}

Given the practical complications associated with shielding, of particular 
interest is understanding the effect of a longitudinal 
magnetic field on single photoelectron outputs. We used the PMT response function 
and fitting procedure described in Section 3 to fit the ADC distributions 
from pixels 45 and 61 and from quads A and B for various longitudinal 
field settings (Figs.~\ref{45_one}, \ref{61_one}, \ref{quad_a_one}, 
\ref{quad_b_one}, bottom panels). Our fits are shown in Figs.~\ref{45_fit}, 
\ref{61_fit}, \ref{quad_a_fit} and \ref{quad_b_fit}. It can be seen that 
the fit was stable up to the highest magnetic field setting for pixels and 
quads allowing a reliable extraction of the fit parameters associated 
with the gain, $mean(SPE)$, and the average number of photoelectrons, $\mu$. 
The fit results are summarized in Fig.~\ref{losses} where we plot, on the 
left panel, \textit{$mean(SPE)$} and, on the right panel, \textit{$\mu$}.

\begin{figure}[htbp]
\vspace*{-0.1in}
\centering
\begin{tabular}{c}
\includegraphics[width=9.8cm]{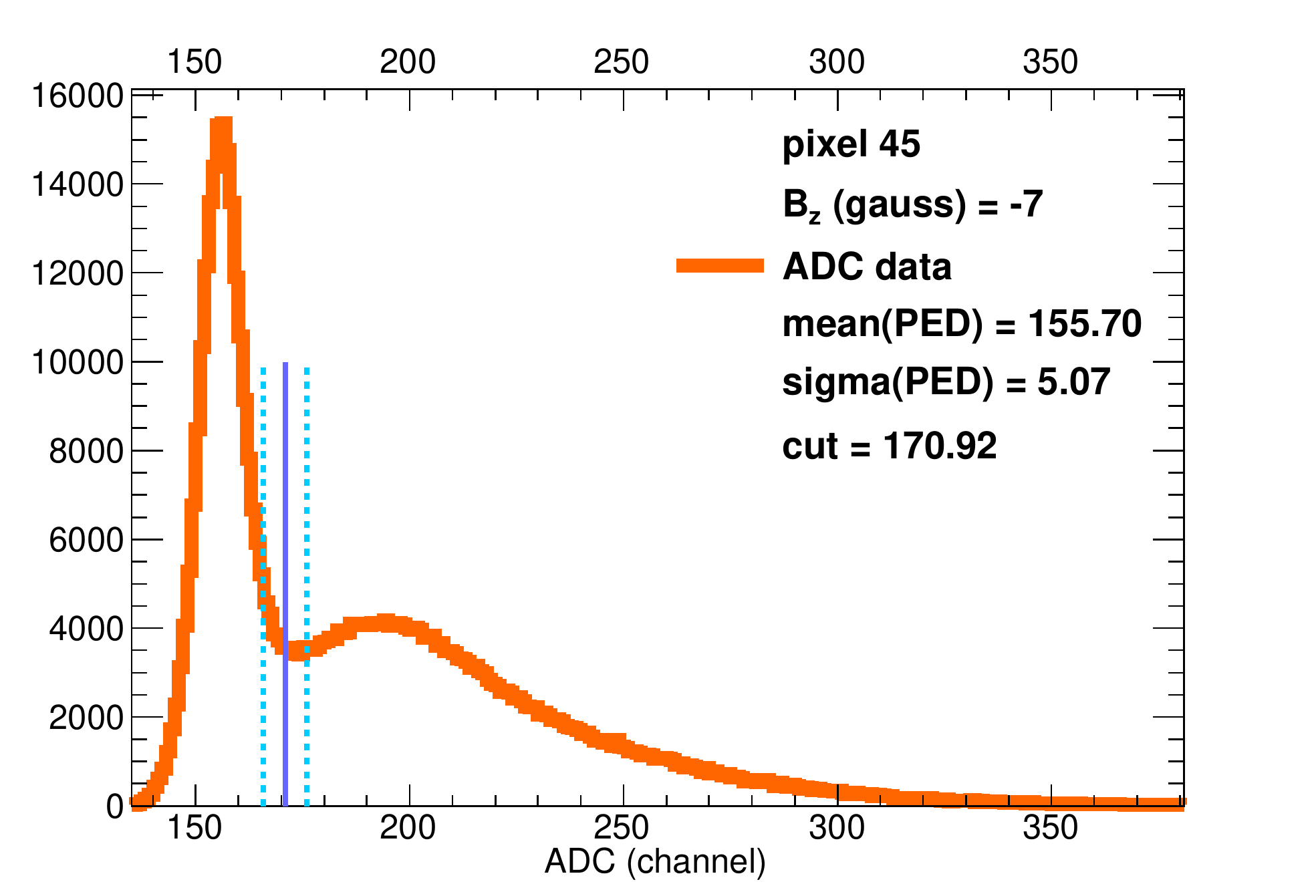}
\end{tabular}
\linespread{0.5}
\caption[]{
{Representative ADC distribution from pixel 45 at a given longitudinal magnetic field setting of 
$-7$ Gauss. The cuts used to separate the signal from the pedestal are shown 
by the full and dashed lines (see text for details).} }
\label{method}
\end{figure}

In addition, we 
used an integration procedure to quantify the loss of single photoelectrons due 
to magnetic field. In Fig.~\ref{method} we show the ADC distribution from 
pixel 45 for a magnetic field setting of $B_z = -7$ Gauss. The 
pedestal distribution has been fit (the mean and standard deviation associated 
with the pedestal are given in Fig.~\ref{method}) and we used the standard 
deviation thus extracted to place a cut 3 standard deviations away from the 
mean to formally separate the signal from pedestal (full blue line in 
Fig.~\ref{method}). Then the ratio of signal to 
total events has been calculated for each field setting as the ratio of 
the ADC distribution integral with the three sigma cut as the lower limit 
and the integral over the full range ADC distribution. We also varied the 
cut by one sigma around the central cut (dashed 
blue lines in Fig.~\ref{method}) and we recalculate the ratios. The results 
for all longitudinal field settings normalized to the no field setting are shown 
as bands for pixels and quads in Fig.~\ref{losses}, right panel. The results obtained when 
using the three sigma cut are plotted as central value with the width of the 
bands given by the one sigma cuts around the central cut. The results obtained 
utilizing the two methods agree within 10\%.

For pixel 45 there is at most a 5\% loss of gain as indicated by the 
red circles in Fig.~\ref{losses} (left panel) and up to 15\% losses 
associated with the photoelectron extracted from the photocathode missing 
the first dynode (right panel). This is also evident in Fig.~\ref{45_one}, 
bottom panels, where the progression of the ADC distribution with a 
longitudinal ($B_{z}$) magnetic field shows that the position of the single photoelectron 
peak with respect to the pedestal remains mostly unchanged 
during the field scan while the fraction of signal to pedestal events 
decreases as the field increases. A very similar pattern is observed 
for pixel 61: there is at most 15\% decrease in overall gain as seen in Fig.~\ref{losses} 
(right panel, blue triangles) but the dominant effect is the misfocusing of the 
photoelectron extracted from the photocathode which can lead to losses 
up to 50\%. This effect is clearly visible in Fig.~\ref{61_one}, bottom 
panel. The same conclusions apply to the quads behavior in a longitudinal 
magnetic field: there is little gain loss, no more than 5\% 
(Fig.~\ref{losses}, left panel, green crosses and black diamonds) but up to 
30\% single photoelectron losses (Fig.~\ref{losses}, right panel).

\begin{figure}[htbp]
\vspace*{-0.1in}
\centering
\begin{tabular}{cc}
\hspace{1.9in}
\includegraphics[width=8.8cm]{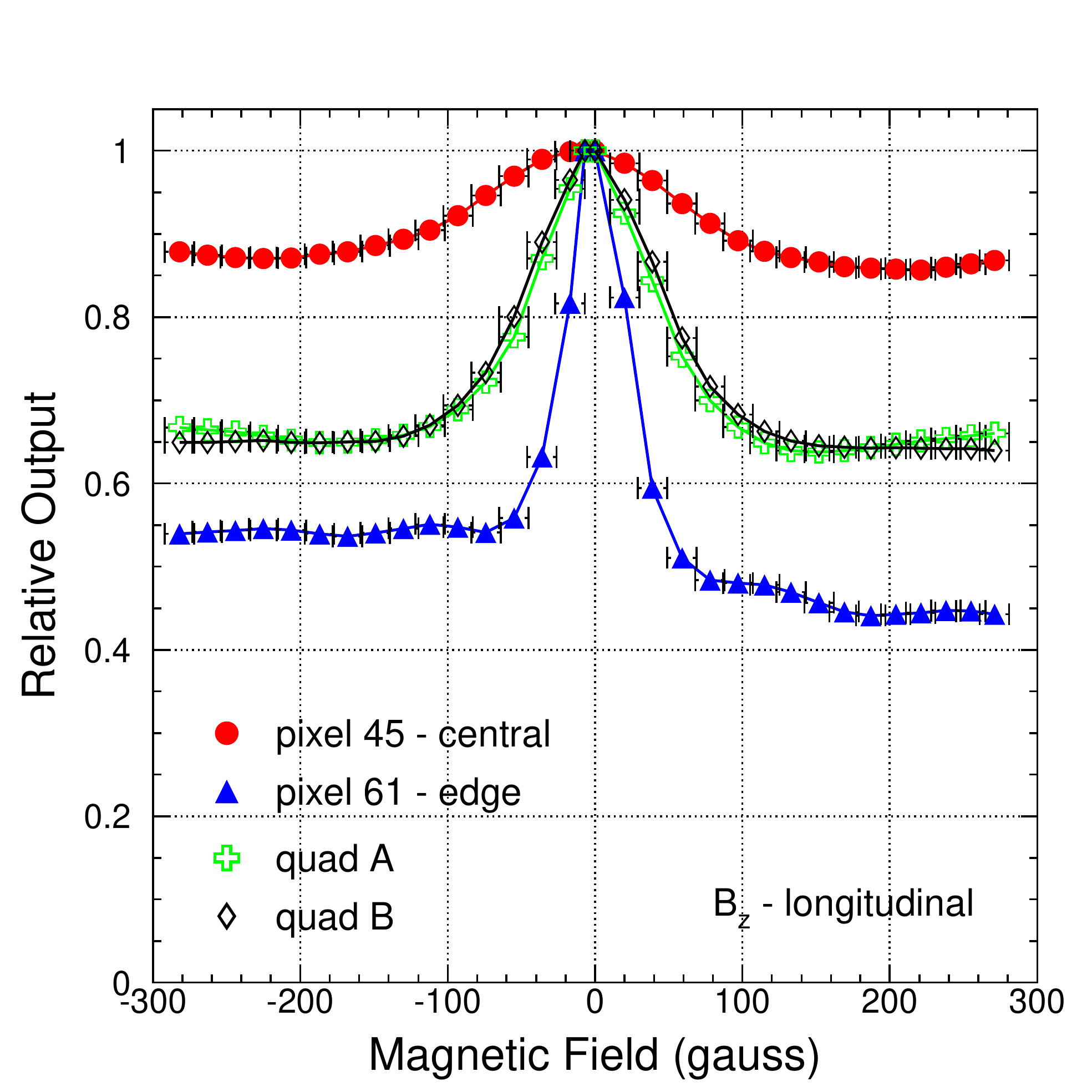}\\
\hspace{-1.5in}
\includegraphics[width=8.8cm]{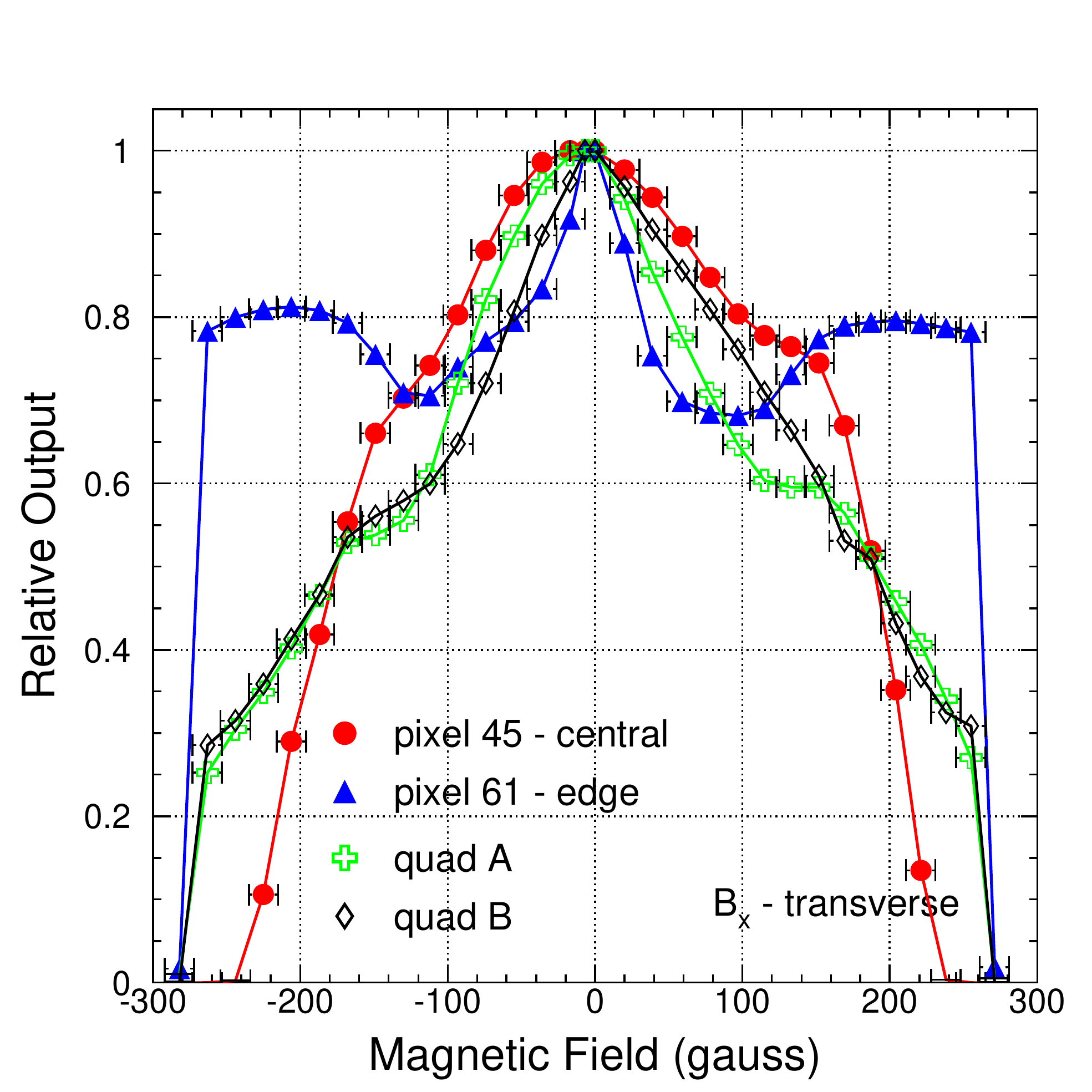}
&
\hspace{-2.in}
\includegraphics[width=8.8cm]{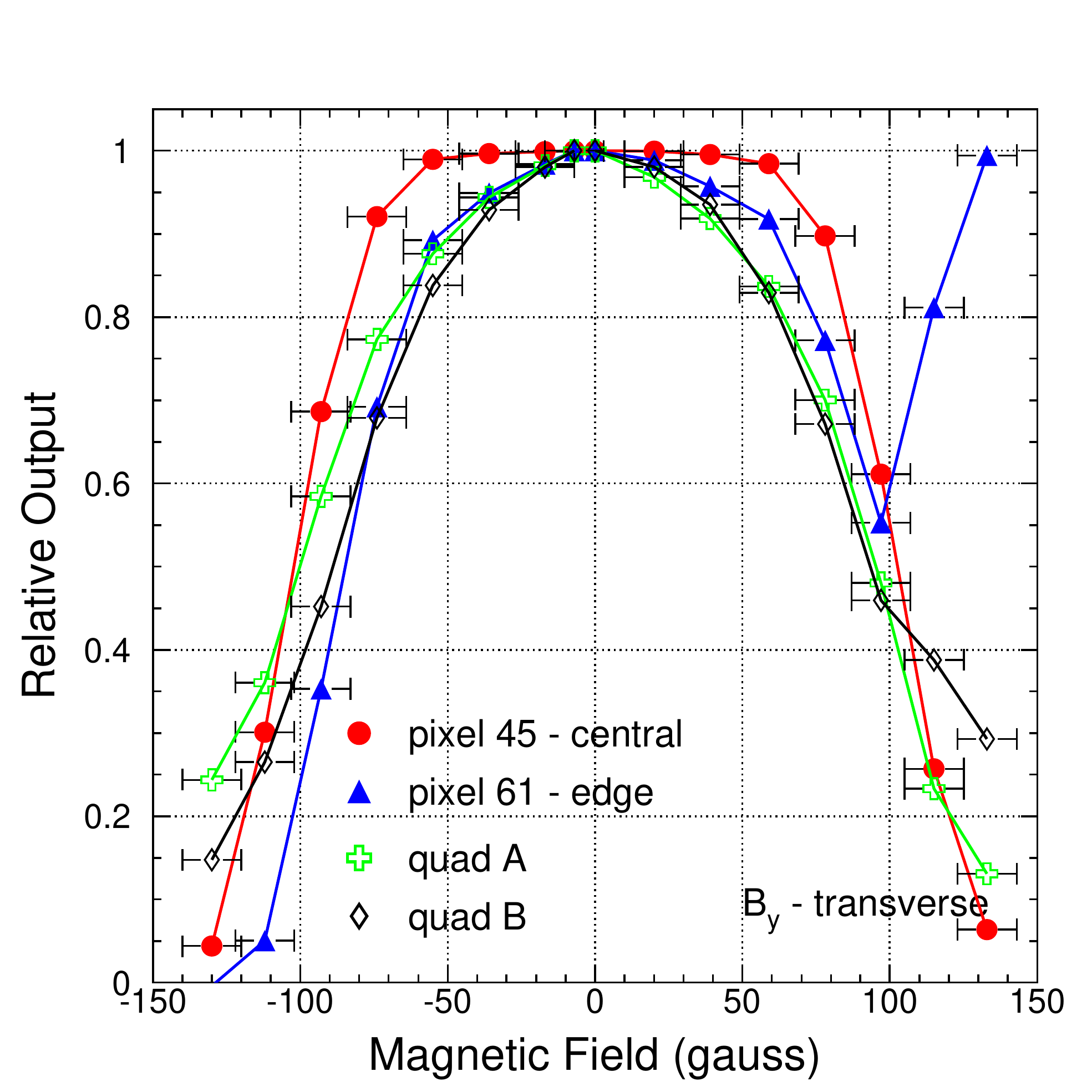} 
\end{tabular}
\linespread{0.5}
\caption[]{
{Relative change in output (\emph{v.s.} no-field configuration) in a longitudinal (top) and 
transverse (bottom) magnetic field for 
pixels 45 and 61 and for quads A and B. The curves are shown to guide the eye.} }
\label{summary_many}
\end{figure}

In a $B_x$ transverse field pixel 45 and the two quads experience a drastic 
loss of single photoelectrons as shown in Figs.~\ref{45_one}, \ref{quad_a_one} 
and \ref{quad_b_one}, top panels. The same qualitative trend is seen here 
as in Fig.~\ref{summary_many}, left bottom panel, where the change of 
large outputs with a $B_x$ transverse field was displayed. In both 
configurations the output from the central pixel is reduced to noise at 
$\approx$ 240 Gauss and quads experience the same reduction at about 
270--280 Gauss. The edge pixel (61) behaves quite differently as shown 
in Fig.~\ref{61_one}, top panel: there is a relatively small loss of 
signal up to a field of $\approx$ 260 Gauss to which both mechanisms 
described above appear to contribute and then there is a sudden drop 
to zero of the output at about 280 Gauss. This is again consistent with 
the results shown in Fig.~\ref{summary_many}, left bottom panel.

The losses in a $B_y$ transverse field are dramatic for all observables. 
Below 50 Gauss there are no losses for the central pixel as seen in 
Fig.~\ref{45_one}, middle panels, but as the field value increases 
the output is lost by $\approx$100 Gauss. This is consistent with the 
study on large signals displayed in Fig.~\ref{summary_many}, bottom 
right panel. The same trend is observed for the edge pixel and 
quads but with a faster drop of signal with field for values below 
50 Gauss. At fields above $+$100 Gauss the edge pixel experiences a 
recovery of signal (see also Fig.~\ref{summary_many}, bottom right 
panel).

\section{Conclusions}\label{sec:conclusions}

We tested one unit of the photomultiplier tube H8500C-03 with the goal of 
extracting its resolution for single photoelectron detection, and to study its 
behavior in magnetic fields of up to 300 Gauss. We measured the output from 
several individual pixels (chosen based on their gain and positioning in the 
multianode matrix), and also the summed output from 2 groups of 16 pixels. We found that 
the resolution for single photoelectron detection is better than one photoelectron 
for both individual pixels and for sum of pixels in spite of the relatively low gain 
of the PMT and pixel-to-pixel gain non-uniformities. This PMT is, therefore, 
suitable for applications where the single photoelectron detection is necessary. 
Our magnetic field tests revealed that a large longitudinal field, up to 300 Gauss, 
induces a relative signal reduction of at most 55\% for an edge pixel and of 
at most 15\% for a central pixel. 
The signal degradation in transverse magnetic fields is significantly more pronounced, 
however, such transverse fields are also the easiest to shield in practice. 
Our studies of single 
photoelectron output reduction with magnetic field point to the 
field induced loss of the photoelectron extracted from the photocathode as 
primary cause of signal loss. With appropriate shielding this PMT 
could function well in high magnetic field environments.

\section*{Acknowledgment}

The collaboration wishes to acknowledge the Detector Group at Jefferson Lab: 
Drew Weisenberger, Jack Mckisson, and Carl Zorn for their 
help with the PMT readout. We would also like to thank the 
Hamamatsu representative Ardavan Ghassemi for useful discussions. 
This work was supported by the U.S. Department of Energy. Jefferson 
Science Associates operates the Thomas Jefferson National Accelerator Facility 
under DOE contract No. DE-AC05-06OR23177. This work was also supported by the U.S. 
Department of Energy under Contract No. DE-FG02-03ER41231.


\bibliographystyle{elsarticle-num}
\bibliography{jlab_duke_malace_mapmt}

\begin{thebibliography}{1}
\expandafter\ifx\csname url\endcsname\relax
  \def\url#1{\texttt{#1}}\fi
\expandafter\ifx\csname urlprefix\endcsname\relax\def\urlprefix{URL }\fi
\expandafter\ifx\csname href\endcsname\relax
  \def\href#1#2{#2} \def\path#1{#1}\fi

\bibitem{solid}
H.~Gao, {\it et al.}, Transverse spin structure of the nucleon through target
  single spin asymmetry in semi-inclusive deep-inelastic (e,e'$\pi^{+(-)}$)
  reaction at jefferson lab, Eur. Phys. J. Plus 126~(2).
\newblock \href {http://dx.doi.org/10.1140/epjp/2011-11002-4}
  {\path{doi:10.1140/epjp/2011-11002-4}}.

\bibitem{bellamy}
E.~H. Bellamy, {\it et al.}, Absolute calibration and monitoring of a
  spectrometric channel using a photomultiplier, Nucl. Instr. and Meth. in
  Phys. Res. A 339 (1994) 468--476.
\newblock \href {http://dx.doi.org/10.1016/0168-9002(94)90183-X}
  {\path{doi:10.1016/0168-9002(94)90183-X}}.

\end{thebibliography}

\onecolumn

\clearpage

\clearpage

\clearpage

\clearpage

\clearpage


\end{document}